\documentclass[
11pt, %
english, %
singlespacing, %
headsepline, %
]{MastersDoctoralThesis} %

\usepackage[utf8]{inputenc} %
\usepackage[T1]{fontenc} %

\usepackage{aas_macros}
\usepackage{amsmath,amssymb}
\usepackage{graphicx}%
\usepackage{dcolumn}%
\usepackage{bm}%
\usepackage{tikz}
\usepackage{todonotes}
\usepackage{subcaption}

\usepackage{palatino} %

\usepackage[backend=bibtex,style=authoryear,natbib=true]{biblatex} %
\usepackage[autostyle=true]{csquotes} %

\usepackage{rotating} %

\usepackage{array}
\newcolumntype{L}{>{\centering\arraybackslash}m{3cm}}

\newcommand{\lcdm}{$\Lambda$CDM}    %
\newcommand{\om}[1]{\Omega_{#1}}         %
\newcommand{\ph}[1]{\phantom{#1}}
\newcommand{\eps}{\epsilon}             %
\newcommand{\tildv}{\tilde{v}}
\newcommand{\brho}{\bar{\rho}}

\newcommand{\Lbeta}{\mathcal{L}_\beta} %
\newcommand{\cD}{\bar{D}}                      %
\newcommand{\cgam}{\bar{\gamma}}   %
\newcommand{\cGam}{\bar{\Gamma}} %
\newcommand{\cA}{\bar{A}}                    %
\newcommand{\cK}{\bar{K}}                  %
\newcommand{\cmR}{\bar{\mR}}           %
\newcommand{\conf}{e^{4\phi}}             %
\newcommand{\iconf}{e^{-4\phi}}             %
\newcommand{\dt}{\frac{d}{dt}} %
\newcommand{\gam}{\gamma}               %
\newcommand{\Gam}{\Gamma}              %
\newcommand{\alp}{\alpha}                     %
\newcommand{\gamud}[2]{\gam^{#1}_{\phantom{#1}#2}}
\newcommand{\gamdu}[2]{\gam_{#1}^{\phantom{#1}#2}}
\newcommand{\del}[2]{\delta^{#1}_{\phantom{#1}#2}} %

\newcommand{\pdo}{\partial_0}            %
\newcommand{\pdt}{\partial_t}              %

\newcommand{\ic}{c^{-1}}                %
\newcommand{\mR}{\mathcal{R}}           %
\newcommand{\mH}{\mathcal{H}}              %
\newcommand{\mD}{\mathcal{D}}            %
\newcommand{\mQ}{\mathcal{Q}}
\newcommand{\mL}{\mathcal{L}}
\newcommand{\mW}{\mathcal{W}}

\newcommand{\avgb}[1]{\langle #1 {\rangle_b}} %
\newcommand{\avgh}[1]{\langle #1 {\rangle_h}} %
\newcommand{\aDb}{a_\mD^b}            %
\newcommand{\aDh}{a_\mD^h}            %
\newcommand{\aD}{a_\mD}            %
\newcommand{\VD}{V_\mD}                %
\newcommand{\VDb}{V_\mD^b}                %
\newcommand{\VDh}{V_\mD^h}                %
\newcommand{\QD}{\mQ_\mD}        %
\newcommand{\LD}{\mL_\mD}         %
\newcommand{\HD}{\mH_\mD}         %

\newcommand{\flrwsolver}{\texttt{FLRWSolver}}
\newcommand{\grhydro}{\texttt{GRHydro}}
\newcommand{\hydrobase}{\texttt{HydroBase}}
\newcommand{\admbase}{\texttt{ADMBase}}
\newcommand{\tmunubase}{\texttt{TmunuBase}}
\newcommand{\mol}{\texttt{MoL}}
\newcommand{\mclachlan}{\texttt{McLachlan}}
\newcommand{\mlbssn}{\texttt{ML\_BSSN}}
\newcommand{\mesc}{\textsc{mescaline}}

\thesistitle{Inhomogeneous cosmology in an anisotropic Universe} %
\supervisorOne{Ass. Prof. Daniel Price} 
\supervisorTwo{Dr. Paul Lasky}
\examiner{} %
\degree{Doctor of Philosophy} %
\author{Hayley J. Macpherson} 
\addresses{} %

\subject{Astrophysics} %
\university{\href{}{Monash University}} %
\department{\href{}{School of Physics and Astronomy}} %
\group{\href{}{Monash Centre for Astrophysics}} %
\faculty{\href{}{Faculty of Science}} %

\hypersetup{pdftitle=\ttitle} %
\hypersetup{pdfauthor=\authorname} %
\hypersetup{pdfkeywords=\keywordnames} %

\begin{document}

\frontmatter %

\pagestyle{plain} %

\begin{titlepage}
\begin{center}

\begin{figure*}[!ht]
	\includegraphics[width=\textwidth]{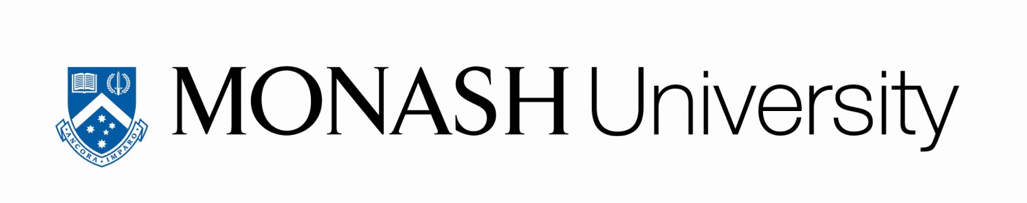}
\end{figure*}

\vspace{15mm}

\HRule \\[0.4cm] %
{\huge \bfseries \ttitle}\\[0.4cm] %

\HRule \\[1.5cm] %

\emph{Author:}\\
{\Large \authorname}\\[1.5cm]

\emph{Supervisors:}\\
{\suponename}  \\ %
{\suptwoname} \\[3cm]

\large \textit{A thesis submitted in fulfillment of the requirements\\ for the degree of \degreename}\\[0.3cm] %
\textit{in the}\\[0.4cm]
\textit{\groupname ,\\\deptname}\\[2cm] %
 
{\large \today}\\[4cm] %

\vfill
\end{center}
\end{titlepage}

\begin{copyrightnotice} 

\copyright \; \authorname \, (2019) \\

\vspace{1cm}

This thesis must be used only under the normal conditions of ``fair dealing'' under the Copyright Act. It should not be copied or closely paraphrased in whole or in part without the written consent of the author. Proper written acknowledgement should be made for any assistance obtained from this thesis. I certify that I have made all reasonable efforts to secure copyright permissions for third-party content included in this thesis and have not knowingly added copyright content to my work without the owner's permission.

\end{copyrightnotice}

\cleardoublepage

\begin{abstract}

With the era of precision cosmology upon us, and upcoming surveys expected to further improve the precision of our observations below the percent level, ensuring the accuracy of our theoretical cosmological model is of the utmost importance. Current tensions between our observations and predictions from the standard cosmological model have sparked curiosity in extending the model to include new physics. Although, some suggestions include simply accounting for aspects of our Universe that are ignored in the standard model. One example acknowledges the fact that our Universe contains significant density contrasts on small scales; in the form of galaxies, galaxy clusters, filaments, and voids. This small-scale structure is \emph{smoothed} out in the standard model, by assuming large-scale homogeneity of the matter distribution, which could have a measurable effect due to the nonlinearity of Einstein's equations. This \emph{backreaction} of small-scale structures on the large-scale dynamics has been suggested to explain the measured accelerating expansion rate of the Universe. 

Current standard cosmological simulations ignore the effects of General Relativity by assuming purely Newtonian dynamics. In this thesis, we take the first steps towards quantifying the backreaction of small-scale structures by performing cosmological simulations that solve Einstein's equations directly. Simulations like these will allow us to quantify potentially important effects on our observations that could become measurable as the precision of these observations increases into the future. 

We begin by testing our computational setup to ensure our results are trustworthy. We then present simulations of a realistic matter distribution with initial conditions inspired by the early moments of our own Universe. Analysing the averaged, large-scale evolution of an inhomogeneous universe in full General Relativity, we find negligible difference from small-scale structures. However, we do find significant effects present on small scales that could potentially influence future observations. While we suggest improvements to our computational framework to validate our results, we conclude that the backreaction of small-scale structures is unlikely to explain the accelerating expansion of the Universe. 

Finally, we suggest future extensions to our analysis to improve the quantification of General-Relativistic effects on our cosmological observations. 

\end{abstract}

\begin{publications}

\begin{enumerate}

    \item \fullcite{macpherson2017a}
    \item \fullcite{macpherson2018b}
    \item \fullcite{macpherson2019a}

\end{enumerate}

\end{publications}

\begin{declaration}

\vspace{1.5cm}

I, \authorname, hereby declare that this thesis contains no material which has been accepted for the award of any other degree or diploma at any university or equivalent institution and that, to the best of my knowledge and belief, this thesis contains no material previously published or written by another person, except where due reference is made in the text of the thesis. 

This thesis includes three original papers published in peer reviewed journals in Chapters~\ref{Chapter3}, \ref{Chapter4}, and \ref{Chapter5}. The core theme of the thesis is cosmological simulations with numerical relativity. The ideas, development and writing up of all the papers in the thesis were the principal responsibility of myself, the student, working within the \deptname\, under the supervision of \suponename\, and \suptwoname. 

The inclusion of co-authors reflects the fact that the work came from active collaboration between researchers and acknowledges input into team-based research. I have renumbered sections of submitted or published papers in order to generate a consistent presentation within the thesis. In the case of Chapters~\ref{Chapter3}, \ref{Chapter4}, and \ref{Chapter5}, my contribution to the work in each case is summarised in the table on the following page.

\vspace{1.5cm}

Student signature: 				   			\hspace{4.5cm} Date: \today

\vspace{1.5cm}

I, \suponename, hereby certify that the above declaration correctly reflects the nature and extent of the student's and co-authors' contributions to this work. In instances where I am not the responsible author I have consulted with the responsible author to agree on the respective contributions of the authors. 

\vspace{1.5cm}

Main Supervisor signature:				  		\hspace{3.0cm} Date: \today

\end{declaration}

\begin{sidewaystable}
\begin{tabular}{| L | L | L | L | L | L |}
	
	\hline
	\textbf{Thesis chapter} & \textbf{Publication title} & \textbf{Status} & \textbf{Nature and \% of student contribution} & \textbf{Co-author names, nature and \% of contribution} & \textbf{Co-authors, Monash student Y/N} \\
	\hline
	
	3 & Inhomogeneous cosmology with numerical relativity & Published (PRD, March 2017) & 80\% -- Concept, coding, analysis of results, manuscript authorship & Paul Lasky: 10\% -- Concept, coding, manuscript input. Daniel Price: 10\% -- Concept, manuscript input. & N \\
	\hline
	4 & Einstein's Universe: Cosmological structure formation in numerical relativity & Published (PRD, March 2019) & 80\% -- Concept, coding, analysis of results, manuscript authorship & Daniel Price: 10\% -- Concept, coding, manuscript input. Paul Lasky: 10\% -- Concept, manuscript input. & N \\
	\hline
	5 & The Trouble with Hubble: Local versus Global Expansion Rates in Inhomogeneous Cosmological Simulations with Numerical Relativity & Published (ApJL, September 2018) & 80\% -- Concept, analysis of results, manuscript authorship & Paul Lasky: 10\% -- Concept, manuscript input. Daniel Price: 10\% -- Concept, manuscript input. & N \\ 
	\hline
\end{tabular}
\end{sidewaystable}

\cleardoublepage

\begin{acknowledgements}

Firstly, I would like to thank my supervisors, Daniel Price and Paul Lasky. From the beginning of this project back in 2015, you have both challenged me and helped me grow, while always being plenty of fun to work with. I wouldn't be the scientist I am today if it weren't for both of your help, encouragement, and breadth of knowledge. I'm inspired by both of your curiosity, without which this work would never have happened. I would like to apologise for being such an expensive student, please accept this thesis as official reimbursement. It's been a grand journey becoming cosmologists together, we made it.

I would also like to thank my collaborators, especially Marco Bruni, Chris Blake, and Julian Adamek, for many valuable discussions and visits that helped me get to where I am today. I have been extremely lucky to travel so much during my PhD, and attend many conferences in some amazing places around the world. To everyone I met --- and explored, ate, drank, and hiked with --- along the way; thank you. 

Over my 7 years at Monash I have made many lifelong friendships. A special thank you to Izzy for keeping me sane over those years, you will be my best friend for life. I would also like to thank all of my office mates, especially Dave, Conrad, and Zac for priceless banter over the past 4 years. The majority of the postgrad cohort in the School of Physics and Astronomy have contributed to my life at Monash in their own way, so to each of you; thank you. Perhaps my largest thanks goes to Jean Pettigrew, for doing basically everything for everyone, all the time. I know I am not alone in saying that you keep us students afloat.

Of course, I would like to thank my parents for their never-ending love and support. Mum, your commitment and success have always inspired me to be the best version of myself I can be, and the care you show for those around you never ceases to amaze me. Hopefully this thesis will be useful in explaining my research to your friends. Dad, your endless encouragement and faith in me are a huge part of my achievements thus far. Thanks for the science chats, South Park and tacos, the good (and the bad) surfs, and of course for being ``a cool nerd... the Ultimate Person'' (Mark Macpherson, 2014). A special thanks to my Grandma, not only for feeding me and chatting MasterChef, but for your unconditional love and for researching any place I'm travelling to better than I ever do. I love you all, thank you.

Lastly, but certainly not least, I would like to thank my girls; Abbey, Meag, and Jess, for always sticking by me and being my best friends and support system since high school, and definitely for the rest of our lives to come. Also, a big thanks to Kym for helping me through the past few months, I couldn't have done it without you. 

This research was supported by an Australian Government Research Training Program (RTP) Scholarship. I also acknowledge the Astronomical Society of Australia for their funding support that helped contribute to this work. 

\end{acknowledgements}

\tableofcontents %

\begin{abbreviations}{ll} %

\textbf{GR} & \textbf{G}eneral \textbf{R}elativity \\
\textbf{LIGO} & \textbf{L}aser \textbf{I}nterferometer \textbf{G}ravitational-Wave \textbf{O}bservatory \\
\textbf{FLRW} & \textbf{F}riedmann-\textbf{L}ema\^itre-\textbf{R}obertson-\textbf{W}alker \\
\textbf{SN1a} & \textbf{S}uper\textbf{N}ova Type \textbf{1a} \\
\textbf{CMB} & \textbf{C}osmic \textbf{M}icrowave \textbf{B}ackground \\
\textbf{\lcdm} & \textbf{$\Lambda$} \textbf{C}old \textbf{D}ark \textbf{M}atter \\
\textbf{NR} & \textbf{N}umerical \textbf{R}elativity \\
\textbf{ADM} & \textbf{A}rnowitt \textbf{D}eser \textbf{M}isner \\
\textbf{BSSN} & \textbf{B}aumgarte-\textbf{S}hapiro-\textbf{S}hibata-\textbf{N}akamura \\
\textbf{CAMB} & \textbf{C}ode for \textbf{A}nisotropies in the \textbf{M}icrowave \textbf{B}ackground \\
\textbf{EOS} & \textbf{E}quation \textbf{O}f \textbf{S}tate \\
\textbf{LSST} & \textbf{L}arge \textbf{S}ynoptic \textbf{S}urvey \textbf{T}elescope \\
\textbf{SKA} & \textbf{S}quare \textbf{K}ilometre \textbf{A}rray \\
\textbf{SDSS} &  \textbf{S}loan \textbf{D}igital \textbf{S}ky \textbf{S}urvey \\
\textbf{2dFGRS} & \textbf{2}-\textbf{d}egree \textbf{F}ield \textbf{G}alaxy \textbf{R}edshift \textbf{S}urvey \\
\textbf{WMAP} & \textbf{W}ilkinson's \textbf{M}icrowave \textbf{A}nisotropy \textbf{P}robe \\
\textbf{BAO} & \textbf{B}aryon \textbf{A}coustic \textbf{O}scillation \\
\textbf{SH0ES} & \textbf{S}upernova \textbf{$H_0$} for the \textbf{E}quation of \textbf{S}tate of dark energy \\
\textbf{RGTC} & \textsc{\textbf{R}iemannian \textbf{G}eometry and \textbf{T}ensor \textbf{C}alculus} \\
\textbf{LTB} & \textbf{L}ema\^itre \textbf{T}olman \textbf{B}ondi \\
\textbf{ET} & \textsc{\textbf{E}instein \textbf{T}oolkit} \\
\textbf{CCZ4} & \textbf{C}onformal and \textbf{C}ovariant \textbf{Z4} \\
\textbf{TVD} & \textbf{T}otal \textbf{V}ariation \textbf{D}iminishing \\ 
\textbf{PPM} & \textbf{P}iecewise \textbf{P}arabolic \textbf{M}ethod \\
\textbf{ENO} & \textbf{E}ssentially \textbf{N}on-\textbf{O}scillatory \\
\textbf{HLLE} & \textbf{H}arten \textbf{L}ax van \textbf{L}eer \textbf{E}infeldt \\
\textbf{RK4} & \textbf{R}unge-\textbf{K}utta \textbf{4}$^{\rm th}$ order \\
\textbf{MoL} & \textbf{M}ethod \textbf{o}f \textbf{L}ines \\

\end{abbreviations}

\dedicatory{For every woman who thought she couldn't}

\mainmatter %

\pagestyle{thesis} %

\chapter{Introduction and Background} %

\label{Chapter1} %

\newcommand{\keyword}[1]{\textbf{#1}}
\newcommand{\tabhead}[1]{\textbf{#1}}
\newcommand{\code}[1]{\texttt{#1}}
\newcommand{\file}[1]{\texttt{\bfseries#1}}
\newcommand{\option}[1]{\texttt{\itshape#1}}

Einstein's general theory of relativity (GR) is the most accurate description of gravity for the Universe. Rather than describing gravity as a force between massive objects --- as in Newton's theory --- gravity is a consequence of geometry. The presence of mass curves spacetime, the ``fabric'' of the Universe, which in turn affects the motion of passing matter. Einstein's theory has proven to better describe dynamics in the local Universe than Newtonian gravity, successfully explaining the precession of the perihelion of Mercury \citep{einstein1916,clemence1947} and the bending of light around the Sun \citep{dyson1920}. More recently, the first detection of gravitational waves with the Laser Interferometer Gravitational-Wave Observatory \citep[LIGO;][]{abbott2016} marked yet another prediction of GR to be confirmed correct. While small-scale physics in our Universe is described with great precision using either the weak- or strong-field limit of GR, the applicability to large-scale physics remains to be thoroughly tested.

Early astronomical observations were limited to stars within our own galaxy --- which have small velocities --- prompting scientists to believe that the Universe was static: neither expanding nor contracting. Einstein's first application of GR to cosmology thus required the cosmological constant, $\Lambda$, to be added into his field equations to counteract the expansion; resulting in a static description of spacetime \citep{einstein1917}. Aleksandr Friedmann and Georges Lema\^itre independently derived expanding solutions to Einstein's equations with no need for the cosmological constant \citep{friedmann1922,friedmann1924,lemaitre1927}. These solutions were largely dismissed until the first observational evidence for an expanding Universe emerged in \citeyear{hubble1929}. Edwin Hubble's observations of extragalactic nebulae showed a positive, linear trend between distance and radial velocity; suggesting that the Universe is expanding \citep{hubble1929}. Einstein later accepted the notion of an expanding Universe, deeming the cosmological constant unnecessary \citep{einstein1931,einsteindesitter1932,straumann2002}.

The basis for the current standard model of cosmology is the Friedmann-Lema\^itre-Robertson-Walker (FLRW) solution to Einstein's equations; describing an expanding, homogeneous, and isotropic spacetime. Application of the FLRW model to our cosmological observations revealed that things were not as expected; leading to an astounding discovery. \citet{riess1998} and \citet{perlmutter1999} discovered the accelerating expansion of the Universe using observations of Type 1a supernovae (SN1a). These SN1a appear fainter than predicted by the FLRW model, implying the expansion of spacetime is accelerating. The return of the cosmological constant $\Lambda$ was imminent, deemed ``dark energy''; a mysterious negative pressure forcing the expansion of the Universe to accelerate at late times. 

Early measurements of the first light after the Big Bang --- the Cosmic Microwave Background (CMB) radiation --- indicate that the Universe has globally flat geometry \citep{jaffe2001}. Constraints that matter only accounted for $\sim 27\%$ of the total energy density of the Universe \citep{bennett2003} require a smoothly-distributed energy to reconcile this result with that of a flat geometry. These measurements fit perfectly with the cosmological constant, $\Lambda$.

The existence of ``dark matter'' --- a type of invisible matter that interacts only gravitationally --- was first inferred from rotation curves of galaxies \citep{freeman1970,rubin1970,rubin1980}. Since then, measurements of the weak gravitational lensing of light \citep[e.g.][]{bacon2000}, and measurements of CMB anisotropy \citep[e.g.][]{bennett2003,hinshaw2013,planck2018a} have strengthened the argument for the existence of dark matter \citep[see, e.g.,][for a review]{bertone2018}.

These discoveries each contributed to the current standard cosmological model; the $\Lambda$ Cold Dark Matter (\lcdm) model, named after the main constituents of the Universe. 

To date, the \lcdm\, model has successfully explained essentially all of our cosmological observations. Notable successes include matching the power spectrum of temperature fluctuations in the CMB \citep[e.g.][]{planck2018a}, the location of the peak separation of large-scale structures --- the baryon acoustic oscillation (BAO) peak --- \citep[e.g.][]{blake2011a}, and the matter power spectrum of the large-scale structure at low redshifts \citep[e.g.][]{reid2010,anderson2014}. 

Aside from its successes, there are some tensions between the \lcdm\, model and what we observe. Most notable are the lack of power at the largest scales in the CMB power spectrum \citep{planck2018a}; the low-multipole ``bump'' visible in the power spectrum data, and the recent $3.7\sigma$ tension between local measurements of the Hubble parameter \citep{riess2018a,riess2018b} compared to that inferred from the CMB \citep{planck2018a}. These tensions may be due to insufficient precision, systematic errors in the measurements, or --- more excitingly --- new physics (see Section~\ref{subsec:curiosities}). 

Existing physics that is currently neglected in the standard model could explain some of these tensions \citep[e.g.][]{buchert2016}. The basis of \lcdm\, is the assumption of a homogeneous, isotropic background spacetime with expansion described by the FLRW model. Current state-of-the-art cosmological simulations model structures evolving under Newtonian gravity on top of an \textit{a-priori} assumed homogeneously expanding background spacetime \citep[e.g.][]{springel2005b,kim2011,genel2014,potter2017}. These simulations are the primary comparison point to our cosmological observations. Newtonian gravity has been shown to be a good approximation for GR on small scales, however, it's applicability to cosmological scales remains uncertain, if only because it is based on instantaneous action-at-a-distance \citep{buchert2016}. Newtonian cosmological simulations thus ignore causality, which could have a significant effect when considering cosmological scales.

The main argument for the assumptions of homogeneity and isotropy underlying the standard model is that our Universe is homogeneous and isotropic on large scales. The Universe is inhomogeneous and anisotropic on small scales, and it is not clear whether \emph{smoothing} over this small-scale structure has a measurable effect on the large-scale dynamics of the Universe. \citet{buchert2000a} showed that the evolution of the average of an inhomogeneous universe does not coincide with the evolution of a homogeneous universe, due to the non-commutation of averaging and time evolution in fully nonlinear GR. Extra mathematical terms contribute to both the expansion of spacetime and the \emph{acceleration} of the expansion of spacetime, the significance of which has been the subject of much debate \citep[e.g.][]{buchert2015,green2016}. 

Upcoming cosmological surveys using state-of-the-art telescopes such as Euclid, the Large Synoptic Survey Telescope (LSST), and the Square Kilometre Array (SKA) are expected to reach percent-level precision \citep{ivezic2008,maartens2015,amendola2016}, at which small differences between Newtonian gravity and GR on cosmological scales could be measurable. Drawing correct conclusions from our observations first requires accurate cosmological simulations; simulations that include any potentially measurable GR effects. The magnitude of the backreaction of structures on the large-scale dynamics of the Universe can only be quantified with a full treatment of GR in a cosmological simulation. 

Numerical relativity (NR) allows us to solve Einstein's field equations numerically, often using a ``3+1'' decomposition. This involves splitting our four-dimensional spacetime into three space dimensions and one time dimension. The ADM formalism, named after its authors \citet{arnowitt1959}, casts Einstein's equations into a weakly hyperbolic form for numerical evolution. This weak hyperbolicity means the system is unstable for long time evolutions. The Baumgarte-Shapiro-Shibata-Nakamura (BSSN) formalism improves on this by instead forming a strongly hyperbolic system \citep{shibata1995,baumgarte1999}, which allows for arbitrarily long, stable time evolutions. \citet{pretorius2005,campanelli2006,baker2006} performed the first successful long-term evolutions of a binary black hole system, including the merger and emission of gravitational waves. Since then, the field of NR has exploded and it is now widely used for simulations of binary mergers of compact objects such as black holes \citep[e.g.][]{baker2006,campanelli2006,buonanno2007,gonzalez2007,hinder2018,huerta2019}, neutron stars \citep[e.g.][]{baiotti2008,paschalidis2011,kastaun2015,chaurasia2018}, stellar collapse and supernovae explosions \citep[e.g.][]{duez2004,montero2012}, and more recently, for cosmology \citep{giblin2016a,giblin2016b,bentivegna2016a,giblin2017b,macpherson2017a,macpherson2019a,macpherson2018b,east2018,daverio2019,barrera-hinojosa2019}.

It is now possible to investigate GR effects on our cosmological observations, test the validity of the assumptions underlying the standard cosmological model, and ensure our cosmological simulations are sufficiently accurate for forthcoming precision cosmological surveys. These are the primary motivations behind this thesis.

We perform three-dimensional cosmological simulations of large-scale structure formation that solve Einstein's equations without approximation using NR. Removing the assumptions underlying current cosmological simulations allows us to fully understand GR's role, and validity, on cosmological scales.

In the remainder of this Chapter we describe the basis of modern cosmology, including the mathematical background for the standard model, observational tests of \lcdm, the current status of state-of-the-art cosmological simulations, cosmological perturbation theory and its current use in these simulations, current observational tensions with the standard model and the resulting suggested extensions to \lcdm. 

In Chapter 2 we outline the methods used to perform our simulations and main analysis. We include an introduction to NR and derive the equations of the BSSN formalism including a discussion of common coordinate choices, a discussion of the workings of the \textsc{Einstein Toolkit} --- the open-source NR code that we use to perform our simulations --- and a description of our post-processing analysis.

In Chapter 3 we describe our computational framework in detail, and show results of two code tests to demonstrate the accuracy of our results. We perform simulations of a homogeneous, isotropic FLRW spacetime and compare to the analytic solution. Perturbing this FLRW spacetime with initially small, simplified inhomogeneities, we show the growth of these perturbations matches the analytic linear solution; but also surpasses this solution into the nonlinear regime of growth where deriving analytic expressions in full GR is not possible.

In Chapter 4 we present NR cosmological simulations of a realistic matter distribution. We generate initial conditions drawn from the temperature anisotropies in the CMB, using the Code for Anisotropies in the Microwave Background \citep[\textsc{CAMB};][]{seljak1996}. We then evolve these initially small inhomogeneities to redshift $z\approx0$, and from the resulting cosmic web we calculate averaged quantities and draw comparisons with an FLRW spacetime.

In Chapter 5 we analyse the effect of local inhomogeneities on an observer's measurement of the Hubble parameter (expansion rate) of the Universe. Our calculations are presented to approximate the expected variance on the local $H_0$ measurement from SN1a in \citet{riess2018a} due to an observer's physical location in an inhomogeneous, anisotropic Universe.

A short summary is presented at the end of each chapter, and a full summary with details of future work is presented in Chapter 6. 

Throughout this thesis, we assume a metric signature $(-,+,+,+)$ unless otherwise stated, and adopt the Einstein summation convention: implied summation over repeated indices. Greek indices represent spacetime indices and take values $(0,1,2,3)$, and Latin indices represent spatial indices and take values $(1,2,3)$.

\section{Einstein's field equations}

Einstein's field equations of GR are
\begin{equation} \label{eq:einstein}
	G_{\mu\nu} \equiv R_{\mu\nu} - \frac{1}{2} \,R \,g_{\mu\nu} = \frac{8 \pi G}{c^4} T_{\mu\nu},
\end{equation}
where $G_{\mu\nu}$ is the Einstein tensor, $g_{\mu\nu}$ is the metric tensor, $G$ is the gravitational constant, $c$ is the speed of light, and $T_{\mu\nu}$ is the stress-energy tensor of matter. The four-dimensional Riemann curvature tensor is
\begin{equation} \label{eq:4Riemdef}
	R^{\alpha}_{\phantom{\alpha}\mu \beta \nu} \equiv \partial_\beta \, ^{(4)}{\Gamma^\alpha_{\mu\nu}} - \partial_\nu \, ^{(4)}{\Gamma^\alpha_{\mu \beta}} + ^{(4)}{\Gamma^\alpha_{\lambda \beta}} ^{(4)}{\Gamma^\lambda_{\mu\nu}} - ^{(4)}{\Gamma^\alpha_{\lambda\nu}} ^{(4)}{\Gamma^\lambda_{\mu \beta}},
\end{equation}
where $\partial_\mu \equiv \partial/\partial x^\mu$ is the partial derivative with respect to coordinate $x^\mu$, and the four-dimensional Christoffel symbols (or connection functions) associated with the metric $g_{\mu\nu}$ are defined as
\begin{equation} \label{eq:christoffel4}
	^{(4)}{\Gamma^\alpha_{\mu\nu}} \equiv \frac{1}{2} g^{\alpha\beta} \left( \partial_\mu g_{\beta \nu} + \partial_\nu g_{\mu\beta} - \partial_\beta g_{\mu\nu} \right).
\end{equation}
We distinguish these four-dimensional objects from their spatial counterparts with $^{(4)}$ (see Section~\ref{subsec:3p1}). The Ricci curvature tensor in Einstein's equations \eqref{eq:einstein} is then the contraction of the Riemann tensor $R_{\mu\nu} \equiv R^{\alpha}_{\ph{\alpha}\mu\alpha\nu}$, and the Ricci scalar is the trace of the Ricci curvature tensor, $R \equiv g^{\mu\nu}R_{\mu\nu}$, where the inverse metric $g^{\mu\nu}$ is defined such that $g^{\mu\nu}g_{\nu\alp} = \del{\mu}{\alp}$, with $\del{\mu}{\alp}$ the Kronecker delta function. Indices of four-dimensional tensors are raised and lowered using the metric tensor, i.e. $A^{\mu}_{\ph{\mu}\nu} = g^{\alp\mu} A_{\alp\nu}$.

The Einstein equations \eqref{eq:einstein} satisfy the contracted Bianchi identities \citep{voss1880},
\begin{equation}
	\nabla_\mu G^{\mu\nu} = 0,
\end{equation}
which consequently implies the conservation of stress-energy,
\begin{equation} \label{eq:conservation_stress_energy}
	\nabla_\mu T^{\mu\nu} = 0,
\end{equation}
from which the equations of General-Relativistic hydrodynamics are derived (see Section~\ref{subsec:grhydro_eqs}). In the above, $\nabla_\mu$ is the covariant derivative associated with the metric $g_{\mu\nu}$, i.e. 
\begin{equation} 
	\nabla_\alp g^{\mu\nu} = \partial_\alp g^{\mu\nu} + ^{(4)}{\Gam^\mu_{\alp\beta}} g^{\beta\nu} + ^{(4)}{\Gam^\nu_{\alp\beta}} g^{\mu\beta} = 0,
\end{equation}
by construction. 

The set of 16 equations \eqref{eq:einstein} --- which reduces to 10 due to symmetries in the metric tensor $g_{\mu\nu}$ --- describe how matter, $T_{\mu\nu}$, interacts with spacetime, $G_{\mu\nu}$. Here, we approximate the matter content of the Universe as a perfect fluid, which has stress-energy tensor
\begin{equation}\label{eq:Tmunu_perfect_fluid}
	T^{\mu\nu} = \rho_0 h u^\mu u^\nu + P \, g^{\mu\nu},
\end{equation}
where $\rho_0$ and $P$ are the rest-mass density and pressure, respectively, and the specific enthalpy is $h=c^2 + \eps + P/\rho_0$, with $\eps$ the internal energy. The dimensionless four velocity of the fluid is defined as %
\begin{equation} \label{eq:4veldef}
	u^\mu \equiv \frac{dx^\mu}{d(c\tau)},
\end{equation}
where $x^\mu=(ct,x^i)$ are the spacetime coordinates, and the four velocity is normalised such that $u_\mu u^\mu = -1$. The proper time, $\tau$, is defined by
\begin{equation}\label{eq:proptime_def}
	c^2 d\tau^2 \equiv - ds^2,
\end{equation}
where $ds$ is the infinitesimal distance between two nearby points in any spacetime,
\begin{equation}\label{eq:ds_def}
	ds^2 \equiv g_{\mu\nu}dx^\mu dx^\nu,
\end{equation}
commonly referred to as the \emph{line element} of a given spacetime.

Often Einstein's equations are written including the cosmological constant $\Lambda$, i.e.
\begin{equation} \label{eq:einstein_lambda}
	R_{\mu\nu} - \frac{1}{2} \,R \,g_{\mu\nu} + \Lambda g_{\mu\nu} = \frac{8 \pi G}{c^4} T_{\mu\nu},
\end{equation}
commonly adopted to describe dark energy; the driving force of the apparent accelerating expansion of the Universe (see Section~\ref{sec:LCDM}). Alternatively, the cosmological constant can be absorbed into the stress-energy tensor as an additional form of matter, using $\rho_R \rightarrow \rho_R + \rho_\Lambda$, and $P\rightarrow P+P_\Lambda$, with $P_\Lambda=-\rho_\Lambda$. Here, $\rho_R$ is the total mass-energy density as measured by an observer comoving with the fluid, and is defined by projecting the stress-energy tensor into the rest frame of the fluid, i.e.
\begin{equation}
	\rho_R c^2 \equiv T_{\mu\nu} u^\mu u^\nu. \label{eq:rhoR_def} 
\end{equation}
Using \eqref{eq:Tmunu_perfect_fluid}, its relation to the rest-mass density $\rho_0$ is therefore
\begin{align}
	\rho_R c^2 &= \rho_0 h u_\mu u^\mu u_\nu u^\nu + P g_{\mu\nu} u^\mu u^\nu, \\
		&= \rho_0 c^2 \left(1 + \frac{\eps}{c^2} \right) \label{eq:rhoR_rho0_relation}.
\end{align}
For the remainder of this thesis, we assume $\rho_R$ contains \emph{all} forms of energy-density --- potentially including a cosmological constant --- unless otherwise stated.
\subsection{Foliation of spacetime}

Einstein's field equations describe the relation between spacetime and matter. In order to simulate the evolution of a four-dimensional spacetime using numerical relativity, we split that spacetime into a series of three-dimensional hypersurfaces (surfaces) that can be evolved in time. Choosing the way the spacetime is split is done via the lapse function, $\alp$, which describes the spacing between subsequent spatial slices in time, and the shift vector, $\beta^i$, which describes how the spatial coordinates are re-labelled between slices. These are known as gauge choices. The surfaces each have dimensionless normal vector
\begin{equation}\label{eq:normal_def}
	n_\mu \equiv -\alp \nabla_\mu x^0, %
\end{equation}
which is normalised such that $n_\mu n^\mu=-1$. We impose $\alp>0$, so that $n_\mu$ is time like. In terms of the lapse and shift functions, the covariant and contravariant normals are $n_\mu = (-\alp,0,0,0)$ and $n^\mu = (1/\alp, - \beta^i/\alp)$, respectively. From \eqref{eq:normal_def} we define the spatial metric tensor induced on the surfaces
\begin{equation}\label{eq:gamab_def}
	\gam_{\mu\nu} \equiv g_{\mu\nu} + n_\mu n_\nu,
\end{equation}
which is used to express four-dimensional objects, such as the Riemann curvature, in terms of purely spatial objects by projecting them onto the spatial surfaces.

The normal vector describes the motion of coordinate observers, and the fluid four velocity describes the motion of observers comoving with the fluid. The Lorentz factor between this pair of observers is
\begin{equation}\label{eq:lorentz_def}
	W \equiv -n_\alp u^\alp = \frac{1}{\sqrt{1 - v^i v_i / c^2}},
\end{equation}
where $v^i$ is the fluid three velocity with respect to an Eulerian observer, which, in terms of the four velocity, lapse, and shift, is
\begin{equation} \label{eq:threevel_def}
	\frac{v^i}{c} = \frac{u^i}{\alp u^0} + \frac{\beta^i}{\alp}.
\end{equation}
In this three-dimensional split, the components of the contravariant four velocity are
\begin{align}
	u^0 &= \frac{dx^0}{d(c \tau)} = \frac{W}{\alp}, \label{eq:u0up}\\
	u^i &= \frac{dx^i}{d(c \tau)} = W \left( \frac{v^i}{c} - \frac{\beta^i}{\alp} \right), \label{eq:uiup}
\end{align}
and the covariant components are
\begin{align} 
	u_0 &= W \left ( \frac{v^i \beta_i}{c} - \alp \right),\label{eq:u0down} \\
	u_i &= W \frac{v_i}{c}. \label{eq:uidown}
\end{align}
Indices of purely spatial objects are raised and lowered using the spatial metric, i.e. $v_i = \gam_{ij} v^j$. For four-dimensional objects this is not the case, i.e. we have $u_i = g_{i\alp} u^\alp \neq g_{ij} u^j$ in general. Full details of the 3+1 foliation and the resulting equations commonly used in numerical relativity are given in Section~\ref{sec:NR}.

\section{Friedmann-Lema\^itre-Robertson-Walker spacetime}\label{sec:FLRW_intro}

The assumptions underlying the FLRW model are that the Universe is both homogeneous --- the same everywhere --- and isotropic --- the same in all directions. This is the \emph{cosmological principle}. 

Applying the cosmological principle to Einstein's equations in spherical polar coordinates results in the line element
\begin{equation}\label{eq:fried_metric_t_withk}
	ds^{2} = -c^2 dt^{2} + a^2(t) \left[ \frac{dr^2}{1 - kr^2} + r^2 \left( d\theta^2 + {\rm sin}^2 \theta d\phi^2 \right) \right],
\end{equation}
where $r$ is the comoving radial distance, and the curvature constant is $k=-1,0,1$ if the universe has an open, flat or closed geometry, respectively. Assuming spatial flatness we can write the above line element in Cartesian coordinates as
\begin{equation} \label{eq:fried_metric_t}
	ds^2 = -c^2 dt^2 + a^2(t) \delta_{ij}dx^i dx^j.
\end{equation}
In the above, $a(t)$ is the spatially homogeneous \emph{scale factor} describing the size of the universe at any time.%

\subsection{About time}

The coordinate time, $t$, in \eqref{eq:fried_metric_t} is also known as the \emph{cosmic time}; the proper time for a clock with zero peculiar velocity in a vacuum FLRW spacetime. The proper time $\tau$, defined in \eqref{eq:proptime_def}, represents the proper time of \emph{any} clock moving along a \emph{general} path $ds$, whereas the cosmic time represents the proper time of a specific clock. 

The conformal time, $\eta$, explicitly describes the time it would take for a photon to travel back to the Big Bang, if the expansion were to suddenly cease, at any time in the Universe's history. For this reason, $\eta$ is not a physically meaningful time, however, the \emph{particle horizon}, defined as $c\eta$, measures the maximum distance any information could have propagated since the Big Bang (in an FLRW spacetime). This horizon is useful in determining causality between different regions in the Universe, and identifying similarities between regions separated by distances larger than $c\eta$ \citep[e.g. the ``horizon problem'' in inflationary cosmology; see][]{dodelson2003}. The conformal time is widely used in cosmology, and is related to the cosmic time via $a(\eta)d\eta=dt$. The FLRW line element \eqref{eq:fried_metric_t} in terms of conformal time is therefore
\begin{equation} \label{eq:fried_metric}
	ds^2 = a^2(\eta) \left( -c^2 d\eta^2 + \delta_{ij}dx^i dx^j \right).
\end{equation}

\subsection{Friedmann equations}\label{subsec:Friedmann_eqs}

With the metric \eqref{eq:fried_metric}, the components of the Einstein tensor $G_{\mu\nu}$ are
\begin{align}
	G_{00} &= \frac{3}{c^2} \left( \frac{a'}{a} \right)^2, \label{eq:G00_FLRW} \\
	G_{ij} &= \frac{1}{c^2} \left[ \left( \frac{a'}{a} \right)^2 - \frac{2a''}{a} \right] \delta_{ij}, \label{eq:Gij_FLRW} \\
	G_{0i} &= 0,
\end{align}
where $a=a(\eta)$, and the $'$ represents a derivative with respect to conformal time, i.e. $a' \equiv \partial_\eta a \equiv \partial a/\partial\eta$. The trace of \eqref{eq:Gij_FLRW} is therefore
\begin{equation}
	G^i_{\ph{i}i} = g^{ij} G_{ij} = \frac{3}{c^2} \left[ \left( \frac{a'}{a} \right)^2 - \frac{2a''}{a} \right]. \label{eq:trG_FLRW}
\end{equation}
The assumption of isotropy underlying the FLRW solution implies the fluid is at rest with respect to the FLRW coordinates, i.e. $u^{\mu}=(1,0,0,0)$, and the assumption of homogeneity implies the density and pressure are functions of time only. Assuming a perfect fluid, the components of the stress-energy tensor are,
\begin{align}
	T_{00} &= \rho_R c^2 a^2, \label{eq:T00_FLRW} \\
	T_{ij} &= P g_{ij}, \label{eq:Tij_FLRW} \\
	T_{0i} &= 0,
\end{align}
and the trace of the spatial components is 
\begin{equation}
	T \equiv T^i_{\ph{i}i} = 3P. \label{eq:trT_FLRW}
\end{equation}

Substituting the time-time components \eqref{eq:G00_FLRW} and \eqref{eq:T00_FLRW} into Einstein's equations \eqref{eq:einstein} gives the first Friedmann equation
\begin{equation}
	\left(\frac{a'}{a}\right)^2 = \frac{8\pi G \rho_R a^2}{3}, \label{eq:fried1}
\end{equation}
and the traces \eqref{eq:trG_FLRW} and \eqref{eq:trT_FLRW}, also with \eqref{eq:fried1}, gives the acceleration equation
\begin{equation}
	\frac{a''}{a} = -\frac{4\pi G a^2}{3} \left( \rho_R + \frac{3 P}{c^2 a^2} \right). \label{eq:fried2}
\end{equation}

From the time component of the conservation law for the stress-energy tensor \eqref{eq:conservation_stress_energy} 
\begin{equation}
	\nabla_\mu T^{\mu0} = \partial_\mu T^{\mu0} + ^{(4)}{\Gam}^\mu_{\alp\mu} T^{\alp0} + ^{(4)}{\Gam}^0_{\alp\mu} T^{\alp\mu} = 0,
\end{equation}
and using the metric \eqref{eq:fried_metric} to calculate the connection functions (see Section~\ref{subsubsec:Ricci_Gams}), we find
\begin{equation}
	\rho_R' = -3 \frac{a'}{a} \left( \rho_R + \frac{P}{c^2} \right), \label{eq:fried_rhodot}
\end{equation}
which describes mass-energy conservation in an FLRW spacetime.

\subsection{Friedmann solutions}

Deriving the time evolution of the FLRW spacetime in most cases requires solving \eqref{eq:fried_rhodot} with \eqref{eq:fried1}, however, to close the system of equations we must first specify an equation of state (EOS) relating the pressure, $P$, to the mass-energy density, $\rho_R$. 

The EOS we choose will depend on the cosmological era we want to describe. At early times in the Universe's history, around the recombination era, radiation dominated the total energy density, with pressure $P=\frac{1}{3}\rho_R c^2$. Substituting this in \eqref{eq:fried_rhodot} gives a density $\rho_R \propto a^{-4}$, meaning the density of radiation dropped off rapidly with time. At later times, matter therefore came to dominate the energy density, which is well approximated as a ``dust'' fluid with $P=0$. For matter domination the Friedmann equations are
\begin{subequations} \label{eqs:friedmann_dust}
	\begin{align}
		\mH^2 &= \frac{8\pi G \rho_R a^2}{3}, \label{eq:fried1_dust} \\ 
		\frac{a''}{a} &= -\frac{4\pi G\rho_R a^2}{3}, \label{eq:fried2_dust}
	\end{align}
\end{subequations}
where we have defined the conformal Hubble parameter to be
\begin{equation}
	\mH(\eta) \equiv \frac{a'}{a}. \label{eq:conf_Hubble_def}
\end{equation}

In some cosmologies the sign of the Hubble parameter $\mH$ is ambiguous, e.g. in bouncing cosmologies \citep[see e.g.][]{novello2008}. To derive the time evolution of the FLRW spacetime for these cases we must instead solve \eqref{eq:fried_rhodot} with the acceleration equation \eqref{eq:fried2}. For the remainder of this thesis, we will only consider cosmologies with positive expansion, i.e. $\mH > 0$, for all time.

Setting $P=0$ in \eqref{eq:fried_rhodot} gives
\begin{align}
	\frac{\rho_R'}{\rho_R} = -3 \frac{a'}{a}, \\
	\Rightarrow \rho_R a^3 = \rho^* \label{eq:rhostar_def}
\end{align}
where $\rho^*\equiv\rho_{R,{\rm init}} a_{\rm init}^3$ is the conserved (constant) rest-frame energy-density, $a_{\rm init}\equiv a(\eta_{\rm init})$ and $\rho_{R,{\rm init}}\equiv\rho_R(\eta_{\rm init})$ are the arbitrary initial values of the scale factor and density, respectively. Substituting \eqref{eq:rhostar_def} into the Friedmann equation \eqref{eq:fried1} gives
\begin{align}
	\frac{a'}{\sqrt{a}} &= \sqrt{ \frac{ 8\pi G \rho^*}{3} }, \\
	\Rightarrow a(\xi) &= a_{\rm init} \xi^2, \label{eq:Fried_dust_asoln}
\end{align}
where we have defined the dimensionless scaled conformal time
\begin{equation}\label{eq:xidef}
	\xi \equiv 1 + \sqrt{\frac{2\pi G\rho^*}{3a_{\rm init}}} \eta,
\end{equation}
and we have set $\eta_{\rm init}=0$ in deriving the above. Using \eqref{eq:Fried_dust_asoln} in \eqref{eq:rhostar_def} we find
\begin{equation}
	\rho_R(\xi) = \frac{\rho_{R,\rm init}}{\xi^6}. \label{eq:Fried_dust_rhosoln}
\end{equation}

According to observations of SN1a, the expansion of the Universe is currently accelerating. This implies a mysterious, negative pressure --- known as ``dark energy'' --- now contributes $\sim 70\%$ of the total energy density, best described via the cosmological constant $\Lambda$ (see Section~\ref{sec:LCDM}). The evolution of the energy density and scale factor in this case is found using an EOS with a combination of dust and dark energy, i.e. $P=-\rho_\Lambda$. 

The FLRW spacetime forms the basis of the current standard cosmological model to describe the large-scale evolution of the Universe, and is adopted in almost all current simulations of cosmological structure formation (see Section~\ref{sec:LCDM}). The FLRW spacetime is also a common choice for a background cosmology in perturbation theory --- a method commonly used to study the growth of structures in the Universe (see Section~\ref{sec:cosmo_perturb}) --- and to estimate the size of General-Relativistic corrections to the standard cosmological model (see Section~\ref{subsec:GR_corrections}).

\section{The Lambda Cold Dark Matter model} \label{sec:LCDM}

The \lcdm\, model became the widely-accepted cosmological model after the discovery of the accelerating expansion of the Universe \citep{riess1998,perlmutter1999}. This model is the adaptation of the FLRW solution of GR to our observable Universe, and is the concordance cosmology supported by many cosmological observations. In this model, the Universe has a flat geometry, a perfect fluid content, and a non-zero cosmological constant $\Lambda$ to explain the accelerating expansion of the Universe. In addition, the majority of matter in the Universe is observed to be a type of slow moving (i.e. not relativistic) matter known as ``cold dark matter'', which only interacts gravitationally and is therefore well approximated as ``dust''.

\subsection{Cosmological parameters}

Considering matter, curvature, and dark energy, the Friedmann equation is derived from Einstein's equations \eqref{eq:einstein_lambda} with $\Lambda\neq0$ and the metric \eqref{eq:fried_metric_t_withk}, with curvature $k\neq0$ in general, giving
\begin{equation} \label{eq:fried1_full}
	\left(\frac{a'}{a}\right)^2 = \frac{8\pi G \rho_R a^2}{3} + \frac{\Lambda c^2}{3} - \frac{k c^2}{a^2}.
\end{equation}
With the definition of the conformal Hubble parameter, \eqref{eq:conf_Hubble_def}, we can write this equation equivalently as
\begin{equation} \label{eq:cosmo_params_FLRW}
	1 = \om{m} + \om{\Lambda} + \om{k},
\end{equation}
where the dimensionless cosmological parameters are
\begin{subequations}\label{eqs:cosmo_params_LCDM}
	\begin{align}
		\om{m} &\equiv \frac{8\pi G \rho_R a^2}{3\mH^2}, \\
		\om{\Lambda} &\equiv \frac{\Lambda c^2}{3 \mH^2}, \\
		\om{k} &\equiv -\frac{k c^2}{a^2 \mH^2}.
	\end{align}
\end{subequations}
These describe the contribution to the total energy-density of the Universe from matter, dark energy, and curvature, respectively. 

Measuring the cosmological parameters using observations allows us to determine the main components of the Universe, under the assumption that the FLRW model is a valid description of our Universe on the scales measured. We discuss the main observations constraining the values of the cosmological parameters in the following sections.

\subsection{Observations} \label{subsec:observations}

The first discovery by \citet{hubble1929} that the Universe is expanding marked the beginning of the standard cosmological model. Early work constraining the energy-density of matter to be $\om{m}<1$ \citep[e.g.][]{efstathiou1990,efstathiou1992,kofman1993,ostriker1995} paved the way for the discovery of the accelerating expansion, implying $\om{\Lambda}>0$ \citep{riess1998,perlmutter1999}. 

The first detection of the CMB radiation revealed a homogeneous and isotropic background glow across the sky. Improvements in instrumentation and better understanding of systematic errors led to the discovery that the CMB radiation was actually \emph{anisotropic}, and even further improvements now allow us to constrain these anisotropies to the percent level. Surveys of SN1a can now constrain cosmological parameters to within $\sim 2\%$ \citep{riess2018a}, and forecasts of upcoming cosmological surveys predict better than percent-level precision \citep{ivezic2008,maartens2015,amendola2016,SKAWG2018,zhan2018}.

Currently, almost all of our cosmological observations are explained extremely well under the standard model. Including the power spectrum of anisotropies in the CMB radiation, the large-scale clustering of galaxies, and the values of the cosmological parameters. Aside from its success, the \lcdm\, model also has its fair share of ``curiosities'', i.e. observations that are \emph{not} explained in the context of \lcdm. These have driven explorations into whether they could be caused by modifications of GR, if the assumptions underlying the standard model are flawed, or if some other exotic physics could explain the discrepancies. We discuss some of these tensions, and some extensions to \lcdm\, proposed to alleviate them, in Section~\ref{subsec:curiosities}.

\subsubsection{Type 1a Supernovae} \label{subsubsec:SN1a_obs}

The luminosity distance $d_L$ of an object is related to the luminosity flux received by an observer, $F$, and the absolute luminosity of the object, $L$, via
\begin{equation}
	F = \frac{L}{4 \pi d_L^2}.
\end{equation}
The flux received by an observer is dependent on $d_L$, which itself depends on the underlying cosmology; and therefore on the energy-density of dark energy and curvature. Measurements of the luminosity distance are used to determine the distance-redshift relation --- i.e. the Hubble diagram ---  and can therefore also be used to determine the cosmological parameters defined in \eqref{eqs:cosmo_params_LCDM}. 

\citet{phillips1993} discovered that SN1a can be used as standardised measures of distances, or ``standard candles'', due to their predictable light curves. Combined with the fact that they have high intrinsic luminosities, and therefore can be observed at large distances, SN1a are perfect candidates for cosmological observation \citep[see][for a review]{leibundgut2000}. The High-z Supernova Search Team \citep{schmidt1998,riess1998} and the Supernova Cosmology Project \citep{perlmutter1999} took advantage of this and measured the Hubble diagram out to redshift $z\sim0.8$ for the first time. The magnitudes of the distant SN1a they found were $\sim15\%$ dimmer than anticipated by a flat FLRW model with zero cosmological constant, implying accelerating expansion with $\Lambda>0$ in the context of FLRW. Alternate explanations for this dimming include intrinsic evolution of the SN1a light curves as a function of redshift \citep{drell2000} and extinction via extragalactic grey dust \citep{aguirre1999a,aguirre1999b}. Further observations of even higher-redshift SN1a, with $z>1$, strongly disfavour these alternate explanations, and suggest that the Universe previously underwent an episode of \emph{deceleration}, and therefore was not always dark-energy dominated \citep{riess2001,riess2004,riess2007}.

SN1a are now widely used as cosmological probes, and many subsequent surveys align with the early results of \citet{riess1998} and \citet{perlmutter1999}, including the new Supernova Cosmology Project \citep{knop2003}, the Supernova Legacy Survey \citep{astier2006}, the ESSENCE supernova survey \citep{miknaitis2006,wood-vasey2007}, the Sloan Digital Sky Survey II, \citep[SDSS][]{kessler2009}, the Union2 compilation \citep{amanullah2010}, the Supernova $H_0$ for the Equation of State of dark energy (SH0ES) project \citep{riess2011,riess2016}, the Dark Energy Survey Supernova Program \citep{abbottDES2019}, and combinations of different SN1a sample sets, for example, Supercal \citep{scolnic2015} and the Pantheon Sample \citep{scolnic2018}.

If we wish to maintain the homogeneous, isotropic FLRW spacetime as the base of the standard cosmological model, the dimming of SN1a can be explained in one of two ways. First is that the Universe contains a significant amount of dark, non-baryonic matter with an effective negative pressure, called ``dark energy''. This is often described as being smoothly distributed through spacetime, as the cosmological constant $\Lambda$. Second is that Einstein's theory of GR breaks down on cosmological scales, and we need an amended theory to explain our Universe on large scales \citep[see][for a review]{clifton2012}. The standard cosmological model assumes that GR is the correct description of the Universe, and adopts the cosmological constant, with observations currently constraining $\om{\Lambda}\approx0.7$. Alternatively, abandoning the assumptions of homogeneity and isotropy has, in some cases, been able to explain the dimming without $\Lambda$ (see Section~\ref{sec:beyond_LCDM}).

Observations of SN1a also allows precise \emph{measurement} of the expansion rate locally via the Hubble parameter, $H_0$. Using the distance modulus relation, and by calibrating the distances to SN1a using Cepheid variable stars in their host galaxies, the most up-to-date measurement of $H_0$ from SN1a has reached 2.3\% precision, with $H_0=73.48\,{\rm km} \,{\rm s}^{-1}\, {\rm Mpc}^{-1}$ and a $1\sigma$ uncertainty of $\pm1.66 \,{\rm km} \,{\rm s}^{-1}\, {\rm Mpc}^{-1}$ \citep{riess2018a}. This measurement of the local expansion rate is in $3.7\sigma$ tension with the model-dependent \emph{inferred} value using the latest CMB data from the \citet{planck2018a}, with $H_0=67.4 \,{\rm km} \,{\rm s}^{-1}\, {\rm Mpc}^{-1}$ and a $1\sigma$ uncertainty of $\pm0.5 \,{\rm km} \,{\rm s}^{-1}\, {\rm Mpc}^{-1}$ We discuss this discrepancy more in Section~\ref{subsubsec:curiosities_Hubble}, and investigate the role of local inhomogeneities on the SN1a measurement in Chapter~\ref{Chapter5}.

\subsubsection{Cosmic Microwave Background} \label{subsubsec:CMB_obs}

The discovery of the CMB radiation revealed a perfectly homogeneous and isotropic background glow across the sky with temperature $\sim 3.5K$ \citep{penzias1965}, providing the first evidence in favour of Big-Bang cosmology \citep{dicke1965}. Increasingly precise measurements of the background radiation revealed tiny anisotropies in temperature \citep{smoot1992}, and recent measurements --- using the Wilkinson Microwave Anisotropy Probe (WMAP) and the Planck satellite --- have uncovered immense amounts of detail in the $\Delta T/\bar{T}\sim 10^{-5}$ anisotropies around the mean $\bar{T}\approx2.7K$ radiation \citep[e.g.][]{komatsu2009,planck2016params}.

The very early Universe was a dense, opaque plasma of baryons and photons. Hot regions in this plasma expanded under high pressure and consequently cooled. As they cooled, the outward pressure was eventually overcome by the gravitational potential, causing these regions to re-collapse, again raising the temperature and pressure to the point at which pressure dominated once again. This interplay between forces caused BAOs; acoustic waves in the dense plasma. As the Universe expanded, less and less Compton scattering occurred, and free electrons bound with protons to form neutral hydrogen; this was the \emph{recombination} era. These free electrons bound in their highest energy state, immediately dropping in energy and emitting the first photons that were able to freely travel through the Universe; this was the photon \emph{decoupling} era \citep[see][]{durrer2008a}.

The small-scale anisotropies we measure in the CMB radiation today are dominated by the acoustic oscillations created during recombination. Measuring the location and amplitude of the resulting peaks in the angular power spectrum allows us to constrain the geometry of the Universe, the energy-density of baryons, and the energy-density of matter \citep{jungman1996}. Temperature anisotropies themselves can be related to perturbations in the metric at recombination, consequently allowing us to constrain the matter perturbations that gave rise to the large-scale structure we see today. 

Measurements of the CMB temperature power spectrum alone only constrain a small subset of the cosmological parameters, and so \citet{planck2018a} combine the temperature power spectrum with measurements of the polarisation power spectrum and the lensing signature of the CMB. Since the CMB photons measured have travelled the entire history of the Universe, they have passed by many massive objects along their path, and therefore there will be a gravitational lensing signature in the radiation we measure \citep{blanchard1987}. Measuring this lensing spectrum requires assumptions about the geometry of the Universe, the depth of gravitational potentials along the photon paths, the radial extent of the potentials, and the average number of potentials any one photon may pass through. These assumptions are made under linear perturbation theory (see Section~\ref{subsec:lin_perturb}), since on the $\sim 1$ arcminute scales considered for CMB lensing studies, weak lensing should be a valid approximation \citep{lewis2006}.

\subsubsection{Large-Scale Structure} \label{subsubsec:LSS_obs}

The initially small temperature anisotropies in the CMB gave rise to the large-scale distribution of galaxy clusters, filaments, and voids we see today. Signatures of the anisotropy in the CMB radiation are therefore present in the measured clustering of galaxies, e.g. the sound horizon introduced in the previous section \citep{eisenstein1998a,eisenstein1998b}. The sound horizon is measured at low redshift as a peak in the matter power spectrum, and the location of this peak (and hence, physical size of the separation) at different redshifts provides intuition into the method of structure formation, a geometrical distance measure, and insight into the expansion history. SDSS \citep{eisenstein2005} and the 2-degree Field Galaxy Redshift Survey \citep[2dFGRS;][]{cole2005} provided the first detections of the predicted peak in the correlation function of the clustering of luminous red galaxies, leading to precise measures of cosmological distance and strong evidence in favour of dark energy. The discovery of the BAO peak also confirms that dark matter must have been present during the recombination era, since the peak is predicted to be different in a baryon-only universe \citep{eisenstein2005}.

Cosmological N-body simulations (see Section~\ref{subsec:Nbody_sims}) generally assume a base-\lcdm\, cosmology to simulate the growth of the large-scale structure from initially small perturbations. These simulations predict a certain distribution and abundance of massive galaxy clusters at a given redshift \citep[e.g.][]{warren2006}. The formation of the dark-matter haloes surrounding these clusters should be dependent only on the geometry of the Universe and the initial fluctuations that gave rise to the gravitational potentials today \citep{haiman2001}. The abundance of these clusters will also be dependent on the growth rate of structure, determined from the total energy density of matter, $\Omega_m$. Including analysis of the \emph{evolution} of the abundance of galaxy clusters can improve the constraints on cosmological parameters \citep{viana1999}. In addition, measuring the weak gravitational lensing of light as it passes by massive clusters allows another method of mapping the dark-matter distribution. This lensing distorts images of distant galaxies lying behind massive clusters along our line of sight, and the degree of this lensing allows us to constrain the mass of the dark-matter halo of the cluster. As with the galaxy cluster abundances, this allows constraints to be put on dark energy's role in cosmological structure formation \citep{frieman2008}. 

In a matter-dominated Universe, on scales where linear perturbation theory is valid, the depth and size of the gravitational potentials are constant in time. When dark energy dominates at later times, the potentials decay due to the accelerated expansion, and photons are lensed as they travel through these potentials. This is the late-time, integrated Sachs-Wolfe effect \citep{rees1968}. Correlating the lensing signal of the CMB with the position of super-voids and super-clusters ($\sim 100$ Mpc in size) --- as measured in the SDSS luminous red galaxies survey --- supports a flat, dark-energy dominated Universe. However, comparisons with \lcdm\, N-body simulations show a $\sim 2\sigma$ deviation from the observations \citep{granett2008}.

\subsubsection{Concordance cosmology}

Most cosmological observations only constrain certain parameters. For example, the CMB radiation does not strongly constrain dark energy, since at recombination the amount of dark energy was negligible compared to matter and radiation. Combinations of cosmological observations are therefore necessary to build a complete picture of the cosmological model that best describes our Universe. The \citet{planck2018a} presented the latest CMB data, making thorough comparisons to not only different kinds of CMB power spectra, but also to independent cosmological probes including BAOs, SN1a, and galaxy clusters, all of which agree on a spatially-flat \lcdm\, cosmology with matter density $\om{m}=0.315\pm0.007$, and dark-energy density $\om{\Lambda}=0.685\pm0.007$, where the dark energy equation of state is $P=-w \rho_\Lambda$ with $w=-1.03\pm0.03$ (consistent with a cosmological constant). See \citet{planck2018a} for the full parameter set describing the currently favoured standard model. 
\subsection{Curiosities in \lcdm}\label{subsec:curiosities}

The \lcdm\, model has successfully predicted and explained most of our cosmological observations. However, alongside its success are several tensions \citep[see, e.g.][for reviews]{bull2016,buchert2016}. These ``curiosities'' have sparked interest in investigating possible extensions or alterations to the standard model, most of which are also strongly motivated by the fact that inflation, dark matter, and dark energy --- the main components of the standard model --- have largely eluded explanation to date.

\subsubsection{CMB power spectrum}

The \citet{planck2018a} measurements of the CMB anisotropy constrain many of the \lcdm\, model parameters to better than percent-level precision, aligning with predictions of the lensing signal present in the CMB radiation and the angular power spectrum at small scales. However, at the largest angular scales measured by the Planck satellite, there is a suspicious ``dip'' in the power spectrum relative to the \lcdm\, prediction, which is also seen in CMB data from WMAP \citep{bennett2003,hinshaw2013}. The fact that \emph{both} satellites independently measure the same dip at the same angular scale essentially rules out instrument systematic errors or foreground structures as causes of the anomaly, instead pointing towards a real feature in the CMB anisotropy \citep{planck2014}. Since small angular-scale data fits the \lcdm\, prediction so well, there is less freedom to move away from the standard model. New physics at the largest angular scales only could be required to explain this discrepancy, however, must maintain the match to observations at small scales. 

\subsubsection{Hubble parameter}\label{subsubsec:curiosities_Hubble}

Arguably, the most significant tension with the latest CMB measurements is that of the Hubble expansion at redshift zero, i.e. $H_0$. The expansion rate inferred from the CMB power spectrum is $H_0=67.4\pm0.5 \,{\rm km} \,{\rm s}^{-1}\, {\rm Mpc}^{-1}$, a highly \emph{model dependent} result, explicitly assuming a base-\lcdm\, model \citep{planck2018a}. As discussed in Section~\ref{subsubsec:SN1a_obs}, SN1a contribute to the cosmic distance ladder at low redshift, providing a model-independent measurement of the local expansion rate of $H_0=73.48\pm1.66 \,{\rm km} \,{\rm s}^{-1}\, {\rm Mpc}^{-1}$ \citep{riess2018a}. This value is in $3.7\sigma$ tension with the inferred expansion rate from Planck data. 

The actual measured values of $H_0$ have not changed significantly from past measurements from SH0ES \citep{riess2011,riess2016} and earlier Planck data \citep{planck2014,planck2016params}, rather the uncertainties on each respective value have tightened considerably. Better understanding of systematic errors, and improvement of distance calibration for SN1a using Cepheid variable stars, have both contributed to tighter constraints, and hence, increasing tension. Systematic errors have been suggested as a cause of the tension, however, re-analyses of the data have not shown any significant difference in the result \citep[e.g.][]{efstathiou2014}. The model dependence of the Planck value has provoked investigation into whether the standard model could be to blame for the discrepancy. Effects from local structure and peculiar velocities have not been shown to be significant enough to explain the tension, however could be important for upcoming precision surveys \citep[e.g.][]{ben-dayan2014,camarena2018}. \citet{bolejko2018a} showed that emerging, globally negative curvature can successfully explain the Hubble tension --- while also reducing the amount of dark energy needed to explain the accelerated expansion --- under the ``silent universe'' approximation \citep[see][]{bruni1995}. In Chapter~\ref{Chapter5} \citep[see also][]{macpherson2018b} we use a fully General-Relativistic treatment to calculate the variance on the local expansion rate in an inhomogeneous Universe. While the approximations we use to calculate $H_0$ need to be improved (see Section~\ref{subsec:FW_raytracing}), we found the effect was below the percent-level, and therefore not enough to explain the tension.

\subsubsection{Low-redshift Universe}

Lithium levels measured in metal-poor stars in our own galaxy are $4-5\sigma$ lower than predicted by \lcdm, suggesting a lower primordial abundance than predicted in Big-Bang nucleosynthesis \citep{cyburt2008}. This tension can be alleviated by assuming the existence of new particles \citep{cyburt2013}, and in some cases supersymmetric particles --- yet undetected --- have also been shown to solve the problem \citep[e.g.][]{jedamzik2004}. Low-redshift clustering of luminous red galaxies on $\sim$Mpc scales differs from \lcdm\, predictions from cosmological simulations by up to $3\sigma$ \citep{wiegand2014}, and predictions of the growth rate of structure --- sensitive to $\om{m}$ --- are significantly higher than those measured with redshift space distortions \citep[see][]{peacock2001}. While the standard model describes the early Universe extremely well, these discrepancies suggest we may need alternative explanations at low redshift. Suggested extensions include higher-order relativistic corrections to Newtonian dynamics (see Section~\ref{subsec:GR_corrections}), or considering that small-scale dynamics \emph{do} contribute to the large-scale evolution of the Universe, and hence the assumptions of homogeneity and isotropy underlying the standard model are not valid (see Section~\ref{subsec:backreaction}).

\subsubsection{Proposed extensions}

\citet{planck2018a} have investigated standard extensions to the \lcdm\, model, including primordial gravitational waves, non-zero spatial curvature, dynamical (evolving) dark energy models \citep[see also][]{DESCollab2017a}, modifications to GR, and different neutrino masses or primordial element abundances. No significant evidence in favour of any of these extensions, as opposed to \lcdm, was found. 

All of the tensions touched on here remain largely unexplained. Given the increasing precision of cosmological observations --- implying tighter constraints on our systematic errors --- we may instead need to turn to inherent flaws in the \lcdm\, model, or new, exotic physics, to explain these curiosities. 

\subsection{Cosmological simulations}\label{subsec:Nbody_sims}

As mentioned in the previous section, the small-amplitude anisotropies in the CMB at recombination seeded the large-scale galaxy structure we see today. As these anisotropies grow over time under the influence of gravity, the dynamics of their evolution becomes increasingly more complicated. While the amplitudes are small (linear), the evolution can be predicted analytically (see Section~\ref{sec:cosmo_perturb}). However, the nonlinear regime of structure formation is only accessible via numerical simulation. 

CMB measurements provide insight into the near-Gaussian anisotropies at recombination, and large-scale galaxy clustering and BAOs measured at different redshifts gives us an idea about the evolution of these perturbations. Ensuring the standard cosmological model correctly predicts the evolution of the large-scale structure, including the nonlinear dynamics at late times, requires cosmological simulations.

These simulations generally adopt a background flat, FLRW spacetime that expands according to the Friedmann equations (with $\Lambda\neq 0$) alongside a purely-Newtonian description for gravity. Initially small perturbations to the density field, based on measurements of the CMB, collapse over time to form a large-scale distribution of galaxy clusters, filaments, and voids that can then be compared to our observations. 

In the early days, matter dynamics were completely approximated by a collisionless, self-gravitating fluid using N-body particle methods --- with each particle of a certain mass representing a collection of physical dark matter particles. While the majority of matter in the Universe is thought to be cold dark matter, which is well approximated as dust, in order to mimic our cosmological observations we need to also consider gas dynamics \citep[e.g. using smoothed particle hydrodynamics, see][]{monaghan1992}. 

The first N-body cosmological simulation was performed by \citet{peebles1970a}, using 300 particles to simulate the formation of the Coma cluster. Since then, advancements in both supercomputing power and improved numerical techniques have allowed the particle number of such simulations to skyrocket to billions \citep[e.g.][]{springel2005,kim2009,kim2011,genel2014} and even trillions \citep[e.g.][]{ishiyama2013,skillman2014,potter2017} over Gpc volumes.

We need our cosmological simulations to sample both very large volumes (of order Gpc or more), while also sampling down to the scales of individual galaxies smaller than the Milky Way. Large-volume simulations are appealing because of the recent rapid increase in sky area sampled by cosmological surveys, and so we need equivalently large simulations to compare with these observations. In addition, to be able to study the prevalence of very high-mass galaxy clusters in the Universe, we need to sample a large enough volume so as to gain a statistically significant sample of these rare objects. Sampling as small scales as possible, while still maintaining a large volume, will allow a complete sampling of the matter power spectrum at a number of redshifts, with the benefit of also probing dynamics on nonlinear scales \citep{kim2011}.

N-body cosmological simulations have provided a wealth of knowledge into the formation and evolution of large-scale structures, and constrained many aspects of the standard cosmological model. However, a key issue is their use of Newtonian gravity. When sampling small-scale dynamics, Newtonian gravity has been shown to be a good approximation to GR in the presence of a weak gravitational field. On cosmological scales, space-like separated events \emph{can} influence one another in Newtonian gravity, since information propagates instantly. In GR, however, information travels at the speed of light, and therefore causality becomes important on sufficiently large scales. In addition, the assumption of a homogeneously expanding background spacetime, that evolves independently of the nonlinear structure formation taking place, is another potential issue with these simulations. In Section~\ref{subsec:backreaction} we discuss the effects on the global expansion rate by small-scale inhomogeneities, the size of which can only be quantified using cosmological simulations that solve Einstein's equations directly (see Chapters~\ref{Chapter3}, \ref{Chapter4}, and \ref{Chapter5}). 
\section{Cosmological perturbation theory}\label{sec:cosmo_perturb}

The Universe is often approximated as being homogeneous and isotropic. However, the mere presence of stars, planets, galaxies, and galaxy clusters shows that in the early Universe there must have been small perturbations that coalesced over time into larger and larger structures. Investigating where these initially small perturbations came from, and how they grew over time into the structure we see today, is a main goal of perturbation theory in cosmology \citep{kodama1984}. 
Perturbation theory uses a \emph{background} cosmological model --- for example, the FLRW model --- and then uses Einstein's equations of General Relativity to describe small perturbations around this background.

\subsection{The gauge problem}

Due to the complete coordinate freedom of GR, perturbations themselves can be dependent on the chosen coordinates. Therefore, solutions to the perturbed Einstein equations may include unphysical ``gauge modes'', as in the pioneering work of \citet{lifshitz1946} and \citet{lifshitz1963}. This means the density perturbation itself is gauge dependent. A physically meaningful perturbation should not be dependent on the coordinates used, which is where the motivation for ``gauge invariant'' formulations of perturbation theory originated. \citet{bardeen1980} wrote the perturbation equations in a completely gauge-invariant way, and analysed the physical interpretation of the scalar, vector, and tensor perturbations. However, even in \citeauthor{bardeen1980}'s formulation, the \emph{density} perturbation remains dependent on the gauge, because it is defined as the difference between the real density and the background density, i.e. $\delta\rho\equiv \rho - \bar{\rho}$. In this definition, we have implicitly defined a particular mapping from the \emph{true} Universe ($\rho$) to the fictitious \emph{background} spacetime ($\bar{\rho}$), and therefore the perturbations are explicitly dependent on the background spacetime, and hence the chosen gauge. The gauge transformation of an arbitrary tensor field perturbation $ \delta X_{\alp} = X_\alp - \bar{X}_\alp$ is \citep{stewart1974}
\begin{equation} \label{eq:gauge_transform_perturb}
	\delta X'_\alp \rightarrow \delta X_\alp + \mL_Y \bar{X}_\alp,
\end{equation}
where $\mL_Y$ is the Lie derivative with respect to the vector field inducing the gauge transformation, $Y$ (see Section~\ref{subsec:ADM}). From \eqref{eq:gauge_transform_perturb} we can see that if the Lie derivative of the background quantity $\bar{X}_\alp$ vanishes, the perturbation is gauge-invariant. Since the Lie derivative of the density in an FLRW spacetime may not vanish (i.e. the time derivative is nonzero), the perturbation $\delta\rho$ is not gauge invariant. \citet{ellisbruni1989} address this by describing the density distribution instead in terms of the density \emph{gradient}, which is zero in the FLRW background, and therefore the variables are gauge invariant in a perturbed-FLRW Universe \citep[see also][]{bruni1992}.

\subsection{Linear perturbation theory}\label{subsec:lin_perturb}

Early in the Universe's history, perturbations to the curvature and the stress-energy tensor were small, allowing us to approximate their evolution using linear perturbation theory. Considering perturbations to the background metric tensor (indicated with an over bar) such that $\delta g_{\mu\nu} \ll \bar{g}_{\mu\nu}$,
\begin{equation}
	g_{\mu\nu} = \bar{g}_{\mu\nu} + \delta g_{\mu\nu},
\end{equation}
and taking the background cosmology as FLRW, the linearly-perturbed line element in the longitudinal gauge is
\begin{equation}\label{eq:metric_lin_perturb}
	ds^2 = a^2(\eta) \left[ - \left(1 + \frac{2\psi}{c^2}\right) c^2 d\eta^2 + \left(1 - \frac{2\phi}{c^2} \right) \delta_{ij} dx^i dx^j \right],
\end{equation}
where $\psi$ and $\phi$ are the first-order scalar perturbations to the metric, relevant for galaxy clustering and light propagation, respectively. Vector and tensor perturbations are subdominant in the linear regime, and so are usually neglected in the context of linear perturbation theory (but see Section~\ref{subsec:weak_field}). In this gauge, the scalar perturbations $\phi$ and $\psi$ coincide with Bardeen's gauge-invariant potentials $\Phi$ and $\Psi$, respectively \citep{bardeen1980}. These small perturbations in the metric tensor are linked to perturbations in the stress-energy tensor via the perturbed Einstein equations
\begin{equation}\label{eq:lin_perturb_Einstein}
	\bar{G}_{\mu\nu} + \delta G_{\mu\nu} = \frac{8\pi G}{c^4}\left( \bar{T}_{\mu\nu} + \delta T_{\mu\nu} \right).
\end{equation}
Solving these equations in linear perturbation theory involves neglecting terms second order or higher (see Section~\ref{subsec:FLRWSolver} for the derivation of the equations used for initial conditions in this thesis). This method is commonly used to describe the high-redshift Universe --- usually in generating initial conditions for cosmological simulations --- and the low-redshift Universe on sufficiently large scales such that fluctuations in the density field are small. 

\subsection{Weak field approximation}\label{subsec:weak_field}

On small enough scales we measure density fluctuations that are $\delta\gg1$, and therefore the assumption of linear perturbations is no longer applicable. Usually, Newtonian dynamics is used to describe the small-scale growth of structure, although cosmological N-body simulations adopting Newtonian gravity also sample cosmological scales, at which point causality can become important \citep{rigopoulos2015}. However, even though density contrasts may be large, the metric perturbation remains small, with amplitude $\phi/c^2\sim10^{-6}-10^{-5}$, from galactic up to cosmological scales.

Linear perturbation theory remains valid for large-scale perturbations (but see Section~\ref{subsec:backreaction} for a discussion of how small-scale nonlinearities could still influence the large-scale dynamics), the dynamics of which are extremely well understood. To bridge the gap between linear perturbation theory on large scales and Newtonian dynamics on small scales, the \emph{weak field} approximation was developed for cosmology \citep{green2011,green2012}. \citeauthor{green2011} consider a general background metric that describes the averaged behaviour of the spacetime, and differences between the actual metric and the background metric are assumed to be small everywhere, neglecting the effects of relativistic objects such as black holes and neutron stars. This assumption does not imply that any derivatives of $\delta g_{\mu\nu}$ must be small. Placing no limit on second derivatives of the metric also implies that matter sources with $\delta\gg 1$ are allowed within this framework, as is commonly observed on galactic and sub-galactic scales. The intention of this framework is to capture both small-scale nonlinear dynamics and the large-scale, averaged evolution of the Universe. 

\citet{adamek2013} applied the weak-field approximation to \textsc{gevolution}; a relativistic N-body code for simulations of cosmological structure formation, by adopting the perturbed metric in the Poisson gauge (here in units with $c=1$)
\begin{equation}
	ds^2 = a^2(\eta) \left[ -\left( 1 + 2\psi \right) d\eta^2 - 2 B_i dx^i d\eta + \left(1 - 2\phi \right)\delta_{ij} dx^i dx^j + h_{ij} dx^i dx^j \right].
\end{equation}
Here, $B_i$ and $h_{ij}$ are the vector and tensor perturbations, respectively, and are kept to first order only since they are, in general, small relative to the scalar perturbations $\phi$ and $\psi$. This is evident in the long-term success of predictions from Newtonian cosmological simulations --- which neglect $B_i$ and $h_{ij}$ --- and in the difficulty of observing these perturbations compared to the scalar potentials \citep[e.g.][]{everitt2011,abbott2016}. 

The scalar perturbations themselves are also only kept to linear order, except when their quadratic terms are multiplied with their spatial derivatives \citep{adamek2016a,adamek2016b}. Higher derivatives of the potentials, corresponding to density fluctuations, are kept to all orders, and velocities are kept to second order. These simulations capture significantly more relativistic effects than Newtonian N-body simulations, however, the weak-field approximation still requires a background spacetime, and hence an explicit description for the averaged evolution of the Universe. Potential issues with this assumption are discussed in Section~\ref{subsec:backreaction}.
\section{Beyond the Standard Model}\label{sec:beyond_LCDM}

According to the standard cosmological model, approximately 95\% of the energy-density of the Universe is made up of the so-called ``dark components'', namely dark matter and dark energy. Neither of these have been directly observed, and the very nature of them both remains a mystery. Not only because of this, but also due to disagreement between some observations and predictions from the standard model, many extensions are now being explored. These include modifications of GR on cosmological scales \citep[see][for a review]{clifton2012}, using relativistic perturbation theory to include effects that are neglected in the standard model (see Section~\ref{subsec:GR_corrections}), and others which question the validity of the underlying assumptions. An example of the latter is discussed in Section~\ref{subsec:backreaction}, where the assumptions of homogeneity and isotropy underlying the \lcdm\, model are called into question. Other assumptions that have been investigated are the Gaussianity of the primordial fluctuations \citep[see, e.g.,][]{verde2000} and the apparent fine-tuning of inflation \citep[see, e.g.,][]{brandenberger2011}. While some ``standard'' extensions have been shown to be disfavoured compared to \lcdm\, \citep[see, e.g.][]{planck2018a}, there is still a huge amount to be explored.
\subsection{General-Relativistic corrections} \label{subsec:GR_corrections}

Large-scale galaxy surveys are interpreted based on cosmological simulations that adopt purely Newtonian dynamics. The clustering of galaxies is explained sufficiently in Newtonian gravity on small enough scales \citep{green2012}, however, there is skepticism regarding the applicability of the Newtonian limit on (or close to) the Hubble scale \citep[e.g.][]{yoo2009}. Some have shown that the Newtonian limit provides the correct dynamics even on large scales \citep[discussed below; see also][]{matarrese1996,hwang2006,chisari2011,oliynyk2014}. Regardless of this, there are still quantities in GR that simply do not exist in the Newtonian approximation \citep{bruni2014}. In addition, relativistic corrections to our observations arise because our observations take place on the past light cone, not over a spatial slice \citep{bertacca2015}. Assessing the size of the GR corrections to the Newtonian limit for large-scale cosmological simulations is therefore important, especially with the increasing precision of cosmological surveys. 

General-Relativistic effects could be important not only in the nonlinear evolution of the density field, but in the setting of initial conditions for simulations. Primordial Gaussian perturbations in the metric tensor are predicted from inflation \citep{maldacena2003}, and N-body simulations are usually initialised using a corresponding Gaussian distribution of density perturbations. This relation is valid based on the Poisson equation, in which the density perturbation and gravitational potential are related linearly. \citet{bruni2014b} used the post-Friedmann expansion \citep{milillo2015} to show that Gaussian fluctuations in the metric correspond to \emph{non-Gaussian} density perturbations on large scales, due to the nonlinearity of Einstein's equations. Cosmological simulations sampling initial conditions above the causal horizon therefore must include this correction to be consistent with GR \citep[see also][]{rampf2013,christopherson2016}. 

Some have calculated GR corrections using linear perturbation theory \citep[e.g.][]{bonvin2011}, while others have extended to second order and higher, including the effects from not only scalar but also vector and tensor perturbations, in addition to primordial non-Gaussianity in some cases \citep[e.g.][]{green2011,bertacca2014,villa2016}. These extensions are intended to be used for corrections to Newtonian N-body initial conditions, but also for the evolution itself, since they provide relativistic corrections to the particle trajectories.

To align particle positions from N-body simulations with observable coordinates we must identify the simulations with a particular gauge \citep{malik2009,yoo2010,challinor2011}. A natural choice is the so-called conformal Newtonian gauge, i.e. a metric equivalent to \eqref{eq:metric_lin_perturb}. Linearising Einstein's equations for this metric yields a Poisson-like equation for the gravitational potential, i.e. \emph{not} equivalent to the Poisson equation in purely Newtonian gravity (see \eqref{eq:perturb_phi_eq_psi_1} in Section~\ref{subsec:FLRWSolver}). This suggests that N-body simulations are not solving the correct dynamics \citep{chisari2011}. \citet{fidler2015} defined the \emph{N-body gauge}, in which the density field calculated when counting particles in N-body simulations aligns with the comoving density field defined in Einstein's equations. This gauge is therefore suggested as a useful gauge to interpret N-body simulations \citep[see also][]{fidler2016}.

Aside from analysing the \emph{dynamic} evolution in N-body simulations, and assessing their relevance in GR, some relativistic effects are exactly zero in the Newtonian limit, and so require a full GR treatment. The frame-dragging potential, gravitational waves, and the difference between the two potentials $\phi$ and $\psi$ in the metric \eqref{eq:metric_lin_perturb} are all examples of relativistic effects that we know exist, but are zero in Newtonian gravity. Gravitational waves have now been observed \citep{abbott2016}, and the frame dragging effect is present in cosmological perturbation theory \citep{bardeen1980} and has also been measured in our own Solar System \citep{everitt2011}. The ``gravitational slip'' $|\phi-\psi|$ (see Chapter~\ref{Chapter3}) can be measured from the integrated Sachs-Wolfe effect, weak gravitational lensing, and in the matter power spectrum \citep{daniel2010a}. It is zero at first order in GR, however becomes non-zero at higher orders in perturbation theory \citep[and in some modified gravity theories, see][]{daniel2008}. 

Post-Friedmann expansion provides an approximation for GR that captures \emph{both} the small-scale nonlinear dynamics and the large-scale linear dynamics \citep{milillo2015}. In the Newtonian limit of this expansion, there is a non-zero vector potential, in addition to the usual scalar potential, in the metric tensor \citep{bruni2014}. This encapsulates the frame-dragging effect, sourced by purely Newtonian terms, and therefore can be calculated from nonlinear N-body simulations. \citet{bruni2014} performed the first calculation of the frame-dragging potential from a purely Newtonian simulation, showing it has small enough magnitude that N-body dynamics should be unaffected, however, could be measurable in weak-lensing cosmological surveys \citep[see also][]{thomas2015,thomas2015b}.

\subsection{Backreaction}\label{subsec:backreaction}

Our Universe is approximated as homogeneous and isotropic on large scales, however it is extremely inhomogeneous and anisotropic on small scales. The process of smoothing over these small-scale structures to achieve large-scale homogeneity is often referred to as the ``averaging problem'' \citep[see][]{clarkson2011,wiltshire2011}. Due to the nonlinearity of Einstein's equations, when averaging an inhomogeneous fluid (e.g., the Universe on small scales) over large scales, there are extra terms governing the evolution of the averaged fluid compared to a homogeneous fluid. The theory of \textit{backreaction} states that the globally-averaged expansion of an inhomogeneous, anisotropic Universe \emph{does not} coincide with the expansion rate of the homogeneous, isotropic model \citep{buchert2000a}. Extra terms appear in the equation for the acceleration of the expansion rate of the Universe, and therefore have been suggested as alternate explanations for dark energy.

Quantifying the size of the backreaction effect ultimately requires simulations that solve Einstein's equations using numerical relativity. The formalism for calculating cosmological averages is explicitly dependent on the chosen slicing conditions, since the averages themselves must be taken over a specified three-dimensional domain. The dependence of backreaction on slicing has been explored in \citet{adamek2019a}, and can show up to a 10\% difference depending on which three-dimensional slices are chosen. It is therefore important to specify a slicing condition that is physically interesting and best represents our measurements of averages in the Universe.

\subsubsection{Averaging comoving domains} \label{subsubsec:backreaction_comoving}

The original formalism of \citet{buchert2000a} is based on averaging over a spatial surface that is comoving with the fluid flow. This is called the synchronous, comoving gauge; a popular choice in relativistic cosmological perturbation theory, both due to its simplicity and the parallels that can be drawn with Newtonian Lagrangian coordinates \citep{bruni1992,bruni2014ac}. However, this spatial surface only exists if the fluid is vorticity free \citep{ehlers1993}. In this gauge, the proper time measured by a comoving observer coincides with the coordinate time on the spatial slices --- i.e. a lapse function $\alp=1$ --- and the coordinate observers follow the fluid flow --- i.e. a shift vector $\beta^i=0$. The normal vector orthogonal to the spatial slices in this case is therefore $n^\mu=(1,0,0,0)$, here coinciding with the four velocity of the fluid $u^\mu$.

We want to study the averaged dynamics of inhomogeneous, anisotropic dust. The kinematical quantities describing the expansion rate, shear, and vorticity of this fluid are defined, respectively, by decomposing the four velocity of the fluid,
\begin{align}
	\Theta &\equiv \nabla_\mu u^\mu, \\
	\sigma_{\mu\nu} &\equiv b^\alp_{\ph{\alp}\mu} b^\beta_{\ph{\beta}\nu} \nabla_{(\alp}u_{\beta)} - \frac{1}{3} \Theta b_{\mu\nu}, \\
	w_{\mu\nu} &\equiv b^\alp_{\ph{\alp}\mu} b^\beta_{\ph{\beta}\nu} \nabla_{[\alp} u_{\beta]},
\end{align}
where in the comoving gauge $w_{\mu\nu}=0$. The projection tensor $b_{\mu\nu} \equiv g_{\mu\nu} + u_\mu u_\nu$ is purely spatial, and in this gauge is equivalent to the metric tensor $\gam_{ij}$ describing the spatial surfaces. In the above, we use rounded and square brackets around indices to denote the symmetric and antisymmetric parts of a tensor, respectively, i.e.
\begin{align} \label{eq:symmetric_antisymmetric_def}
	A_{(ij)} &\equiv \frac{1}{2} \left( A_{ij} + A_{ji} \right), \\
	A_{[ij]} &\equiv \frac{1}{2} \left( A_{ij} - A_{ji} \right).
\end{align}
When considering three-dimensional slices embedded in four-dimensional spacetime, Einstein's equations are split into the Hamiltonian and momentum constraint equations, and a system of evolution equations for the metric and extrinsic curvature of the slices (see Section~\ref{subsec:ADM}). In terms of the kinematical quantities above, the Hamiltonian constraint equation can be written as (with $\Lambda=0$)
\begin{equation}\label{eq:ham_constraint_buchert}
	\frac{1}{2}\mR + \frac{1}{3}\Theta^2 - \sigma^2 = \frac{8\pi G}{c^2} \rho,
\end{equation}
where $\sigma^2\equiv\frac{1}{2}\sigma^{ij}\sigma_{ij}$, and $\mR$ is the Ricci scalar describing the intrinsic curvature of the surfaces. Here, the density $\rho$ is the projection of the stress-energy tensor into the spatial surfaces.

Raychaudhuri's equation governs the evolution of the expansion scalar, $\Theta$, and is derived from the trace of the evolution equation for the extrinsic curvature \citep{raychaudhuri1957,matarrese1996},
\begin{equation}\label{eq:Raychaudhuri}
	\pdo \Theta + \frac{1}{3}\Theta^2 + 2\sigma^2 + \frac{4\pi G}{c^2} \rho = 0,
\end{equation}
where $\partial_0\equiv\partial/\partial x^0 = \ic \partial_t$, with $t$ the coordinate time (which in the comoving gauge coincides with the proper time). 

The average of a scalar $\chi$, which here is a function of Lagrangian (comoving) coordinates and time, over some arbitrary domain $\mD$ (lying on the spatial slice) is defined as \citep{buchert2000a},
\begin{equation}\label{eq:avg_def}
	\avgb{\chi(t,x^i)} \equiv \frac{1}{\VDb} \int_\mD \chi(t,x^i) \sqrt{b} \, d^3 x,
\end{equation}
where $b$ is the determinant of the projection tensor $b_{ij}$, the volume element is $dV=\sqrt{b}\,d^3 x$, and the volume of the domain is defined as
\begin{equation}\label{eq:VD_def}
	\VDb \equiv \int_\mD \sqrt{b} \, d^3 x.
\end{equation}
The dimensionless, effective scale factor is defined from the volume,
\begin{equation}\label{eq:aD_def}
	\aDb \equiv \left( \frac{\VDb(t)}{\VDb(t_{\rm init})} \right)^{1/3},
\end{equation}
and describes the expansion rate of the domain, where $t_{\rm init}$ is the initial time. We can then write the expansion scalar in terms of the effective scale factor,
\begin{equation} \label{eq:theta_aD_relation}
	\avgb{\Theta} = \frac{\pdo \VDb}{\VDb} = 3 \frac{\pdo \aDb}{\aDb}.
\end{equation}

To study the averaged dynamics of the fluid, we want to study the averaged Raychaudhuri equation and Hamiltonian constraint. First taking the time derivative of \eqref{eq:avg_def}, we get
\begin{align}
	\pdo \avgb{\chi} &= \pdo \left(\frac{1}{\VDb}\right) \int_\mD \chi \sqrt{b} \, d^3 x + \frac{1}{\VDb} \int_\mD \left( \sqrt{b} \, \pdo \chi + \chi \pdo \sqrt{b} \right) \, d^3 x, \\
			&= - \frac{\pdo\VDb}{\VDb} \avgb{\chi} + \avgb{\pdo \chi } + \frac{1}{\VDb} \int_\mD \chi \, \pdo \left(\sqrt{b}\right)\, d^3 x. \label{eq:avgchidot_step2}
\end{align}
Using the evolution equation for the projection tensor $b_{ij}$ (see Section~\ref{subsec:ADM}), we can show \citep[see][]{buchert2000a},
\begin{equation} \label{eq:dtsqrtb_buchert}
	\pdo \sqrt{b} = \Theta \sqrt{b},
\end{equation}
so \eqref{eq:avgchidot_step2} becomes
\begin{equation}
	\pdo \avgb{\chi} = - \frac{\pdo \VDb}{\VDb}\avgb{\chi} + \avgb{\pdo \chi} + \avgb{\Theta\chi}.
\end{equation}
From this, we substitute \eqref{eq:theta_aD_relation} and find the commutation rule to be
\begin{equation}\label{eq:commutation_buchert}
	\pdo \avgb{\chi} - \avgb{\pdo \chi} = \avgb{\Theta\chi} - \avgb{\Theta}\avgb{\chi}.
\end{equation}

Now taking the average of Raychaudhuri's equation \eqref{eq:Raychaudhuri} by averaging each individual term, i.e.
\begin{equation}\label{eq:avg_raychaudhuri_step0}
	\avgb{\pdo\Theta} + \frac{1}{3}\avgb{\Theta^2} + 2\avgb{\sigma^2} + \frac{4\pi G}{c^2} \avgb{\rho} = 0,
\end{equation}
and using the commutation rule \eqref{eq:commutation_buchert}, we have
\begin{equation}
	\avgb{\pdo\Theta} =  \pdo \avgb{\Theta} - \avgb{{\Theta^2}} + {\avgb{\Theta}}^2.
\end{equation}
Substituting the above, along with 
\begin{align}
	\pdo \avgb{\Theta} &= 3 \, \pdo \left(\frac{\pdo \aDb}{\aDb}\right), \\
		&= 3 \, \frac{\pdo^2\aDb}{\aDb} - 3 \left(\frac{\pdo\aDb}{\aDb}\right)^2,
\end{align}
and \eqref{eq:theta_aD_relation}, into \eqref{eq:avg_raychaudhuri_step0} gives
\begin{align}
	-\frac{2}{3}\avgb{\Theta^2} + \frac{2}{3}\avgb{\Theta}^2 + 3 \frac{\pdo^2 \aDb}{\aDb} + 2\avgb{\sigma^2} + \frac{4\pi G}{c^2} \avgb{\rho} = 0,
\end{align}
which we rearrange to get the averaged Raychaudhuri equation,
\begin{equation}\label{eq:avg_Raychaudhuri}
	3 \frac{\pdo^2\aDb}{\aDb}  + \frac{4\pi G}{c^2} \avgb{\rho} = \QD.
\end{equation}
Here, we have defined the kinematical backreaction term
\begin{equation}\label{eq:QD_def}
	\QD \equiv \frac{2}{3} \left( \avgb{\Theta^2} - \avgb{\Theta}^2 \right) - 2\avgb{\sigma^2},
\end{equation}
which exists precisely because the commutation rule \eqref{eq:commutation_buchert} does not vanish. We now take the average of the Hamiltonian constraint \eqref{eq:ham_constraint_buchert} in the same way,
\begin{equation}
	\frac{1}{2}\avgb{\mR} + \frac{1}{3}\avgb{\Theta^2} - \avgb{\sigma^2} = \frac{8\pi G}{c^2} \avgb{\rho},
\end{equation}
and substituting \eqref{eq:QD_def} and \eqref{eq:theta_aD_relation} quickly gives the averaged Hamiltonian constraint,
\begin{equation}\label{eq:avg_han_constraint}
	\frac{3}{c^2} \left( \frac{\pdt\aDb}{\aDb} \right)^2 - \frac{8\pi G}{c^2} \avgb{\rho} + \frac{1}{2}\avgb{\mR} + \frac{1}{2} \QD = 0.
\end{equation}

We define the effective Hubble parameter in the domain $\mD$ from the effective scale factor,
\begin{equation}\label{eq:HD_def}
	\HD \equiv \frac{\pdt\aDb}{\aDb},
\end{equation}
and using this we can rewrite \eqref{eq:avg_han_constraint} to give
\begin{equation}
	\om{m} + \om{R} + \om{Q} = 1,
\end{equation}
where we have defined the dimensionless cosmological parameters,
\begin{equation}\label{eq:cosmo_params_buchert}
	\om{m} \equiv \frac{8\pi G\avgb{\rho}}{3 \HD^2}, \quad \om{R} \equiv -\frac{\avgb{\mR}c^2}{6 \HD^2}, \quad \om{Q} \equiv - \frac{\QD c^2}{6 \HD^2},
\end{equation}
describing, respectively, the matter, curvature, and backreaction content of an averaged inhomogeneous Universe.

This system is analogous to the FLRW model discussed in Section~\ref{sec:FLRW_intro}, however the cosmological parameters \eqref{eq:cosmo_params_buchert} are different to those in the FLRW model \eqref{eqs:cosmo_params_LCDM} since here they are derived in full GR, rather than under the assumptions of homogeneity and isotropy.

\subsubsection{Averaging general foliations} \label{subsubsec:general_foliation}

While the synchronous, comoving gauge is a useful representation of observers following the fluid flow, in practice it presents computational difficulties (see Section~\ref{subsec:coordinates}). It is therefore useful to generalise the above averaging formalism for any slicing condition, i.e. for any form of the lapse function or shift vector. 

Several generalisations of the \citet{buchert2000a} averaging formalism to arbitrary coordinates have been proposed \citep{brown2009a,larena2009,gasperini2010}, with differences in these formalisms stemming mainly from the definition of the Hubble expansion, $\HD$. In the previous section, we defined the Hubble parameter from the expansion scalar $\Theta$, which measures the divergence of the four velocity of the fluid. In the specific case of the synchronous, comoving gauge, the four velocity of the fluid coincides with the normal vector orthogonal to the spatial surfaces, i.e. $u^\mu=n^\mu$. In the case of a general foliation, this may not be the case, and we will generally have $u^\mu\neq n^\mu$. In \citet{brown2009a} and \citet{gasperini2010}, the Hubble parameter is defined as the divergence of the normal vector, $\HD\propto \nabla_\mu n^\mu$, representing the expansion of \textit{coordinate observers}. Similarly, the volume element in these works is defined using the spatial metric $h_{\mu\nu}\equiv g_{\mu\nu} + n_\mu n_\nu$, which is used to project four-dimensional objects onto the spatial slices orthogonal to $n_\mu$ (in the general case, $h_{\mu\nu}$ is distinct from $b_{\mu\nu}$ used in the previous section). \citet{larena2009} defines the Hubble parameter from the divergence of the fluid four velocity, hence analysing a more physically interesting expansion rate, since we, as observers, will measure the expansion rate of the fluid and \textit{not} that of our coordinates. For this reason, here, we follow the generalised averaging formalism of \citet{larena2009}.

In general coordinates, the normal vector orthogonal to the spatial surfaces is given by $n^\mu=(1/\alp, -\beta^i/\alp)$. The four velocity of the fluid is again decomposed into its expansion rate, shear, and vorticity, respectively
\begin{align}
	\theta &\equiv h^{\mu\nu} \nabla_\mu u_\nu, \label{eq:theta_larena_def} \\
	\sigma_{\mu\nu} &\equiv h^{\alp}_{\ph{\alp}\mu} h^{\beta}_{\ph{\beta}\nu} \nabla_{(\alp} u_{\beta)} - \frac{1}{3} \theta h_{\mu\nu}, \label{eq:sigma_def_larena}\\
	\omega_{\mu\nu} &\equiv h^{\alp}_{\ph{\alp}\mu} h^{\beta}_{\ph{\beta}\nu} \nabla_{[\alp} u_{\beta]}.
\end{align}
We also decompose the normal vector and the Eulerian velocity $v^i$ in a similar way, giving
\begin{align}
	\Sigma_{\mu\nu} &\equiv h^{\alp}_{\ph{\alp}\mu} h^{\beta}_{\ph{\beta}\nu} \nabla_{(\alp} n_{\beta)} + \frac{1}{3} K h_{\mu\nu},  \\
	\beta_{\mu\nu} &\equiv h^{\alp}_{\ph{\alp}\mu} h^{\beta}_{\ph{\beta}\nu} \nabla_{(\alp} v_{\beta)} - \frac{1}{3} \kappa h_{\mu\nu}, \label{eq:betaij_larena} \\
	M_{\mu\nu} &\equiv h^{\alp}_{\ph{\alp}\mu} h^{\beta}_{\ph{\beta}\nu} \nabla_{[\alp} v_{\beta]},
\end{align}
where
\begin{align}
	\kappa &\equiv h^{\alp\beta} \nabla_\alp v_\beta, \label{eq:kappa_larena} \\
	K & = - h^{\alp\beta} \nabla_\alp n_\beta. \label{eq:trK_larena}
\end{align}
Here, K is the trace of the extrinsic curvature of the spatial hypersurfaces (see Section~\ref{subsec:ADM}).

The Hamiltonian constraint can be written in terms of the above variables,
\begin{equation} \label{eq:ham_constraint_larena}
	W^2 \mR - 2\sigma^2 - 2\sigma_B^2 + \frac{2}{3}\left(\theta + \theta_B \right)^2 - \frac{16\pi G}{c^2} W^2 \rho = 0,
\end{equation}
where $W$ is the Lorentz factor describing the motion between normal observers and observers comoving with the fluid, and we have used the following for convenience
\begin{align}
	\theta_B &\equiv - W \kappa - W^3 B, \label{eq:thetaB} \\
	\sigma_{B\mu\nu} &\equiv - W \beta_{\mu\nu} - W^3 \left( B_{(\mu\nu)} - \frac{1}{3} B h_{\mu\nu} \right), \label{eq:sigmaBij}
\end{align}
where we have also defined 
\begin{align}
	\sigma^2 &\equiv \frac{1}{2} \sigma^{ij} \sigma_{ij}, \\
	\sigma_B^2 &\equiv \frac{1}{2} \sigma_{Bij} \sigma_B^{ij} + \sigma_{ij}\sigma_B^{ij}. \label{eq:sigmaB2}
\end{align}
We have also introduced the tensor
\begin{equation}
	\begin{aligned}
		B_{\mu\nu} &\equiv \frac{1}{3}\kappa \left(v_\mu n_\nu + v_\mu v_\nu \right) + \beta_{\alp\mu} v^\alp n_\nu + \beta_{\alp\mu} v^\alp v_\nu \\
		&+ M_{\alp\mu} v^\alp n_\nu + M_{\alp\mu} v^\alp v_\nu, \label{eq:Bmunu_larena}
	\end{aligned}
\end{equation}
and $B=\frac{1}{3} \kappa v^\mu v_\mu + \beta_{\mu\nu} v^\mu v^\nu$ is its trace.

The averaging procedure for the non-comoving formalism is defined in a similar way, however we define the volume element here instead using the projection tensor $h_{\mu\nu}$
\begin{equation} \label{eq:VDh_def}
	\VDh \equiv \int_\mD \sqrt{h} \, d^3 x,
\end{equation}
where $h\equiv{\rm det}| h_{\mu\nu}|$, so that the average of a function $\chi$ is
\begin{equation}
	\avgh{\chi(t,x^i)} \equiv \frac{1}{\VDh} \int_\mD \chi(t,x^i) \sqrt{h} \, d^3 x.
\end{equation}

The effective Hubble parameter can then be defined from the expansion scalar \eqref{eq:theta_larena_def} \citep{larena2009,umeh2011},
\begin{equation} \label{eq:HDh_def}
	\HD^h \equiv \frac{c}{3} \avgh{\alp\theta},
\end{equation}
which describes the expansion rate of the fluid as seen by an observer on the hypersurface defined by $h_{\mu\nu}$ \citep{larena2009}. The effective scale factor defined in \eqref{eq:aD_def} describes the expansion of the volume element of the domain, $\VD$. In the case of a comoving slice, $u^\alp = n^\alp$, this expansion corresponds to the expansion of the fluid flow. In this case, we have in general $u^\alp \neq n^\alp$ and so the effective \emph{volume} scale factor here is
\begin{equation}\label{eq:aDV_larena}
	\aD^V \equiv \left( \frac{\VDh(t)}{\VDh(t_{\rm init})} \right)^{1/3},
\end{equation}
which describes the expansion of the coordinate observers themselves. We can also define the effective \emph{fluid} scale factor from the Hubble expansion, by setting
\begin{equation} \label{eq:aDh_def}
	\HD^h = \frac{\pdt \aDh}{\aDh}.
\end{equation}
In this general formalism, using the evolution equation for the metric $h_{ij}$ (see Section~\ref{subsec:ADM}), we can show\footnote{We note an error in equation (30) in \citet{larena2009} --- corresponding to our equation \eqref{eq:dtsqrth_larena} --- and consequently (31) and (34) --- our equations \eqref{eq:aDvdot_on_aDV} and \eqref{eq:scale_factors_relation_larena}, respectively. In the paper, the author has $-\kappa$ in place of $\theta_B$ in our expressions. See Appendix~\ref{sec:appx_aD_typo} for details.}
\begin{equation}\label{eq:dtsqrth_larena}
	\frac{1}{\sqrt{h}} \pdo \sqrt{h} = \frac{\alp}{W} \left( \theta + \theta_B \right) + D_i \beta^i,
\end{equation}
which is equivalent to \eqref{eq:dtsqrtb_buchert} in the case of a comoving formalism, i.e. $\alp=1$, $W=\theta_B=\beta^i=0$. From \eqref{eq:dtsqrth_larena}, the rate of change of the volume is then
\begin{align}
	\frac{\pdo \VDh}{\VDh} &= 3 \frac{\pdo \aD^V}{\aD^V}, \\
		&= \avgh{\frac{\alp}{W} \left(\theta + \theta_B \right) + D_i \beta^i}, \label{eq:aDvdot_on_aDV}
\end{align}
and using this with \eqref{eq:aDh_def}, we can show the relation between the two effective scale factors is
\begin{equation}\label{eq:scale_factors_relation_larena}
	\aDh = \aD^V \, {\rm exp}\left( -\frac{c}{3} \int_{t_{\rm init}}^t \avgh{ \frac{\alp}{W} \left(\theta + \theta_B \right) - \alp\theta + D_i \beta^i } dt \right),
\end{equation}
see Appendix~\ref{sec:appx_expn} for more details.

Averaging each term in \eqref{eq:ham_constraint_larena}, we find
\begin{equation}\label{eq:avg_ham_larena}
	\begin{aligned}
		\avgh{W^2\mR} &- 2\avgh{\sigma^2} - 2\avgh{\sigma_B^2} + \frac{2}{3}\avgh{\theta^2} + \frac{4}{3} \avgh{\theta\theta_B} + \frac{2}{3}\avgh{\theta_B^2} \\
			&- \frac{16\pi G}{c^2} \avgh{W^2 \rho} = 0.
	\end{aligned}
\end{equation}
The kinematical backreaction term adapted for this generalised foliation is
\begin{equation}\label{eq:QD_def_larena}
	\QD^h \equiv \frac{2}{3} \left( \avgh{\theta^2} - \avgh{\theta}^2 \right) - 2\avgh{\sigma^2},
\end{equation}
and the additional backreaction term due to the non-zero coordinate velocity is
\begin{equation}\label{eq:LD_def_larena}
	\LD \equiv 2\avgh{\sigma_B^2} - \frac{2}{3}\avgh{\theta_B^2} - \frac{4}{3} \avgh{\theta\theta_B}.
\end{equation}
With these definitions in \eqref{eq:avg_ham_larena}, we arrive at the averaged Hamiltonian constraint
\begin{equation}
	\om{m} + \om{R} + \om{Q}  + \om{L} = 1.
\end{equation}
Here, the cosmological parameters are
\begin{align}
	\om{m} &\equiv \frac{8\pi G \avgh{W^2 \rho}}{3 {\HD^h}^2}, \quad \om{R} \equiv - \frac{\avgh{W^2\mR}c^2}{6 {\HD^h}^2}, \\
	\om{Q} &\equiv - \frac{\QD^h c^2}{6 {\HD^h}^2}, \quad\quad\quad \om{L} \equiv \frac{\LD c^2}{6 {\HD^h}^2},
\end{align}
which describe the content of an averaged, inhomogeneous universe as calculated by a general, non-comoving observer.

\subsubsection{Improved general formalism} \label{subsubsec:improved_general_averaging}

The above formalism describes the averaged cosmological dynamics of the fluid, as seen by normal observers, by projecting properties of the fluid into the hypersurfaces defined by $h_{\mu\nu}$. However, the volume element \eqref{eq:VDh_def} is propagated along the normal vector to the hypersurfaces, rather than along the fluid four-velocity vector. This means that matter is free to flow into and out of the domain over time --- implying mass is not conserved within the domain --- and the evolution of the averaged quantities within the domain is therefore dependent on the chosen coordinates and slicing.

\citet{buchert2018b} propose a new, coordinate-independent averaging scheme in which the volume element is mass preserving --- by propagating the averaging domain along the fluid flow lines. The averaged quantities in this case are projections from the normal frame into the fluid-comoving frame, and therefore represent the averaged dynamics of the fluid as seen by \emph{comoving} observers, rather than that seen by normal observers as in \citet{larena2009} \citep[and see also][]{brown2009b,gasperini2010}.

The \emph{proper} volume element comoving with the fluid is defined in the same way as $\VDb$ in \eqref{eq:VD_def}, and the averaging operator appears the same as \eqref{eq:avg_def}, however the domain $\mD$ in this case lies in the non-comoving hypersurfaces, rather than the comoving hypersurfaces as in \citet{buchert2000a}. The proper volume element is related to the Riemannian volume element $\VDh$ defined in \eqref{eq:VDh_def}, via
\begin{equation}
	\VDb = \avgb{W} \VDh.
\end{equation}
The averaged Hamiltonian constraint in this formalism is (for $\Lambda=0$)
\begin{equation} \label{eq:rescaled_avg_ham_constraint}
	\frac{3\mH_\mD^2}{c^2} - \frac{8\pi G}{c^2} \avgb{\tilde{\rho}} + \frac{1}{2}\avgb{\tilde{\mR}_b} + \frac{1}{2} \tilde{\mQ}_\mD = 0
\end{equation}
where the effective Hubble parameter is defined in \eqref{eq:HD_def}. The tilde represents rescaled kinematic fluid variables,
\begin{align}
	\tilde{\rho} &\equiv \frac{\alp^2}{W^2} \rho, \quad \tilde{\mR}_b \equiv \frac{\alp^2}{W^2} \mR_b, \\
	\tilde{\Theta} &\equiv \frac{\alp^2}{W^2} \Theta, \quad \tilde{\sigma}^2 \equiv \frac{\alp^2}{W^2} \sigma^2, \quad \tilde{w}^2 \equiv \frac{\alp^2}{W^2} w^2,
\end{align}
where $w^2\equiv \frac{1}{2} w^{ij} w_{ij}$ is the vorticity scalar, and the rescaled kinematic backreaction term in \eqref{eq:rescaled_avg_ham_constraint} is defined as
\begin{equation}
	\tilde{\mQ}_\mD \equiv \frac{2}{3} \left( \avgb{ \tilde{\Theta}^2 } - \avgb{ \tilde{\Theta} }^2 \right) - 2 \avgb{\tilde{\sigma}^2} + 2 \avgb{ \tilde{w}^2 }.
\end{equation}
Here, $\mR_b$ is distinct from the Ricci scalar of the spatial hypersurfaces, $\mR$, and instead represents the fluid rest-frame spatial curvature \citep{buchert2018b}.

\subsubsection{Quantifying $\QD$} \label{subsubsec:quantifying_QD}

The basic result that averaged properties of an inhomogeneous fluid do not satisfy Einstein's equations has sparked many investigations into the size of the resultant effect. Some attempts have been able to completely explain the accelerating expansion with $\Lambda=0$, while others have shown that backreaction itself is a small effect, however, still can be relevant for upcoming precision cosmological surveys. In contrast, some argue that whether or not $\QD$ provides acceleration is irrelevant, and actual connection to \emph{observables} in the Universe is more important \citep{ishibashi2006,green2014}. Because of these strikingly different results, the amplitude of $\QD$ itself is still heavily debated \citep[see, e.g.][]{buchert2015,green2015,green2016}. 

\citeauthor{wiltshire2008}'s ``timescape'' cosmology considers virialised objects to be spatially flat, and the void regions surrounding them to be negatively curved \citep[see][]{wiltshire2007a,wiltshire2007b,wiltshire2008,wiltshire2009}. In the context of \citeauthor{buchert2000a}'s averaging scheme, observers located in the dense ``walls'' surrounding virialised objects measure an \emph{apparent} cosmic expansion when the fraction of the total volume occupied by voids reaches $\sim 0.6$. The difference is largely due to the fact that the clocks of observers located in dense regions will tick differently to the globally-averaged clock. The timescape model has also been shown to fit SN1a light-curve data as well as --- or better than --- \lcdm\, in some cases \citep{dam2017,smale2011}. 

Backreaction as calculated from purely Newtonian simulations has also been shown to be significant; in some cases explaining cosmic acceleration. \citet{roukema2017} used peculiar velocity gradients from Newtonian N-body simulations to calculate the backreaction parameter, and hence the differential expansion due to structure formation \citep[see also][]{rasanen2006b}. In this model, an accelerating global expansion was found when considering $\sim 2.5{\rm Mpc}/h$ averaging regions in calculating $\QD$. Using a similar method, \citet{racz2017} calculated the local expansion rate of smoothed regions from N-body simulations using the Friedmann equations (i.e., not considering any relativistic quantities such as curvature or backreaction). With this method, again on a certain coarse-graining scale, the modified simulations provide an extremely close fit to the SN1a data, while also resolving the tension between the locally-measured Hubble expansion and that from the CMB. Both of these approaches have used purely Newtonian simulations to find a \emph{global} backreaction. However, the original \citet{buchertehlers1997} averaging scheme clearly showed that there can be no global backreaction effect in Newtonian simulations with periodic boundary conditions, since $\QD$ itself manifests as a pure boundary term in Newtonian dynamics \citep[see][for direct comments on these works]{buchert2018a,kaiser2017}. Backreaction can still be studied in the context of these Newtonian simulations on sub-periodicity scales, however in this case the measurement is of \emph{cosmic variance} from peculiar velocities, rather than a pure GR effect \citep{buchert2000,buchert2012}.

Studying backreaction in the context of perturbation theory is potentially problematic since it still requires a background spacetime. Regardless, perturbation theory has still provided some constraints on the size of the backreaction effect in this context. High-order terms in the perturbative series get progressively smaller for the early Universe, however, at redshifts $z\lesssim 1$ these high-order terms have been shown to have similar magnitudes --- i.e. the series does not converge --- and therefore can contribute to accelerating expansion \citep{rasanen2004,notari2006}. Second-order perturbation theory has shown that super-horizon fluctuations generated at inflation could be responsible for the apparent accelerating expansion \citep{barausse2005,kolb2006}. However, these works have been criticised since the higher-order terms neglected in the perturbation series can be shown to cancel these second-order effects \citep{hirata2005}. 

Second-order analysis of the scale dependence of backreaction shows that averaged curvature effects can reach $\sim 10\%$ at 80 Mpc scales --- just below the homogeneity scale of the Universe --- and fall to order $1\%$ at $\sim 200$ Mpc scales, with $\QD$ becoming important inside $\sim 30$ Mpc scales \citep{li2007,li2008}. The weak-field approximation improves on perturbation theory (as discussed in Section~\ref{subsec:weak_field}), in which only the metric perturbations are assumed small. In the weak-field limit, backreaction is small \citep{adamek2015,adamek2019a}. However, relativistic effects in the Hubble diagram \citep{adamek2018b} and in redshift-space distortions on Gpc scales \citep{adamek2018a} have been suggested to be important. 

Backreaction itself is an attractive explanation to the accelerating expansion without introducing any new, exotic physics, but simply by considering Einstein's GR in full. However, any suggestions that propose the standard cosmological model --- which has been accepted as correct for decades --- as incorrect or flawed will be subject to a necessary amount of scrutiny.

\subsection{Exact inhomogeneous cosmology}

Exact solutions to Einstein's equations have proven to be extremely useful in analysing the behaviour and evolution of simple objects in astrophysics, for example, black holes using the \citet{schwarzschild1916} and \citet{kerr1963} solutions. For inhomogeneous cosmology, there are not many exact solutions to choose from \citep[see, e.g.][for reviews]{bolejko2009book,bolejko2011}. Commonly adopted are the Lema\^itre-Tolman-Bondi (LTB) model \citep{lemaitre1933,tolman1934,bondi1947}, a spherically-symmetric dust solution, the Szekeres model \citep{szekeres1975}, a general non-symmetric dust solution, and ``Swiss Cheese'' models \citep[e.g.][]{einsteinstraus1945}, which are often groups of LTB or Szekeres solutions on a homogeneous background \citep{kai2007}. 

The LTB model has been used to suggest that a nearby, large-scale inhomogeneity is causing an apparent accelerating expansion. Measurements by an observer located at the centre of such an inhomogeneity align with the SN1a data without the need for dark energy \citep{celerier2000}, while still describing the position of the first peak in the angular power spectrum of the CMB anisotropy \citep{alnes2006} and observed BAO data \citep{garcia-bellido2008}. Even though under-dense regions of the correct radius have been detected \citep{frith2003}, the measured density does not match the minimum requirement for accelerating expansion \citep{alexander2009}.

The LTB solution has proven useful as a toy model for inhomogeneous cosmology, however, its inherent symmetries call for the use of more general models to validate the results. Szekeres models are inhomogeneous, exact solutions to Einstein's equations for a dust fluid containing no gravitational radiation; a type of ``silent'' Universe \citep[see][]{bruni1995,bolejko2018a}. \citet{ishak2008} used a Szekeres model with zero curvature at large distances from the observer and negative curvature nearby, alongside $\Lambda=0$, to fit the SN1a data competitively with \lcdm, while still satisfying the requirement of spatial flatness at CMB scales. Improving further on these models are ``Swiss Cheese'' models, considering multiple LTB \citep[e.g.][]{biswas2008} or Szekeres \citep[e.g.][]{bolejko2011} holes in a homogeneous background cheese. In some cases these inhomogeneities have been shown to produce a percent-level effect on our observations \citep{bolejko2012,fleury2013}. Apparent accelerating expansion only arises if observers are located in a large $\sim 500$ Mpc void  \citep{marra2007,alexander2009,bolejko2011}, which has been ruled out with CMB constraints in the case of Swiss cheese models \citep{valkenburg2009}. \citet{moss2011} found inconsistencies between LTB void models and observational data, finding the approximate models have very low expansion rates, Universe ages inconsistent with observations, and much smaller local matter fluctuations than measured.

Another family of exact inhomogeneous cosmological models commonly used to address backreaction are black-hole lattices \citep{lindquistwheeler1957}. These models involve regular grids of Schwarzschild masses in an otherwise vacuum spacetime \citep[see][for a review]{bentivegna2018}. Cosmological averaging in these spacetimes exhibits large backreaction effects for small numbers of masses, however the global expansion approaches FLRW for large numbers of masses  \citep[e.g.][]{clifton2012b,bentivegna2013}. Regardless of their inherent large-scale homogeneity and global FLRW expansion, some optical properties measured in these spacetimes --- such as the luminosity distance --- \emph{do not} match the prediction from the FLRW model \citep{bentivegna2017}. While black-hole lattices are useful toy models to study exact inhomogeneous cosmology in the presence of strong-field objects, the mass distribution is extremely simplified and the solutions themselves are static; missing the dynamic aspect of inhomogeneous cosmology.

\subsection{Numerical, General-Relativistic cosmology}

Significant progress has been made towards quantifying the backreaction effect in inhomogeneous cosmologies. In some cases described above, the effect has been significant enough to explain the accelerating expansion, and in other cases the effect is either percent level or completely negligible. All the methods described above have their own respective drawbacks, be it due to simplifying assumptions or symmetries. In order to be able to fully quantify the effect of backreaction in our own Universe, we must solve Einstein's equations in full for the complex, nonlinear matter distribution we observe. Advances in numerical relativity and computational resources over the past two decades (see Section~\ref{sec:NR}) now allow for the stable simulation of relativistic objects such as black holes and neutron stars. Application of numerical relativity to large-scale, inhomogeneous cosmological simulations has emerged over the past few years \citep{giblin2016a,bentivegna2016a,macpherson2017a}, and the field is rapidly advancing with new codes \citep{bentivegna2016b,mertens2016,daverio2017,east2018}, measurements of observables from fully relativistic simulations \citep{giblin2016b,giblin2017b}, and the application of particle-based methods alongside numerical relativity \citep[as opposed to purely mesh-based codes; see][]{giblin2018,daverio2019,barrera-hinojosa2019}. As of yet, no significant backreaction effect has been measured in full numerical relativity \citep{bentivegna2016a,macpherson2019a}, however, a lack of virialisation and periodic boundary conditions may potentially explain these results (see Chapter~\ref{Chapter4}). 

These recent advancements in cosmology with numerical relativity will not only allow a full and final quantification of the backreaction effect, but will in general significantly improve the accuracy of large-scale cosmological simulations. This will allow us to assess and quantify all General-Relativistic effects on our observations, test Einstein's GR on cosmological scales, and improve our understanding of the Universe.

\chapter{Methods} %

\label{Chapter2} %

\vspace{10mm}

In this Chapter, we outline the methods used for the simulations and analysis presented in this thesis. We introduce a brief history of numerical relativity, the formalism and system of equations, and how these are implemented in \textsc{Cactus} and the \textsc{Einstein Toolkit} (ET); the numerical relativity code used in this thesis. We derive the system of equations used to develop the initial conditions for our simulations, and the methods employed in \mesc\, to perform post-simulation analysis.

\section{Numerical Relativity}\label{sec:NR}

Solving Einstein's equations numerically allows the study of dynamics of strong-field gravitating objects, which are otherwise inaccessible with analytic methods. Binary black holes and neutron stars --- or black hole, neutron star pairs --- and the resulting gravitational-wave emission and propagation, tidal disruption of a star by a black hole, accretion disks around spinning black holes, supernova explosions, and cosmological structure formation are all examples of the highly nonlinear, relativistic phenomena which can now be studied in detail due to the advancement of this field.

The pioneers of numerical relativity knew the importance of using numerical techniques to study problems that were difficult to solve analytically. \citet{hahn1964} numerically evolved the merger of two ends of a wormhole to study the two-body problem, \citet{eppley1977} created initial data for evolving source-free gravitational radiation, and \citet{smarr1978} studied coordinate choices in the numerical evolution of Einstein's equations, relating these choices to different families of observers. 

In the decades following development of the ADM formalism \citep[][see the next section]{arnowitt1959}, the advancement of numerical-relativity simulations was hindered both by the lack of computational power and stability issues with the form of the evolution equations themselves. Early works focused on the study of pure gravitational waves and stellar collapse \citep{nakamura1987}, and inhomogeneous inflationary cosmology \citep{laguna1991}. 

The main obstacle standing in the way of long-term simulations of compact objects was an efficient method for dealing with singularities. Many singularity-avoiding coordinates were proposed, for example ``maximal slicing'' \citep[][and see Section~\ref{sssec:maximal}]{smarr1978a}, in which time is \emph{stopped} in the vicinity of the singularity, but continues moving forward in other regions. While this avoids issues in evolving the singularity itself, it induces strong gradients in the metric which generally cause the code to fail. In some cases, these slices have been found to be successful in spherical symmetry \citep[e.g.][]{bona1995} and for evolving brief periods of black-hole mergers \citep{anninos1995,brugmann1999}. Alternatively, boundary conditions on the event-horizon edge can completely excise the singularity from the simulation, while still evolving all observable regions \citep[e.g.][]{bardeen1983a,thornburg1987}. This improved the situation, although still only short-lived simulations were possible \citep[e.g.][]{seidel1992,brandt2000,alcubierre2001,thornburg2004}.

Simulations of the head-on collision of non-rotating black holes, including the extraction of gravitational waves, were limited to axially-symmetric, two-dimensional models \citep{smarr1977,seidel1992,anninos1993,bernstein1994}. As computing power and memory continued to increase, parallelisation of codes became possible. \citet{anninos1995} used excision to perform the first fully three-dimensional simulation in Cartesian coordinates of a single black hole, evolved for several light-crossing times. The ``puncture'' method --- i.e., placing the singularity away from the points on the computational grid where the variables are evaluated \citep{brandt1997} --- allowed for longer evolutions of both distorted and colliding black holes \emph{without} the need for excision \citep{alcubierre2003}. 

Even with the improvement of stability via the BSSN formalism \citep{baumgarte1999,shibata1995}, at this time, all binary black-hole and neutron-star \citep[e.g.][]{oohara1999,miller1999} simulations adopted some kind of symmetry, usually to reduce the computational memory required. The first fully three-dimensional, non-axisymmetric binary black-hole simulations --- including the merger and ring-down stages and the emission of gravitational waves --- were performed by \citet{pretorius2005,campanelli2006,baker2006}. Since this ground-breaking work, the field of numerical relativity has exploded. 

These early works paved the way for more recent advancement in numerical relativity, touching many aspects of relativistic astrophysics such as black holes \citep[e.g.][]{baker2006,campanelli2006,buonanno2007,gonzalez2007,hinder2018,huerta2019}, neutron stars \citep[e.g.][]{baiotti2008,paschalidis2011,kastaun2015,chaurasia2018}, stellar collapse and supernovae explosions \citep[e.g.][]{duez2004,montero2012}, and more recently, for cosmology \citep[e.g.][]{giblin2016a,bentivegna2016a,macpherson2017a,east2018,daverio2019,barrera-hinojosa2019}.

\subsection{3+1 Foliation} \label{subsec:3p1}

In the 3+1 foliation of spacetime, Einstein's four-dimensional equations may be written in terms of purely spatial objects constructed from the three-dimensional metric tensor of the embedded spatial hypersurfaces (surfaces), which are then evolved forward in time. Foliations of this type began with the ADM formalism \citep{arnowitt1959}.

Evolutions using this formalism begin with the construction of an initial Cauchy hypersurface --- i.e. specifying the metric tensor $g_{\mu\nu}$ and its first time derivative $\pdt g_{\mu\nu}$ at every point on the initial space-like surface. To then evolve these surfaces forwards in time we thus need to specify the second time derivatives $\partial^2_t g_{\mu\nu}$, which, due to the symmetry of the metric tensor $g_{\mu\nu}$, implies we need ten equations in total to evolve the system. These second time derivatives will be present in \emph{some} components of the Ricci tensor $R_{\mu\nu}$ in the field equations (and hence the Einstein tensor $G_{\mu\nu}$), and so we need to identify which components are relevant for time evolution. 

The contracted Bianchi identities give 
\begin{align}
	\nabla_\nu G^{\mu\nu} &= \partial_\nu G^{\mu\nu} +^{(4)}{\Gam^{\mu}_{\nu\alp}}G^{\nu\alp} + ^{(4)}{\Gam^{\nu}_{\nu\alp}}G^{\alp\mu} = 0, \\
	\Rightarrow \pdo G^{\mu t} &= - \partial_i G^{\mu i} - ^{(4)}{\Gam^{\mu}_{\nu\alp}}G^{\nu\alp} - ^{(4)}{\Gam^{\nu}_{\nu\alp}}G^{\alp\mu},
\end{align}
where the right hand side contains only second time derivatives (in the Ricci tensor $R_{\mu\nu}$). This implies that $G^{\mu t}$ cannot contain any second time derivatives itself, and the four components of Einstein's equations
\begin{equation} \label{eq:Gmut_dof}
	G^{\mu t} = \frac{8\pi G}{c^4} T^{\mu t},
\end{equation}
therefore cannot contribute to the evolution of the metric tensor, and instead act as constraint equations that must be satisfied on every surface during the evolution. We now have only the remaining six components of Einstein's equations
\begin{equation}
	G^{ij} = \frac{8\pi G}{c^4} T^{ij}.
\end{equation}
Since we require ten equations in total to evolve the system, this leaves us with four extra degrees of freedom for the evolution. The lapse function $\alp$ describes how much proper time elapses between surfaces, and the shift vector $\beta^i$ describes how the spatial coordinates $x^i$ transform from one surface at time $t$ to the next at time $t+dt$. The remaining four degrees of freedom for the evolution are in the second time derivatives $\partial_t^2 \alp$ and $\partial_t^2 \beta^i$. 

The initial data we must specify in $g_{\mu\nu}$ and $\partial_t g_{\mu\nu}$ comprises in total twenty independent choices. Using the constraint equations \eqref{eq:Gmut_dof} these choices are reduced to sixteen. Due to the coordinate invariance of GR, any metric which satisfies Einstein's equations in some coordinate system $x^\mu$ must also satisfy Einstein's equations in some other coordinate system $x^{\mu '}$. This means our choice of the coordinates $x^\mu$ constrain the physical form of the metric $g_{\mu\nu}$, and therefore eliminate four degrees of freedom from our choice of initial data. The \emph{evolution} of our chosen coordinates is defined via the lapse, $\alp$, and the shift, $\beta^i$, including their first time derivatives $\partial_t \alp$ and $\partial_t \beta^i$. Since $\alp$ and $\beta^i$ form part of the metric tensor itself (see \eqref{eq:gdown_alp_beta} below), these eight gauge choices imply that overall, twelve of our total twenty degrees of freedom in the initial conditions are purely coordinate-based choices. Considering also the four constraint equations, we therefore are left with only two \emph{physical} degrees of freedom in the metric, and two in its first time derivative. 

We want to represent the four-dimensional Einstein equations in terms of purely spatial objects that exist on the chosen surfaces. This is done by defining projection operators using the spatial metric \eqref{eq:gamab_def} induced on the surfaces and the normal vector \eqref{eq:normal_def}. Raising one index of the spatial metric, $\gam^{\mu}_{\ph{\mu}\nu}$, is useful for projecting four-dimensional tensors into the spatial surfaces, while the normal projector $-n^\mu n_\nu$ extracts the time-like part of a tensor. Every free index of the tensor being projected must be contracted with one of these projection operators. This projection is also used to define the covariant derivative with respect to the spatial metric, by projecting the four-dimensional covariant derivative into the surfaces, e.g. for a scalar $\chi$,
\begin{equation}
	D_\mu \chi \equiv \gamdu{\mu}{\nu} \nabla_\nu \chi,
\end{equation}
for a vector (one-form) $F^\nu$,
\begin{equation}
	D_\mu F^\nu \equiv \gamdu{\beta}{\nu} \gamdu{\mu}{\alp} \nabla_\alp F^\beta,
\end{equation}
and for a tensor $P^\nu_{\ph{\nu}\beta}$,
\begin{equation}
	D_\mu P^\nu_{\ph{\nu}\beta} \equiv \gamdu{\beta}{\eps} \gamdu{\delta}{\nu} \gamdu{\mu}{\alp} \nabla_\alp P^\delta_{\ph{\delta}\eps},
\end{equation}
see \citet{baumgarteshapiro2010}.

The Riemann curvature tensor associated with the spatial metric, i.e. the curvature of the spatial surfaces, is
\begin{equation} \label{eq:3Riemdef}
	\mR^{k}_{\ph{k}i l j} = \partial_l \Gam^k_{ij} - \partial_j \Gam^k_{i l} + \Gam^k_{m l} \Gam^m_{ij} - \Gam^k_{m j} \Gam^m_{i l},
\end{equation}
where the spatial connection functions are
\begin{equation} \label{eq:3christoffeldef}
	\Gam^{k}_{i j} = \frac{1}{2}\gam^{k m} \left(\partial_i \gam_{j m} + \partial_j \gam_{m i} - \partial_m \gam_{i j}\right).
\end{equation}
Both \eqref{eq:3Riemdef} and \eqref{eq:3christoffeldef} can therefore be constructed entirely from the spatial metric and its spatial derivatives. The spatial Ricci curvature tensor and its trace are $\mR_{ij} \equiv \mR^k_{\ph{k}ikj}$ and $\mR \equiv \mR^i_{\ph{i}i}$, respectively. 

The connection functions \eqref{eq:3christoffeldef} differ from the \emph{spatial components} of the four-dimensional connection functions \eqref{eq:christoffel4}, denoted by $^{(4)}{\Gamma^k_{ij}}$. The contravariant four-dimensional metric in the 3+1 foliation is \citep{baumgarteshapiro2010}
\begin{align}
	g^{\mu\nu} &= \gam^{\mu\nu} - n^\mu n^\nu \\
	 &= 
	\begin{bmatrix}
	 	-\alp^{-2} & \alp^{-2} \beta^i \\
        	 	\alp^{-2} \beta^j & \gam^{ij} - \alp^{-2} \beta^i \beta^j
  	\end{bmatrix},
\end{align}
and the covariant metric is
\begin{equation} \label{eq:gdown_alp_beta}
	g_{\mu\nu} =
	\begin{bmatrix}
	 	-\alp^2 + \beta_k\beta^k & \beta_i \\
        	 	\beta_j & \gam_{ij}
  	\end{bmatrix},
\end{equation}
from which we can see that $g_{ij}=\gam_{ij}$, however, in general $g^{ij}\neq \gam^{ij}$. This implies the spatial components of \eqref{eq:christoffel4} will not be equivalent to the spatial connection functions \eqref{eq:3christoffeldef}, except in the case of $\beta^i=0$. 

The curvature \eqref{eq:3Riemdef} does not contain \textit{all} of the information about the four-dimensional spacetime curvature, since its four-dimensional relative \eqref{eq:4Riemdef} contains time derivatives, and $\mR^k_{\ph{k}i l j}$ can be constructed purely from spatial derivatives of the spatial metric. It therefore describes only the curvature of the embedded surfaces themselves, and contains no information about how these surfaces are embedded in the four-dimensional manifold. This information is encoded in the extrinsic curvature, $K_{ij}$, which measures the gradient of the normal vectors of the embedded surfaces, and hence describes how these surfaces are placed in the four-dimensional manifold. It is defined as the gradient of the normal vector projected onto the surfaces,
\begin{equation} \label{eq:Kijdef}
	K_{\mu\nu} \equiv - \gam_\mu^{\ph{\mu}\alp} \gam_\nu^{\ph{\nu}\beta} \nabla_\alp n_\beta,
\end{equation}
and in the 3+1 decomposition can also be related to the time derivative of the spatial metric (see Section~\ref{subsec:ADM}). The extrinsic curvature measures local deviations in the direction of the normal vector, and thus describes how the surfaces are deformed on each spatial slice.

\subsection{ADM Formalism} \label{subsec:ADM}

The line element, in Cartesian coordinates, in the 3+1 decomposition is
\begin{equation} \label{eq:3+1lineelement}
	ds^2 = -\alp^2 c^2 dt^2 + \gam_{ij}\left(dx^i + \beta^i c dt\right)\left(dx^j + \beta^j c dt\right).
\end{equation}
In the ADM formalism, Einstein's equations are decomposed into four constraint equations and a set of evolution equations for the spatial metric and the extrinsic curvature.

\subsubsection{Constraint equations}

The constraint equations are derived by relating the four-dimensional Riemann tensor $R^\alp_{\ph{\alp}\mu\beta\nu}$ to its three-dimensional counterpart $\mR^k_{\ph{k}ilj}$, which lives on the surfaces. This is done via the full spatial projection of the four-dimensional Riemann tensor, i.e. $\gamud{\mu}{a} \gamud{\nu}{b} \gamud{\alp}{c} \gamud{\beta}{d} R_{\mu\nu\alp\beta}$. Contracting the resulting relation and eliminating the four-dimensional Riemann tensor using Einstein's equations \eqref{eq:einstein} results in the Hamiltonian constraint,
\begin{equation} \label{eq:ham_constraint}
	\mR + K^2 - K_{ij}K^{ij} - \frac{16\pi G}{c^2}\rho = 0,
\end{equation}
where the mass-energy density measured by an observer moving along the normal to the surfaces is defined as
\begin{equation}\label{eq:rho_def}
	\rho\,c^2 \equiv T_{\mu\nu} n^\mu n^\nu.
\end{equation}

Taking a spatial projection of the four-dimensional Riemann tensor with one index projected in the normal (time) direction, i.e. $\gamud{\mu}{a} \gamud{\nu}{b} \gamud{\alp}{c} n^\beta R_{\mu\nu\alp\beta}$, which is related to spatial derivatives of the extrinsic curvature, and contracting the result gives the momentum constraint
\begin{equation} \label{eq:mom_constraint}
	D_j K^j_{\ph{j} i} - D_i K - \frac{8\pi G}{c^3} S_i = 0,
\end{equation}
where $K=\gam^{ij}K_{ij}$ is the trace of the extrinsic curvature, and the momentum density is defined as 
\begin{equation}
	S_i \equiv -\gam_{i \mu} n_\nu T^{\mu\nu}. \label{eq:Si_def}
\end{equation}
For the full derivation of the constraint and evolution equations, see \citet{baumgarteshapiro2010}.

If the constraint equations are satisfied on this initial surface, and the relevant evolution equations are also satisfied, then the constraint equations will be satisfied exactly on any future surface. Numerical evolution of the initial surface forwards in time will result in nonzero constraint violation due to truncation errors. We discuss this further in Section~\ref{subsec:ET}.

\subsubsection{Evolution equations}

The evolution equation for the spatial metric is derived from the definition of the extrinsic curvature \eqref{eq:Kijdef}, giving
\begin{equation} \label{eq:gamij_evolution}
	\dt \gam_{ij} = -2\alp K_{ij},
\end{equation}
where the time derivative is defined as 
\begin{equation}\label{eq:dt_def}
	\dt \equiv \frac{\partial}{\partial x^0} - \Lbeta,
\end{equation}
and $\Lbeta$ is the Lie derivative associated with the shift vector $\beta^i$. The Lie derivative describes the change in a tensor, vector, or scalar along a vector congruence --- in this case the shift vector $\beta^i$ --- and is defined independent of the metric on the surface. For example, the Lie derivative of a covariant tensor $X_{ij}$ along $\beta^i$ is
\begin{equation}\label{eq:Liederiv_def1}
	\Lbeta X_{ij} \equiv \beta^k \partial_k X_{ij} + X_{ik} \partial_j \beta^k + X_{kj} \partial_i \beta^k,
\end{equation}
and for a contravariant tensor is
\begin{equation}\label{eq:Liederiv_def2}
	\Lbeta X^{ij} \equiv \beta^k \partial_k X^{ij} - X^{ik} \partial_k \beta^j - X^{kj} \partial_k \beta^i.
\end{equation}
So long as the connection functions associated with the spatial metric are symmetric in their lower indices, i.e. $\Gam^i_{jk} = \Gam^i_{(jk)}$, then we can interchange the partial derivatives in the Lie derivative with covariant derivatives to obtain a coordinate-free expression \citep[see][]{baumgarteshapiro2010}.

A tensor is defined by its transformation law, however, if instead an object transforms slightly differently, specifically by picking up a power of the Jacobian during the coordinate transform, it is instead a \emph{tensor density}, with ``weight'' $\mW$ equal to the power of the Jacobian in its coordinate transformation \citep[see][]{baumgarteshapiro2010}. In the above, if $X_{ij}$ were instead a tensor density, we would have an extra term $\mW X_{ij}\partial_k \beta^k$ added to the respective Lie derivative (this will be used in the next section).

We can write the evolution equation for the contravariant spatial metric (which will be of use later), starting with the spatial metric
\begin{align}
	\gam_{ij}\gam^{ij} &= \del{i}{i} = 3, \\
	\Rightarrow \dt \left( \gam_{ij}\gam^{ij} \right) &= 0, \\
	\gam^{ij} \dt \gam_{ij} &= -\gam_{ij} \dt \gam^{ij},
\end{align}
from this we can substitute \eqref{eq:gamij_evolution} into the left-hand side, and deduce that
\begin{equation}
	\dt \gam^{ij} = 2\alpha K^{ij}.
\end{equation}

The evolution equation for the extrinsic curvature is derived from a projection of the four-dimensional Riemann tensor, with two indices projected along the normal direction, i.e. $n^\mu n^\alp \gamud{\beta}{a} \gamud{\nu}{b} R_{\mu\nu\alp\beta}$, giving
\begin{equation} \label{eq:Kij_evolution}
	\begin{aligned} 
		\dt K_{ij} &= \alp\left[ \mR_{ij} - 2 K_{ik} K^k_{\ph{k}j} + KK_{ij}\right] - D_i D_j \alp \\
			&- \frac{8\pi G}{c^4} \alp \left[S_{ij} - \frac{1}{2}\gam_{ij} (S-\rho\,c^2)\right],
	\end{aligned}
\end{equation}
where we have defined the spatial stress and its trace to be, respectively, 
\begin{equation}
	S_{ij} \equiv \gam_{i\mu} \gam_{j\nu} T^{\mu\nu}, \quad S \equiv \gam^{ij} S_{ij}.
\end{equation}

For a system of equations to be ``well posed'', when treated as a Cauchy problem, the solution to such a system must be bounded by an exponential function that is independent of the initial data. The solution must not be able to grow unbounded. In terms of hyperbolic partial differential equations, such as Einstein's equations, the system would be said to be \textit{weakly hyperbolic} if it were not well posed \citep[see][for more detail]{alcubierre2008}.

Einstein's equations in the ADM formalism are only weakly hyperbolic, and are therefore not expected to remain well behaved for long time evolutions \citep[see][]{kidder2001}. The hyperbolicity is spoiled by the presence of mixed second derivatives in the spatial Ricci tensor in the right-hand side of the evolution equation \eqref{eq:Kij_evolution}. Without these terms, the evolution equations could be written as a set of wave equations.
To strengthen the hyperbolicity of the system these mixed derivative terms can be removed, as is done in the BSSN formalism \citep{baumgarte1999,shibata1995}. There are other methods of stabilising the evolution of the ADM system, including abandoning the 3+1 foliation altogether \citep[see][and Sections~\ref{sssec:harmonic} and \ref{subsubsec:Z4} for more details of some of these alternatives]{baumgarteshapiro2010}.

\subsection{The BSSN Formalism} \label{subsec:BSSN}

The BSSN formalism \citep{shibata1995,baumgarte1999} re-casts the ADM equations into a form that is strongly hyperbolic, and hence allows for arbitrarily long, stable evolutions of Einstein's equations.

\subsubsection{Conformal decomposition}

First, we define the \textit{conformal metric} $\cgam_{ij}$, defined by decomposing the spatial metric into two parts using a conformal factor
\begin{equation} \label{eq:confmetricdef}
	\gam_{ij} = \conf \cgam_{ij}, \quad {\rm or,} \quad \gam^{ij} = \iconf \cgam^{ij}.
\end{equation}
In general, conformal decompositions of the metric involve an arbitrary factor $\psi$. The BSSN formalism corresponds to setting $\psi=e^{4\phi}$, which turns out to be a convenient choice when deriving the evolution equations. Quantities written with an over bar are associated with the conformal metric $\cgam_{ij}$, and those associated with the spatial metric $\gam_{ij}$ are written without an over bar. Indices of conformal quantities are raised and lowered with the conformal metric. In Cartesian coordinates, we choose $e^{4\phi} = {\rm det}(\gam_{ij})^{1/3}\equiv\gam^{1/3}$, so that $\cgam = 1$. This convenient choice makes the conformal factor $\phi$ a tensor density of weight $\mW=1/6$, implying the covariant conformal metric $\cgam_{ij}$ is a tensor density of weight $\mW=-2/3$.

The traceless part of the extrinsic curvature is defined as
\begin{equation}\label{eq:Kij_trace_traceless}
	A_{ij} \equiv K_{ij} - \frac{1}{3}\gam_{ij} K,
\end{equation}
and the conformal traceless part of the extrinsic curvature is
\begin{equation}
	A_{ij} = \conf \cA_{ij}, \quad A^{ij} = \iconf \cA^{ij}.
\end{equation}
Substituting the conformal metric \eqref{eq:confmetricdef} into the definition of the spatial connection functions \eqref{eq:3christoffeldef} gives their conformal transformation
\begin{equation}\label{eq:conformal_connection}
	\Gam^i_{jk} = \cGam^i_{jk} + 2\left( \del{i}{j} \partial_k \phi + \del{i}{k} \partial_j \phi - \cgam_{jk}\cgam^{il} \partial_l \phi \right),
\end{equation}
which we substitute into the contraction of \eqref{eq:3Riemdef} to get the conformal transformation of the spatial Ricci tensor 
\begin{equation} \label{eq:Rijconf}
	\begin{aligned}
		\cmR_{ij} &= \mR_{ij} + 2 \left( \bar{D}_i \bar{D}_j \phi + \cgam_{ij} \cgam^{lm} \bar{D}_l \bar{D}_m \phi \right)  \\
	        	           &- 4 \left[ \bar{D}_i (\phi) \bar{D}_j (\phi) - \cgam_{ij} \cgam^{lm} \bar{D}_l (\phi) \bar{D}_m (\phi) \right],
	\end{aligned}
\end{equation}
or
\begin{equation}\label{eq:conf_Ricci_vs_nonconf_Ricci}
	\cmR_{ij} = \mR_{ij} - \cmR_{ij}^\phi,
\end{equation}
where $\cmR_{ij}^\phi$ is the part of \eqref{eq:Rijconf} that only depends on the conformal factor $\phi$. From \eqref{eq:Rijconf} we can write the conformal transformation of the Ricci scalar to be
\begin{equation}\label{eq:3R_conformal}
	\cmR \equiv \cgam^{ij} \cmR_{ij} = \conf \mR + 8 e^{-\phi} \cD^2 e^\phi,
\end{equation}
where we have used
\begin{equation}
	\cD^2 e^\phi = e^ \phi \left[ \cgam^{ij} \cD_i \cD_j \phi + \cgam^{ij} \cD_i (\phi) \cD_j (\phi) \right],
\end{equation}
and $\cD^2\equiv\cgam^{ij}\cD_i \cD_j$ is the conformal, spatial, covariant Laplacian.

\subsubsection{Constraint equations}

In the following section, we will slightly alter the form of the ADM evolution equations to strengthen the hyperbolicity of the system. We do not need to do this for the constraint equations since they are not evolved. However, see Section~\ref{subsec:CCZ4} for some formalisms that \emph{do} evolve the constraint equations to restrict the violations in numerical-relativity simulations. Here, we write the constraint equations in terms of the conformal variables. For the Hamiltonian constraint \eqref{eq:ham_constraint} we first use \eqref{eq:Kij_trace_traceless}, which gives
\begin{equation}
	K_{ij} K^{ij} = A_{ij} A^{ij} + \frac{1}{3} K^2 = \cA_{ij} \cA^{ij} + \frac{1}{3} K^2,
\end{equation}
and using the conformal decomposition of the spatial Ricci scalar \eqref{eq:3R_conformal}, we can write \eqref{eq:ham_constraint} as
\begin{equation}\label{eq:BSSN_ham_constraint}
	\frac{e^\phi}{8} \cmR - \cD^2 e^\phi - \frac{e^{5\phi}}{8} \cA_{ij} \cA^{ij} + \frac{e^{5\phi}}{12} K^2 - \frac{2\pi G}{c^2} e^{5\phi} \rho = 0.
\end{equation}

The first term in the momentum constraint \eqref{eq:mom_constraint}, using \eqref{eq:Kij_trace_traceless}, is
\begin{equation}\label{eq:BSSN_mom_constraint_step0}
	D_j K^j_{\ph{j}i} = D_j \cA^j_{\ph{j}i} + \frac{1}{3} \cD_i K,
\end{equation}
since $A^j_{\ph{j}i} = \cA^j_{\ph{j}i}$, $D_i K = \cD_i K$, and $\bar{K}=K$. We now need to relate the covariant derivative $D_i$ in the above to its conformal counterpart $\cD_i$, which we do using the relation between the conformal and non-conformal spatial connection functions. First we expand the spatial covariant derivative
\begin{equation}
	D_j \cA^j_{\ph{j}i} = \partial_j \cA^j_{\ph{j}i} + \Gam^j_{kj} \cA^k_{\ph{k}i} - \Gam^k_{ji} \cA^j_{\ph{j}k},
\end{equation}
and now using \eqref{eq:conformal_connection} for the two right-most terms above, we find
\begin{equation}
	D_j \cA^j_{\ph{j}i} = \cD_j \cA^j_{\ph{j}i}  + 6 \cA^j_{\ph{j}i} \cD_j \phi.
\end{equation}
Substituting this expression into \eqref{eq:BSSN_mom_constraint_step0}, and then the result into \eqref{eq:mom_constraint} gives the conformal momentum constraint
\begin{equation} \label{eq:BSSN_mom_constraint}
	e^{-6\phi} \cD_k \left( e^{6\phi} \cA^k_{\ph{k}i} \right) - \frac{2}{3} \cD_i K - \frac{8\pi G}{c^3}S_i = 0.
\end{equation}

\subsubsection{Evolution equations}

Introducing the conformal factor $\phi$ in the BSSN formalism means we now require an additional equation. We derive this by first substituting the conformal metric \eqref{eq:confmetricdef} into \eqref{eq:gamij_evolution}, giving
\begin{equation}
	\dt \cgam_{ij} + 4 \cgam_{ij} \dt \phi = -2\alp \cK_{ij},
\end{equation}
taking the trace gives
\begin{equation}\label{eq:dtphi_step2}
	\cgam^{ij} \dt \cgam_{ij} + 12 \dt \phi = -2\alp K.
\end{equation}
We now use the identity \citep[see][]{carroll1997}
\begin{equation} \label{eq:dt_lndetgam}
	\dt {\rm ln}\gam^{1/2} = \frac{1}{2} \gam^{ij} \dt \gam_{ij},
\end{equation}
which for the conformal metric is,
\begin{equation} \label{eq:dt_lndetgam_conf}
	\dt {\rm ln} \cgam^{1/2}  = \frac{1}{2}\cgam^{ij} \dt \cgam_{ij} = 0,
\end{equation}
since $\cgam=1$ in Cartesian coordinates, and so \eqref{eq:dtphi_step2} becomes
\begin{equation} 
	\dt \phi = - \frac{1}{6} \alp K. \label{eq:phi_evolution}
\end{equation}
The evolution equation for the trace of the extrinsic curvature is found by first taking the trace of \eqref{eq:Kij_evolution}, which results in
\begin{equation}
	\gam^{ij} \dt K_{ij} = \alp \left( \mR - 2 K_{ij} K^{ij} + K^2 \right) - D^2 \alp + \frac{4\pi G}{c^4} \alp \left( S - 3\rho c^2 \right),
\end{equation}
where $D^2 \equiv \gam^{ij} D_i D_j$ is the covariant Laplacian associated with the spatial metric. We then eliminate $\mR$ using the Hamiltonian constraint \eqref{eq:ham_constraint}, giving
\begin{equation} \label{eq:dtK_step2}
	\gam^{ij} \dt K_{ij} =  - \alp K_{ij} K^{ij} - D^2 \alp + \frac{4\pi G}{c^4} \alp \left( S + \rho c^2 \right).
\end{equation}
We can then expand the derivative of the trace of the extrinsic curvature to give
\begin{align}
	\dt K &= \dt (\gam^{ij} K_{ij}), \\
		&= \gam^{ij} \dt K_{ij} + 2\alp K_{ij} K^{ij},
\end{align}
where we have used \eqref{eq:gamij_evolution}. Substituting this into \eqref{eq:dtK_step2}, and using \eqref{eq:Kij_trace_traceless}, we find
\begin{equation} \label{eq:trK_evolution}
	\dt K = \alp \left( \cA_{ij} \cA^{ij} + \frac{1}{3} K^2 \right) - D^2 \alp + \frac{4\pi G}{c^4} \alp \left(S + \rho \,c^2\right).
\end{equation}
We can find the traceless part of the evolution equations by subtracting \eqref{eq:phi_evolution} and \eqref{eq:trK_evolution} from the ADM evolution equations \eqref{eq:gamij_evolution} and \eqref{eq:Kij_evolution}. These are
\begin{equation} \label{eq:dtconfgamma}
	\dt \cgam_{ij} = -2\alp\cA_{ij}, \\
\end{equation}
and
\begin{equation} \label{eq:dtAij}
	\begin{aligned} 
		\dt \cA_{ij} &= \iconf \left[ - (D_i D_j \alp)^{{\rm TF}} + \alp \left(\mR_{ij}^{\rm TF} - \frac{8\pi G}{c^4} S_{ij}^{\rm TF} \right) \right] \\
	 	                      &+ \alp \left(K \cA_{ij} - 2 \cA_{ik} \cA^k_{\ph{k}j} \right),
	\end{aligned}
\end{equation}
where the superscript ${\rm TF}$ represents the trace-free part of a tensor. That is, we define,
\begin{align}
	S_{ij}^{\rm TF} &\equiv S_{ij} - \frac{1}{3}\gam_{ij}S, \\
	\mR_{ij}^{\rm TF} &\equiv \mR_{ij} - \frac{1}{3}\gam_{ij}\mR, \label{eq:TF_Ricci} \\
	\left(D_i D_j \alp\right)^{\rm TF} &\equiv D_i D_j \alp - \frac{1}{3}\gam_{ij} D^2 \alp.
\end{align}
We can use the conformal decomposition of the Ricci tensor, \eqref{eq:conf_Ricci_vs_nonconf_Ricci}, and the Ricci scalar, \eqref{eq:3R_conformal}, to relate the trace-free Ricci tensor \eqref{eq:TF_Ricci} to its conformal counterpart,
\begin{equation}
	\cmR_{ij}^{TF} \equiv \cmR_{ij} - \frac{1}{3}\cgam_{ij}\cmR.
\end{equation}
We find
\begin{equation}\label{eq:RijTF_confRijTF}
	\mR_{ij}^{TF} = 	\cmR_{ij}^{TF} + \cmR_{ij}^\phi + \frac{8}{3} \cgam_{ij} e^{-\phi} \cD^2 e^\phi,
\end{equation}
where $\cmR_{ij}^\phi$ is defined from \eqref{eq:Rijconf}.

Computing the Ricci tensor $\mR_{ij}$ in \eqref{eq:dtAij} would again introduce similar mixed derivative terms which spoil the hyperbolicity of the ADM equations. To avoid this, we introduce the contracted conformal connection functions
\begin{equation} \label{eq:conf_gamma}
	\cGam^i \equiv \cgam^{jk} \cGam^i_{jk},
\end{equation}
where the $\cGam^i_{jk}$ are the connection functions associated with the conformal metric. The conformal Ricci tensor in terms of these conformal connection functions is then
\begin{equation}\label{eq:cRij_Gami}
	\cmR_{ij} = -\frac{1}{2} \cgam^{lm} \partial_m \partial_l \cgam_{ij} + \cgam_{k(i} \partial_{j)} \cGam^k + \cGam^k \cGam_{(ij)k} + \cgam^{lm} \left( 2\cGam^k_{l(i}\cGam_{j)km} + \cGam^k_{im} \cGam_{klj} \right),
\end{equation}
where $\Gam_{ijk} \equiv \gam_{im}\Gam^m_{jk}$ for both conformal and non-conformal connection functions, and round brackets around indices denote the symmetric parts of a tensor, defined in \eqref{eq:symmetric_antisymmetric_def}.

There are several ways to ensure the mixed derivatives in $\mR_{ij}$ are eliminated in the evolution equations. One of these is to make \eqref{eq:conf_gamma} a gauge choice, and choose $\cGam^i=0$ so that the mixed derivative terms vanish completely; the ``Gamma-driver'' condition (see Section~\ref{sssec:gamdriver}). However, this reduces the gauge freedom of the system, and may lead to undesirable coordinates that could form coordinate singularities. Another method --- used in the BSSN formalism --- is to evolve \eqref{eq:conf_gamma} as a new variable; which both eliminates the mixed derivative terms \textit{and} retains the gauge freedom of the system via the lapse and the shift vector. 

In Cartesian coordinates, where we have chosen $\cgam = 1$, we can write \eqref{eq:conf_gamma} as
\begin{equation}
	\cGam^i = -\partial_j \cgam^{ij},
\end{equation}
using the identity \eqref{eq:dt_lndetgam_conf}. Taking the time derivative of the above gives
\begin{equation} \label{eq:dtGami_start}
	\pdo \cGam^i = - \partial_j \pdo \cgam^{ij}.
\end{equation}
Expanding the Lie derivative in \eqref{eq:dtconfgamma} gives
\begin{align}
	2\alp \cA^{ij} &= \pdo \cgam^{ij} - \Lbeta \cgam^{ij}, \\
	\Rightarrow \pdo \cgam^{ij} &= 2\alp \cA^{ij} + \beta^k \partial_k \cgam^{ij} - 2\cgam^{k(i} \partial_k \beta^{j)} + \frac{2}{3} \cgam^{ij} \partial_k \beta^k,
\end{align}
and then substituting this expression into \eqref{eq:dtGami_start} we find
\begin{equation} \label{eq:dt_cGam_step2}
	\begin{aligned}
		\pdo \cGam^i &= -2\cA^{ij}\partial_j \alp - 2\alp \partial_j \cA^{ij} + \beta^k\partial_k \cGam^i - \cGam^k\partial_k\beta^i \\
				&+ \frac{1}{3} \cgam^{ij} \partial_j \partial_k \beta^k + \cgam^{kj} \partial_k \partial_j \beta^i + \frac{2}{3} \cGam^i \partial_k \beta^k,
	\end{aligned}
\end{equation}
where we have used the definition \eqref{eq:conf_gamma} to simplify the expression. We can eliminate the derivative term $\partial_j \cA^{ij}$ in the above with the momentum constraint \eqref{eq:mom_constraint}, which along with \eqref{eq:Kij_trace_traceless} gives
\begin{equation} \label{eq:djAij_step1}
	\partial_j A^{ij} = \frac{8\pi G}{c^3} S^i + \frac{2}{3} \gam^{ij} \partial_j K - \Gam^i_{jk} A^{kj} - \Gam^j_{jk} A^{ik}.
\end{equation}
We now substitute the conformal trace-free extrinsic curvature $\cA^{ij}$ and conformal metric $\cgam^{ij}$ into the above expression, and after some simplification we find
\begin{equation} \label{eq:djAij_step2}
	\partial_j \cA^{ij} = \frac{8\pi G}{c^3} \cgam^{ij} S_j + 4\cA^{ij}\partial_j\phi - \left(\Gam^i_{jk} \cA^{kj} + \Gam^j_{jk} \cA^{ik}\right) + \frac{2}{3} \cgam^{ij} \partial_j K.
\end{equation}
The only non-conformal objects remaining in this expression are the connection functions, which we can relate to their conformal counterparts using \eqref{eq:conformal_connection}, so that the connection function terms in \eqref{eq:djAij_step2} become
\begin{equation}
	\Gam^i_{jk} \cA^{kj} + \Gam^j_{jk} \cA^{ik} = \cGam^i_{jk} \cA^{kj} + 10 \cA^{ij} \partial_j \phi + \cGam^j_{jk} \cA^{ik},
\end{equation}
and using \eqref{eq:dt_lndetgam} we can show that $\cGam^j_{jk} = 0$. Substituting this expression back into \eqref{eq:djAij_step2} gives
\begin{equation}
	\partial_j \cA^{ij} = \frac{8\pi G}{c^3} \cgam^{ij} S_j - 6\cA^{ij}\partial_j\phi - \cGam^i_{jk} \cA^{kj} + \frac{2}{3} \cgam^{ij} \partial_j K.
\end{equation}
Finally, to arrive at the evolution equation for the contracted connection coefficients we substitute the above expression for $\partial_j\cA^{ij}$ into \eqref{eq:dt_cGam_step2}, which gives
\begin{equation} \label{eq:cGami_evol}
	\begin{aligned}
		\dt \cGam^i &= -2 \cA^{ij} \partial_j \alp + 2\alp\left( \cGam^i_{jk} \cA^{kj} - \frac{2}{3} \cgam^{ij} \partial_j K - \frac{8\pi G}{c^3} \cgam^{ij} S_j + 6 \cA^{ij} \partial_j \phi \right) \\
		&+ \frac{2}{3} \cGam^i \partial_j \beta^j + \frac{1}{3} \cgam^{li} \partial_l \partial_j \beta^j + \cgam^{lj} \partial_j \partial_l \beta^i.
	\end{aligned}
\end{equation}
This equation, together with equations \eqref{eq:phi_evolution}, \eqref{eq:trK_evolution}, \eqref{eq:dtconfgamma}, and \eqref{eq:dtAij}, form the full BSSN system of evolution equations, summarised in Table~\ref{tab:equations_used_summary}. Evolving the contracted connection functions as independent functions means that the definition \eqref{eq:conf_gamma} acts as a new constraint equation together with the Hamiltonian \eqref{eq:BSSN_ham_constraint} and momentum constraints \eqref{eq:BSSN_mom_constraint}.

The matter source terms $\rho, S_i$, and $S_{ij}$ appearing in both the ADM and BSSN formalisms above are projections of the stress-energy tensor into the \emph{spatial surfaces}. These quantities are distinct from those measured in the rest frame of the fluid. From \eqref{eq:rho_def} we can relate $\rho$ to the mass-energy density measured in the \emph{fluid} rest frame, $\rho_R$,
\begin{align}
	\rho c^2 &\equiv T_{\mu\nu} n^\mu n^\nu, \\
		&= \rho_0 h u_\mu n^\mu u_\nu n^\nu + P g_{\mu\nu} n^\mu n^\nu, \\
		&= \rho_R c^2 W^2 + P \left( W^2 - 1 \right), \label{eq:rho_rhoR_relation}
\end{align}
where we have used $n_\mu n^\mu = -1$ and the definition of the Lorentz factor \eqref{eq:lorentz_def}.

\subsection{Coordinate choices}\label{subsec:coordinates}

As mentioned in Section~\ref{subsec:3p1}, there exist four degrees of freedom alongside the evolution and constraint equations derived in Section~\ref{subsec:ADM} and \ref{subsec:BSSN}. These freedoms are encompassed in choosing the lapse function and the shift vector. The lapse function $\alp$ describes the time slicing; the relation between proper time and coordinate time between, and across, spatial surfaces. The shift vector $\beta^i$ describes the spatial gauge; describing how the spatial coordinates are translated from one surface to the next. These functions are traditionally chosen in ways that avoid (or prevent) coordinate or physical singularities, with some coordinate choices made purely for simplicity, and others developed to avoid numerical issues with these simpler choices, a few of which we will discuss briefly below.

\subsubsection{Geodesic slicing} \label{subsubsec:geodesic_slicing}

Geodesic slicing is the simplest choice for the spatial gauge and time slicing, in which we have $\alp=1$ and $\beta^i=0$. The choice of zero shift implies that coordinate observers coincide with normal observers, and the choice of $\alp=1$ implies that these observer's proper time coincides with the coordinate time of the surfaces. While this choice simplifies the evolution equations, it also can form coordinate singularities when evolving nonlinear problems. To show this, we can take the trace of \eqref{eq:Kij_evolution}, which gives \citep[see][]{smarr1978},
\begin{equation}
	\dt K = - \gam^{ij} D_i D_j \alp + \alp \left[K_{ij} K^{ij} + \frac{4\pi G}{c^4}(\rho c^2 + S) \right],
\end{equation}
and substituting the conditions for geodesic slicing, $\alp=1$, $\beta^i=0$, gives
\begin{equation}
	\pdo K = K_{ij} K^{ij} + \frac{4\pi G}{c^4}(\rho c^2 + S),
\end{equation}
and for a perfect fluid (using \eqref{eq:Tmunu_perfect_fluid}) we can write the trace of the spatial stress as
\begin{align}
	S \equiv \gam^{ij} S_{ij} &= \gam^{ij} \gam_{i\mu} \gam_{j\nu} T^{\mu\nu} \\
	&= \gam^j_{\ph{j}\mu} \gam_{j\nu} u^\mu u^\nu (\rho c^2 + P) + \gam^j_{\ph{j}\mu} \gam_{j\nu} g^{\mu\nu} P \\
	&= (\rho c^2 + P) u^j u_j + 3P
\end{align}
and in geodesic slicing we have $u_j=n_j=0$, so we have
\begin{equation}
	\pdo K = K_{ij} K^{ij} + \frac{4\pi G}{c^4}(\rho c^2 + 3 P).
\end{equation}
So long as the strong energy condition for a perfect fluid is satisfied,
\begin{equation}
	\rho c^2 + 3P \geq 0,
\end{equation}
then both terms on the right hand side are positive, implying that $K$ will increase without limit. From \eqref{eq:Kijdef}, we can write
\begin{equation} \label{eq:trKdivn}
	K = -\nabla_\mu n^\mu,
\end{equation}
which shows that in geodesic slicing the normal vectors will therefore converge and create caustics, and hence coordinate singularities. This is expected as these choices correspond to observers freely falling along geodesics (i.e. with no acceleration), and geodesics will converge during gravitational collapse. So long as we are not simulating nonlinear gravitational collapse, e.g. simulating evolution of an FLRW spacetime, geodesic slicing is well suited. However, in simulations of large-scale cosmological structure formation, geodesics will cross once nonlinear structures begin to form.

\subsubsection{Maximal slicing}\label{sssec:maximal}

Maximal slicing provides a condition that prevents the convergence of coordinate observers seen in geodesic slicing \citep{smarr1978}. From \eqref{eq:trKdivn}, an obvious choice to stop the focusing of normal observers is $K=0$, and choosing also $\pdt K=0$ ensures this will be true on all subsequent time slices. Equation \eqref{eq:trK_evolution} can then be simplified into an elliptic equation for the lapse
\begin{equation} \label{eq:maximal_elliptic_lapse_zeroK}
	D^2\alp = \alp K_{ij} K^{ij} + \frac{4\pi G}{c^4} \alp \left(S + \rho \,c^2\right).
\end{equation}
In practice, the maximal slicing condition will only be satisfied approximately, i.e. $K \neq 0$, due to truncation errors in the simulation. Even if $\pdt K=0$ is enforced throughout, the maximal slicing condition will still be violated. Instead, a condition is specified to drive $K$ back towards zero,
\begin{equation}
	\pdo K = - m K,
\end{equation}
where $m$ is a positive constant with dimensions of inverse length. Again using \eqref{eq:trK_evolution} we find a new elliptic equation for the lapse,
\begin{equation}\label{eq:elliptic_lapse}
	D^2\alp = \alp \left[ K_{ij} K^{ij} + \frac{4\pi G}{c^4} \left(S + \rho \,c^2\right) \right] + \beta^i \partial_i K + m K,
\end{equation}
which is essentially correcting \eqref{eq:maximal_elliptic_lapse_zeroK} for violations of the maximal slicing condition. The lapse function that satisfies \eqref{eq:elliptic_lapse} describes the maximal slices. Elliptic equations are computationally expensive to invert in three dimensions, so to make this equation cheaper to solve numerically, we convert it into a parabolic form using a derivative of the lapse \citep[i.e. similar to relaxation methods for solving elliptic equations;][]{press1986}. We introduce an arbitrary time coordinate $\lambda=\eps t$, where $\eps$ is some constant with dimension $L$. We then set $\partial_\lambda \alp$ equal to \eqref{eq:elliptic_lapse}, which gives a parabolic equation for the lapse
\begin{equation}
	\pdo \alp = \eps D^2 \alp - \eps \alp \left[ K_{ij} K^{ij} + \frac{4\pi G}{c^4} \left(S + \rho \,c^2\right) \right] - \eps \beta^i \partial_i K - \eps m K,
\end{equation}
or, more simply
\begin{equation} \label{eq:alp_Kdriver}
	\pdo \alp = -\eps \left( \pdo K + m K \right),
\end{equation}
which is referred to as a ``K-driver'' condition \citep{balakrishna1996}. The maximal slicing condition is then satisfied in the limit $\eps\rightarrow\infty$, however a very large $\eps$ would require an extremely small time step, which can significantly increase the computational cost of the simulation. 

\subsubsection{Harmonic coordinates and slicing} \label{sssec:harmonic}

Writing Einstein's equations in harmonic coordinates was first done to avoid some of the computational issues associated with the ADM formalism \citep[e.g.][]{fischer1973}, by abandoning the 3+1 decomposition altogether and keeping Einstein's equations in a four-dimensional form. We first define the four-dimensional contracted connection functions, similar to in Section~\ref{subsec:BSSN}, 
\begin{equation} \label{eq:4gam_contracted}
	{^{(4)}\Gam}^\alp \equiv g^{\mu\nu} {^{(4)}\Gam}^\alp_{\mu\nu}.
\end{equation}
If we choose for these contracted connection functions to vanish,
\begin{equation} \label{eq:4gam_contracted_harmonic}
	{^{(4)}\Gam}^\alp = 0,
\end{equation}
then the coordinates themselves satisfy the wave equation, $\nabla^\alp\nabla_\alp x^\mu = 0$, and are therefore harmonic functions \citep[e.g.][]{york1979}. As discussed in Section~\ref{subsec:BSSN}, defining the contracted connection functions simplifies the form of the Riemann --- and hence Ricci --- tensor by absorbing second derivatives of the metric into first derivatives of the functions \eqref{eq:4gam_contracted}. The harmonic coordinate choice \eqref{eq:4gam_contracted_harmonic} will obviously simplify this even further, and reduces Einstein's equations to a set of nonlinear wave equations, which provides a mathematical advantage since the behaviour of equations of this type is extremely well understood \citep{baumgarteshapiro2010}.

An issue with harmonic coordinates arises because there is no strict requirement for the time coordinate to remain time-like throughout the simulation, and this can cause numerical problems \citep{garfinkle2002}. A way around this is to introduce source functions to the wave equation for the coordinates, i.e. $\nabla^\alp\nabla_\alp x^\mu = H_\mu$. The function $H_\mu$ can then be used to control the behaviour of the coordinates in the simulation, known as ``generalised harmonic coordinates'' \citep{pretorius2005b}.

Using completely harmonic coordinates is not normally done in numerical simulations because the choices of initial data and coordinates are not as clear as in a 3+1 decomposition. Usually, the coordinates are chosen via the lapse and shift, which both have a clear geometric interpretation in relation to the spatial surfaces. In harmonic coordinates, $H_\mu$ encompasses the coordinate choices, but does not have a clear geometric interpretation, making it difficult to form a desirable coordinate system \citep{pretorius2005b}. Because the spacetime is not split into a series of spatial surfaces when using harmonic coordinates, this also makes the generation of initial data more difficult. The regular 3+1 constraint equations, along with the evolution equations, can instead be used to generate the initial data necessary for the harmonic evolution equations; specifying the four-metric $g_{\mu\nu}$ and its time derivative at some initial instant in time. 

Harmonic slicing of the lapse is much more common in numerical simulations, and involves setting only the time component of the contracted connection functions to be zero, i.e. ${^{(4)}\Gam}^0 = 0$. Alongside a zero shift vector, this gives the evolution equation for the lapse to be
\begin{equation} \label{eq:lapse_evol_harmonic}
	\pdo \alp = -\alp^2 K,
\end{equation}
which can be integrated by first substituting a contraction of the ADM evolution equation \eqref{eq:gamij_evolution}, namely,
\begin{equation}
	\alp K = D_i \beta^i - \pdo {\rm ln} \gam^{1/2},
\end{equation}
again with zero shift, which gives
\begin{equation}
	\pdo {\rm ln} \alp = - \pdo {\rm ln} \gam^{1/2}.
\end{equation}
Solving the above gives the general form of the lapse function in harmonic slicing to be
\begin{equation}\label{eq:lapse_soln_harmonic}
	\alp = F(x^i) \gam^{1/2}
\end{equation}
where $F(x^i)$ is an arbitrary, dimensionless, purely spatial function.

\subsubsection{Generalised slicing form}

The appeal of the evolution of the lapse function using \eqref{eq:lapse_soln_harmonic} is that there is no need to invert an elliptic equation at each time step, as with maximal slicing. However, the singularity avoidance of harmonic slicing is not as strong as in maximal slicing \citep{shibata1995}. 

The Bona-Masso family of slicing conditions \citep{bona1995} are a generalisation of the evolution equation for the lapse \eqref{eq:lapse_evol_harmonic},
\begin{equation}\label{eq:bonamassolapse}
	\pdo \alp = -\alp^2 f(\alp) K,
\end{equation}
where $f(\alp)$ is a positive, dimensionless, arbitrary function (which may be a function of the lapse, or a constant). Choosing $f=1$ reduces the slicing condition to harmonic slicing, and $f=0$ reduces it to geodesic slicing. Another popular choice is $f=2/\alp$, which results in the lapse
\begin{equation}
	\alp = 1 + {\rm ln}\gam,
\end{equation}
known as ``1+log'' slicing, which has been proven to have better singularity avoidance than maximal slicing \citep{alcubierre2008}.

\subsubsection{Gamma-driver condition} \label{sssec:gamdriver}

The Gamma-driver condition is related to the minimal distortion shift conditions \citep[see][]{baumgarteshapiro2010}, which was developed to limit the time evolution of the conformal metric in order to reduce spurious coordinate modes in $\cgam_{ij}$ \citep{smarr1978}. We can limit the conformal metric's evolution by setting the time derivative of the conformal, contracted connection functions to zero,
\begin{equation} \label{eq:Gamdriver_zero}
	\pdo \cGam^i = 0,
\end{equation}
which, combined with \eqref{eq:cGami_evol}, gives a set of coupled elliptic equations for the shift vector,
\begin{equation} \label{eq:elliptic_shift}
	\begin{aligned}
		 \beta^j \partial_j \cGam^i &- \cGam^j \partial_j \beta^i + \frac{2}{3} \cGam^i \partial_j \beta^j + \frac{1}{3} \cgam^{li} \partial_l \partial_j \beta^j + \cgam^{lj} \partial_j \partial_l \beta^i \\
		 &= 2 \cA^{ij} \partial_j \alp - 2\alp\left( \cGam^i_{jk} \cA^{kj} - \frac{2}{3} \cgam^{ij} \partial_j K - \frac{8\pi G}{c^3} \cgam^{ij} S_j + 6 \cA^{ij} \partial_j \phi \right).
	\end{aligned}
\end{equation}
As in Section~\ref{sssec:maximal}, we want to drive away any potential violations of \eqref{eq:Gamdriver_zero}, so we instead choose
\begin{equation} \label{eq:Gamdriver_etaGam}
	\pdo \cGam^i = - \eta \cGam^i.
\end{equation}
We then convert \eqref{eq:elliptic_shift} into parabolic form in the same way as in maximal slicing, giving the ``Gamma-driver'' condition for the shift \citep{alcubierre2001,duez2003}
\begin{equation} \label{eq:Gamdriver_shift_evol}
	\pdo \beta^i = k \left( \pdo \cGam^i + \eta \cGam^i \right),
\end{equation}
where $k$ and $\eta$ are arbitrary constants with dimensions length and inverse length, respectively. 

\subsection{Constraint violation management} \label{subsec:CCZ4}

The constraint equations \eqref{eq:ham_constraint} and \eqref{eq:mom_constraint} are zero analytically, however, finite differencing errors introduce a non-zero constraint violation into the evolution. In the ADM formalism, since the constraint equations are not evolved, any local constraint violation remains where it is and can grow, which may become unstable. In deriving the evolution equations for the BSSN formalism, specifically the evolution equation for the contracted connection functions \eqref{eq:cGami_evol}, we brought the momentum constraint into the evolution equations. This means that local violation of the momentum constraint can now propagate and move off the grid (if coupled with suitable boundary conditions). This stabilises the evolution and damps the constraint violation in the simulation. 

\subsubsection{Z4 Formulation}\label{subsubsec:Z4}

In the original ADM formalism, the 3+1 decomposition of spacetime splits Einstein's equations into separate evolution and constraint equations, the latter of which are only enforced on the initial data, and not constrained during the simulation. Any resulting constraint violations are not invariant under coordinate transforms, breaking the general covariance of the system. The Z4 formulation \citep{bona2003} is a covariant extension to Einstein's equations, which involves introducing a new four-vector $Z_\mu$, such that the field equations become
\begin{equation} \label{eq:Z4_Einstein}
	G_{\mu\nu} + \nabla_\mu Z_\nu + \nabla_\nu Z_\mu = \frac{8\pi G}{c^4} T_{\mu\nu},
\end{equation}
with the constraint $Z_\mu=0$ now acting as a numerical check on the accuracy of a simulation. A 3+1 decomposition of \eqref{eq:Z4_Einstein} results in a system of equations with no static constraints, with $Z_\mu$ evolved as a part of the system, maintaining general covariance. The regular 3+1 constraint equations \eqref{eq:ham_constraint} and \eqref{eq:mom_constraint} can be solved to generate initial data for the Z4 system, since the two formulations are equivalent in the case $Z_\mu=0$. The extent to which the solution is matching Einstein's equations can then be monitored via $Z_\mu$ \citep{bona2003}. The addition of this new four-vector is analogous to the addition of the contracted connection functions in harmonic coordinates, which also results in a general covariant system, however the two systems share similar drawbacks, as discussed in Section~\ref{sssec:harmonic} \citep{garfinkle2002,pretorius2005b}.

\subsubsection{Damped Z4 system}

The $\lambda-$system is a method of constraint-violation damping by adding dynamical variables into the system, which act as time derivatives of each constraint expression \citep{brodbeck1999}. Additional terms are brought into the evolution equations of these extra variables such that the variables themselves, and hence the constraints, are damped during the evolution. The Z4 system is a $\lambda-$system with no damping terms; the $Z_\mu$ four-vector acts as the additional dynamical variables related to the constraints. The ``damped Z4 system'' \citep{gundlach2005} takes advantage of this by adding damping terms to \eqref{eq:Z4_Einstein},
\begin{equation} \label{eq:CCZ4}
	\begin{aligned}
		G_{\mu\nu} &+ \nabla_\mu Z_\nu + \nabla_\nu Z_\mu \\
			 &+ \kappa_1 \left[ n_\mu Z_\nu + n_\nu Z_\mu - (1+\kappa_2)g_{\mu\nu} n^\sigma Z_\sigma \right] = \frac{8\pi G}{c^4} T_{\mu\nu},
	\end{aligned}
\end{equation}
where $\kappa_1, \kappa_2$ are free parameters to control the level of damping.

\subsubsection{Conformal and covariant Z4 system}\label{subsubsec:CCZ4}

The conformal and covariant Z4 system \citep[CCZ4;][]{alic2012} is an extension to the damped Z4 system, by performing a 3+1 decomposition of \eqref{eq:CCZ4} and casting the equations into conformal trace-free form, as in the BSSN formalism. The aim is to combine the benefits of the BSSN and generalised harmonic formalisms. The BSSN formalism is appealing numerically because of its gauge freedom in the form of the lapse function and the shift vector, which allow for singularity avoidance without the need to completely excise a region of spacetime from the simulation. In addition, the conformal decomposition allows for potentially singular terms to be absorbed into the conformal factor, rather than the metric. The appeal of the generalised harmonic formalism is the evolution of the constraints is included, meaning initial data that satisfies the constraints will satisfy them at all times \citep{pretorius2005b}. In the case of initially small, inhomogeneous constraint violations, the CCZ4 system will constrain the growth of these violations during the simulation with appropriately chosen values for $\kappa_i$ \citep[see][]{gundlach2005}.

\subsection{General-Relativistic Hydrodynamics}\label{subsec:grhydro_eqs}

Many astrophysical (and general hydrodynamical) phenomena produce shock waves, and the numerical modelling of these nonlinear waves poses several issues. High resolution shock-capturing methods cast the evolution equations into conservative form to ensure that mass, momentum, and energy are conserved across the shock boundaries during the evolution of the shock. This introduces difficulties of its own, since physically interesting qualities of the fluid are described by the \textit{primitive} variables; mass density, pressure, velocity, and internal energy, while the \textit{conserved} variables are the ones used for evolution. In addition, the primitive variables themselves may be required to calculate the source terms to evolve the conserved variables. Solving for the primitive variables from the conserved variables can be difficult for particular systems, since a simple analytic relation between the two may not exist. This must be performed at every time step, adding significant computational time to the calculation. 

The equations of General-Relativistic hydrodynamics are derived from the conservation of rest-mass and energy-momentum \citep[see][]{banyuls1997}
\begin{equation}
	\nabla_\mu (\rho_0 u^\mu) = 0, \quad \nabla_\mu T^{\mu\nu} = 0.
\end{equation}
These equations can be written in flux-conservative form as
\begin{equation} \label{eq:FC_GRHydro}
	\pdo {\bf U} + \partial_i {\bf F}^i = {\bf S},
\end{equation}
where ${\bf U}=[ D, S^c_j, \tau^c ]$ are the conserved variables, here defined in Eulerian coordinates as \citep{wilson1972,font2008} 
\begin{subequations}\label{eqs:GRHydro_conserved}
	\begin{align} 
		D &\equiv \sqrt{\gam} \rho_0 W, \label{eq:Dcons} \\
		S^c_i &\equiv \sqrt{\gam} \rho_0 h W^2 \frac{v_i}{c}, \label{eq:Sjcons} \\
		\tau^c &\equiv \sqrt{\gam}  \left( \rho_0 h W^2 - P \right) - D c^2. \label{eq:taucons}
	\end{align}
\end{subequations}
Here, $h = c^2 + \eps + P/\rho_0$ is the specific enthalpy, and $\eps$ is the specific internal energy. The Lorentz factor, $W$, and the fluid three velocity, $v^i$, are defined in \eqref{eq:lorentz_def} and \eqref{eq:threevel_def}, respectively. 

The vectors ${\bf F}^i$ and ${\bf S}$ in \eqref{eq:FC_GRHydro} are the fluxes and source terms, respectively, defined by %
\begin{subequations} \label{eqs:GRHydro_flux_source}
	\begin{align}
		{\bf F}^i &= \left[ \alp D \tildv^i, \alp \left( S^c_j \, \tildv^i + \delta^i_j P \right) , \alp \left( \tau^c \tildv^i + P v^i \right) \right], \label{eq:GRHydroFi} \\
		{\bf S} &= \left[ 0, T^{\mu\nu} \left( \partial_\mu g_{\nu j} - {^{(4)}}{\Gam^\lambda_{\mu\nu}} g_{\lambda j} \right), \alp \left( T^{\mu 0} \partial_\mu {\rm ln}\alp - T^{\mu\nu} {^{(4)}}{\Gam^0_{\mu\nu}} \right) \right], \label{eq:GRHydroS}
	\end{align}
\end{subequations}
where 
\begin{equation}
	\tildv^i \equiv \frac{v^i}{c} - \frac{\beta^i}{\alp}.
\end{equation}
Defining an EOS for the fluid in question --- i.e. a description of the pressure of the fluid in terms of rest-mass density and internal energy --- closes the system.

While the conserved variables $D,S^c_j$, and $\tau^c$ are the variables actually evolved in the simulation, the primitive variables $\rho_0,P,v^i$, and $\eps$ are required to calculate the source terms ${\bf S}$ for the evolution of the conserved variables (via the stress-energy tensor). From \eqref{eqs:GRHydro_conserved} we can see that converting primitive to conservative variables is straightforward analytically, but the reverse is not. We discuss briefly a few different methods for converting conservative to primitive variables in Section~\ref{subsec:GRHydro}.

\section{The \textsc{Einstein Toolkit}} \label{subsec:ET}

The \textsc{Cactus} code was first written in 1992 by Ed Seidel and his group at the Max Planck Institute for Gravitational Physics (Albert Einstein Institute), initially to allow for a collaborative, parallel platform for numerical-relativity simulations. The \textsc{Cactus} framework consists of a central core, ``flesh'', and application modules, ``thorns'', which communicate with each other via the \textsc{Cactus} flesh, allowing for thorns to be developed and maintained independently from one another. The \textsc{Cactus} framework was later generalised for other computational scientists requiring large-scale collaborative computing \citep{cactus}, and the numerical-relativity capabilities of the \textsc{Cactus} code were collected into the \textsc{Einstein Toolkit} (ET) \citep{loffler2012}. 

The ET\footnote{\url{http://www.einsteintoolkit.org}} itself consists of about 100 thorns used for relativistic astrophysics, including vacuum spacetime solvers (e.g. \mclachlan), General-Relativistic hydrodynamics (e.g. \grhydro), adaptive mesh refinement, analysis thorns, and thorns for different initial conditions. We briefly introduce the main thorns used in this thesis, and discuss the relevant equations being solved.

\subsection{Base thorns}

The main appeal of the ET's structure is the ability to be used collaboratively, and for different parts of the code to be used in different ways. Main evolution thorns (e.g. for the hydrodynamics and spacetime) are written in a way such that supplementary thorns can be easily substituted in their place. A large part of this structure being able to work is through the use of several ``base'' thorns, which store the sets of variables common amongst different methods for numerical evolution of particular systems, and therefore the variables that are common among different thorns. For example, hydrodynamic evolution thorns communicate directly with the base thorn \hydrobase, which stores the \textit{primitive} hydrodynamical variables (see Section~\ref{subsec:grhydro_eqs}), and spacetime evolution thorns communicate with \admbase, which stores the variables evolved using a 3+1 decomposition of spacetime (as discussed in Section~\ref{subsec:ADM} and \ref{subsec:BSSN}). The base thorns \hydrobase\, and \admbase\, also act as an interface to specify initial conditions, and for performing analyses \citep[see][for a detailed discussion of \textsc{Cactus} and ET structure]{zilhao2013}. 

Spacetime evolution thorns in the ET evolve only the left hand side of Einstein's equations and therefore must be sourced by the stress-energy tensor if matter is present. However, calculating the stress-energy tensor \eqref{eq:Tmunu_perfect_fluid} requires \textit{both} matter and spacetime variables. The thorn \tmunubase\, builds the stress-energy tensor $T_{\mu\nu}$ by communicating separately with the hydrodynamic and spacetime thorns, and can then feed it back into the relevant spacetime evolution thorn. This means that the evolution of the hydrodynamics and spacetime are completely independent, and so different thorns can be easily substituted.

\subsection{\mclachlan} \label{subsec:McLachlan}

The \mclachlan\footnote{\url{https://www.cct.lsu.edu/~eschnett/McLachlan/}} group of thorns is a code for solving the left hand side of Einstein equations using a 3+1 conformal decomposition \citep{brownD2009}. The code itself is generated by \texttt{Kranc}\footnote{\url{http://kranccode.org}} --- a \textsc{Mathematica} program that converts a system of partial differential equations into \textsc{Cactus} code. While the code itself solves the vacuum Einstein equations, it can be linked to seperate thorns that solve the hydrodynamical system coupled to the spacetime, such as \grhydro, explained in the next section. The spacetime variables are discretised on a grid, with options for adaptive mesh refinement via the \texttt{Carpet} driver thorn, which also handles memory, parallelisation, input and output, and time evolution \citep{schnetter2004}. 

The \mclachlan\, code implements both the conformal trace-free BSSN (as described in Section~\ref{subsec:BSSN}) and CCZ4 (as described in Section~\ref{subsec:CCZ4}) formalisms, via two thorns \mlbssn\, and \texttt{ML\_CCZ4}. The variables evolved in \mlbssn\, are the conformal factor $\phi$, the trace of the extrinsic curvature $K$, the conformal metric $\cgam_{ij}$, the conformal trace-free extrinsic curvature $\cA_{ij}$, and the contracted conformal connection functions $\cGam^i$ (see Table~\ref{tab:equations_used_summary} for the full system). \texttt{ML\_CCZ4} extends this system with additional damping terms related to the four-vector $Z_\mu$, as in \eqref{eq:CCZ4},  which are controlled using the parameters $\kappa_1,\kappa_2$, and also by evolving the quantity $\hat{\cGam}^i\equiv \cGam^i + 2\cgam^{ij} Z_j$, along with the projection of $Z_\mu$ along the normal direction; $n_\mu Z^\mu$ \citep[see][]{alic2012}. Both thorns adopt a generalised Bona-Masso slicing of the lapse function, using \eqref{eq:bonamassolapse}, and evolve the shift vector under the ``Gamma-driver'' condition, using \eqref{eq:Gamdriver_shift_evol}.

\subsection{\grhydro} \label{subsec:GRHydro}

\grhydro\, is the main hydrodynamical evolution thorn in the ET, evolving the equations of ideal General-Relativistic hydrodynamics (or magnetohydrodynamics), and was built from the public \texttt{Whisky}\footnote{\url{http://www.whiskycode.org/}} code \citep{baiotti2005,hawke2005,giacomazzo2007,baiotti2008,mosta2014}. The method of lines thorn \mol\, is used for time evolution, implementing a numerical method for solving partial differential equations, in which spatial derivatives are discretised and time derivatives are left continuous. This then allows use of a regular numerical method for ordinary differential equations.

Discretising a fluid on a computational grid results in artificial discontinuities in the fluid across cell boundaries. This is often dealt with by averaging the primitive variables across a cell boundary in order to calculate the flux of the fluid across the boundary, which in itself requires reconstructing the primitive variables between neighbouring cells. Performing this reconstruction at high accuracy can result in spurious numerical oscillations when near a shock (according to Godunov's theorem). To achieve monotonicity, reconstruction methods such as total variation diminishing (TVD), the piecewise parabolic method \citep[PPM;][]{colella1984}, and essentially non-oscillatory \citep[ENO;][]{harten1987} methods are required; each of which are implemented in \grhydro. Once the primitive variables have been reconstructed on the cell boundaries, they are used as initial conditions for the chosen Riemann solver in \grhydro, which may be the Harten-Lax-van Leer-Einfeldt \citep[HLLE;][]{harten1983,einfeldt1988}, Roe \citep{roe1981}, or Marquina \citep{marquina1992} solver. 

The EOS of the fluid is specified and handled separately in the thorn \texttt{EOS\_Omni}, in which ``Polytype'' EOS, $P=P(\rho_0)$ (including a polytropic EOS), or general EOS, $P=P(\rho_0,\eps)$ (including a gamma-law and hybrid EOS), are implemented \citep[see][for full details of all those available]{loffler2012}.

To simplify the calculations performed in \grhydro, time derivatives of the spatial metric in the source terms ${\bf S}$ in \eqref{eq:GRHydroS} are eliminated using the evolution equation \eqref{eq:gamij_evolution}. Time components of the four-dimensional connection functions are written in terms of spatial and time derivatives of the spatial metric, lapse, and shift. The time derivatives of the lapse and shift are specified in the chosen gauge, and explicit spatial derivatives of the spatial metric are eliminated using its spatial covariant derivative, which is zero by construction, i.e.
\begin{equation}
	D_i \gam^{jk} = \partial_i \gam^{jk} + 2 \gam^{lk} \Gam^j_{il} = 0.
\end{equation}
Hence, spatial derivatives of $\gam_{ij}$ can be written in terms of the connection functions. 

The conversion from primitive to conservative variables is simple analytically, as can be seen in \eqref{eqs:GRHydro_conserved}, however the reverse is not as straightforward. In \grhydro\, this conversion is performed using a Newton-Raphson iteration, however, the specific method is dependent on the user-chosen EOS. In the case of a general EOS, i.e. $P=P(\rho,\eps)$, the root of the function $f = \bar{P} - P(\brho,\bar{\eps})$ is found using approximate guesses for $\bar{P},\brho,\bar{\eps}$. In this case the pressure is a function of both the density and internal energy, so the derivatives $dP/d\rho$ and $dP/d\eps$ (required to find the root) are supplied from the relevant EOS thorn. In the case of a ``Polytype'' EOS, the root of the function $f = \brho\bar{W} - D/\sqrt{\gam}$ is found using a similar method \citep[see][and the \grhydro\, documentation]{loffler2012}.

\subsection{Initial data and \flrwsolver}\label{subsec:FLRWSolver}

The ET contains several thorns for initialising different setups in both spacetime and matter. These include binary neutron stars, binary black holes, magnetised neutron stars, Minkowski and Kasner spacetimes, Kerr and Schwarzschild spacetimes (in several coordinate systems), and linear gravitational waves. The ET is not used extensively for cosmology, with only a few tests of exact cosmological spacetimes having being previously performed \citep{vulcanov2002}. 

We developed an initial-condition thorn for linearly-perturbed FLRW spacetimes; \flrwsolver\, \citep[see][]{macpherson2017a}. Around the same time, a group of thorns \textsc{CTThorns} (\textsc{CosmoToolkit}) was released and added to the public release of the ET, to both initialise cosmological spacetimes and evolve them with a new hydrodynamic evolution thorn for dust \citep[see][]{bentivegna2016a}. 

To set up initial data, \flrwsolver\, communicates directly with the base thorns \hydrobase\, and \admbase, by filling the initial data for the primitive hydrodynamic variables --- the rest-mass density, pressure, velocity, and internal energy --- and the spacetime variables --- the spatial metric, extrinsic curvature, lapse, and shift.

In \flrwsolver, we currently only consider small perturbations around the flat, dust FLRW model, under the assumption that linear perturbation theory is valid, i.e.,
\begin{equation}
	\frac{|\phi|}{c^2}, \, \frac{|\psi|}{c^2}, \, |\delta|, \, \frac{|v^i|}{c} \ll 1,
\end{equation}
for the metric, density, and velocity perturbations, respectively. 

From the linearly perturbed Einstein equations \eqref{eq:lin_perturb_Einstein}, we have 
\begin{equation} \label{eq:background_einstein}
	\bar{G}_{\mu\nu} = \frac{8\pi G}{c^4} \bar{T}_{\mu\nu},
\end{equation}
and
\begin{equation} \label{eq2:perturbed_einstein} %
	\delta G_{\mu\nu} = \frac{8\pi G}{c^4} \delta T_{\mu\nu},
\end{equation}
solving \eqref{eq:background_einstein} results in the Friedmann equations derived in Section~\ref{subsec:Friedmann_eqs}. In the following derivation, we solve \eqref{eq2:perturbed_einstein} to find analytic evolution equations for the metric, density, and velocity perturbations. We neglect any terms that are second order or higher. We use the resulting equations to generate initial conditions for the cosmological simulations in Chapters~\ref{Chapter3}, \ref{Chapter4}, and \ref{Chapter5}.

We use the Riemannian Geometry and Tensor Calculus (RGTC) package\footnote{Written by Sotirios Bonanos, see: \url{http://library.wolfram.com/infocenter/MathSource/4484/}} for \textsc{Mathematica} to calculate the components of the Einstein tensor, $G_{\mu\nu}$, for the metric \eqref{eq:metric_lin_perturb}. The time-time and time-space components are, to linear order,
\begin{align}
	G_{00} &= \frac{3\mH^2}{c^2} + \frac{2\,\partial^2\phi}{c^2} -  \frac{6 \mH \phi'}{c^4},  \label{eq:G00_linear}\\
	G_{0i} &= \frac{2}{c} \left( \mH \frac{\partial_i\psi}{c^2} + \frac{\partial_i \phi'}{c^2} \right),\label{eq:G0i_linear}
\end{align}
where $\partial^2\equiv\delta^{ij}\partial_i\partial_j$. The spatial components are
\begin{equation} \label{eq:Gij_linear}
	\begin{aligned}
		G_{ij} &= \left[ \frac{1}{c^2} \left( \mH^2 - \frac{2a''}{a} \right) \left(1 - \frac{2\psi}{c^2} - \frac{2\phi}{c^2} \right) + \frac{2 \mH}{c^2} \left(\frac{\psi'}{c^2} + \frac{2\phi'}{c^2} \right) \right]\delta_{ij} \\
		&+ \left( \frac{2\phi''}{c^4} + \frac{\partial^2\psi}{c^2} - \frac{\partial^2\phi}{c^2} \right) \delta_{ij} - \frac{\partial_i\partial_j\psi}{c^2} + \frac{\partial_i\partial_j\phi}{c^2}.
	\end{aligned}
\end{equation}
For dust, i.e. using \eqref{eq:Tmunu_perfect_fluid} with $P=0$, and zero shift, the time-time component of the stress-energy tensor is
\begin{equation}
	\begin{aligned}
		T_{00} &= \rho_R c^2 W^2\alp^2, \\
			&= \rho_R c^2 W^2 a^2 \left(1 + \frac{2\psi}{c^2} \right), 
	\end{aligned}
\end{equation}
where we have used \eqref{eq:u0down}. The time-space components are
\begin{equation}
	\begin{aligned}
	T_{0i} &= - \rho_R c^2 W^2 \frac{v_i}{c}  \alp, \\
		&= - \rho_R c^2 W^2 \frac{v_i}{c} a \left(1 + \frac{\psi}{c^2}\right),
	\end{aligned}
\end{equation}
where we have used \eqref{eq:uidown} and the linear approximation $\alp=a\sqrt{1+2\psi/c^2} \approx a(1 + \psi/c^2)$. The spatial components are
\begin{equation}
	T_{ij} = \rho_R W^2 v_i v_j.
\end{equation}
Expanding the density and velocity in terms of the background and the linear perturbations gives
\begin{align}
	\rho_R &= \brho_R( 1 + \delta ), \\
	v^i &= \bar{v}^i + \delta v^i,
\end{align}
where we have introduced the fractional density perturbation $\delta\equiv \delta\rho/\brho_R = (\rho_R - \brho_R)/\brho_R$. For FLRW, $\bar{v}^i=0$, so from here on we denote $v^i = \delta v^i$. The Lorentz factor is $W\approx 1$ to linear order, and with the above perturbations the components of the stress-energy tensor become
\begin{align}
	T_{00} &= \brho_R c^2 a^2 + \brho_R c^2 a^2 \left( \delta + \frac{2\psi}{c^2} \right), \label{eq:T00_linear}\\
	T_{0i} &= - \brho_R c^2 a \frac{v_i}{c},\label{eq:T0i_linear} \\
	T_{ij} &= 0. \label{eq:Tij_linear}
\end{align}

\subsubsection{Linearly perturbed equations}

The time-time component of \eqref{eq2:perturbed_einstein} is, using \eqref{eq:T00_linear} and \eqref{eq:G00_linear},
\begin{equation}
	\frac{2\,\partial^2\phi}{c^2} - \frac{6 \mH \phi'}{c^4} = \frac{8\pi G}{c^4} \left[ \brho_R c^2 a^2 \left( \delta + \frac{2\psi}{c^2} \right) \right],
\end{equation}
which gives
\begin{equation}
	\partial^2 \phi - 3 \mH \left( \frac{\phi'}{c^2} + \frac{\mH\psi}{c^2} \right) = 4\pi G \brho_R \delta a^2.
\end{equation}
The time-space components of \eqref{eq2:perturbed_einstein} are, using \eqref{eq:T0i_linear} and \eqref{eq:G0i_linear},
\begin{equation}
	\mH \partial_i\psi + \partial_i \phi'  = - 4\pi G \brho_R a v_i.
\end{equation}
Now considering the spatial components, we first take the trace of \eqref{eq:Gij_linear}
\begin{equation} \label{eq:trG_spatial}
	\begin{aligned}
		G^k_{\ph{k}k} &= \frac{3}{a^2 c^2}\left( \mH^2 - \frac{2a''}{a} \right) \left(1 - \frac{2\psi}{c^2} - \frac{2\phi}{c^2} \right) \\
		&+ \frac{6\mH}{a^2 c^2} \left( \frac{\psi'}{c^2} + \frac{2\phi'}{c^2} \right) + \frac{6\phi''}{a^2 c^4} + \frac{2}{a^2} \left(\frac{\partial^2\psi}{c^2} - \frac{\partial^2\phi}{c^2} \right),
	\end{aligned}
\end{equation}
then considering the perturbed part of this equation, with $T=g^{ij}T_{ij}=0$, we find
\begin{equation}
	\phi'' + \mH (\psi' + 2\phi') + \left(\frac{2a''}{a} - \mH^2 \right) \left(\phi + \psi\right) + \frac{c^2}{3} \partial^2 \left(\psi - \phi \right) = 0.
\end{equation}
Next, we consider the trace-free part of the spatial components by subtracting the spatial trace \eqref{eq:trG_spatial} from the full spatial components \eqref{eq:Gij_linear}, i.e. by defining
\begin{align}
	G_{ij}^{\rm TF} &\equiv G_{ij} - \frac{1}{3} g_{ij} G^k_{\ph{k}k} = G_{ij} - \frac{1}{3}a^2 \delta_{ij} G^k_{\ph{k}k}, \\
	T_{ij}^{\rm TF} &\equiv T_{ij} - \frac{1}{3} g_{ij} T = 0,
\end{align}
which hold for an FLRW background, and solving
\begin{equation}
	G_{ij}^{\rm TF} = \frac{8\pi G}{c^4} T_{ij}^{\rm TF}.
\end{equation}
This gives
\begin{align}
	G_{ij}^{\rm TF} = \left[ \left(\partial_i \partial_j - \frac{1}{3}\delta_{ij} \partial^2 \right) \left(\phi - \psi \right) \right] &= 0, \\
	\Rightarrow \partial_{\langle i} \partial_{j \rangle} \left(\phi - \psi \right) &= 0, \label{eq:phi_minus_psi_zero}
\end{align}
where $\partial_{\langle i} \partial_{j \rangle} \equiv \partial_i \partial_j - 1/3\,\delta_{ij} \partial^2$. Equation \eqref{eq:phi_minus_psi_zero} implies, in the linear regime, the temporal and spatial perturbations of the metric are equal, i.e. $\phi = \psi$. Our full system of equations therefore simplifies to
\begin{subequations} \label{eqs:perturb_phi_eq_psi}
	\begin{align}
		\partial^2 \phi - 3 \mH \left( \frac{\phi'}{c^2} + \frac{\mH\phi}{c^2} \right) &= 4\pi G \brho_R \delta a^2, \label{eq:perturb_phi_eq_psi_1} \\
		\mH \partial_i\phi + \partial_i \phi' &= - 4\pi G \brho_R a  v_i, \label{eq:perturb_phi_eq_psi_2} \\
		\phi'' + 3 \mH \phi' &= 0, \label{eq:perturb_phi_eq_psi_3}
	\end{align}
\end{subequations}
where we have used $2a''/a - \mH^2 = 0$, which can be shown using the Friedmann equations \eqref{eq:fried2} and \eqref{eq:fried1} with $P=0$.

\subsubsection{Linearly perturbed solutions}

For a flat, matter-dominated FLRW universe, the expansion rate follows \eqref{eq:Fried_dust_asoln}, giving the analytic form of $\mH$. This means we can solve \eqref{eq:perturb_phi_eq_psi_3} to arrive at an analytic expression for the metric perturbation,
\begin{equation} \label{eq:phi_solution_general}
	\phi(\xi) = f(x^i) - \frac{g(x^i)}{5 \, \xi^5},
\end{equation}
where $f,g$ are arbitrary, time-independent functions, and $\xi$ is the scaled conformal time defined in \eqref{eq:xidef}. We now substitute \eqref{eq:phi_solution_general} into the Hamiltonian constraint \eqref{eq:perturb_phi_eq_psi_1} to derive the analytic form for the fractional density perturbation, $\delta$. We find
\begin{equation} \label{eq:delta_fandg}
	\delta(\xi) = C_1 \xi^2 \partial^2 f - \frac{2}{c^2} f - \frac{C_1}{5} \xi^{-3} \partial^2 g - \frac{3}{5\,c^2} \xi^{-5} g,
\end{equation}
where we have used the Friedmann equation \eqref{eq:fried1_dust}, and the solutions \eqref{eq:Fried_dust_rhosoln} and \eqref{eq:Fried_dust_asoln}, for the background density, $\brho_R$, and scale factor, $a$, respectively. We also define $C_1\equiv a_{\rm init}/(4\pi G\rho^*)$, as in \citet{macpherson2017a}.

We now substitute \eqref{eq:phi_solution_general} into \eqref{eq:perturb_phi_eq_psi_2} to derive the analytic form of the velocity perturbation
\begin{equation} \label{eq:3vel_full}
	v^i = \frac{C_3}{a_{\rm init}} \xi^{-1} \partial^i f + \frac{3\, C_3}{10\, a_{\rm init}} \xi^{-6} \partial^i g,
\end{equation}
where $C_3\equiv-\sqrt{a_{\rm init} / (6\pi G \rho^*)}$. 

The system of equations \eqref{eq:phi_solution_general}, \eqref{eq:delta_fandg}, and \eqref{eq:3vel_full} describes the evolution of linear perturbations to the FLRW metric, so long as our initial assumptions about the magnitude of the perturbations themselves remain valid. 

The general evolution equation for the fractional density perturbation \eqref{eq:delta_fandg} contains both growing and decaying modes. Since we are interested in analysing the \textit{growth} of structure in the Universe, we choose $g=0$ and extract only the growing mode. The analytic solutions governing the evolution of the metric, density, and velocity perturbations in the linear regime then become
\begin{subequations} \label{eqs:growingmode_perturb_analytic}
	\begin{align}
		\phi &= f, \label{eq:growingmode_phi} \\
		\delta &= C_1 \xi^2 \partial^2 f - \frac{2}{c^2} f, \label{eq:growingmode_density} \\
		v^i &= \frac{C_3}{a_{\rm init}} \xi^{-1} \partial^i f. \label{eq:growingmode_velocity}
	\end{align}
\end{subequations}
We therefore have $\phi'=0$, i.e. the spatial distribution of the metric perturbation is constant in the linear regime. These analytic solutions provide the initial conditions in \flrwsolver, for different choices of $f$. 

The fluid three velocity with respect to the Eulerian observer \eqref{eq:growingmode_velocity}, defined in \hydrobase\, via \eqref{eq:threevel_def}, implies here (for zero shift)
\begin{equation}\label{eq:vi_dxdt_relation}
	\frac{dx^i}{d t} = \alp v^i.
\end{equation}
The velocity $v^i$ decays over time, as per \eqref{eq:growingmode_velocity}, whereas the coordinate velocity $dx^i/ dt\propto \xi$ in the linear regime, when $\alp\approx a$.

\subsubsection{Single-mode perturbation}

We can choose a simple form of $\phi$ to be a single-mode, sinusoidal perturbation with dimensionless amplitude $\phi_0 \ll 1$,
\begin{equation}
	\frac{\phi}{c^2} = \phi_0 \sum_{i=1}^3 {\rm sin} \left(\frac{2\pi x^i}{L} \right),
\end{equation}
where $L$ is the wavelength of the perturbation with dimension of length. With this form of the metric perturbation, the corresponding density and velocity perturbations (initially, i.e. $\xi=1$) are then
\begin{align}
	\delta &= - \left( \frac{4 \pi^2 c^2 C_1}{L^2} + 2 \right) \phi_0  \sum_{i=1}^3 {\rm sin} \left(\frac{2\pi x^i}{L} \right), \\
	\frac{v^i}{c} &= \frac{2\pi c \, C_3}{L \, a_{\rm init}} \,\phi_0\, {\rm cos} \left(\frac{2\pi x^i}{L} \right).
\end{align}
We evolve these initial perturbations in the ET in Chapter~\ref{Chapter3}, Section~\ref{sec:linear}, and compare the numerical growth to the analytic solutions \eqref{eqs:growingmode_perturb_analytic}. 

\subsubsection{Multi-mode perturbations}

We can analyse the growth of perturbations similar to those in our own Universe by drawing our initial conditions from the spectrum of perturbations in the CMB. The Code for Anisotropies in the Microwave Background \citep[\textsc{CAMB};][]{seljak1996,lewis2002}\footnote{\url{https://camb.info}} is a cosmology code primarily used for generating linear power spectra of the fluctuations in the CMB. \textsc{CAMB} is written in the synchronous (comoving) gauge, i.e. $\alp=1$, $\pdt\alp=0$ and $\beta^i=0$, which differs to the longitudinal gauge, as used in \flrwsolver. However, for the scales we currently sample, the matter power spectra in the synchronous and longitudinal gauges are almost identical, and so we expect negligible difference in the generation of initial conditions on $\lesssim 5-10$ Gpc scales.

To generate the CMB-like initial conditions for \flrwsolver\, we use parameters consistent with \citet{planck2016params}. We use the matter power spectrum output from \textsc{CAMB} as the spectrum of initial fluctuations in the density perturbation $\delta$. We invert \eqref{eq:growingmode_density} in Fourier space to solve for the corresponding metric perturbation $\phi$, from which we specify the velocity field $v^i$. The full method of generating the initial conditions is discussed in Chapter~\ref{Chapter4}, Section~\ref{subsec:cmbfluc}.

\subsection{Setup for this thesis}\label{subsec:our_code_setup}

For the work presented in this thesis, we use our thorn \flrwsolver\, to initialise several cosmological spacetimes, described in detail in Chapters \ref{Chapter3} and \ref{Chapter4}. Aside from this thorn, we use \mclachlan, specifically \mlbssn, to evolve the spacetime variables using the BSSN formalism, with shift vector $\beta^i=0$ and the generalised slicing condition \eqref{eq:bonamassolapse} with $f(\alp)=1/3$, i.e.,
\begin{equation}
	\pdo \alp = -\frac{1}{3}\alp^2 K,
\end{equation}
which for FLRW gives the conformal time parameterisation, i.e. $\alp=a$. For the evolution of the hydrodynamical variables we use \grhydro, with PPM reconstruction \citep{colella1984} and the HLLE Riemann solver \citep{harten1983,einfeldt1988}. In \texttt{EOS\_Omni} we use the ``Polytype'' equation of state, specifically a polytrope with pressure defined by
\begin{equation}
	P = K_{\rm poly} \rho_0^2,
\end{equation}
since \grhydro\, currently cannot handle $P=0$ (dust). We choose $K_{\rm poly}=0.1$, which sufficiently satisfies $P\ll\rho_0$. For this EOS, the internal energy is not evolved, and is instead set directly from the rest-mass density, i.e. $\eps=\rho_0$. We use periodic boundary conditions on a regular Cartesian mesh (i.e., no adaptive mesh refinement), and use Runge-Kutta fourth order (RK4) time integration in the thorn \mol, with the condition $\Delta t = 0.5 \Delta x$, where $\Delta t, \Delta x$ are the time step and grid spacing, respectively. We expect second-order convergence of our solutions, since the spatial order of \grhydro\, is second order. The full system of equations solved in our simulations is summarised in Table~\ref{tab:equations_used_summary}.

Using initial conditions describing a flat, dust, FLRW spacetime, and seperate initial conditions describing small perturbations to this spacetime (see Section~\ref{sec:cosmo_perturb}), in \flrwsolver\, and evolving with the above setup, we matched the analytic evolution for the homogeneous scale factor and density to within $10^{-6}$, and linear perturbations to the density, velocity, and metric to within $10^{-4}$. We tested the convergence of our errors with increasing resolution, including the Hamiltonian and momentum constraints, and saw the expected fourth-order convergence for FLRW (only time derivatives) and second-order convergence for the linear perturbations (both time and space derivatives). For more details of the computational tests performed for this setup, see Chapter~\ref{Chapter3} (and Chapter~\ref{Chapter4} for more complex perturbations).

\begin{table}
	\centering	
	\begin{tabular}{| l | c |}
      
      \hline
      \textbf{BSSN Equations} & \textbf{Eq. number} \\ \hline
      
      \parbox{4cm}{\begin{align*}
      	\dt \phi &= - \frac{1}{6} \alp K
	\end{align*}} & \eqref{eq:phi_evolution} \\ \hline

      \parbox{4cm}{\begin{align*} 
      	\dt K &= \alp \left( \cA_{ij} \cA^{ij} + \frac{1}{3} K^2 \right) - D^2 \alp + \frac{4\pi G}{c^4} \alp \left(S + \rho \,c^2\right)
	\end{align*}} & \eqref{eq:trK_evolution} \\ \hline

      \parbox{4cm}{\begin{align*}
      	\dt \cgam_{ij} &= -2\alp\cA_{ij}
	\end{align*}} & \eqref{eq:dtconfgamma} \\ \hline

      \parbox{4cm}{\begin{align*}
      \dt \cA_{ij} &= \iconf \left[ - (D_i D_j \alp)^{{\rm TF}} + \alp \left(\mR_{ij}^{\rm TF} - \frac{8\pi G}{c^4} S_{ij}^{\rm TF} \right) \right] \\
      	&+ \alp \left(K \cA_{ij} - 2 \cA_{ik} \cA^k_{\ph{k}j} \right)
	\end{align*}} & \eqref{eq:dtAij} \\ \hline
	
      \parbox{4cm}{\begin{align*}
       \mR_{ij}^{\rm TF} &= \cmR_{ij} - \frac{1}{3}\cgam_{ij}\cmR + \cmR_{ij}^\phi + \frac{8}{3} \cgam_{ij} e^{-\phi} \cD^2 e^\phi
       \end{align*}} & \eqref{eq:RijTF_confRijTF} \\ \hline

      \parbox{4cm}{\begin{align*}
       \cmR_{ij} &= -\frac{1}{2} \cgam^{lm} \partial_m \partial_l \cgam_{ij} + \cgam^{lm} \left( 2\cGam^k_{l(i}\cGam_{j)km} + \cGam^k_{im} \cGam_{klj} \right) \\
       		&+ \cgam_{k(i} \partial_{j)} \cGam^k + \cGam^k \cGam_{(ij)k} 
       \end{align*}} & \eqref{eq:cRij_Gami} \\ \hline

      \parbox{4cm}{\begin{align*}
      \dt \cGam^i &= -2 \cA^{ij} \partial_j \alp + 2\alp\left( \cGam^i_{jk} \cA^{kj} - \frac{2}{3} \cgam^{ij} \partial_j K\right) + \frac{2}{3} \cGam^i \partial_j \beta^j \\
      	& - 2\alp\left(\frac{8\pi G}{c^3} \cgam^{ij} S_j + 6 \cA^{ij} \partial_j \phi \right) + \frac{1}{3} \cgam^{li} \partial_l \partial_j \beta^j + \cgam^{lj} \partial_j \partial_l \beta^i
	\end{align*}} & \eqref{eq:cGami_evol} \\ \hline

	\textbf{Gauge conditions} & \textbf{Eq. number} \\ \hline
	
	\parbox{4cm}{\begin{align*}
		\beta^i = 0
		\end{align*}} & N/A \\ \hline
	
	\parbox{4cm}{\begin{align*}
		\pdo \alp = \frac{1}{3}\alp^2 K
		\end{align*}} & \eqref{eq:bonamassolapse} \\ \hline

	\textbf{Equations of Hydrodynamics} & \textbf{Eq. number} \\ \hline
	
	\parbox{4cm}{\begin{align*}
		\pdo D + \partial_i \left(\alp D \tildv^i\right) = 0
		\end{align*}} & \eqref{eq:Dcons}, \eqref{eqs:GRHydro_flux_source} \\ \hline

	\parbox{4cm}{\begin{align*}
		\pdo S^c_j &+ \partial_i \left[\alp \left( S^c_j \, \tildv^i + \delta^i_j P \right) \right] = T^{\mu\nu} \left( \partial_\mu g_{\nu j} - {^{(4)}}{\Gam^\lambda_{\mu\nu}} g_{\lambda j} \right)
		\end{align*}} & \eqref{eq:Sjcons}, \eqref{eqs:GRHydro_flux_source} \\ \hline
	
	\parbox{4cm}{\begin{align*}
		\pdo \tau^c &+ \partial_i \left[\alp \left( \tau^c \tildv^i + P\tildv^i \right) \right] =\alp \left( T^{\mu 0} \partial_\mu {\rm ln}\alp - T^{\mu\nu} {^{(4)}}{\Gam^0_{\mu\nu}} \right)
		\end{align*}} & \eqref{eq:taucons}, \eqref{eqs:GRHydro_flux_source} \\ \hline
	
	\end{tabular}
	
	\caption{System of equations solved in our cosmological simulations using the \textsc{Einstein Toolkit}.}
	\label{tab:equations_used_summary}

\end{table}

\section{Post-processing analysis: \mesc}\label{sec:mescaline}

The ET has some built in thorns specifically for analysis, which calculate quantities such as the trace of the extrinsic curvature, the Ricci scalar, the determinant of the spatial metric, the metric and extrinsic curvature in different coordinates, while some thorns include routines for locating black hole horizons and calculating constraint violation \citep{loffler2012,zilhao2013}. While these thorns are useful for generating output of physically interesting quantities while the simulation is running, to analyse General-Relativistic effects in inhomogeneous cosmology, there are many more quantities that we are interested in. \textsc{Mescaline} is a post-processing analysis code to read in three-dimensional \textsc{Cactus} data in HDF5 format and calculate quantities such as the spatial Ricci tensor and its trace, the trace of the extrinsic curvature, the expansion rate and shear of the fluid, and spatially-averaged quantities over both the entire volume and in sub-domains within the volume. In \mesc\, we adopt geometric units, $G=c=1$, and so for this section we present all equations in these units.

\subsection{Key calculations}

\subsubsection{Ricci tensor and connection functions} \label{subsubsec:Ricci_Gams}

The spatial Ricci tensor is the contraction of the spatial Riemann curvature tensor of the spatial surfaces, \eqref{eq:3Riemdef},
\begin{equation}
	\mR_{ij} = \mR^k_{\ph{k}ikj} = \partial_k \Gamma^{k}_{ij} - \partial_j\Gamma^{k}_{ik} + \Gamma^{k}_{lk}\Gamma^{l}_{ij} - \Gamma^{k}_{jl}\Gamma^{l}_{ik}.
\end{equation}
In \mesc\, we calculate $\mR_{ij}$ directly from ET output using the spatial metric and its spatial derivatives --- via the spatial connection functions --- with either a second-order or fourth-order approximation of the derivative. The Ricci scalar is then the trace of the Ricci tensor, $\mR \equiv \gam^{ij}\mR_{ij}$. We calculate the trace of the extrinsic curvature $K\equiv\gam^{ij}K_{ij}$ using direct output of $K_{ij}$ from the ET. When calculating the trace, we assume the rank-2, covariant tensor in question is symmetric in its indices; true for all cases in \mesc. 

It is useful to write the time components of the four-dimensional connection functions in terms of purely spatial objects, some of which we use to calculate the expansion scalar $\theta$. These are
\begin{subequations} \label{eqs:4Gam_time_components}
	\begin{align}
		^{(4)}{\Gam}^0_{00} &= - \frac{1}{3} \alp K, \quad ^{(4)}{\Gam}^0_{0i} = \frac{1}{\alp} \partial_i \alp, \quad ^{(4)}{\Gam}^0_{ij} = -\frac{1}{\alp} K_{ij}, \label{eq:4Gam_time_line1} \\
		^{(4)}{\Gam}^i_{00} &= \alp \gam^{ij} \partial_j \alp, \quad ^{(4)}{\Gam}^i_{0k} = - \gam^{ij} \alp K_{kj}, \label{eq:4Gam_time_line2}
	\end{align}
\end{subequations}
and the spatial components of the four-dimensional connection functions are here equal to the spatial connection functions, since $g_{ij}=\gamma_{ij}$ (always), and $g^{0i}=0$, $g_{0i}=0$, and $g^{ij}=\gam^{ij}$ for $\beta^i=0$.

\subsubsection{Constraint violation}

We calculate the violation of the Hamiltonian and momentum constraint equations via \eqref{eq:ham_constraint} and \eqref{eq:mom_constraint}, respectively. The violation in the Hamiltonian constraint is
\begin{equation}
	H \equiv \mR + K^2 - K_{ij}K^{ij} -16\pi \rho,
\end{equation}
where $\rho$ is the total mass-energy density projected into the normal frame, which we can relate to the rest-mass density (as output from the ET, and therefore read into \mesc) via \eqref{eq:rho_rhoR_relation}, which for dust gives,
\begin{equation}
	\rho = \rho_R W^2 = \rho_0 \left(1 + \eps \right) W^2,
\end{equation}
where we have used \eqref{eq:rhoR_rho0_relation}. For our chosen EOS we have $\eps=\rho_0$ (see Section~\ref{subsec:our_code_setup}), and so
\begin{align}
	\rho &= \rho_0 W^2 + \rho_0^2 W^2, \\
		&\approx \rho_0 W^2,
\end{align}
since (in code units) we set $\rho_0 \approx 6\times10^{-9}$, and so $\rho_0^2 \approx 10^{-17}$ (see Chapter~\ref{Chapter3}).

The violation in the momentum constraint is
\begin{equation}
	M_i \equiv D_j K^j_{\ph{j}i} - D_i K - 8\pi S_i,
\end{equation}
where the momentum density is defined in \eqref{eq:Si_def}, which we can write as
\begin{equation}
	S_i = -\gam_{ij} n_0 T^{0j},
\end{equation}
since $n_i=0$ and $\gam_{i0}=0$. Using \eqref{eq:Tmunu_perfect_fluid} with $P=0$, and four velocity \eqref{eq:u0up} and \eqref{eq:uiup}, we find
\begin{equation}
	S_i = \gam_{ij} v^j W^2 \rho_R,
\end{equation}
where $\rho_R$ is the projection of the stress-energy tensor into the rest frame of the fluid \eqref{eq:rhoR_def}, and is related to the rest-mass density via \eqref{eq:rhoR_rho0_relation}. The magnitude of the momentum constraint violation is $M^2\equiv\gam^{ij}M_i M_j$, and for both violations we calculate the $L_1$ error at each time,
\begin{align}
	L_1(H) &= \frac{1}{N} \sum_{a=1}^N | H_a |, \\
	L_1(M) &= \frac{1}{N} \sum_{a=1}^N | M_a |,
\end{align}
where $N$ is the total number of grid cells, and $H_a, M_a$ are the Hamiltonian violation and magnitude of the momentum violation at grid cell $a$, respectively. The above $L_1$ errors quantify the raw constraint violation in each case, but to calculate the relative violation we define the ``energy scales'' as the sum of the squares of the individual terms in each violation \citep[as in][]{mertens2016}, i.e.
\begin{align}
	[H] &\equiv \sqrt{ \mR^2 + \left( K^2 \right)^2 + \left( K_{ij}K^{ij} \right)^2 + \left( 16\pi\rho \right)^2 }, \\
	[M] &\equiv \sqrt{ \left( D_j K^j_{\ph{j}i} \right) \left( D_j K^{ji}_{\ph{i}} \right) + D_i \left( K \right) D^i \left( K \right) + \left( 8 \pi \right)^2 S_i S^i },
\end{align}
and calculate the relative $L_1$ violations as
\begin{equation}
	L_1\left( H/[H] \right) = \frac{ \frac{1}{N} \sum_{a=1}^N | H_a |}{ \frac{1}{N} \sum_{a=1}^N [H]_a },
\end{equation}
and
\begin{equation}
	L_1\left( M/[M] \right) = \frac{ \frac{1}{N} \sum_{a=1}^N | M_a |}{ \frac{1}{N} \sum_{a=1}^N [M]_a }.
\end{equation}

\subsubsection{Expansion scalar}

We implement the generalised averaging scheme described in Section~\ref{subsubsec:general_foliation} \citep{larena2009,umeh2011}. To calculate the backreaction terms $\QD^h$ \eqref{eq:QD_def_larena} and $\LD$ \eqref{eq:LD_def_larena}, and hence the cosmological parameters, we first must calculate the expansion scalar, $\theta$, the shear tensor $\sigma_{\mu\nu}$, as well as the additional scalars and tensors related to the divergence of the peculiar velocity field $v^i$, at every coordinate point in the domain.

The expansion scalar is defined as the divergence of the fluid four velocity projected into the surface defined by $h_{\mu\nu}$,
\begin{align}
	\theta &\equiv h^{\mu\nu} \nabla_\mu u_\nu, \\
		&= h^{ij} \nabla_i u_j,
\end{align}
since $h_{\mu\nu}$ is purely spatial and so $h^{00}=0$ and $h^{0i}=0$. Expanding the covariant derivative gives
\begin{align}
	\theta &= h^{ij} \left( \partial_i u_j - ^{(4)}{\Gam}^\alp_{ij} u_\alp \right) \\
		&= h^{ij} \left( \partial_i u_j - ^{(4)}{\Gam}^0_{ij} u_0 - \Gam^k_{ij} u_k \right)
\end{align}
since $\nabla_i$ is the spatial component of the covariant derivative associated with the metric $g_{\mu\nu}$, the connection functions involved are still the four-dimensional connection functions, and hence we have a term involving the time component $^{(4)}{\Gam}^0_{ij}$, which we substitute from \eqref{eq:4Gam_time_line1}. The expansion scalar is then
\begin{equation}
	\theta = h^{ij} \partial_i u_j - W K - h^{ij} \Gam^k_{ij} u_k,
\end{equation}
where we have also used \eqref{eq:u0down}.

\subsubsection{Shear tensor}

The definition of the shear tensor is
\begin{align}
	\sigma_{\mu\nu} &\equiv h^{\alp}_{\ph{\alp}\mu} h^{\beta}_{\ph{\beta}\nu} \nabla_{(\alp} u_{\beta)} - \frac{1}{3} \theta h_{\mu\nu}, \\
	\Rightarrow \sigma_{ij} &= \nabla_{(i} u_{j)} - \frac{1}{3} \theta h_{ij},
\end{align}
i.e., it is purely spatial. Expanding the covariant derivative, and again using \eqref{eq:4Gam_time_line1}, we find
\begin{equation}
	\sigma_{ij} = \partial_{(i} u_{j)} - W K_{ij} - \Gam^k_{ij} u_k - \frac{1}{3} \theta h_{ij},
\end{equation}
where we have again substituted $^{(4)}{\Gam}^0_{ij}$ from \eqref{eq:4Gam_time_line1}. We then calculate the rate of shear as $\sigma^2\equiv \frac{1}{2} \sigma_{ij}\sigma^{ij}$.

\subsubsection{Other tensors}

To be able to form the cosmological parameters we also need to calculate the scalars $\sigma_B^2$ \eqref{eq:sigmaB2} and $\theta_B$ \eqref{eq:thetaB}. The scalar $\theta_B$ is built from the divergence of the peculiar velocity, $\kappa$ \eqref{eq:kappa_larena}, and the trace of the tensor $B_{ij}$ \eqref{eq:Bmunu_larena}, which itself depends on the tensor $\beta_{ij}$ \eqref{eq:betaij_larena}. From their definitions, we can write these quantities as
\begin{align}
	\kappa &= h^{ij} \partial_i v_j - h^{ij} \Gam^k_{ij} v_k, \\
	\beta_{ij} &= \partial_{(i} v_{j)} - \Gam^k_{ij} v_k - \frac{1}{3}\kappa h_{ij}, \\
	B &= \frac{1}{3} \kappa v_i v^i + \beta_{ij} v^i v^j,
\end{align}
where we have used the fact the peculiar velocity is purely spatial, i.e. $v_0=0$. The scalar $\theta_B$ is then
\begin{equation}
	\theta_B = -W \kappa - W^3 B.
\end{equation}

To calculate $\sigma_B^2$ we first must calculate the tensor $\sigma_{Bij}$ using \eqref{eq:sigmaBij}, which we can write as
\begin{equation}
	\sigma_{Bij} = - W \beta_{ij} - W^3 \left( B_{(ij)} - \frac{1}{3} B h_{ij} \right)
\end{equation}
where we have
\begin{equation}
	B_{ij} = \frac{1}{3}\kappa v_i v_j + \beta_{ki} v^k v_j + \partial_{(k} v_{i)} v^k v_j.
\end{equation}

\subsubsection{Effective scale factors}

We calculate the volume scale factor $\aD^V$ using \eqref{eq:aDV_larena} and the definition of the volume element \eqref{eq:VDh_def}. To find the scale factor describing the expansion of the fluid, $\aDh$, we follow \citet{larena2009b} and use the expressions for the Hubble parameter \eqref{eq:aDh_def} and \eqref{eq:HDh_def}, along with the evolution of the volume element \eqref{eq:aDvdot_on_aDV}, to get
\begin{equation} 
	\frac{\pdt \aD^V}{\aD^V} - \frac{\pdt \aDh}{\aDh} = \frac{1}{3} \avgh{A},
\end{equation}
where $A\equiv \frac{\alp}{W} \left(\theta-\kappa\right) - \alp\theta$ for simplicity\footnote{We note an error in \citet{larena2009b} in this relation, which is corrected with $-\kappa\rightarrow\theta_B$ in $A$. We have not corrected it here since we use this form in our analysis, however, see Appendix~\ref{sec:appx_aD_typo} for a re-analysis with the error fixed.}. This implies
\begin{align}
	\pdt {\rm ln} \left( \frac{\aD^V}{\aDh} \right) = \frac{1}{3} \avgh{A}.
\end{align}
Using a second-order approximation for the time derivative, we have
\begin{equation}
	\frac{ {\rm ln} \left. \frac{\aD^V}{\aDh} \right\rvert_{n} - {\rm ln} \left. \frac{\aD^V}{\aDh} \right\rvert_{n-2}}{2\Delta t} \approx \frac{1}{3} \left. \avgh{A}\right\rvert_{n-1}
\end{equation}
where $n$, $n-1$, and $n-2$ represent times $t$, $t-\Delta t$, and $t-2\Delta t$, respectively. We therefore calculate the fluid scale factor from the volume scale factor using
\begin{equation}
	\left. \frac{\aDh}{\aD^V} \right\rvert_n \approx {\rm exp} \left( - \frac{2 \Delta t}{3} \left. \avgh{A}\right\rvert_{n-1} \right) \left. \frac{\aDh}{\aD^V} \right\rvert_{n-2}.
\end{equation}

\subsection{Averaging}\label{subsec:mesc_averaging}

As discussed previously, we calculate the expansion scalar and shear tensor at every cell in the computational domain. However, to investigate the effect of inhomogeneities on the global evolution of the expansion rate and cosmological parameters, we must calculate averages of these quantities. The averaging domain $\mD$ is entirely arbitrary, and choosing it depends on the physical problem we are interested in. See Chapter~\ref{Chapter4} to see the results of averaging an inhomogeneous cosmological simulation with numerical relativity using \mesc.

\subsubsection{Global averages}

The simplest choice of the averaging domain is the entire computational grid. This is useful to study the large-scale (global) effects of backreaction, and to look into the potential for these effects to explain the accelerating expansion of the Universe; i.e. to explain dark energy (see Section~\ref{subsec:backreaction}). Global averages, however, can be susceptible to boundary problems since the boundary is included in the averaging process. It has not yet been studied how the use of periodic boundary conditions affects the size of the backreaction effect measured globally in full GR simulations. In Newtonian simulations with periodic boundary conditions, the global backreaction vanishes identically \citep[see][]{buchertehlers1997,buchert2018a}. We discuss this more in Chapter~\ref{Chapter6}.

\subsubsection{Subdomain averaging}

Rather than studying the global dynamics, we can analyse the average properties on smaller scales using subdomains located within the computational domain. This will largely remove any spurious boundary effects, simply by not including the boundary in the averaging. In \mesc\, we perform this subdomain averaging using an arbitrary number, $N$, of randomly placed spherical domains of an arbitrary radius, $r_\mD$, within the global domain. The number of spheres and each sphere's radius (where all $N$ spheres are given the same radius) are specified by the user before compiling. From these, we randomly generate $N$ sets of $x^i=(x,y,z)$ coordinates lying within the computational grid, representing the origins of each individual sphere. These origins are generated such that the edge of the outermost sphere is not allowed to exit the computational domain, and the spheres are allowed to overlap with one another.

An issue with the averaging formalism used here, briefly discussed in Section~\ref{subsubsec:improved_general_averaging}, is that the domain $\mD$ does not conserve mass during evolution. That is, if we choose to calculate averages within subdomains as a function of time, the mass contained within these domains is free to move into and out of the sphere itself. This issue is present in the averaging in \mesc\, because the coordinate positions of the sphere's origins stay fixed during the evolution, and are not propagated along with the fluid flow. In order to address this, not only the origin of the sphere but the edges of the sphere also need to be propagated along the fluid four-velocity vector, i.e., we would need to allow the sphere to deform as the fluid evolves. The computational overhead of this propagation is outside the scope of this thesis, and so long as the velocities in the simulation are $v^i\ll c$ (as is the case here), the approximation of stationary origins is sufficiently valid.

\subsection{Assumptions}

In writing \mesc\, we have adopted a few key assumptions to simplify the code, namely:
\begin{enumerate}
	\item Regular Cartesian grid,
	\item Zero shift vector,
	\item Periodic boundary conditions,
	\item Matter-dominated fluid, i.e. $P=0$,
	\item Geometric units, i.e. $G=c=1$,
\end{enumerate}
each of which are common choices in cosmological simulations with numerical relativity \citep[see e.g.][]{bentivegna2016a,giblin2016a,macpherson2017a}. The assumption of a regular Cartesian grid is implemented when defining the grid spacing, specifically for computing spatial derivatives using a finite difference approximation, i.e. using (for fourth-order derivatives)
\begin{equation}
	\frac{\partial f}{\partial x^i} \approx \frac{-f(x^i+2\Delta x^i) + 8 f(x^i+\Delta x^i) - 8  f(x^i-\Delta x^i) +  f(x^i-2\Delta x^i)}{12 \Delta x^i}
\end{equation}
where we set $\Delta x^i = \Delta x = \Delta y = \Delta z$. 

Assuming a zero shift vector, i.e. $\beta^i=0$, simplifies the expressions for the time components of the four-dimensional connection functions, and the time derivative \eqref{eq:dt_def}, since for $\beta^i=0$ we have $\mL_\beta=0$ and hence $\dt = \pdt$ (both used in calculating the backreaction terms). It also simplifies \eqref{eq:uiup} relating the four velocity and the Eulerian three velocity, the latter of which is output from the ET. In addition, adopting a zero shift affects the way we raise and lower indices of four-dimensional objects, i.e. for the spatial components of the four velocity,
\begin{align}
	u_i &= g_{i\mu} u^\mu, \\
	&= g_{i0} u^0 + g_{ij} u^j, \\
	&= g_{ij} u^j.
\end{align}

Periodic boundary conditions are implemented in our approximation of spatial derivatives only, and are the most reasonable choice for cosmological simulations without needing to simulate the entire past light cone of an observer. 

In \mesc\, we assume the matter content is dust ($P=0$) in relating densities $\rho$ and $\rho_0$ measured in the normal frame and fluid rest-frame, respectively, and in calculating the momentum density, $S_i$, for the momentum constraint violation. However, since \grhydro\, is not equipped for $P=0$, we instead use $P\ll\rho$ in the simulations themselves (see Section~\ref{subsec:our_code_setup}), which we found to be sufficient to match the dust evolution for FLRW and small perturbations to this background (see Chapter~\ref{Chapter3}).

\chapter{Inhomogeneous cosmology with numerical relativity} %

\label{Chapter3} %

\vspace{10mm}

Published in:\\
\citet{macpherson2017a}. Physical Review D, \textbf{95.6, 064028}.

\section*{Abstract}
We perform three-dimensional numerical relativity simulations of homogeneous and inhomogeneous expanding spacetimes, with a view towards quantifying non-linear effects from cosmological inhomogeneities. We demonstrate fourth-order convergence with errors less than one part in $10^{6}$ in evolving a flat, dust Friedmann-Lema\^itre-Roberston-Walker (FLRW) spacetime using the \textsc{Einstein Toolkit} within the \textsc{Cactus} framework. We also demonstrate agreement to within one part in $10^{3}$ between the numerical relativity solution and the linear solution for density, velocity and metric perturbations in the Hubble flow over a factor of $\sim350$ change in scale factor (redshift). We simulate the growth of linear perturbations into the non-linear regime, where effects such as gravitational slip and tensor perturbations appear. We therefore show that numerical relativity is a viable tool for investigating nonlinear effects in cosmology.

\section*{A note on notation}
We have altered the notation throughout this chapter, including Appendix~\ref{appx:newt_gauge}, to be consistent with Chapters~\ref{Chapter1} and \ref{Chapter2}, unless explicitly stated otherwise. For these exceptions, we maintain the notation of the publication for consistency with figures in their published form. Aside from these changes, this chapter is consistent with the accepted version of \citet{macpherson2017a}.

\pagebreak

\section{\label{sec:intro}Introduction}
 Modern cosmology relies on the cosmological principle --- that the Universe is sufficiently homogeneous and isotropic on large scales to be described by a Friedmann-Lema\^itre-Robertson-Walker (FLRW) model. Cosmological N-body simulations \citep[e.g.][]{genel2014,springel2005,kim2011} encode these assumptions by prescribing the expansion to be that of the FLRW model, governed by the Friedmann equations, while employing a Newtonian approximation for gravity.
 
The transition to cosmic homogeneity begins on scales $\sim 80 h^{-1}$ Mpc \cite[e.g.][]{yadav2010,scrimgeour2012}, but is inhomogeneous and anisotropic on smaller scales.
Upcoming cosmological surveys utilising Euclid, the Square Kilometre Array (SKA) and the Large Synoptic Survey Telescope (LSST)  \citep{amendola2016,maartens2015,ivezic2008} will reach a precision at which nonlinear General-Relativistic effects from these inhomogeneities could be important. A more extreme hypothesis \citep{rasanen2004,kolb2005,kolb2006, notari2006, rasanen2006a,rasanen2006b,li2007,li2008, larena2009,buchert2015,green2016,bolejkolasky2008} is that such inhomogeneities may provide an alternative explanation for the accelerating expansion of the Universe, via \emph{backreaction} \citep[see][for a review]{buchert2008,buchert2012}, replacing the role assigned to dark energy in the standard $\Lambda$CDM model \citep{riess1998,perlmutter1999,parkinson2012,samushia2013}. 

Quantifying the General-Relativistic effects associated with nonlinear structures ultimately requires solving Einstein's equations. Post-Newtonian approximations are a worthwhile approach \citep{matarrese1996,rasanen2010,green2011,green2012,adamek2013,adamek2016a,adamek2016b,sanghai2015,oliynyk2014,noh2004}, however the validity of these must be checked against a more precise solution since the density perturbations themselves are highly nonlinear.

 An alternative approach is to use numerical relativity, which has enjoyed tremendous success over the past decade \citep{pretorius2005,campanelli2006,baker2006}.

Cosmological modelling with numerical relativity began with evolutions of planar and spherically symmetric spacetimes using the Arnowitt-Deser-Misner (ADM) formalism \citep{arnowitt1959}, including Kasner and matter-filled spacetimes \citep{centrella1979}, the propagation and collision of gravitational wave perturbations \citep{centrella1980, centrella1982} and linearised perturbations to a homogeneous spacetime \citep{centrella1983,centrella1984}. More recent work has continued to include symmetries to simplify the numerical calculations \citep[e.g.][]{rekier2015,torres2014}.

Simulations free of these symmetries have only emerged within the last year. \citet{giblin2016a} studied the evolution of small perturbations to an FLRW spacetime, exploring observational implications in \citep{giblin2016b}. \citet{bentivegna2016a} showed differential expansion in an inhomogeneous universe, and quantified the backreaction parameter from \citep{buchert2000a} for a single mode perturbation. These works all indicate that the effects of nonlinear inhomogeneities may be significant.

In this work, we perform a feasibility study of numerical solutions to the full Einstein equations for inhomogeneous cosmologies by simulating the growth of structure in a model three-dimensional universe and comparing to known analytic solutions. Our approach is similar to \citet{giblin2016a,bentivegna2016a,giblin2016b}, with differences in the generation of initial conditions and numerical methods. We use the freely-available \textsc{Einstein Toolkit}, based on the \textsc{Cactus} infrastructure \citep{loffler2012,zilhao2013}. We benchmark our three-dimensional numerical implementation on two analytic solutions of Einstein's equations relevant to cosmology: FLRW spacetime and the growth of linear perturbations. We also present the growth of perturbations into the nonlinear regime, and analyse the resulting gravitational slip \citep{daniel2008,daniel2009} and tensor perturbations. 

In Section \ref{sec:method_chp3} we describe our numerical methods, including choices of gauge (\ref{sec:gauge_chp3}) and an overview of the derivations of the linearly perturbed Einstein equations used for our initial conditions (\ref{sec:initial_conditions}). In Section \ref{sec:FLRW} we describe the setup (\ref{sec:FLRW_setup}) and results (\ref{sec:FLRW_results}) of our evolutions of a flat, dust FLRW universe. The derivation of initial conditions for linear perturbations to the FLRW model are described in \ref{sec:linear_setup}, with results presented in \ref{sec:linear_results}. The growth of the perturbations to nonlinear amplitude is presented in \ref{sec:nonlinear}, with analysis of results and higher order effects in \ref{sec:nonlinear_results} and \ref{sec:nonlinear_highorder} respectively. In this chapter, we adopt geometric units with $G=c=1$, Greek indices run from 0 to 3 while Latin indices run from 1 to 3, with repeated indices implying summation.

\section{\label{sec:method_chp3}Numerical Method}
We integrate Einstein's equations with the \textsc{Einstein Toolkit}, a free, open-source code for numerical relativity \citep{loffler2012}. This utilises the \textsc{Cactus} infrastructure, consisting of a central core, or ``flesh'', with application modules called ``thorns'' that communicate with this flesh \citep{cactus}. The \textsc{Einstein Toolkit} is a collection of thorns for computational relativity, used extensively for simulations of binary neutron star and black hole mergers \citep[e.g][]{kastaun2015,radice2015,baiotti2005}. Numerical cosmology with the \textsc{Einstein Toolkit} is a new field \citep{bentivegna2016a}. We use the \texttt{McLachlan} code \citep{brownD2009} to evolve spacetime using the Baumgarte-Shapiro-Shibata-Nakamura (BSSN) formalism \citep{shibata1995, baumgarte1999}, and the \texttt{GRHydro} code to evolve the hydrodynamical system \citep{baiotti2005, giacomazzo2007, mosta2014}; a new setup for cosmology with the \textsc{Einstein Toolkit}.

We use the fourth-order Runge-Kutta method, adopt the Marquina Riemann solver and use the piecewise parabolic method for reconstruction on cell interfaces. \texttt{GRHydro} is globally second order in space due to the coupling of hydrodynamics to the spacetime \citep{hawke2005, mosta2014}. We therefore expect fourth-order convergence of our numerical solutions for the spatially homogeneous FLRW model. Once perturbations are introduced to this model we expect our solutions to be second-order accurate. 

We have developed a new thorn, \texttt{FLRWSolver}, to initialise an FLRW cosmological setup with optional linear perturbations. We evolve our simulations in a cubic domain on a uniform grid with periodic boundary conditions with $x^{i}$ in [-240,240]. Our domain sizes are $20^{3}, 40^{3}$ and $80^{3}$, respectively using 70 (8 cores), 380 (8 cores) and 790 (16 cores) CPU hours.

\subsection{\label{sec:gauge_chp3}Gauge}
The gauge choice corresponds to a choice of the lapse function, $\alpha$, and shift vector, $\beta^{i}$. The metric written in the $(3+1)$ formalism is
\begin{equation}\label{eq:3p1_metric}
	ds^{2} = -\alpha^{2}dt^{2} + \gamma_{ij}(dx^{i} + \beta^{i}dt)(dx^{j} + \beta^{j}dt), 
\end{equation}
where $\gamma_{ij}$ is the spatial metric. Previous cosmological simulations with numerical relativity adopt the synchronous gauge, corresponding to $\alpha=1, \;\beta^{i}=0$ \citep{giblin2016a, bentivegna2016a}. We instead utilise the general spacetime foliation of \citet{bona1995},
\begin{equation}
	\partial_{t}\alpha = -\alpha^{2} \, f(\alpha) \, K, \label{eq:harmonic_gauge}
\end{equation}
where $f(\alpha)>0$ is an arbitrary function, and $K=\gamma^{ij}K_{ij}$. We set the shift vector $\beta^{i}=0$.
Harmonic slicing uses $f$ = const., while $f=1/\alpha$ corresponds to the ``1+log'' slicing common in black hole binary simulations. We choose $f=0.25$ to maintain the stability of our evolutions, as in \citet{torres2014}. This allows for longer evolutions for the same computational time, compared to ``1+log'' slicing, due to the increased rate of change of the lapse. We adopt this gauge for numerical convenience, and acknowledge possible alternative methods include using synchronous gauge with adaptive time-stepping.

We use \eqref{eq:harmonic_gauge} for evolution only. We scale to the gauge described in the next section for analysis.

\subsection{\label{sec:initial_conditions}Perturbative Initial Conditions}
Bardeen's formalism of cosmological perturbations \citep{bardeen1980} was developed with the intention to connect metric perturbations to physical perturbations in the Universe. This connection is made clear by defining the perturbations as gauge-invariant quantities in the longitudinal gauge.

The general line element of a perturbed, flat FLRW universe, including scalar ($\Phi, \Psi$), vector ($B_{i}$) and tensor ($h_{ij}$) perturbations takes the form
\begin{equation}\label{eq:perturbed_metric_svt}
	\begin{aligned}
		ds^{2} = a^{2}(\eta)[-(1 &+ 2\Psi)d\eta^{2} - 2B_{i}dx^{i}d\eta \\
		&+ (1-2\Phi)\delta_{ij} dx^{i}\,dx^{j}+ h_{ij}dx^{i}\,dx^{j}], 
	\end{aligned}
\end{equation}
where $\eta$ is conformal time, $a(\eta)$ is the FLRW scale factor and $\delta_{ij}$ is the identity matrix. 
We derive initial conditions from the linearly perturbed Einstein equations, implying negligible vector and tensor perturbations \citep{adamek2013}. This is valid as long as our simulations begin at sufficiently high redshift that the Universe may be approximated by an FLRW model with small perturbations. 
Considering only scalar perturbations the metric becomes
\begin{equation}
	ds^{2} = a^{2}(\eta)[-(1+2\Psi)d\eta^{2} + (1-2\Phi)\delta_{ij}dx^{i}dx^{j}], \label{eq:perturbed_metric}
\end{equation}
where $\Phi$ and $\Psi$ here coincide with Bardeen's gauge-invariant scalar potentials\footnote{These perturbations are equivalent to $\phi$ and $\psi$ used in Chapter~\ref{Chapter1} and Section~\ref{subsec:FLRWSolver}.} \citep{bardeen1980}. Here we see that $\Psi$, the Newtonian potential, will largely influence the motion of non-relativistic particles; where the time-time component of the metric dominates the motion. The Newtonian potential plays the dominant role in galaxy clustering. Relativistic particles will \emph{also} be affected by the curvature potential $\Phi$, and so both potentials influence effects such as gravitational lensing \citep{bertschinger2011, bardeen1980}.

The metric perturbations are coupled to perturbations in the matter distribution via the stress-energy tensor. We approximate the homogeneous and isotropic background as a perfect fluid in thermodynamic equilibrium, giving
\begin{equation}
	T_{\mu\nu} = \left(\rho + P\right)u_{\mu}u_{\nu} + P\,g_{\mu\nu}, \label{eq:stress_energy_full}
\end{equation}
where $\rho$ is the total energy density\footnote{The energy-density $\rho$ used throughout this Chapter is the total \emph{rest-frame} energy-density, and is equivalent to $\rho_R$ used in Chapters~\ref{Chapter1} and \ref{Chapter2}.}, $P$ is the pressure and $u^{\mu}$ is the four-velocity of the fluid. 
We assume a dust universe, implying negligible pressure ($P\ll\rho$), and we solve the perturbed Einstein equations,
\begin{equation}
	\delta G_{\mu\nu} = 8\pi \, \delta T_{\mu\nu}, \label{eq:perturbed_einstein_chp3}
\end{equation}
using linear perturbation theory. From the time-time, time-space, trace and trace free components of \eqref{eq:perturbed_einstein_chp3}, we obtain the following system of equations \citep{sachs1967,adamek2013}

\begin{subequations} \label{eqs:perturbed_einstein_chp3}
	\begin{align}
		\partial^2\Phi - 3\mH\left(\Phi' + \mH \Psi\right) &= 4\pi  \bar{\rho}\,\delta a^{2}, \label{eq:einstein_1_chp3} \\ 
		\mH \partial_{i}\Psi + \partial_{i}\Phi' &= -4\pi \bar{\rho} \,a^{2} \delta_{ij}\delta v^{j}, \label{eq:einstein_2_chp3} \\ 
		\Phi'' + \mH \left(\Psi' + 2\Phi' \right) &= \frac{1}{3}\partial^2(\Phi - \Psi), \label{eq:einstein_3_chp3} \\ 
		\partial_{\langle i}\partial_{j\rangle} \left(\Phi - \Psi\right) &= 0. \label{eq:einstein_4_chp3}
	\end{align}
\end{subequations}
Here $\mH\equiv a'/a$ is the Hubble parameter, $\partial_{i} \equiv \partial / \partial x_{i}$, $\partial^2\equiv\partial^{i}\partial_{i}$, $\partial_{\langle i}\partial_{j\rangle}\equiv \partial_{i}\partial_{j} - 1/3\,\delta_{ij}\partial^2$, and $'$ represents a derivative with respect to conformal time, $\eta$. The quantity $|\Phi-\Psi|$ is known as the gravitational slip \citep{daniel2008,daniel2009,bertschinger2011}, which is zero in the linear regime and in the absence of anisotropic stress. At higher orders in perturbation theory, the gravitational slip is non-zero, and $\Phi\neq\Psi$ \citep[see e.g.][]{ballesteros2012}.

We perturb the density and coordinate three velocity\footnote{The velocity used throughout this Chapter is $v^i\equiv dx^i/dt$, which differs from the velocity used in Chapter~\ref{Chapter2} via \eqref{eq:vi_dxdt_relation}.} by making the substitutions
\begin{subequations} \label{eqs:matter_perturb}
	\begin{align}
		\rho &= \bar{\rho}\,(1+\delta), \\ 
		v^{i} &= \delta v^{i},
	\end{align}
\end{subequations}
where $\bar{\rho}$ represents the background FLRW density, and $\bar{v}^{i}=0$.
We derive the relativistic fluid equations from the components of the energy-momentum conservation law,
\begin{equation}
	\nabla_{\alpha} T_{\mu}^{\phantom{\alpha}\alpha} = 0,
\end{equation}
where $\nabla_{\alpha}$ is the covariant derivative associated with the 4-metric. The resulting continuity and Euler equations are, 
\begin{subequations} %
	\begin{align}
		\partial_t \delta &= 3 \partial_t \Phi - \partial_{i}v^{i}, \label{eq:rel_continuity} \\
		\mH v^{i} + \partial_t v^{i} &= -\partial^{i}\Psi. \label{eq:rel_euler}
	\end{align}
\end{subequations}

\begin{figure*}[!ht]
	\includegraphics[width=\textwidth]{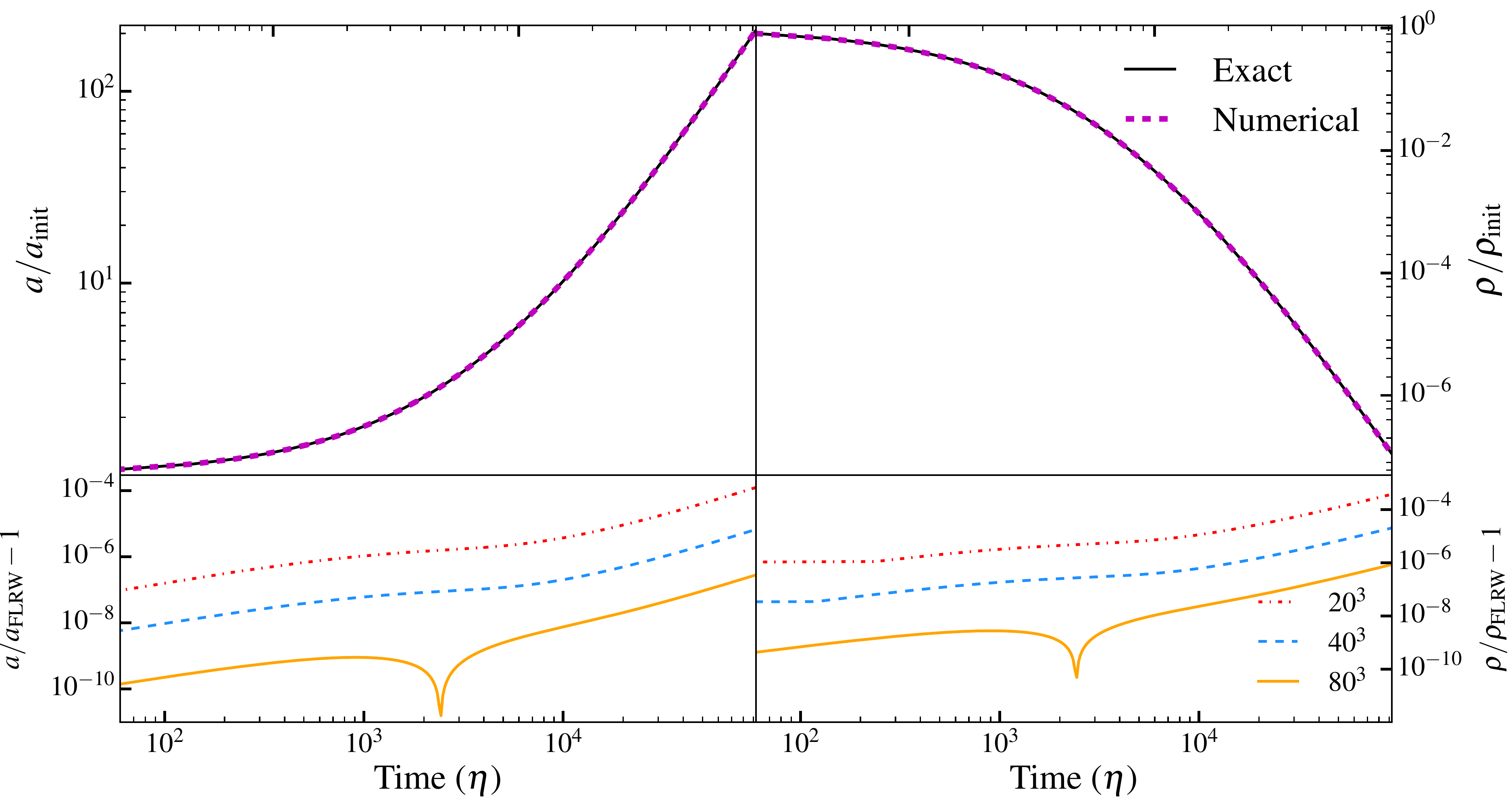}%
	\caption{\label{fig:FLRW_a_rho} Comparison between our numerical simulations (magenta) and the exact solutions (black) for a dust FLRW universe. Top: evolution of the scale factor, $a$ (left) and the density, $\rho$ (right), relative to their initial values $a_{\mathrm{init}}$ and $\rho_{\mathrm{init}}$, as a function of conformal time $\eta$. Bottom: errors in the FLRW scale factor (left) and density (right) at domain sizes $20^{3}, \,40^{3}$ and $80^{3}$.}
\end{figure*}
\begin{figure*}[!ht]
	\includegraphics[width=\textwidth]{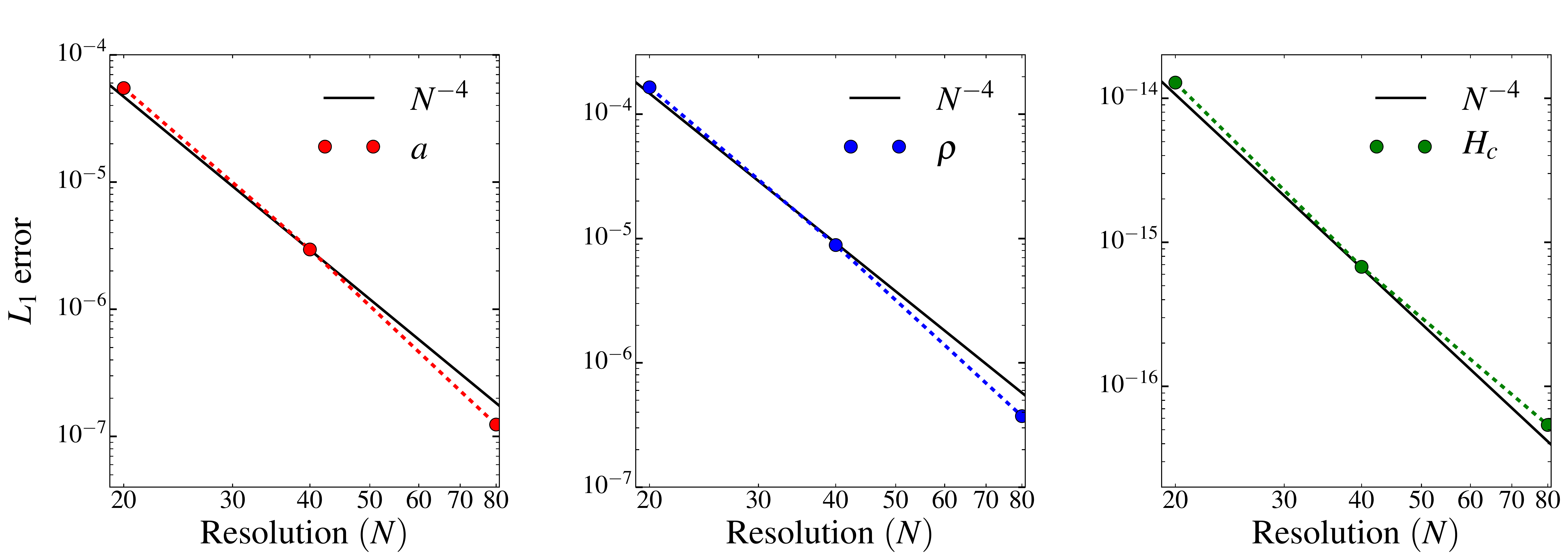}%
	\caption{\label{fig:FLRW_RMS}Fourth-order convergence in the FLRW calculations, showing $L_{1}$ error as a function of resolution for the scale factor (left), density (middle), and Hamiltonian constraint (right). $N$ refers to the number of grid points along one spatial dimension. Filled circles indicate data points from our simulations, dashed lines join these points, and black solid lines indicate the expected $N^{-4}$ convergence.}
\end{figure*}

\section{\label{sec:FLRW}FLRW spacetime}
We test our thorn \texttt{FLRWSolver} together with the \textsc{Einstein Toolkit} on two analytic solutions to Einstein's equations relevant to cosmology. Our first and simplest test is the flat, dust FLRW model. Here we initialise a homogeneous and isotropic matter distribution and spatial metric, and evolve in the gauge outlined in Section~\ref{sec:gauge_chp3}. While the \textsc{Einstein Toolkit} has been previously tested on FLRW and Kasner cosmologies \citep{loffler2012, vulcanov2002}, this is an important first test of \texttt{FLRWSolver} and its interaction with the evolution thorns.

\subsection{\label{sec:FLRW_setup}Setup}
The line element for a spatially homogeneous and isotropic FLRW spacetime is given by
\begin{equation}
	ds^{2} = - dt^{2} + a^2(t) \left[ \frac{dr^2}{1 - kr^2} + r^2 \left( d\theta^2 + {\rm sin}^2 \theta d\phi^2 \right) \right],
\end{equation}
where $k=-1,0,1$ if the universe is open, flat or closed respectively.
Assuming homogeneity and isotropy Einstein's equations reduce to the Friedmann equations \citep{friedmann1922, friedmann1924}, 
\begin{subequations}\label{eq:Friedmann}
	\begin{align}
		\left(\frac{a'}{a}\right)^{2} &= \frac{8\pi\rho \,a^{2}}{3} - k, \label{eq:friedmann_1}\\
		\rho' &= -3\frac{a'}{a}\left(\rho + P\right). \label{eq:friedmann_2}
	\end{align}
\end{subequations}
In the remainder of this chapter we assume a flat spatial geometry, supported by combined Planck and Baryon Acoustic Oscillation data \citep{planck2016params}. The flat ($k=0$), dust ($P\ll\rho$) solution to \eqref{eq:Friedmann} is
\begin{equation}
	\frac{a}{a_{\mathrm{init}}} = \xi^{2}, \quad
	\frac{\rho}{\rho_{\mathrm{init}}} = \xi^{-6},
\end{equation}
where $a_{\mathrm{init}}, \rho_{\mathrm{init}}$ are the values of $a, \rho$ at $\eta=0$ respectively, and we have introduced the scaled conformal time coordinate
\begin{equation}
	\xi\equiv 1 + \sqrt{\frac{2\pi\rho^{*}}{3a_{\mathrm{init}}}}\,\eta,
\end{equation}
where $\rho^{*}=\rho\,a^{3}$ is the conserved (constant) comoving density for an FLRW universe. The familiar $\hat{t}^{2/3}$ solution for the scale factor arises in the Newtonian gauge with $ds^{2} = -d\hat{t}^{2} + \gamma_{ij}dx^{i}dx^{j}$ (for a flat spacetime; see Appendix \ref{appx:newt_gauge}).

We initialise a homogeneous and isotropic matter distribution by specifying constant density $\rho_{\mathrm{init}}=10^{-8}$ and zero velocity in \texttt{FLRWSolver}, with $a_{\mathrm{init}}=1$. The \textsc{Einstein Toolkit} then initialises the stress-energy tensor, coupled to our homogeneous and isotropic spacetime, characterised by the spatial metric, $\gamma_{ij} = a^{2}(\eta)\delta_{ij}$, and extrinsic curvature, also set in \texttt{FLRWSolver}. We define the extrinsic curvature via the relation
\begin{equation}
 	\frac{d}{dt}\gamma_{ij} = -2\alpha K_{ij}, \label{eq:excurv_def}
\end{equation}	
where $d/dt \equiv \partial / \partial t - \mathcal{L}_{\beta}$, and $\mathcal{L}_{\beta}$ is the Lie derivative with respect to the shift vector. Since we choose $\beta^{i}=0$, we have $d/dt = \partial / \partial t$. The extrinsic curvature for our FLRW setup is therefore
\begin{equation}
	K_{ij} = -\frac{\partial_t (a) a}{\alpha}\delta_{ij}.
\end{equation}
We evolve the system until the domain volume has increased by one million, corresponding to a change in redshift of $\sim100$. 

To analyse our results we scale the time from the metric \eqref{eq:3p1_metric} to the longitudinal gauge \eqref{eq:perturbed_metric} using the coordinate transform $t=t(\eta)$. 
This gives
\begin{equation}
	\frac{dt}{d\eta} = \frac{a(\eta)}{\alpha(t)}, 
\end{equation}
which we integrate to find the scaled conformal time in terms of $t$ to be
\begin{equation}
	\xi(t) = \left(\sqrt{6\pi\rho_{\mathrm{init}}} \,\int\alpha(t)\,dt + 1\right)^{1/3},
\end{equation}
where we numerically integrate the lapse function $\alpha$ using the trapezoidal rule. This coordinate transformation allows us to simulate longer evolutions for less computational time, while still performing our analysis in the longitudinal gauge to extract physically meaningful results.

\subsection{\label{sec:FLRW_results}Results}
Figure~\ref{fig:FLRW_a_rho} compares our numerical relativity solutions with the exact solutions to the Friedmann equations. The top panels show the time evolution of $a$ and $\rho$ (dashed magenta curves) relative to their initial values, which may be compared to the exact solutions, $a_{\mathrm{FLRW}}$ and $\rho_{\mathrm{FLRW}}$ (black solid curves). The bottom panels show the residuals in our numerical solutions at resolutions of $20^{3}$, $40^{3}$ and $80^{3}$. The error can be seen to decrease when the spatial resolution is increased. The increase in spatial resolution causes the time step to decrease via the Courant condition. To quantify this, we compute the $L_{1}$ error, given by (e.g. for the scale factor)
\begin{equation}
	L_{1}(a) = \frac{1}{n} \sum_{i=1}^{n} \left|\frac{a}{a_{FLRW}}-1\right|,
\end{equation}
where $n$ is the total number of time steps.
As outlined in Section \ref{sec:method_chp3}, we expect fourth-order convergence due to the spatial homogeneity. Figure~\ref{fig:FLRW_RMS} demonstrates this is true for the scale factor (left), density (middle) and the Hamiltonian constraint (right), 
\begin{equation}
	H \equiv \,\mR + K^{2} - K_{ij}K^{ij} -16\pi\rho = 0, \label{eq:Hamiltonian}
\end{equation}
where $\mR$ is the 3-Riemann scalar and $K=\gamma^{ij}K_{ij}$\footnote{The density in \eqref{eq:Hamiltonian} is technically the density observed in the \emph{normal} frame, equivalent to $\rho$ in Chapter~\ref{Chapter2}. However, for dust in FLRW and linear perturbation theory, we have $\rho=\rho_R$, see \eqref{eq:rho_rhoR_relation}.}. For the FLRW model this reduces to the first Friedmann equation \eqref{eq:friedmann_1}.

The results of this test demonstrate that the \textsc{Einstein Toolkit}, in conjunction with our initial-condition thorn \texttt{FLRWSolver}, produces agreement with the exact solution for a flat, dust FLRW spacetime, with relative errors less than $10^{-6}$, even at low spatial resolution ($80^{3}$). 

\begin{figure*}[!ht]
	\includegraphics[width=\textwidth]{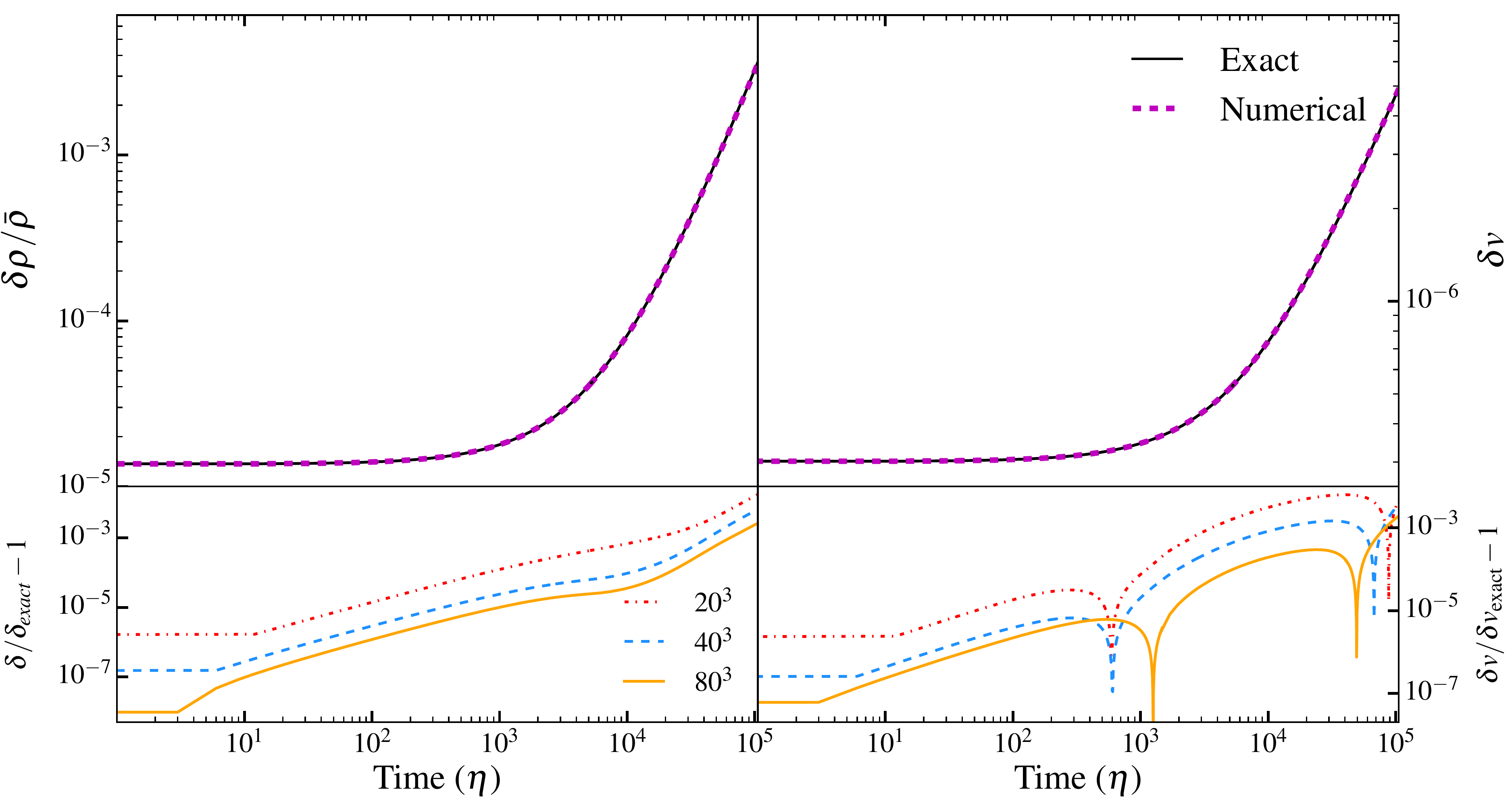}
	\caption{\label{fig:perturb_deltas}Comparison between our numerical relativity solutions and exact solutions for the linear perturbations to a dust FLRW model. We show the conformal time ($\eta$) evolution of the fractional density perturbation (top left) and the velocity perturbation (top right) computed from one-dimensional slices along the $x$ axis of our domain. Bottom: relative errors for calculations at $20^{3}$, $40^{3}$ and $80^{3}$. }
\end{figure*}
\begin{figure*}[!ht]
	\includegraphics[width=\textwidth]{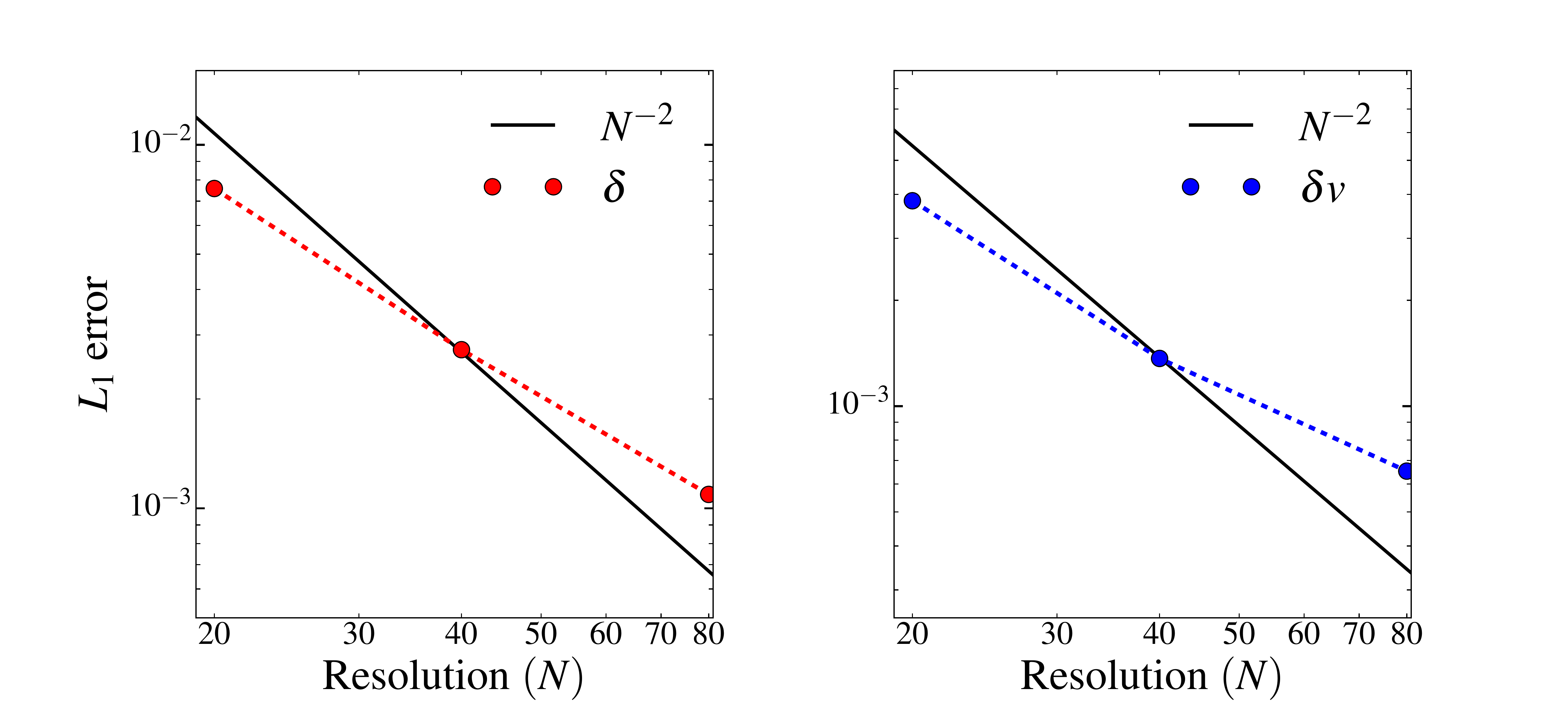}%
	\caption{\label{fig:perturb_converge}Second-order convergence of our numerical solutions to the exact solutions for a linearly perturbed FLRW spacetime, showing $L_{1}$ errors in the density (left) and velocity perturbations (right). $N$ refers to the number of grid points along one spatial dimension. Filled circles indicate data points from our simulations, dashed lines join these points, and black solid lines indicate the expected $N^{-2}$ convergence.}
\end{figure*}

\section{\label{sec:linear}Linear Perturbations}
For our second test we introduce small perturbations to the FLRW model. The evolution of these perturbations in the linear regime can be found by solving the system of equations \eqref{eqs:perturbed_einstein_chp3}. We use these solutions (derived below) to set the initial conditions.

\subsection{\label{sec:linear_setup}Setup}
In the absence of anisotropic stress we have $\Psi=\Phi$. Equation \eqref{eq:einstein_3_chp3} then becomes purely a function of $\Phi$ and the FLRW scale factor $a$. Solving this gives
\begin{equation}\label{eq:phi_full}
	\Phi = f(x^{i}) - \frac{g(x^{i})}{5\,\xi^{5}}, 
\end{equation}
where $f, g$ are functions of only the spatial coordinates. We substitute \eqref{eq:phi_full} into the Hamiltonian constraint, Equation \eqref{eq:einstein_1_chp3}, to give the fractional density perturbation $\delta \equiv \delta\rho / \bar{\rho}$, in the form
\begin{equation}\label{eq:delta_full}
	\begin{aligned} 
		\delta = C_{1}\, \xi^{2}\, \partial^2&f(x^{i}) - 2 \,f(x^{i}) \\
		&- C_{2} \,\xi^{-3}\,\partial^2g(x^{i}) - \frac{3}{5} \xi^{-5} g(x^{i}),
	\end{aligned}
\end{equation}
where we have defined
\begin{equation}
	C_{1}\equiv \frac{a_{\mathrm{init}}}{4\pi\rho^{*}},\quad C_{2}\equiv \frac{a_{\mathrm{init}}}{20\pi\rho^{*}}. \label{eq:c1c2}
\end{equation}
Using the momentum constraint, Equation \eqref{eq:einstein_2_chp3}, the velocity perturbation $\delta v^{i}$ is therefore
\begin{equation}\label{eq:vel_full}
	\delta v^{i} = C_{3}\,\xi\, \partial^{i}f(x^{i}) + \frac{3}{10}C_{3}\,\xi^{-4}\, \partial^{i}g(x^{i})
\end{equation}
where we have
\begin{equation}
	C_{3}\equiv-\sqrt{\frac{a_{\mathrm{init}}}{6\pi\rho^{*}}}. \label{eq:c3c4}
\end{equation}
Equation~\eqref{eq:delta_full} demonstrates both a growing and decaying mode for the density perturbation \citep{bardeen1980, mukhanov1992}. We set $g(x^{i})=0$ to extract only the growing mode, giving
\begin{subequations} \label{eqs:exact_solns_growing}	
	\begin{align}
		\Phi &= f(x^{i}), \label{eq:phi} \\
		\delta &= C_{1}\, \xi^{2}\, \partial^2f(x^{i}) - 2 \,f(x^{i}), \label{eq:delta}\\
		\delta v^{i} &= C_{3}\,\xi\, \partial^{i}f(x^{i}), \label{eq:vel}
	\end{align}
\end{subequations}
from which we set our initial conditions. 
We choose
\begin{equation}
	\Phi = \Phi_{0} \sum_{i=1}^{3} \mathrm{sin}\left(\frac{2\pi x^{i}}{L}\right),
\end{equation}
where $L$ is the length of one side of our computational domain. We require the amplitude $\Phi_{0}\ll1$ so that our assumptions of linearity are valid, and so we set $\Phi_{0}=10^{-8}$.
This choice then sets the form of our density and velocity perturbations, as per \eqref{eq:delta} and \eqref{eq:vel}. At $\eta=0$ ($\xi=1$) these are,
\begin{align}
	\delta &= - \left[ \left(\frac{2\pi}{L}\right)^{2}C_{1} + 2\right] \Phi_{0} \sum_{i=1}^{3} \mathrm{sin}\left(\frac{2\pi x^{i}}{L}\right),\label{eq:initial_delta}\\
	\delta v^{i} &= \frac{2\pi C_{3}}{L}\, \Phi_{0} \mathrm{cos}\left(\frac{2\pi x^{i}}{L}\right), \label{eq:initial_deltav}
\end{align}
and the choice of $\Phi_{0}$ results in amplitudes of $\delta\sim 10^{-5}$ and $\delta v^{i} \sim 10^{-7}$.
We set these matter perturbations in \texttt{FLRWSolver}, implementing negligible pressure and again using \eqref{eq:excurv_def} to define the extrinsic curvature. For a linearly perturbed FLRW spacetime with $\Psi=\Phi$ and $\Phi'=0$ we have
\begin{equation}
	K_{ij} = -\frac{\partial_t (a) a}{\alpha}(1-2\Phi)\delta_{ij}.
\end{equation}
We evolve these perturbations in the harmonic gauge until the volume of the domain has increased by 125 million, $(\Delta a)^{3} \sim 1.25\times10^{8}$, corresponding to a factor of 500 change in redshift.

\subsection{\label{sec:linear_results}Results}
Dashed magenta curves in Figure~\ref{fig:perturb_deltas} show the conformal time evolution of the fractional density perturbation, $\delta\equiv \delta\rho / \bar{\rho}$ (top left), and the velocity perturbation, $\delta v$ (top right). Solid black curves show the solutions \eqref{eq:delta} for $\delta_{\mathrm{exact}}$ and \eqref{eq:vel} for $\delta v_{\mathrm{exact}}$. Bottom panels show the relative errors for three different resolutions. 
Figure~\ref{fig:perturb_converge} shows the $L_{1}$ error as a function of resolution, demonstrating the expected second-order convergence.
\begin{figure*}[!ht]
	\begin{centering}
	\includegraphics[width=\textwidth]{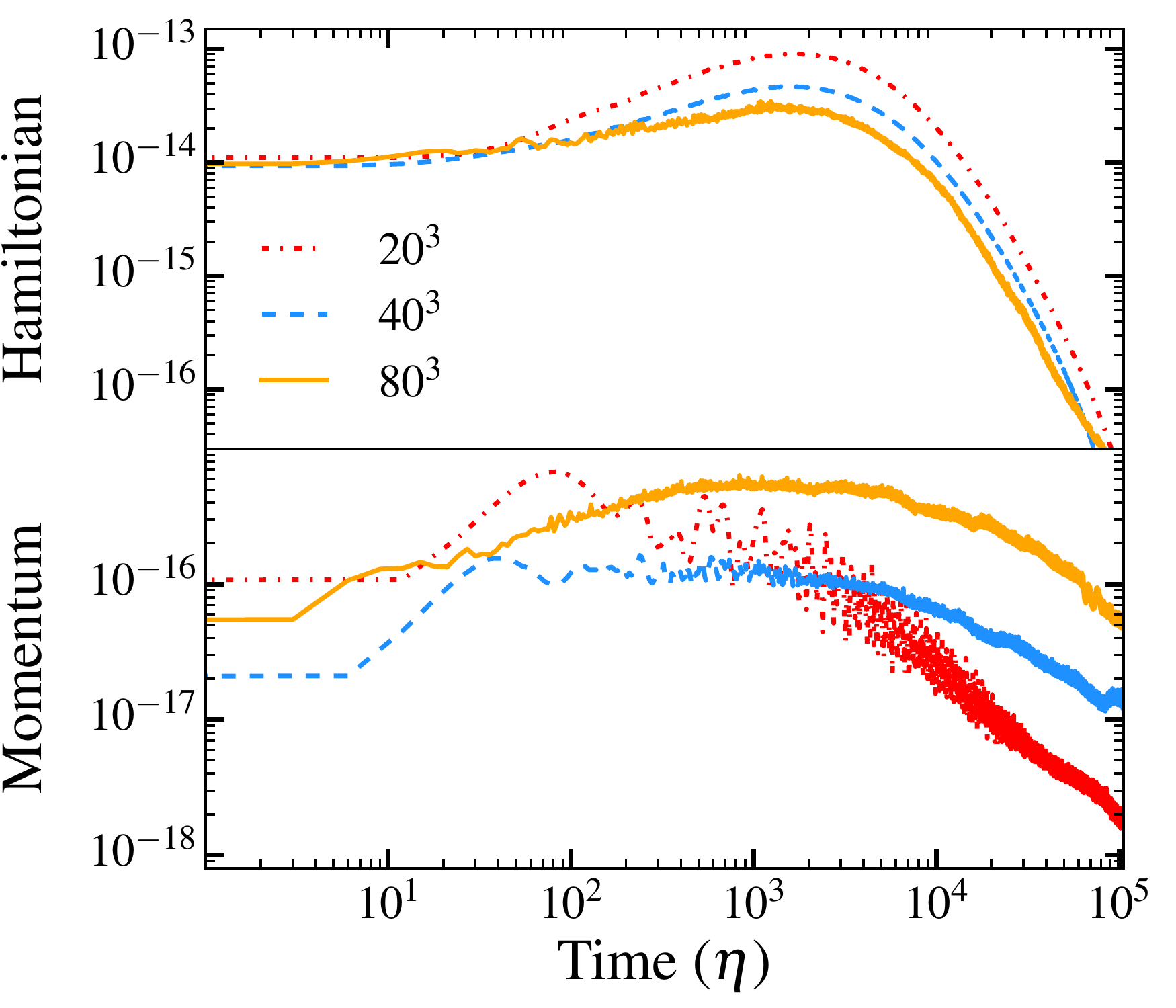}
	\caption{\label{fig:perturb_ham_mom}Maximum (anywhere in the domain) of the Hamiltonian (top panel) and momentum (bottom panel) constraints in a linearly perturbed FLRW spacetime. We show evolution over conformal time $\eta$ at resolutions $20^{3}$, $40^{3}$ and $80^{3}$.}
	\end{centering}
\end{figure*}
Figure~\ref{fig:perturb_ham_mom} shows the Hamiltonian (top), and momentum (bottom) constraints as a function of conformal time at our three chosen resolutions. The Hamiltonian constraint was defined in Equation \eqref{eq:Hamiltonian}. For our linearly perturbed FLRW spacetime this reduces to Equation \eqref{eq:einstein_1_chp3}.
The momentum constraint is
\begin{equation}
	M_{i} \equiv D_{j}K^{j}_{\phantom{j}i} - D_{i}K - S_{i} = 0,
\end{equation}
where $D_{j}$ is the covariant derivative associated with the 3-metric, and the matter source $S_{i} = -\gamma_{i\alpha}n_{\beta}T^{\alpha\beta}$, with $n_{\beta}$ the normal vector \citep{baumgarte1999}.
For linear perturbations this constraint reduces to Equation \eqref{eq:einstein_2_chp3}.

Figure~\ref{fig:perturb_ham_mom} shows a better preservation of the Hamiltonian constraint with increasing resolution. The momentum constraint shows the opposite. We attribute this to the momentum constraint being preserved to of order the roundoff error, which will become larger with an increase in resolution. Even at the highest resolution the momentum constraint is preserved to within $10^{-15}$.

This second test has demonstrated a match to within $\sim10^{-3}$ of our numerical relativity solutions to the exact solutions for the linear growth of perturbations, while exhibiting the expected second-order convergence.

\begin{figure*}[!ht]
	\includegraphics[width=\textwidth]{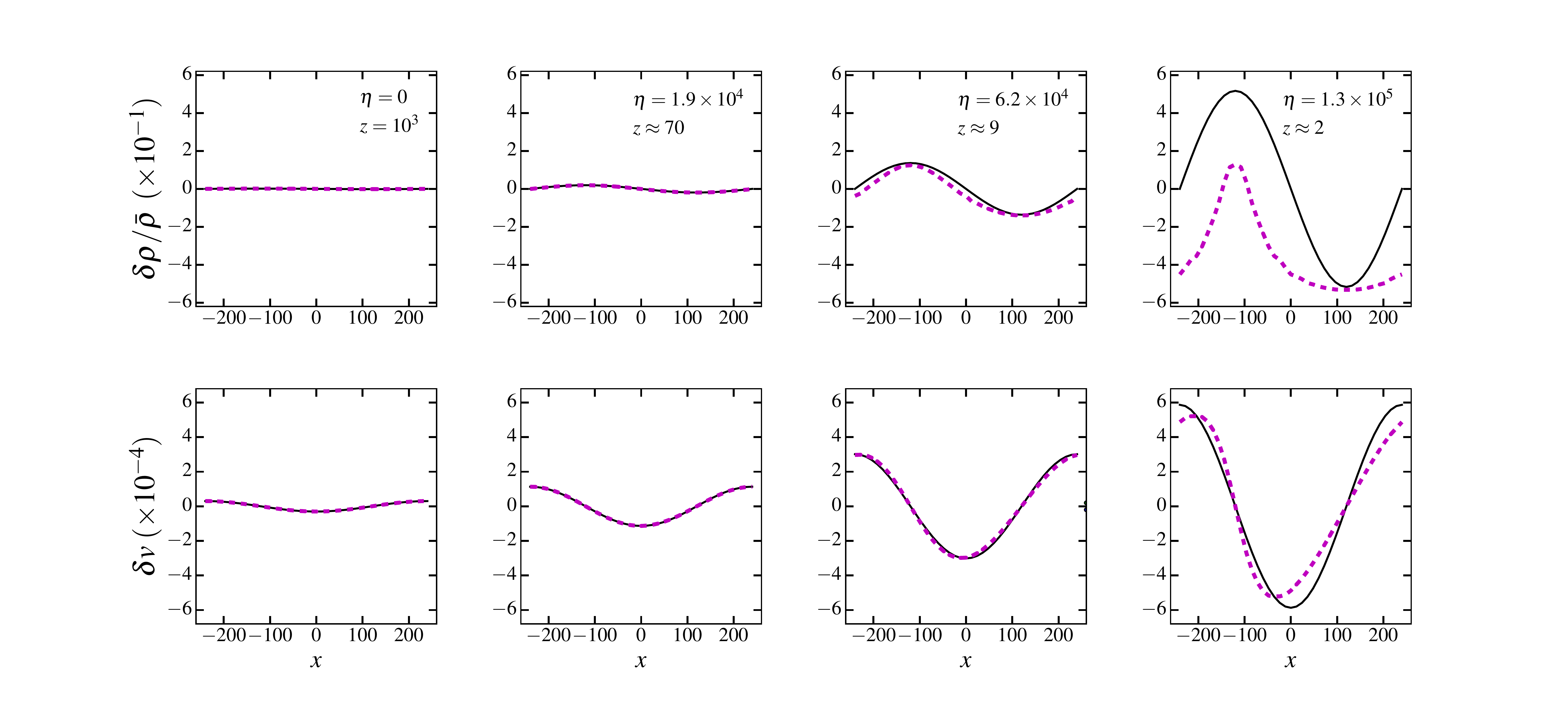} 
	\caption{\label{fig:nonlinear_x}Density (top row) and velocity (bottom row) perturbations as a function of position within our domain. Here we show one-dimensional slices of our $40^{3}$ domain during the conformal time ($\eta$) evolution. All quantities are shown in code units. Dashed magenta curves show our numerical solutions, and black solid curves show the exact solutions for the linear regime. Initial data ($\eta=0$; first column) and $\eta=1.9\times10^{4}$ (second column) match linear theory. We see a clear deviation from linear theory at $\eta=6.2\times 10^{4}$ (third column) and $\eta=1.3\times10^{5}$ (fourth column). Simulation redshifts are shown as an indicator of the \emph{change} in redshift.}
\end{figure*}
\begin{figure*}[!ht]
	\includegraphics[width=\textwidth]{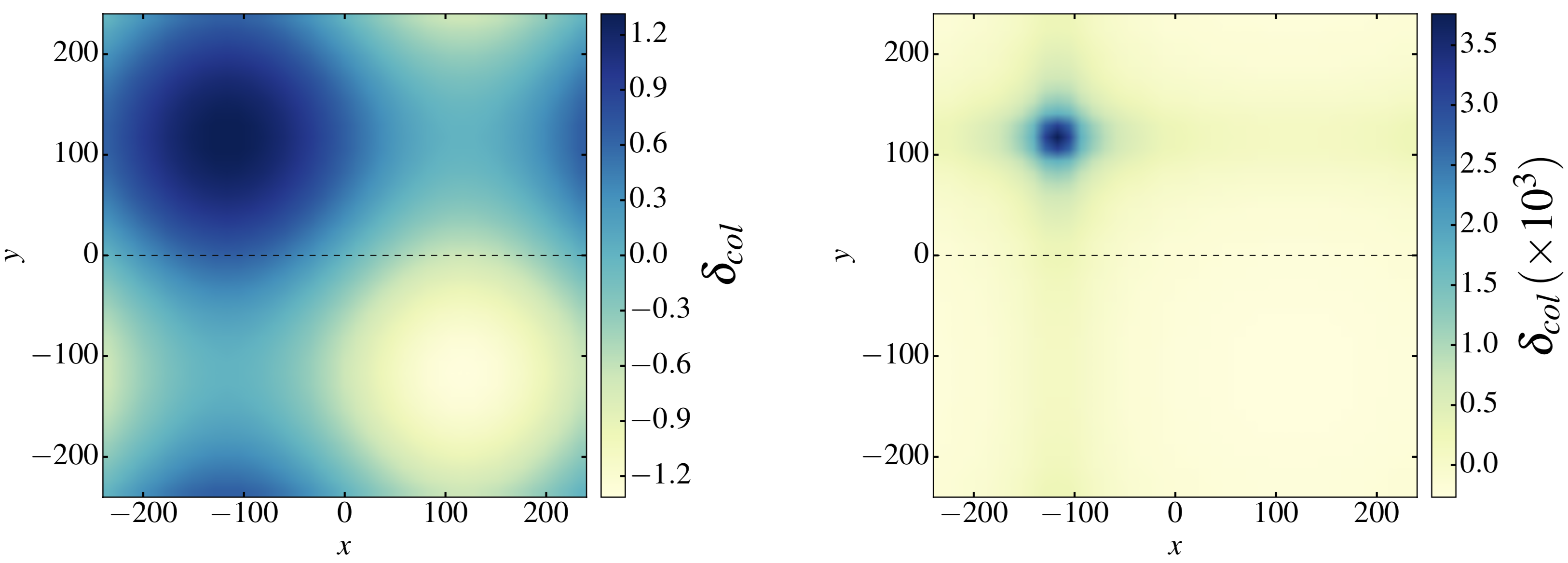}
	\caption{\label{fig:nonlinear_2D_delta} Column-integrated density perturbation showing the gravitational collapse of an overdense region. The two panels correspond to the left and right panels of Figure~\ref{fig:nonlinear_x} respectively, at conformal times of $\eta=0$ and $\eta=1.3\times10^{5}$. All quantities are shown in code units for our $40^{3}$ simulation. Grey dashed lines indicate the position of the one-dimensional slices shown in Figure~\ref{fig:nonlinear_x}.}
\end{figure*}

\begin{figure*}[!ht]
	\includegraphics[width=\textwidth]{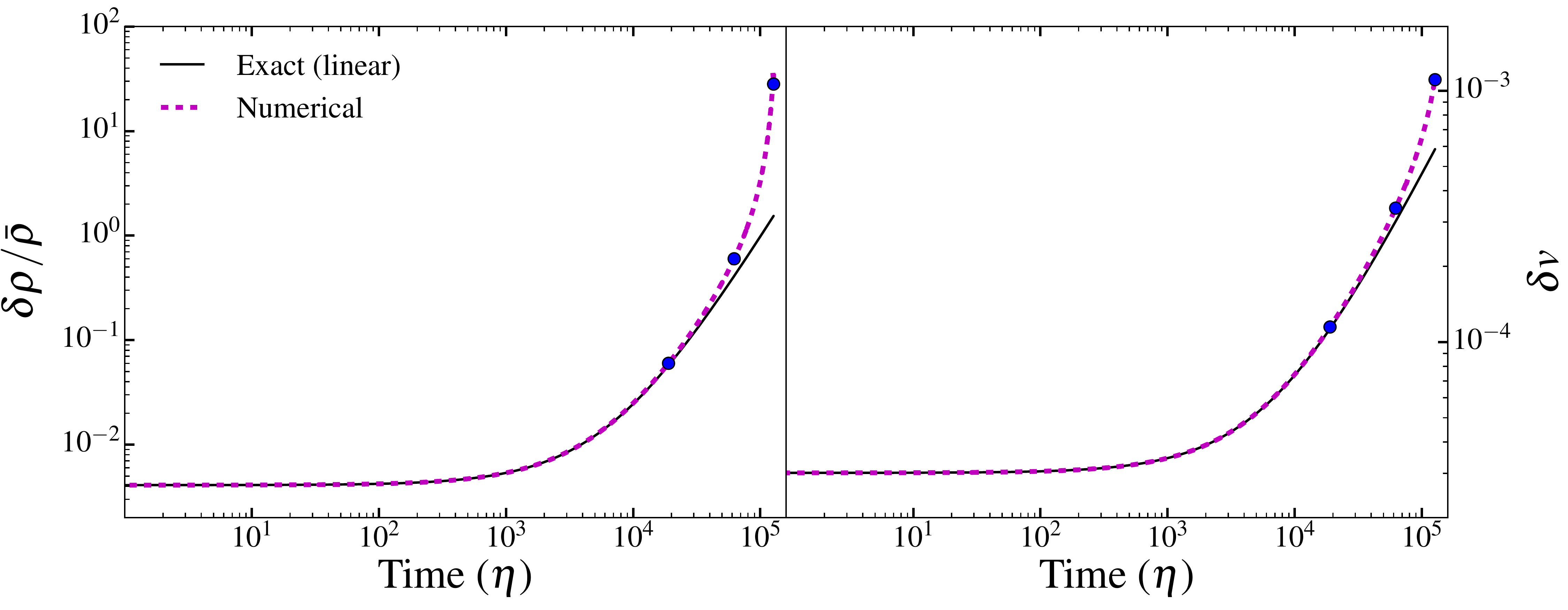}
	\caption{\label{fig:nonlinear_dr_dv}Nonlinear growth of the density (left) and velocity (right) perturbations. Dashed magenta curves show the maximum value within the domain as a function of conformal time $\eta$ (in code units), while black solid curves show the analytic solutions for linear growth. Here we show the simulation with domain size $40^{3}$. Blue circles represent the times of the ($\eta>0$) panels shown in Figure \ref{fig:nonlinear_x}.}
\end{figure*}

\section{\label{sec:nonlinear}Nonlinear evolution}
In order to evolve our perturbations to nonlinear amplitude in a reasonable computational time, we increase the size of our initial perturbations to $\Phi_{0}=10^{-6}$, which in turn gives $\delta\sim 10^{-3}$ and $\delta v^{i}\sim 10^{-5}$. The linear approximation remains valid.

We choose the starting redshift to be that of the cosmic microwave background (CMB). That is, we set $z=1000$, such that our initial density perturbation is roughly consistent with the amplitude of temperature fluctuations in the CMB ($\sim 10^{-5}$) \citep{bennett2013}. We emphasise that this redshift, and all redshifts shown in figures, should not be taken literally; its purpose is to assign an approximate \textit{change} in redshift, calculated directly from the FLRW scale factor.

\subsection{\label{sec:nonlinear_results}Results}
Figure~\ref{fig:nonlinear_x} shows a series of one-dimensional slices through the origin of the $y$ and $z$ axes at four different times. Dashed magenta curves show solutions for the density (top row) and velocity (bottom row) perturbations, which may be compared to the black solid curves showing the analytic solutions for linear perturbations. At $\eta=0$ and $\eta=1.9\times10^{4}$ (first and second columns respectively) the solutions are linear, while at $\eta=6.2\times10^{4}$  (third column) both the density and velocity perturbations deviate from linear theory. The perturbations are nonlinear at $\eta=1.3\times10^{5}$ (fourth column) where matter collapses towards the overdensity, indicated by the shift in the maximum velocity. 

The final column shows an apparent decrease in the average density. This is simply an artefact of taking a one-dimensional slice through a three-dimensional box. Figure~\ref{fig:nonlinear_2D_delta} shows the column-density perturbation, $\delta_{col}$, computed by integrating the density perturbation along the $z$ axis. Panels show $\eta=0$ and $\eta=1.3\times10^{5}$ respectively. The right panel shows an increase of $\sim 3000$ times in the column-density perturbation at $x,y\approx-120,120$. A corresponding void can be seen in the lower right of Figure~\ref{fig:nonlinear_2D_delta}, explaining the underdensity along the $y$ axis seen in the final column of Figure~\ref{fig:nonlinear_x}.

Figure~\ref{fig:nonlinear_dr_dv} shows the maximum value of the density (left) and velocity (right) perturbations as a function of time. Dashed magenta curves show the numerical solutions, which may be compared to the black curves showing the linear analytic solutions. 
Perturbations can be seen to deviate from the linear approximation at $\eta\approx3\times10^{4}$, when $\delta\rho/\bar{\rho} \approx 0.1$. 
At $\eta\approx10^{5}$, the maximum of the density and velocity perturbations have respectively grown 25 and 2 times larger than the linear solutions.

\subsection{\label{sec:nonlinear_highorder}Gravitational slip and tensor perturbations}
Gravitational slip is defined as the difference between the two potentials $\Phi$ and $\Psi$ \citep{daniel2008,daniel2009}, which is zero in the linear regime, see equation \eqref{eq:einstein_4_chp3}, but nonzero in the nonlinear regime \citep[see e.g.][]{ballesteros2012}.

We reconstruct $\Phi$ and $\Psi$ from the metric components, although we note the interpretation of these potentials becomes unclear in the nonlinear regime. From \eqref{eq:perturbed_metric_svt} the spatial metric is
\begin{equation}
	\gamma_{ij} = a^{2}\left[(1-2\Phi)\delta_{ij} + h_{ij}\right],
\end{equation}
and we adopt the traceless gauge condition $\delta^{ij}h_{ij}=0$ \citep{green2012,adamek2013}. The potential $\Phi$ is then
\begin{equation}
	\Phi=\frac{1}{2}\left(1 - \frac{\delta^{ij}\gamma_{ij}}{3\,a^{2}}\right), \label{eq:phi_reconstruct}
\end{equation}
which holds for all times the metric \eqref{eq:perturbed_metric_svt} applies. The potential $\Psi$ is more complicated: our gauge choice implies lapse evolution according to \eqref{eq:harmonic_gauge}, where we have set $f(\alpha)=1/4$, and 
\begin{equation}
	K=-3\frac{\partial_t a}{a\alpha},
\end{equation}
in the linear regime, which gives
\begin{equation}
	\partial_t \alpha = \frac{3}{4}\frac{\partial_t (a) \alpha}{a}.
\end{equation}
Integrating this results in a lapse evolution of
\begin{equation}
	\frac{\alpha}{\alpha_{\mathrm{init}}} = D(x^{i}) \left(\frac{a}{a_{\mathrm{init}}}\right)^{3/4},
\end{equation}
where $D(x^{i})$ is a function of our spatial coordinates. According to the metric \eqref{eq:perturbed_metric}, and with $\alpha_{\mathrm{init}}=a_{\mathrm{init}}=1$ this implies \begin{equation}
	\alpha = \sqrt{1+2\Psi}\,a^{3/4},
\end{equation}
from which we reconstruct the potential $\Psi$ to be
\begin{equation}
	\Psi = \frac{1}{2}\left[\left(\frac{\alpha}{a^{3/4}}\right)^{2} - 1\right], \label{eq:psi_reconstruct}
\end{equation}
valid in the linear regime. Our gauge choice $\beta^{i}=0$ implies that in the nonlinear regime we expect additional modes to be present in this reconstruction of $\Psi$.

\begin{figure*}[!ht]
	\includegraphics[width=\textwidth]{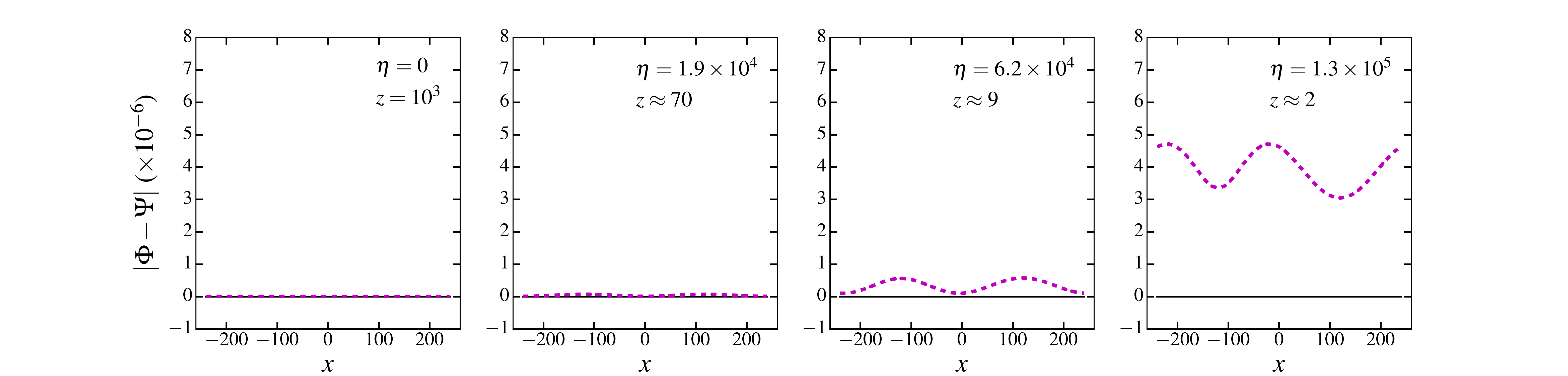} 
	\caption{\label{fig:nonlinear_slip_x}Time evolution of a one-dimensional slice of the gravitational slip. Dashed magenta curves show numerical solutions for a $40^{3}$ domain, while black solid lines show the solution for linear perturbation theory; zero. The potentials $\Phi$ and $\Psi$ are reconstructed according to \eqref{eq:phi_reconstruct} and \eqref{eq:psi_reconstruct} respectively. Initial data ($\eta=0$) is shown in the left column, and time increases towards the right as indicated by timestamps. We show the simulation redshift as an indicator of the approximate \textit{change} in redshift only, and all quantities here are shown in code units.}
\end{figure*}

We use an FLRW simulation for the scale factor $a$ in \eqref{eq:phi_reconstruct} and \eqref{eq:psi_reconstruct}, from which we calculate the gravitational slip $|\Phi - \Psi|$.
This is potentially problematic once the perturbations become nonlinear, as the gauges of the two simulations will differ.
Figure~\ref{fig:nonlinear_slip_x} shows one-dimensional slices of the gravitational slip at the same times as was shown in Figure~\ref{fig:nonlinear_x}. Dashed curves show the numerical results, with black lines showing the linear solution; zero gravitational slip. In the fourth panel ($\eta=1.3\times10^{5}$) we see a positive shift of the gravitational slip to $\approx 4\times10^{-6}$ for this one-dimensional slice, with the maximum value in the three-dimensional domain being $6.5\times10^{-6}$ at this time. The Newtonian potential $\Psi$ has a positive average value at $\eta=1.3\times10^{5}$, due to the majority of the domain being underdense (see Figure~\ref{fig:nonlinear_2D_delta}), and the potential $\Phi$ takes a negative average value. This can be interpreted as an overall positive contribution to the expansion, from the metric \eqref{eq:perturbed_metric_svt}.

Relativistic corrections to one-dimensional N-body simulations in \citep{adamek2013} resulted in a gravitational slip of $4\times10^{-6}$. We show a gravitational slip of the same amplitude, including the full effects of General Relativity in a three-dimensional simulation, for a time when our density perturbation is comparable in size to that of \citet{adamek2013}.
Gravitational slip is a measurable effect that can be quantified by combining weak gravitational lensing and galaxy clustering \citep{bertschinger2011}. 
Our simulations show tentative evidence for the importance of gravitational slip due to nonlinear gravitational effects. However, robust predictions require higher resolution simulations with more realistic initial conditions.

\begin{figure}[h]
	\includegraphics[width=\columnwidth]{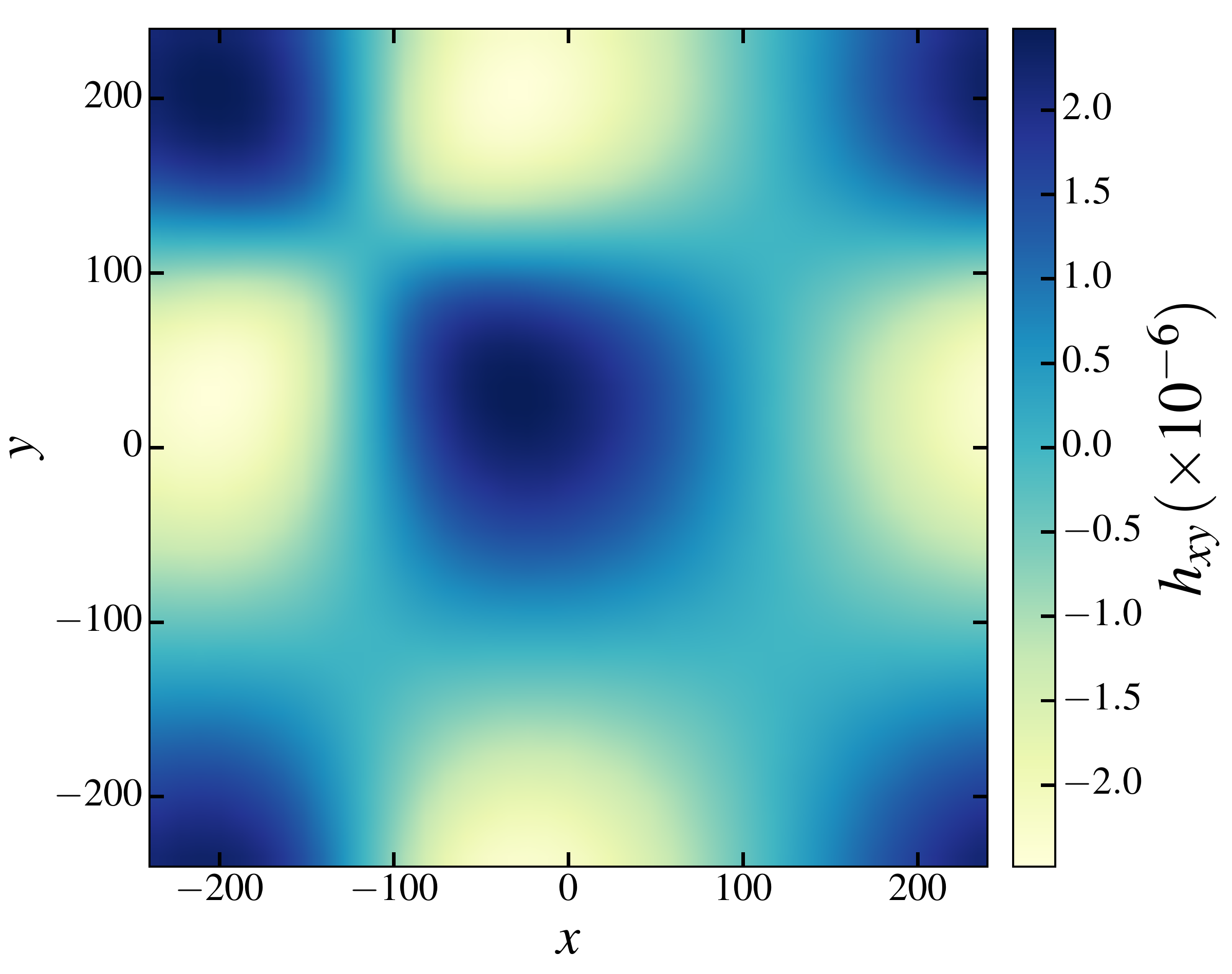}
	\caption{\label{fig:hxy_2D}Two-dimensional slice of the $xy$ component of the tensor perturbation $h_{ij}$ at $\eta=1.3\times10^{5}$. We use \eqref{eq:hij_recon} to calculate $h_{xy}$ using the off-diagonal metric component $g_{xy}$. All quantities are shown in code units for our $40^{3}$ simulation. }
\end{figure}

In our initial conditions we neglected vector and tensor perturbations in the perturbed FLRW metric \eqref{eq:perturbed_metric_svt}, since in the linear regime the scalar perturbations dominate. These higher order perturbations appear in the nonlinear regime. The tensor perturbation can be extracted from the off-diagonal, spatial components of the metric,
\begin{equation}\label{eq:hij_recon}
	\gamma_{ij} = a^{2}h_{ij} \quad \mathrm{for}\,i\neq j,
\end{equation}
however, details of these tensor modes may be dependent on the choice of gauge. 
We calculate $h_{ij}$ using the value of $a$ as per the scalar perturbations.

Figure~\ref{fig:hxy_2D} shows a two-dimensional cross-section of the $xy$ component of the tensor perturbation $h_{ij}$. All other components are identical. The cross-section is shown at $\eta=1.3\times10^{5}$, corresponding to the right panel of Figures~\ref{fig:nonlinear_x}, \ref{fig:nonlinear_2D_delta} and \ref{fig:nonlinear_slip_x}. While the maximum amplitude of the tensor perturbation is small ($\sim 2\times10^{-6}$), 
an asymmetry develops in $h_{xy}$, corresponding to the location of the overdensity in Figure~\ref{fig:nonlinear_2D_delta}.  We also see a diffusion of the tensor perturbation in the void, indicating the beginning of growth of higher order perturbations.

\section{\label{sec:conclusion}Discussion and Conclusions}
We have demonstrated the feasibility of inhomogeneous cosmological simulations in full General Relativity using the \textsc{Einstein Toolkit}. The overall approach is similar to other recent attempts \citep{giblin2016a,bentivegna2016a}, with the main difference being in the construction of initial conditions which allows us to simulate a pure growing mode, instead of a mix of growing and decaying modes \citep[see][]{daverio2017}. We also use a different code to \citep{giblin2016a}, allowing for independent verification. As with the other studies we were able to demonstrate the evolution of a density perturbation into the nonlinear regime. 

As this is a preliminary study, we have focused on the numerical accuracy and convergence, rather than a detailed investigation of physical effects such as backreaction. Our main conclusions are:
\begin{enumerate}
\item We demonstrate fourth-order convergence of the numerical solution to the exact solution for a flat, dust FLRW universe with errors $\sim 10^{-5}$ even at low spatial resolution ($40^{3}$).
\item We demonstrate second-order convergence of the numerical solutions for the growth of linear perturbations, matching the analytic solutions for the cosmic evolution of density, velocity and metric perturbations to within $\sim10^{-3}$.
\item We show that numerical relativity can successfully be used to follow the formation of cosmological structures into the nonlinear regime. We demonstrate the appearance of non-zero gravitational slip and tensor modes once perturbations are nonlinear with amplitudes of $\sim4\times10^{-6}$ and $\sim2\times10^{-6}$ respectively. \end{enumerate}

The main limitation to our study is that we have employed only low-resolution simulations compared to current Newtonian N-body cosmological simulations \citep[e.g.][]{genel2014,springel2005,kim2011}, and used only simple initial conditions rather than a more realistic spectrum of perturbations \citep[but see][]{giblin2016a}. Representing the density field on a grid means our simulations are limited by the formation of shell-crossing singularities. The relative computational expense means that General-Relativistic simulations are unlikely to replace the Newtonian approach in the near future. However, they are an important check on the accuracy of the approximations employed.

\chapter[Einstein's Universe]{Einstein's Universe: Cosmological structure formation in numerical relativity} %

\label{Chapter4} %

\vspace{10mm}

Published in:\\
\citet{macpherson2019a}. Physical Review D, \textbf{99.6, 063522}.

\section*{Abstract}
We perform large-scale cosmological simulations that solve Einstein's equations directly via numerical relativity. Starting with initial conditions sampled from the cosmic microwave background, we track the emergence of a cosmic web without the need for a background cosmology. We measure the backreaction of large-scale structure on the evolution of averaged quantities in a matter-dominated universe. Although our results are preliminary, we find the global backreaction energy density is of order $10^{-8}$ compared to the energy density of matter in our simulations, and is thus unlikely to explain accelerating expansion under our assumptions. Sampling scales above the homogeneity scale of the Universe ($100-180\,h^{-1}$Mpc), in our chosen gauge, we find $2-3\%$ variations in local spatial curvature.

\section*{A note on notation}
We have altered the notation throughout this chapter, including Appendices~\ref{sec:appx_avg}, \ref{sec:appx_expn}, \ref{appx:constraints}, and \ref{appx:convergence}, to be consistent with Chapters~\ref{Chapter1} and \ref{Chapter2}, unless explicitly stated otherwise. For these exceptions, we maintain the notation of the publication for consistency with figures in their published form. Aside from these changes, this chapter is consistent with the accepted version of \citet{macpherson2019a}.

\pagebreak

\section{Introduction}
Modern cosmology derives from the Friedmann-Lema\^itre-Robertson-Walker (FLRW) metric --- an exact solution to Einstein's equations that assumes homogeneity and isotropy. The formation of cosmological structure means that the Universe is neither homogeneous nor isotropic on small scales. The Lambda Cold Dark Matter ($\Lambda$CDM) model assumes the FLRW metric, and has been the leading cosmological model since the discovery of the accelerating expansion of the Universe \citep{riess1998,perlmutter1999}. Since then it has had many successful predictions, including the location of the baryon acoustic peak \citep[e.g.][]{kovac2002,eisenstein2005,cole2005,blake2011a,ata2018}, the polarisation of the cosmic microwave background (CMB) \citep{planck2016params,hinshaw2013}, galaxy clustering, and gravitational lensing \citep[e.g.][]{bonvin2017,hildebrandt2017,DESCollab2017a}. Despite these successes, tensions with observations have arisen. Most notable is the recent $3.8\sigma$ tension between measurements of the Hubble parameter, $H_{0}$, locally \citep{riess2018b} and the value inferred from the CMB under $\Lambda$CDM \citep{planck2016params}. 

The assumptions underlying the standard cosmological model are based on observations that our Universe is, \textit{on average}, homogeneous and isotropic. However, the averaged evolution of an inhomogeneous universe does not coincide with the evolution of a homogeneous universe \citep{buchertehlers1997,buchert2000a}. Additional ``backreaction'' terms exist, but their significance has been debated \citep[e.g.][]{rasanen2006a,rasanen2006b,li2007,li2008,larena2009,clarkson2011a,wiltshire2011,wiegand2012,green2012,buchert2012,green2014,buchert2015,green2016,bolejko2017d,roukema2017,kaiser2017,buchert2018a}.

State-of-the-art cosmological simulations currently employ the FLRW solution coupled with a Newtonian approximation for gravity \citep{springel2005,kim2011,genel2014}. These simulations have proven extremely valuable to furthering our understanding of the Universe. However, General-Relativistic effects on our observations cannot be fully studied when the formation of large-scale structure has no effect on the surrounding spacetime. Whether or not these effects are significant can only be tested with numerical relativity, which allows us to fully remove the assumptions of homogeneity and isotropy. Initial works have shown emerging relativistic effects such as differential expansion \citep{bentivegna2016a}, variations in proper length and luminosity distance relative to FLRW \citep{giblin2016a,giblin2016b}, and the emergence of tensor modes and gravitational slip \citep{macpherson2017a}. A comparison between Newtonian and fully General-Relativistic simulations found sub-percent differences in the weak-field regime \citep{east2018}, in agreement with post-Friedmannian N-body calculations \citep{adamek2013,adamek2014b}. 

In this work, we present cosmological simulations with numerical relativity, using realistic initial conditions, evolved over the entire history of the Universe. Here we use a fluid approximation for dark matter, however, this is one more step along the road to fully relativistic cosmological N-body calculations. We focus on the global backreaction of cosmological structures on averaged quantities, including the matter, curvature, and backreaction energy densities, and how these averages vary as a function of physical size of the averaging domain. We test the global and local effects on the expansion rate, including the potential for backreaction to contribute to the accelerating expansion of the Universe. In a companion paper we examine whether local variations in the Hubble expansion rate can explain the discrepancy between local and global measurements of the Hubble constant \citep[see][and Chapter~\ref{Chapter5}]{macpherson2018b}.

In Section~\ref{sec:compsetup} we describe our computational setup, in Section~\ref{sec:ics} we describe the derivation and implementation of initial conditions drawn from the CMB, in Section~\ref{sec:gauge} and \ref{sec:averaging} we describe our choice of gauge and averaging scheme respectively, and in Section~\ref{sec:results} we present our simulations and averaged quantities. We discuss our results in Section~\ref{sec:discussion} and conclude in Section~\ref{sec:conclude}. Unless otherwise stated, we adopt geometric units with $G=c=1$, where $G$ is the gravitational constant and $c$ is the speed of light. Greek indices take values 0 to 3, and Latin indices from 1 to 3, with repeated indices implying summation.

\section{Computational Setup} \label{sec:compsetup}
\subsection{\textsc{Cactus} and FLRWSolver}
To evolve a fully General-Relativistic cosmology we use the open-source \textsc{Einstein Toolkit} \citep{loffler2012}, a collection of codes based on the \textsc{Cactus} framework \citep{cactus}. Within this toolkit we use the \texttt{ML\_BSSN} thorn \citep{brownD2009} for evolution of the spacetime variables using the BSSN formalism \citep{shibata1995,baumgarte1999}, and the \texttt{GRHydro} thorn for evolution of the hydrodynamics \citep{baiotti2005,giacomazzo2007,mosta2014}. In addition, we use our initial-condition thorn, \texttt{FLRWSolver} \citep{macpherson2017a}, to initialise linearly-perturbed FLRW spacetimes with perturbations of either single-mode or CMB-like distributions.

We assume a dust universe, implying pressure $P=0$, however \texttt{GRHydro} currently has no way to implement zero pressure for hydrodynamical evolution. Instead we set $P\ll\rho$, with a polytropic equation of state,
\begin{equation}
	P = K_\mathrm{poly} \rho\,^{2},
\end{equation}
where $K_\mathrm{poly}$ is the polytropic constant, which we set $K_\mathrm{poly}=0.1$ in code units. We have found this to be sufficient to match the evolution of a homogeneous, isotropic, matter-dominated universe. Deviations from the exact solution for the scale factor evolution, at $80^{3}$ resolution, are within $10^{-6}$ \citep[see][]{macpherson2017a}. 

We perform a series of simulations with varying resolutions, $64^{3}, 128^{3}$, and $256^{3}$, and comoving physical domain sizes, $L=100$ Mpc, 500 Mpc, and 1 Gpc, to study different physical scales. We simulate all three domain sizes at $64^{3}$ and $128^{3}$ resolution, and only the $L=1$ Gpc domain size at $256^3$ resolution due to computational constraints. During the evolution we do not assume a cosmological background, and for convenience, since we have not yet implemented a cosmological constant in the \textsc{Einstein Toolkit}, we assume $\Lambda=0$.

Post-processing analysis is performed using the \textsc{mescaline} code, which we introduce and describe in Section~\ref{subsec:mescaline_chp4}.

\subsection{Length unit}
We choose the comoving length unit of our simulation domain to be 1 Mpc, implying a domain of $L=100$ in code units is equivalently $L=100$ Mpc. 

In geometric units $c=1$, and so we can relate our length unit, $l=1$ Mpc, and our time unit, $t_c$, via the speed of light (in physical units)

\begin{equation} \label{eq:tunit}
	t_c = \frac{l}{c}. %
\end{equation}
To find our background FLRW density we use $H(z=0)=H_0$, with units of s$^{-1}$. This implies
\begin{equation} \label{eq:H0units}
H_{0,\mathrm{code}} \times \frac{1}{t_c} = H_{0,\mathrm{phys}},
\end{equation}
where $H_{0,\mathrm{code}}$ and $H_{0,\mathrm{phys}}=100\,h\,\mathrm{km\,s^{-1}\,Mpc^{-1}}$ are the Hubble parameter expressed in code units and physical units, respectively. We use \eqref{eq:tunit} together with \eqref{eq:H0units} and the Friedmann equation for a flat, matter-dominated model
\begin{equation}
	H = \frac{\dot{a}}{a} = \sqrt{\frac{8\pi \bar{\rho}}{3}},
\end{equation}
where an overdot represents a derivative with respect to proper time, $\bar{\rho}$ is the homogeneous density, $a$ is the FLRW scale factor, and we have $G=1$ in code units. We find the background FLRW density, evaluated at $z=0$, in code units, to be
\begin{equation} \label{eq:rho0_a01}
	\bar{\rho}_{0,\mathrm{code}} = 1.328\times10^{-8}\,h^{2}.
\end{equation}

For computational reasons we adopt the initial FLRW scale factor $a_\mathrm{init}=a(z=1100)=1$, whilst the usual convention in cosmology is to set $a_0=a(z=0)=1$. The density \eqref{eq:rho0_a01} was calculated using the Hubble parameter $H_{0,\mathrm{phys}}$ evaluated with $a_0=1$. The comoving (constant) FLRW density is $\rho^*=\bar{\rho}\,a^3=\bar{\rho}_0\,a_0^3$, and so \eqref{eq:rho0_a01} is the comoving density $\rho^*$. We choose $h=0.704$, and our choice $a_\mathrm{init}=1$ implies our initial background density is the comoving FLRW density.

\subsection{Redshifts}
Simulations are initiated at $z=1100$ and evolve to $z=0$. We quote redshifts computed from the value of the FLRW scale factor at a particular conformal time,
\begin{equation}
	a(\eta) = \frac{z_\mathrm{cmb} + 1}{z(\eta) + 1},
\end{equation}
where $z_\mathrm{cmb}=1100$. Since we set $a_\mathrm{init}=1$, we have $a_0=1101$. The evolution of the FLRW scale factor in conformal time is
\begin{equation}
	a(\eta) = a_\mathrm{init} \, \xi^2, \label{eq:confa}
\end{equation}
where $\xi$ is the scaled conformal time defined in Section~\ref{subsec:linperturb}. Importantly, the redshifts presented throughout this chapter are indicative only of the amount of coordinate time that has passed, and are not necessarily indicative of redshifts measured by observers in an inhomogeneous universe.

\section{Initial conditions} \label{sec:ics}

\subsection{Linear Perturbations} \label{subsec:linperturb}
We solve the linearly-perturbed Einstein equations to generate our initial conditions. Assuming only scalar perturbations, the linearly-perturbed FLRW metric in the longitudinal gauge is
\begin{equation} \label{eq:longitudinal_metric_chp4}
	ds^{2} = -a^{2}(\eta)\left(1+2\psi\right)d\eta^{2} + a^{2}(\eta)\left(1-2\phi\right)\delta_{ij}dx^{i}dx^{j}.
\end{equation}
In this gauge the metric perturbations $\phi$ and $\psi$ are the Bardeen potentials \citep{bardeen1980}. These are related to perturbations in the matter distribution via the linearly perturbed Einstein equations
\begin{equation}\label{eq:perturbed_einstein_chp4}
	\bar{G}_{\mu\nu} + \delta G_{\mu\nu} = 8\pi\left(\bar{T}_{\mu\nu} + \delta T_{\mu\nu}\right),
\end{equation}
where an over-bar represents a background quantity, and $\delta X$ represents a small perturbation in the quantity $X$, with $\delta X\ll X$. A matter-dominated (dust) universe has stress-energy tensor
\begin{equation}
	T_{\mu\nu} = \rho \,u_{\mu}u_{\nu},%
\end{equation}
where $\rho$ is the mass-energy density\footnote{The density $\rho$ used throughout this Chapter is the total \emph{rest-frame} mass-energy density, and is equivalent to $\rho_R$ used in Chapters~\ref{Chapter1} and \ref{Chapter2}.}, $u^{\mu}=dx^{\mu}/d\tau$ is the four-velocity of the fluid, and $\tau$ is the proper time. Assuming small perturbations to the matter we have
\begin{align}
	\rho &= \bar{\rho} + \delta\rho = \bar{\rho}(1+\delta), \\
	v^{i} &= \delta v^{i},
\end{align}
where the fractional density perturbation is $\delta\equiv\delta\rho/\bar{\rho}$, and $v^{i}=dx^{i}/dt$ is the three-velocity\footnote{The velocity used throughout this Chapter differs from the velocity used in Chapter~\ref{Chapter2}, they are related via \eqref{eq:vi_dxdt_relation}.}, with $t$ the coordinate time.%

Solutions to \eqref{eq:perturbed_einstein_chp4} are found by taking the time-time, time-space, trace and trace-free components, given by
 \begin{subequations} \label{eqs:perturbed_einstein}
	\begin{align}
		\partial^{2}\phi - 3 \mathcal{H}\left(\phi' + \mathcal{H} \psi\right) &= 4\pi  \bar{\rho}\,\delta a^{2}, \label{eq:einstein_1_chp4} \\ 
		\mathcal{H} \partial_{i}\psi + \partial_{i}\phi' &= -4\pi \bar{\rho} \,a^{2} \delta_{ij}v^{j}, \label{eq:einstein_2_chp4} \\ 
		\phi'' + \mathcal{H}\left(\psi' + 2\phi' \right) &= \frac{1}{3}\partial^{2}(\phi - \psi), \label{eq:einstein_3_chp4} \\ 
		\partial_{\langle i}\partial_{j\rangle} \left(\phi - \psi\right) &= 0, \label{eq:einstein_4_chp4}
	\end{align}	
\end{subequations}
respectively, where we have assumed all perturbations are small such that second-order (and higher) terms can be neglected. Here, $\partial_{i} \equiv \partial / \partial x^{i}$, $\partial^{2}=\partial^{i}\partial_{i}$, $\partial_{\langle i}\partial_{j\rangle}\equiv \partial_{i}\partial_{j} - 1/3\,\delta_{ij}\partial^{2}$, a $'$ represents a derivative with respect to conformal time, and $\mathcal{H}\equiv a'/a$ is the conformal Hubble parameter. Solving these equations, we find
\begin{subequations} \label{eqs:linear_solns}
    \begin{align}
    	\psi &= \phi = f(x^{i}) - \frac{g(x^{i})}{5\,\xi^{5}}, \\
     	\delta &= C_{1}\, \xi^{2}\, \partial^{2}f(x^{i}) - 2 \,f(x^{i}) - C_{2} \,\xi^{-3}\,\partial^{2}g(x^{i}) - \frac{3}{5} \xi^{-5} g(x^{i}), \\
     	v^{i} &= C_{3}\,\xi\, \partial^{i}f(x^{i}) + \frac{3}{10}C_{3}\,\xi^{-4}\, \partial^{i}g(x^{i}),
    \end{align}
\end{subequations}
where $f, g$ are arbitrary functions of spatial position, we introduce the scaled conformal time coordinate
\begin{equation}
	\xi \equiv 1 + \sqrt{\frac{2\pi\rho^{*}}{3\,a_\mathrm{init}}}\eta,
\end{equation}
and we have defined
\begin{equation} 
	C_{1}\equiv \frac{a_{\mathrm{init}}}{4\pi\rho^{*}},\quad C_{2}\equiv \frac{a_{\mathrm{init}}}{20\pi\rho^{*}}, \quad C_{3}\equiv-\sqrt{\frac{a_{\mathrm{init}}}{6\pi\rho^{*}}}.
\end{equation}
Equations \eqref{eqs:linear_solns} contain both a growing and decaying mode for the density and velocity perturbations. We choose $g=0$ to extract only the growing mode of the density perturbation, and our solutions become
\begin{subequations} \label{eqs:linear_solnsg0}
    \begin{align}
    	\psi &= \phi = f(x^{i}), \\
     	\delta &= C_{1}\, \xi^{2}\, \partial^{2}f(x^{i}) - 2 \,f(x^{i}), \\
     	v^{i} &= C_{3}\,\xi\, \partial^{i}f(x^{i}),
    \end{align}
\end{subequations}
implying $\phi'=0$ in the linear regime.

\begin{figure*}
	\begin{centering}
	\includegraphics[width=0.8\textwidth]{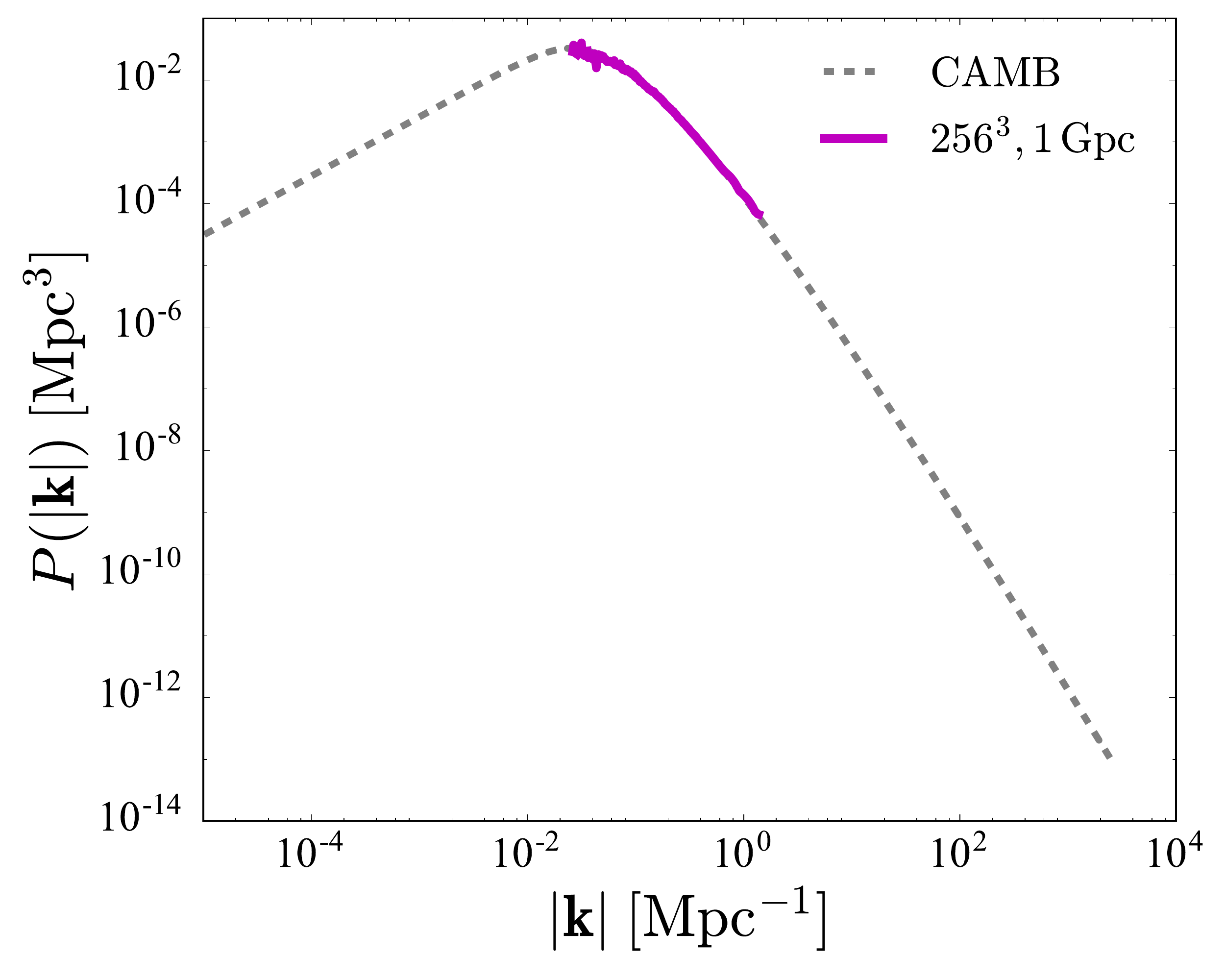}
    \caption{Matter power spectrum of our initial conditions. Grey dashed curve shows the power spectrum produced with the Code for Anisotropies in the Microwave Background (\textsc{CAMB}). We show the power as a function of wavenumber $|\textbf{k}|=\sqrt{k_x^2+k_y^2+k_z^2}$. The magenta curve shows the section of the power spectrum we sample when using a domain size of $L=1$ Gpc with resolution $256^{3}$.}
    \label{fig:powerspectrum}
	\end{centering}
\end{figure*}

\begin{figure*}
	\begin{centering}
	\includegraphics[width=\textwidth]{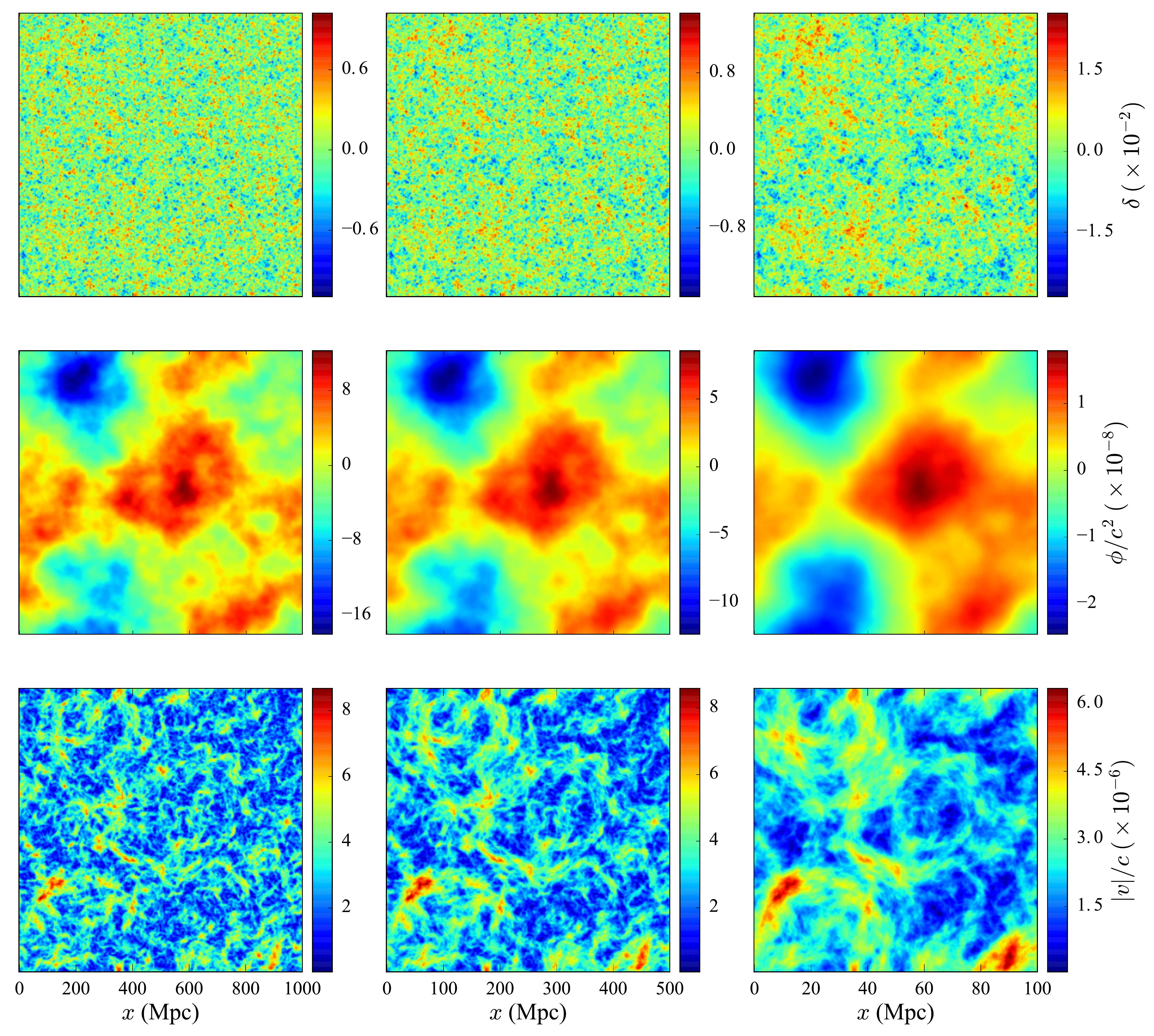}
    \caption{Initial conditions drawn from the cosmic microwave background power spectrum. Here we show initial conditions for the density (top row), metric (middle row), and velocity (bottom row) perturbations for three different physical domain sizes. Left to right shows domain sizes $L=1$ Gpc, 500 Mpc, and 100 Mpc. We show a two-dimensional slice through the midplane of each domain. All initial conditions shown here are at $256^{3}$ resolution, and all quantities are shown in code units -- normalised by the speed of light for the metric and velocity perturbations. The magnitude of the velocity is $|v|=\sqrt{v_x^{2} + v_y^{2} + v_z^{2}}$.}
    \label{fig:256ICs}
	\end{centering}
\end{figure*}
\subsection{Cosmic Microwave Background fluctuations} \label{subsec:cmbfluc}
We use \eqref{eqs:linear_solns} along with the Code for Anisotropies in the Microwave Background \citep[\textsc{CAMB};][]{lewis2002} to generate the matter power spectrum at $z=1100$, with parameters consistent with \citet{planck2016params} as input. Figure~\ref{fig:powerspectrum} shows the matter power spectrum from \textsc{CAMB} (grey curve), as a function of wavenumber $|\textbf{k}|=\sqrt{k_x^{2} + k_y^{2} + k_z^{2}}$. We use the Python module \texttt{c2raytools} \footnote{https://github.com/hjens/c2raytools} to generate a 3-dimensional Gaussian random field drawn from the \textsc{CAMB} power spectrum. This provides the initial density perturbation. The magenta curve in Figure~\ref{fig:powerspectrum} shows the region of the matter power spectrum sampled in our highest resolution ($256^{3}$), largest domain size ($L=1$ Gpc) simulation. The smallest $\textbf{k}$ component sampled represents the largest wavelength of perturbations --- approximately the length of the box, $L$ --- and the largest $\textbf{k}$ component sampled represents the smallest wavelength of perturbations --- two grid points. To relate the initial density perturbation to the corresponding velocity and metric perturbations, we transform \eqref{eqs:linear_solns} into Fourier space. Initially, $\xi=1$ which gives a density perturbation of the form
\begin{equation}
	\delta(\textbf{k}) = -\left( C_1 |\textbf{k}|^{2} + 2 \right) \,\phi(\textbf{k}),
\end{equation}
where $\textbf{k}=(k_x, k_y, k_z)$, so we can define an arbitrary function $\delta(\textbf{k})$, and construct the metric perturbation and velocity, respectively, using

\begin{subequations}\label{eqs:Fspace}
    \begin{align} 
    	\phi(\textbf{k}) &= -\frac{\delta(\textbf{k})}{C_1 |\textbf{k}|^{2} + 2}, \\
    	\textbf{v}(\textbf{k}) &= C_3 \,i\,\textbf{k}\,\phi(\textbf{k}),
    \end{align}
\end{subequations}
where $i^{2}=-1$. With the Fourier transform of the Gaussian random field as $\delta(\textbf{k})$, we calculate the velocity and metric perturbations in Fourier space using \eqref{eqs:Fspace}, and then use an inverse Fourier transform to convert the perturbations to real space. The density perturbation $\delta$ is already dimensionless, and we normalise by the speed of light, $c$, to convert $v^{i}$ and $\phi$ to code units. Figure~\ref{fig:256ICs} shows initial conditions at $256^{3}$ resolution for box sizes $L=1$ Gpc, 500 Mpc, and 100 Mpc in the left to right columns, respectively. The top row shows the density perturbation $\delta$, the middle row shows the normalised metric perturbation $\phi/c^2$, and the bottom row shows the magnitude of the velocity perturbation normalised to the speed of light $|v|/c$. These initial conditions are sufficient to describe a linearly-perturbed FLRW spacetime in \texttt{FLRWSolver}.

We assume a flat FLRW cosmology for the initial instance only. Simulations begin with small perturbations at the CMB, and so the assumption of a linearly-perturbed FLRW spacetime is sufficiently accurate.

\section{Gauge} \label{sec:gauge}

The (3+1) decomposition of Einstein's equations \citep{arnowitt1959} results in the metric
\begin{equation} \label{eq:3p1metric}
	ds^{2} = -\alpha^{2}dt^{2} + \gamma_{ij}\left( dx^{i} + \beta^{i}dt \right)\left(dx^{j} + \beta^{j}dt\right),
\end{equation}
where $\gamma_{ij}$ is the spatial metric, $\alpha$ is the lapse function, $\beta^{i}$ is the shift vector, and $x^{i}$ are the spatial coordinates. The lapse function determines the relationship between proper time and coordinate time from one spatial slice to the next, while the shift vector determines how spatial points are relabelled between slices. In cosmological simulations with numerical relativity the comoving synchronous gauge (geodesic slicing) is a popular choice \citep[e.g.][]{bentivegna2016a,giblin2016a,giblin2016b,giblin2017a,giblin2017b}, which involves fixing $\alpha=1$, $\beta^{i}=0$, and $u^{\mu}=(1,0,0,0)$, or $u^{\mu}=(1/a,0,0,0)$ for conformal time, throughout the simulation. This gauge choice can become problematic at low redshifts when geodesics begin to cross, and can form singularities. 
We choose $\beta^{i}=0$ and evolve the lapse according to the general spacetime foliation
\begin{equation} \label{eq:lapseevol}
	\partial_{t}\alpha = - f(\alpha) \,\alpha^{2} K,
\end{equation}
where $f(\alpha)$ is a positive and arbitrary function, and $K=\gamma^{ij}K_{ij}$ is the trace of the extrinsic curvature.  We choose $f=1/3$, and use the relation from the (3+1) ADM equations \citep{shibata1995}
\begin{equation}
	\partial_{t} \,\mathrm{ln}(\gamma^{1/2}) = -\alpha K,
\end{equation}
where $\gamma$ is the determinant of the spatial metric. Integrating \eqref{eq:lapseevol} gives
\begin{equation}
	\alpha = C(x^{i})\,\gamma^{1/6},
\end{equation}
where $C(x^{i})$ is an arbitrary function of spatial position.

For our initial conditions we have $\gamma_{ij}=a^{2}(1-2\phi)\delta_{ij}$, implying $\gamma^{1/6}=a\,\sqrt{1-2\phi}$. We therefore choose
\begin{equation}
	C(x^{i}) = \frac{\sqrt{1+2\psi}}{\sqrt{1-2\phi}},
\end{equation}
on the initial hypersurface, so that $\alpha=a\,\sqrt{1+2\psi}$, as in the metric \eqref{eq:longitudinal_metric_chp4}.

\section{Averaging scheme} \label{sec:averaging}

We adopt the averaging scheme of \citet{buchert2000a} generalised for an arbitrary coordinate system \citep{larena2009b,brown2009a,brown2009b,clarkson2009,gasperini2010,umeh2011}  \footnote{During the review of this paper, \citet{buchert2018b} raised some concerns regarding the averaging formalism of \citet{larena2009b}. We aim to investigate the proposed alterations in a later work (see Sections~\ref{subsubsec:improved_general_averaging} and \ref{subsec:FW_improved_averaging}).}. The average of a scalar quantity $\psi(x^{i},t)$ is defined as

\begin{equation}\label{eq:avgdef}
	\langle\psi\rangle = \frac{1}{V_{\mathcal{D}}} \int_{\mathcal{D}}\psi \sqrt{\gamma}\;d^{3}x,
\end{equation}
where the average is taken over some domain $\mathcal{D}$ lying within the chosen hypersurface, and $V_{\mathcal{D}} = \int_{\mathcal{D}} \sqrt{\gamma} d^{3}x$
 is the volume of that domain. The normal vector to our averaging hypersurface is $n_{\mu}=(-\alpha,0,0,0)$, corresponding to the four-velocity of observers within our simulations. These  observers are not comoving with the fluid, implying $n_\mu \neq u_{\mu}$, and the tilt between these two vectors results in additional backreaction terms due to nonzero peculiar velocity $v^{i}$. As in \citet{larena2009b,clarkson2009,brown2013}, we define the Hubble expansion of a domain $\mathcal{D}$ to be associated with the expansion of the fluid, $\theta$,
 \begin{equation} \label{eq:hubble_def}
 	\mathcal{H_{D}} \equiv \frac{1}{3}\langle \theta \rangle,
 \end{equation}
 where
 \begin{equation} \label{eq:theta}
 	\theta\equiv h^{\alpha\beta}\nabla_{\alpha}u_{\beta},
\end{equation}
is the projection of the fluid expansion onto the three-surface of averaging, with the projection tensor $h_{\alpha\beta} \equiv g_{\alpha\beta} + n_{\alpha}n_{\beta}$. In our case, this represents the expansion of the fluid as observed in the gravitational rest frame \citep{umeh2011}. 

Averaging Einstein's equations in this frame, with $P=\Lambda=0$, gives the averaged Hamiltonian constraint
\begin{equation} \label{eq:buchert1}
	6\mathcal{H_D}^{2} = 16\pi\langle W^{2}\rho\rangle - \mathcal{R_D} - Q_\mathcal{D} + \mathcal{L_D},
\end{equation}
where $W$ is the Lorentz factor, $\mathcal{R_D}$ is the averaged Ricci curvature scalar, $Q_\mathcal{D}$ is the kinematical backreaction term, and $\mathcal{L_D}$ is the additional backreaction term due to nonzero peculiar velocities in our gauge. For definitions of these terms, see Appendix~\ref{sec:appx_avg}.

\begin{figure*}
	\includegraphics[width=\textwidth]{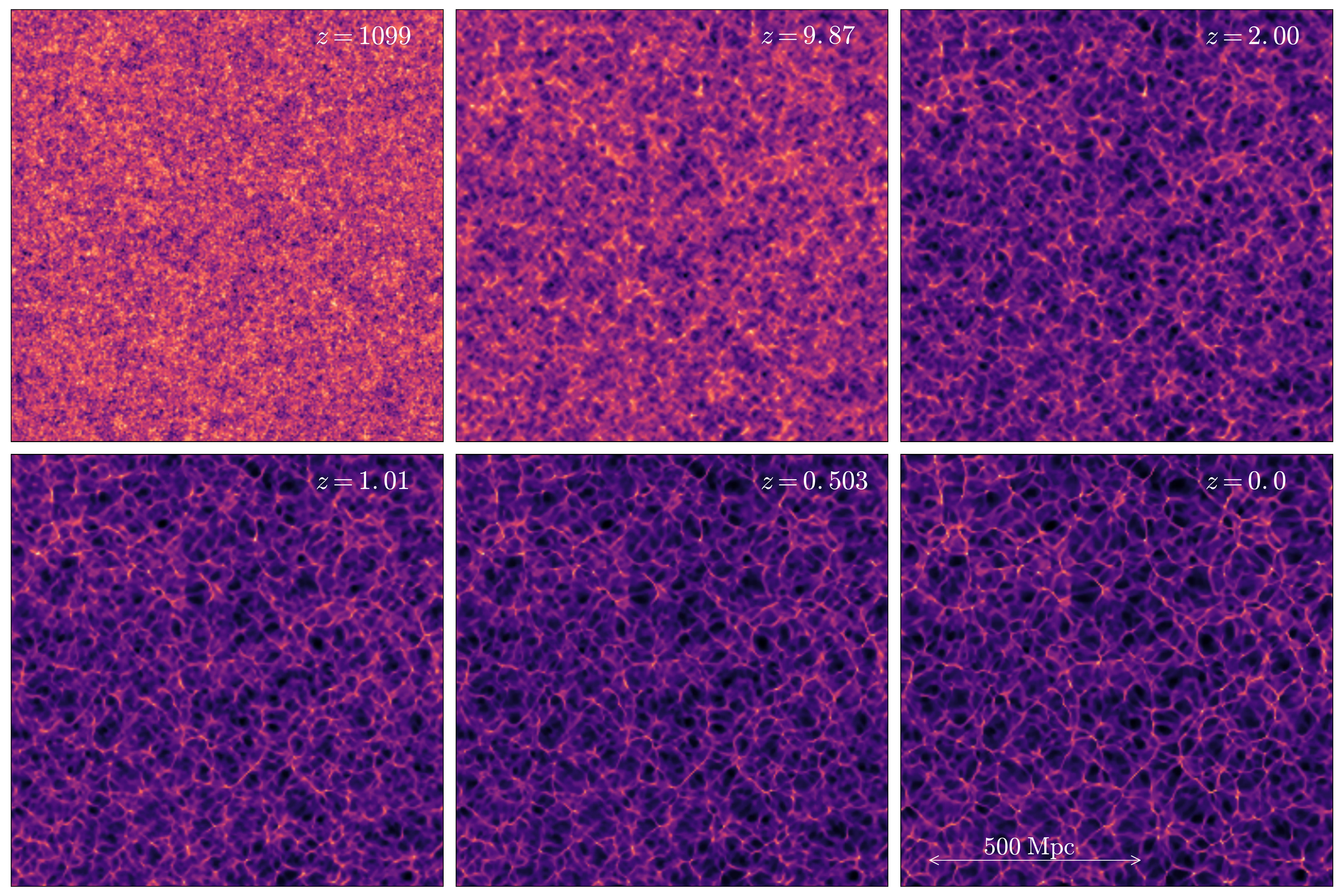}
    \caption{Evolution of a fully General Relativistic cosmic web. Here we show a $256^{3}$ simulation, in an $L=1$ Gpc domain. This simulation has evolved from the cosmic microwave background ($z=1100$; top left) until today ($z=0$; bottom right). Each panel shows a two-dimensional slice of the density perturbation in the midplane of the domain. We can see the familiar web structure of modern cosmological N-body simulations using Newtonian gravity, however this cosmic web contains all of the corresponding General-Relativistic information. The standard deviations of the fractional density perturbation $\delta$ for each panel (progressing in time) are $\sigma_{\delta}=0.0026, 0.15, 0.6, 1.11, 1.89$, and $3.92$, respectively.}
    \label{fig:256evol}
\end{figure*}

We define the effective scale factor, $a_\mathcal{D}$\footnote{The fluid scale factor $\aD$ here is equivalent to $\aDh$ used in Chapter~\ref{Chapter1} and Section~\ref{sec:mescaline}.}, describing the expansion of the fluid, via the Hubble parameter
 \begin{equation} \label{eq:aDdef}
 	\mathcal{H_{D}} = \frac{a'_{\mathcal{D}}}{a_{\mathcal{D}}}.
\end{equation}
This is related to the effective scale factor describing the expansion of the coordinate grid (volume)
\begin{equation} \label{eq:aDVdef}
	a_{\mathcal{D}}^{V}\equiv \frac{V'_\mathcal{D}}{V_\mathcal{D}} = \left(\frac{V_\mathcal{D}(\eta)}{V_\mathcal{D}(\eta_\mathrm{init})}\right)^{1/3},
\end{equation}
via 
\begin{equation}\label{eq:aDaDV}
	a_{\mathcal{D}} = a_{\mathcal{D}}^{V} \;\mathrm{exp}\left(-\frac{1}{3}\int \langle \frac{\alpha}{W}(\theta - \kappa) - \alpha\,\theta\,\rangle \;d\eta\right).
\end{equation}
See Appendix~\ref{sec:appx_expn} for details\footnote{After the publication of this paper, we noticed an error in \eqref{eq:aDaDV} from \citet{larena2009b}. In Appendix~\ref{sec:appx_aD_typo} we show this error makes negligible difference to our results.}.

\subsection{Cosmological parameters}
The dimensionless cosmological parameters describe the content of the Universe. From \eqref{eq:buchert1} we define
\begin{subequations} \label{eqs:cosmic_qrtet}
	\begin{align}
		\Omega_{m} &= \frac{8\pi \langle W^{2}\rho\rangle}{3 \mathcal{H_{D}}^{2}}, \quad \Omega_{R} = -\frac{\mathcal{R_{D}}}{6 \mathcal{H_{D}}^{2}}, \\
		 \Omega_{Q} &= - \frac{Q_{\mathcal{D}}}{6 \mathcal{H_{D}}^{2}}, \quad \Omega_{L} = \frac{\mathcal{L_{D}}}{6\mathcal{H_{D}}^{2}},
	\end{align}
\end{subequations}
giving the Hamiltonian constraint in the form
\begin{equation} \label{eq:content}
	\Omega_{m} + \Omega_{R} + \Omega_{Q} + \Omega_{L} = 1.
\end{equation}
We require this to be satisfied at all times. Here, $\Omega_m$ is the matter energy density, $\Omega_R$ is the curvature energy density, $\Omega_Q+\Omega_L$ is the energy density associated with the backreaction terms; a purely General-Relativistic effect. For a standard $\Lambda$CDM cosmology, these cosmological parameters are $\Omega_m=0.308\pm0.012$, $|\Omega_R|=|\Omega_k|<0.005$, $\Omega_{Q}=0$, and $\Omega_{L}=0$ \citep{planck2016params}.

\begin{figure*}
	\includegraphics[width=\textwidth]{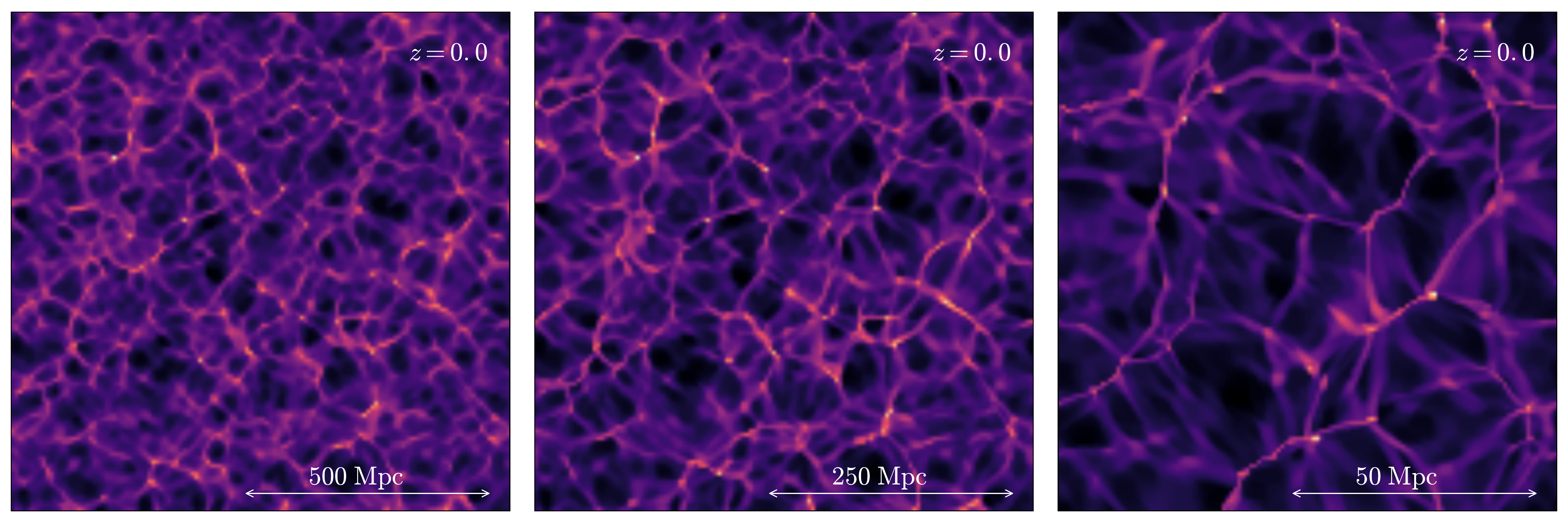}
    \caption{Scale dependence of the cosmic web. Three separate simulations computed at a resolution of $128^{3}$ (left to right) with domain sizes $L=1$ Gpc, 500 Mpc, and 100 Mpc, respectively. All snapshots show a two-dimensional density slice in the midplane of the simulation domain at redshift $z=0$.}
    \label{fig:128allsim_z0}
\end{figure*}
\begin{figure*}
	\includegraphics[width=\textwidth]{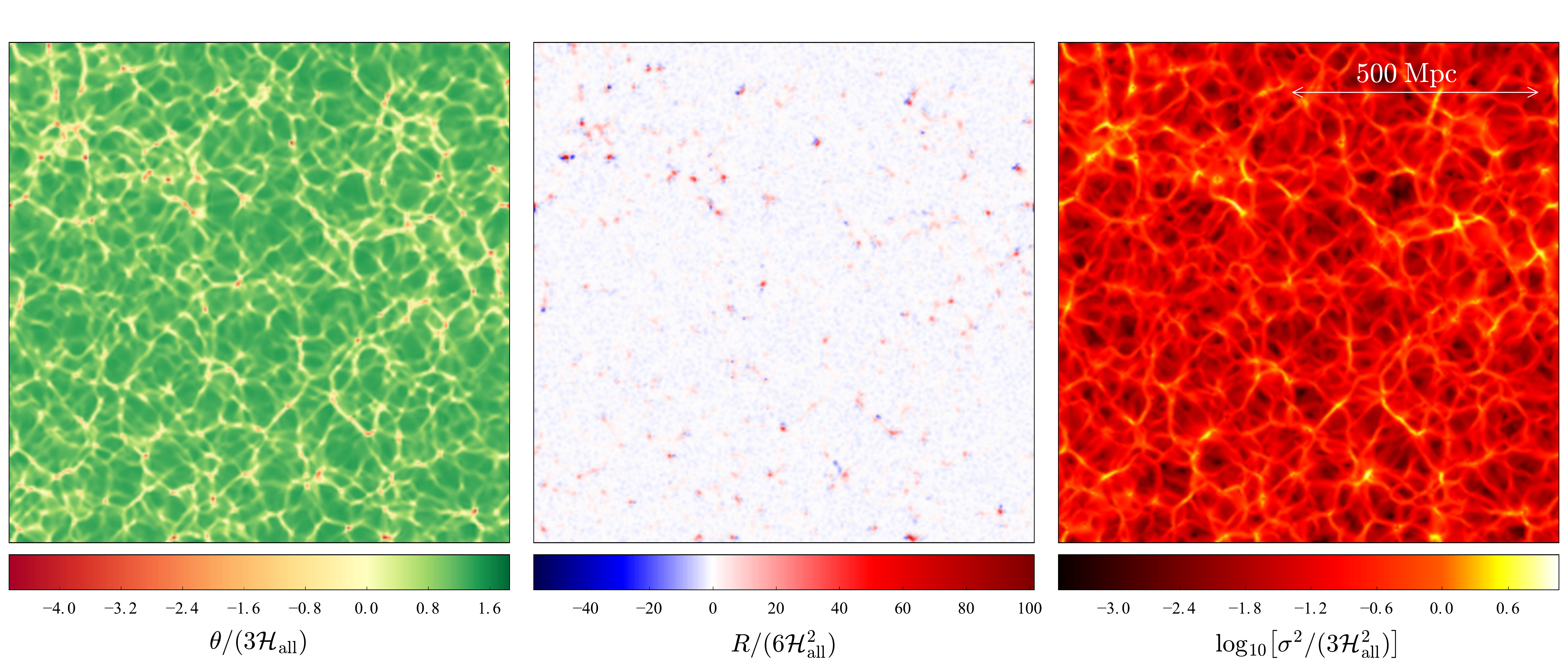} 
    \caption{General-Relativistic attributes of an inhomogeneous, anisotropic universe. Panels (left to right) show the matter expansion rate $\theta$, the spatial Ricci curvature $\mR$, and the shear $\sigma^{2}$, respectively, each relative to the global Hubble expansion $\mathcal{H}_{\rm all}$. Each panel shows a two-dimensional slice at $z=0$ through the midplane of the $L=1$ Gpc domain at $256^{3}$ resolution.}
    \label{fig:GRstuff}
\end{figure*}

\subsection{Post-simulation analysis} \label{subsec:mescaline_chp4}
The Universe is measured to be homogeneous and isotropic on scales larger than $\sim 80-100 h^{-1} \mathrm{Mpc}$ \cite{scrimgeour2012}. Above these scales it is unclear whether the evolution of the average of our inhomogeneous Universe coincides with the FLRW (or $\Lambda$CDM) equivalent. In attempt to address this, we calculate averages over our entire simulation domain, but also over subdomains within the simulation to sample a variety of physical scales. We measure averages over spheres of varying radius $r_{\mathcal{D}}$ embedded in the total volume, from which we calculate the dimensionless cosmological parameters \eqref{eqs:cosmic_qrtet}, the Hubble parameter \eqref{eq:hubble_def}, and consequently the effective matter expansion $a_\mathcal{D}$.

The spatial Ricci tensor $\mR_{ij}$ is the contraction of the Riemann tensor. We calculate this directly from the metric using
\begin{equation} \label{eq:ricci4Ddef}
	\mR_{ij} = \partial_k \Gamma^{k}_{ij} - \partial_j\Gamma^{k}_{ik} + \Gamma^{k}_{lk}\Gamma^{l}_{ij} - \Gamma^{k}_{jl}\Gamma^{l}_{ik},
\end{equation}
where the spatial connection coefficients are
\begin{equation} \label{eq:christoffeldef}
	\Gamma^{k}_{ij} \equiv \frac{1}{2}\gamma^{kl} \left(\partial_i \gamma_{jl} + \partial_j \gamma_{li} - \partial_l \gamma_{ij}\right).
\end{equation}

We use our analysis code \textsc{mescaline}, written to analyse three-dimensional HDF5 data output from our simulations. The code reads in the spatial metric $\gamma_{ij}$, the lapse $\alpha$, the extrinsic curvature $K_{ij}$, the density $\rho$, and the velocity $v^{i}$ from the \textsc{Einstein Toolkit} three-dimensional output. From these quantities we calculate the spatial Ricci tensor $\mR_{ij}$ from the spatial metric, and hence the Ricci scalar via $\mR=\gamma^{ij}\mR_{ij}$. We take the trace of the extrinsic curvature $K=\gamma^{ij}K_{ij}$ and with the set of equations defined in Appendix~\ref{sec:appx_avg} we calculate averages and the resulting backreaction terms. We also use \textsc{mescaline} to calculate the Hamiltonian and momentum constraint violation, discussed in Appendix~\ref{appx:constraints}. We compute derivatives using centred finite difference operators, giving second order accuracy in both space and time, the same order as the \textsc{Einstein Toolkit}'s spatial discretisation.

\section{Results} \label{sec:results}
Figure~\ref{fig:256evol} shows time evolution of a two-dimensional slice of the density $\rho$ through the midplane of the $L=1$ Gpc domain at $256^{3}$ resolution. We show the growth of structures from $z=1100$ (top left) through to $z=0$ (bottom right). The $1\sigma$ variance in $\delta$ evolves from $\sigma_{\delta}=0.0026$ (top left) to $\sigma_{\delta}=3.92$ (bottom right). 

Figure~\ref{fig:128allsim_z0} shows two-dimensional slices through the midplane of three $128^{3}$ resolution simulations with domain size $L=1$ Gpc, 500 Mpc, and 100 Mpc (left to right), at redshift $z=0$. As we sample smaller scales we see a more prominent web structure forming. Our fluid treatment of dark matter implies over-dense regions continue to collapse towards infinite density, rather than forming virialised structures. This should, in general, yield a higher density contrast on small scales than we expect in the Universe.

Figure~\ref{fig:GRstuff} shows (left to right) the matter expansion rate $\theta$, the spatial Ricci curvature $\mR$, and the shear $\sigma^{2}$, respectively, at $z=0$. Each quantity is normalised to the global Hubble expansion $\mathcal{H}_{\rm all}$. The curvature and shear panels are normalised to correspond to the respective density parameters: $\Omega_R$ defined in \eqref{eqs:cosmic_qrtet}, and $\Omega_\sigma = \langle \sigma^2 \rangle / (3 \mathcal{H}_{\rm all}^2)$ defined in \citet{montanari2017}. We calculate $\theta$ using \eqref{eq:theta}, $\sigma^2$ using \eqref{eq:sigma_def} and \eqref{eq:sigma2}, and $R$ using the definitions \eqref{eq:ricci4Ddef} and \eqref{eq:christoffeldef}. Each panel shows a two-dimensional slice through the midpoint of the $L=1$ Gpc domain at $256^{3}$ resolution. Our relativistic quantities can be seen to closely correlate with the density distribution at the same time, shown in the bottom right panel of Figure~\ref{fig:256evol}.

\subsection{Global averages}
\begin{figure*}
	\begin{centering}
	\includegraphics[width=0.95\textwidth]{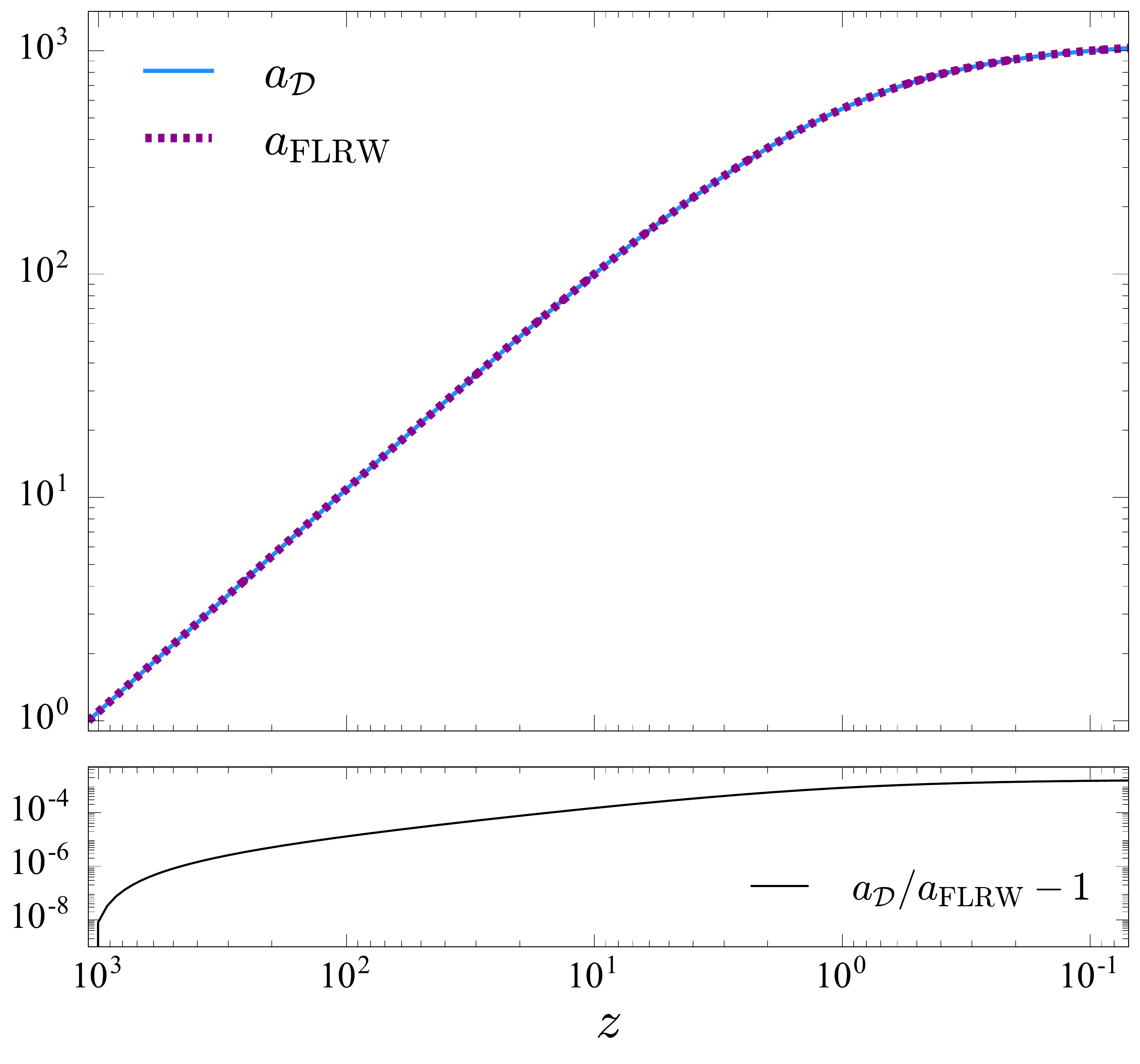}
    \caption{Globally, our expansion coincides with that of FLRW. The blue curve in the top panel shows the effective scale factor $a_\mathcal{D}$, calculated over the entire $L=1$ Gpc domain. The dashed magenta curve shows the equivalent FLRW solution (with $\Omega_m=1$), as a function of redshift. The bottom panel shows the residual error for this $256^3$ resolution calculation.}
    \label{fig:aDaFLRW}
	\end{centering}
\end{figure*}

Figure~\ref{fig:aDaFLRW} shows the global evolution of the effective scale factor, $a_\mathcal{D}$. The blue curve shows $a_\mathcal{D}$ calculated over the whole $L=1$ Gpc, $256^{3}$ resolution domain with \eqref{eq:aDaDV}. The purple dashed curve in the top panel shows the corresponding FLRW solution for the scale factor, $a_\mathrm{FLRW}$, found by solving the Hamiltonian constraint for a flat, matter-dominated, homogeneous, isotropic Universe in the longitudinal gauge, 
\begin{equation}
	\frac{a'}{a} = \sqrt{\frac{8\pi G\bar{\rho}\,a^2}{3}},
\end{equation}
giving the solution \eqref{eq:confa}. The bottom panel of Figure~\ref{fig:aDaFLRW} shows the residual error between the two solutions, which remains below $10^{-3}$ for the evolution to $z=0$.

Analysing the cosmological parameters as an average over the entire simulation domain we find agreement with the corresponding FLRW model in our chosen gauge. Globally, at $z=0$, we find $\Omega_m \approx 1$, $\Omega_R \approx 10^{-8}$, and $\Omega_Q + \Omega_L \approx 10^{-9}$. Systematic errors on these values are discussed in Appendix~\ref{appx:convergence}.

\begin{figure*}
	\includegraphics[width=\textwidth]{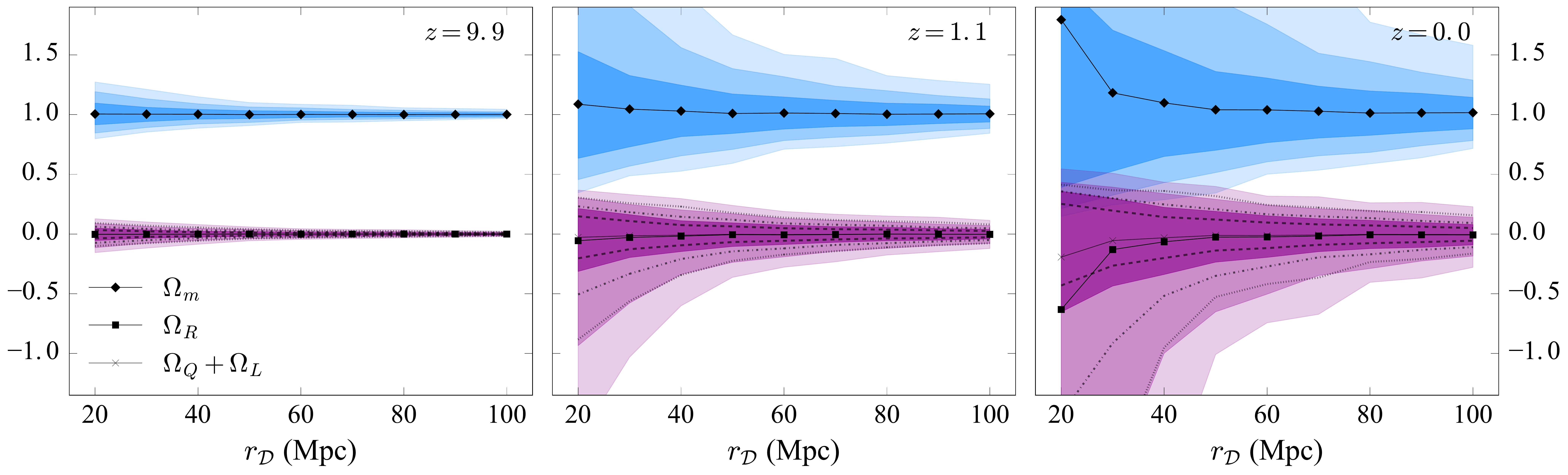}
    \caption{Growing inhomogeneity in matter, curvature, and backreaction. Here we show the cosmological parameters for spheres with various radii $r_{\mathcal{D}}$, randomly placed within an $L=1$ Gpc domain at $256^{3}$ resolution. Black points show mean values over 1000 spheres at each radius, progressively lighter blue and purple shaded regions show the 68\%, 95\%, and 99.7\% confidence intervals for $\Omega_{m}$ and $\Omega_{R}$, respectively. Crosses show the mean contribution from backreaction terms $\Omega_{Q}+\Omega_{L}$, while dashed, dot-dashed, and dotted lines show the 68\%, 95\%, and 99.7\% confidence intervals, respectively. Left to right panels are redshifts $z = 9.9,1.1$, and $0.0$, respectively.}
    \label{fig:omegas_big}
\end{figure*}
\begin{figure*}
	\begin{centering}
	\includegraphics[width=\textwidth]{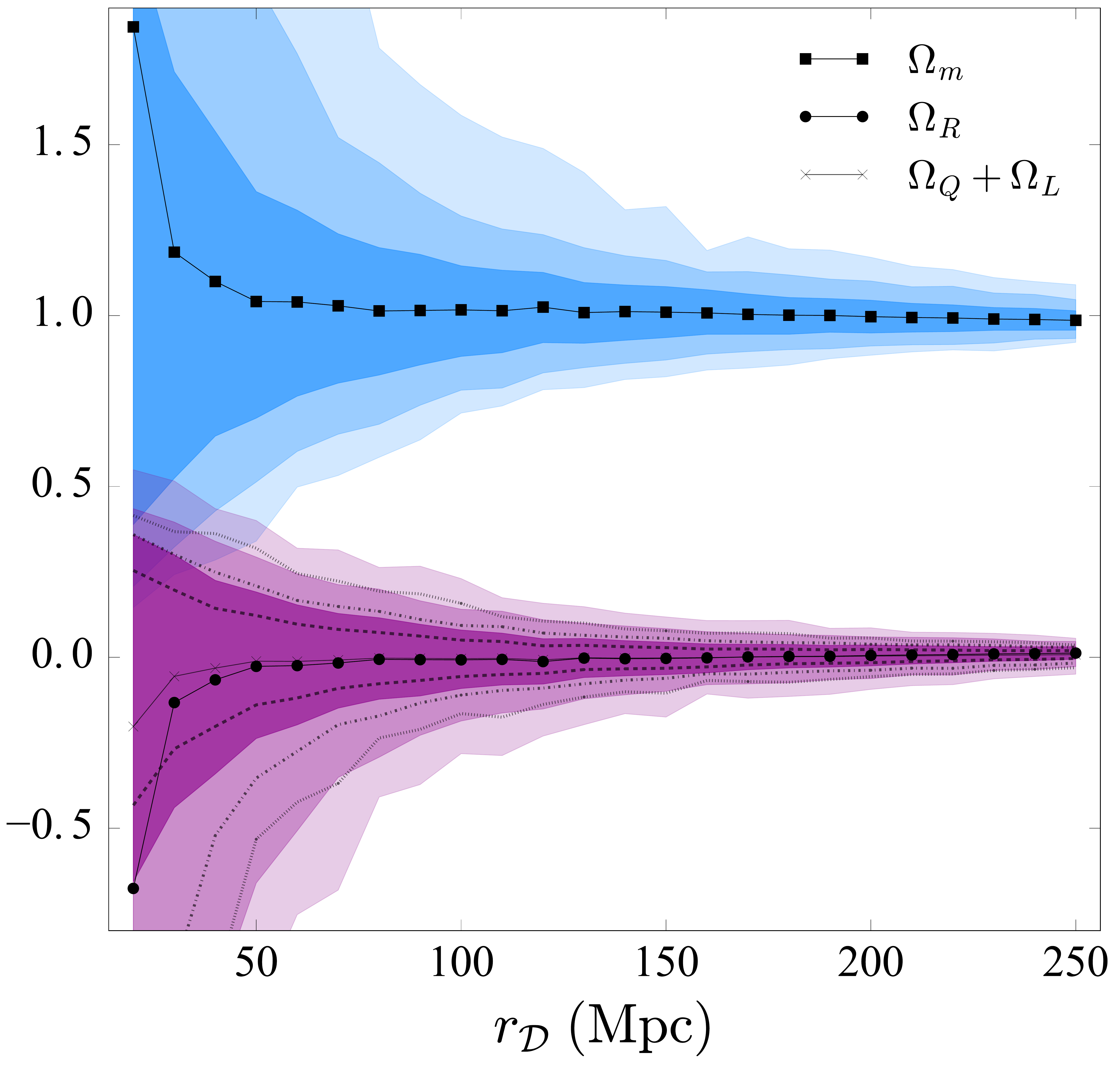}
    \caption{We approach homogeneity when averaging over larger scales. Here we show the right-most panel of Figure~\ref{fig:omegas_big} extending to averaging radius $r_\mathcal{D}=250$ Mpc. Black points show the mean $\Omega_m$, $\Omega_R$, and $\Omega_Q+\Omega_L$ over the 1000 spheres at each radius. Progressively lighter blue and purple shaded regions show the 68\%, 95\%, and 99.7\% confidence intervals for $\Omega_m$ and $\Omega_R$, while dashed, dot-dashed, and dotted lines show these for $\Omega_Q+\Omega_L$.}
    \label{fig:omegas_larger}
	\end{centering}
\end{figure*}
\subsection{Local averages}
\subsubsection{Cosmological parameters}
Figure~\ref{fig:omegas_big} shows cosmological parameters calculated within spheres of various averaging radii, $r_{\mathcal{D}}$, within an $L=1$ Gpc domain at $256^{3}$ resolution. Left to right panels correspond to increasing time (decreasing $z$), showing $z=9.9,1.1$, and $0$, respectively. Black points show the mean value over 1000 spheres at the corresponding averaging radius, showing filled circles for $\Omega_{m}$, filled squares for $\Omega_{R}$, and crosses for $\Omega_{Q} + \Omega_{L}$. Over these 1000 spheres we also show the 68\%, 95\%, and 99.7\% confidence intervals for $\Omega_{m}$ and $\Omega_{R}$ as progressively lighter blue and purple shaded regions, respectively. The same confidence intervals for the contribution from the backreaction terms, $\Omega_{m} + \Omega_{L}$, are shown as dashed, dot-dashed, and dotted lines respectively.  Figure~\ref{fig:omegas_larger} shows the same calculation of the cosmological parameters at $z=0$, extending averaging radii to $r_\mathcal{D}=250$ Mpc. 

At redshift $z=0$, considering averaging radii corresponding to the approximate homogeneity scale of the Universe \citep{scrimgeour2012}, $80<r_\mathcal{D}<100\,h^{-1}$Mpc, we find $\Omega_{m}=1.01\pm0.09 $, $\Omega_{R}=-0.006\pm0.06$, and $\Omega_{Q}+\Omega_{L}=-0.004\pm0.04$. These are the mean values over all spheres with $r_\mathcal{D}=80-100\,h^{-1}$Mpc; 3000 spheres in total. Variations are the 68\% confidence intervals of the distribution.

Below the measured homogeneity scale, with $r_\mathcal{D}<100\,h^{-1}$Mpc, we use 13 individual radii each with a sample of 1000 spheres. We find $\Omega_m=1.1^{+0.12}_{-0.31}$, $\Omega_R=-0.08^{+0.21}_{-0.06}$, and $\Omega_{Q}+\Omega_{L}=-0.03^{+0.11}_{-0.06}$.

Considering scales above this homogeneity scale, we use $100<r_\mathcal{D}<180\,h^{-1}$Mpc with a total of 11 radii sampled and 1000 spheres each. On these scales we find $\Omega_m=0.997\pm0.05$, $\Omega_R=0.005\pm0.03$, and $\Omega_{Q}+\Omega_{L}=0.003\pm0.02$.

Systematic errors in all quoted cosmological parameters here are discussed in Appendix~\ref{appx:convergence}.

\begin{figure*}
	\begin{centering}
	\includegraphics[width=\textwidth]{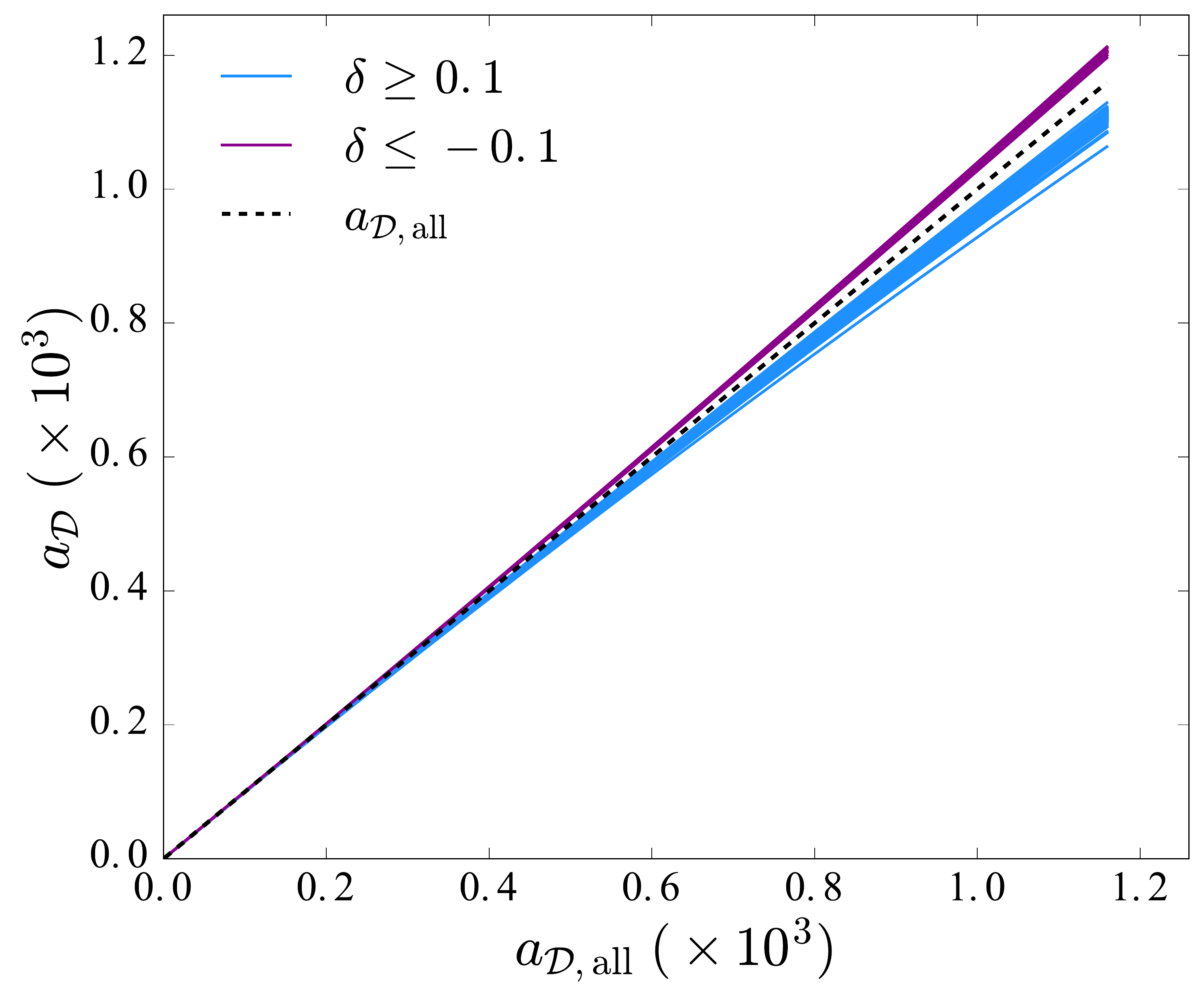}
    \caption{Inhomogeneous expansion as a function of time, showing the effective scale factor $a_\mathcal{D}$ calculated in spheres of radius 100 Mpc as a function of global expansion $a_{\mathcal{D},\mathrm{all}}$. We calculate $a_\mathcal{D}$ in an $L=500$ Mpc simulation at $128^{3}$ resolution. Blue curves show overdense regions with $\delta\geq0.1$, while purple curves show underdense regions with $\delta\leq-0.1$. The black dashed line shows the mean expansion over the whole domain.}
    \label{fig:aDr100}
	\end{centering}
\end{figure*}
\begin{figure*}
	\includegraphics[width=\textwidth]{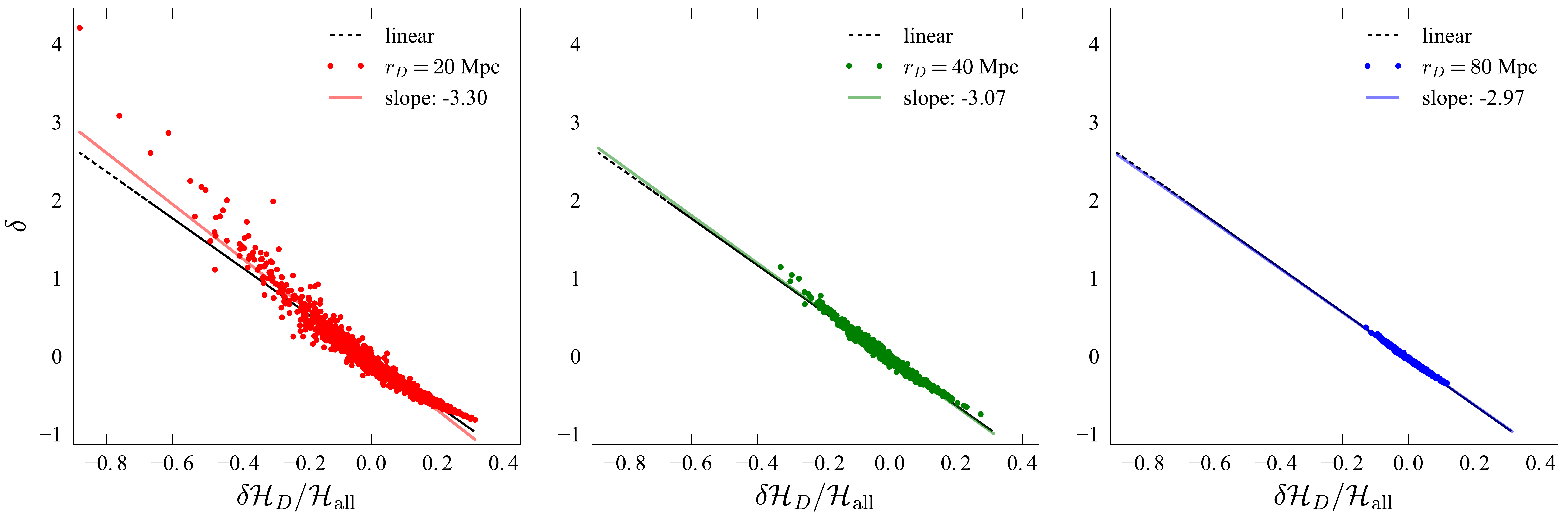}
    \caption{Relation between the fractional density perturbation, $\delta$, and the deviation in the Hubble parameter, $\delta \mathcal{H_D}/\mathcal{H}_\mathrm{all}$, for averaging radii $r_\mathcal{D}=20,40$, and 80 Mpc (left to right), respectively. Points in each panel represent individual spheres of 1000 sampled at each radius, and the solid line of the same colour is the best-fit linear relation, with slope indicated in each panel. The black dashed line is the prediction from linear theory.}
    \label{fig:dHvsdelta}
\end{figure*}

\subsubsection{Scale factor}
Figure~\ref{fig:aDr100} shows the evolution of the effective scale factor calculated within spheres of $r_\mathcal{D}=100$ Mpc, relative to the global value $a_{\mathcal{D},\mathrm{all}}$, which we use as a proxy for time. The dashed line shows the global average, blue curves show $a_\mathcal{D}$ for overdense regions with $\delta\geq0.1$, and purple curves for underdense regions with $\delta\leq-0.1$. In total, we sample 1000 spheres with randomly placed (fixed) origins within an $L=1$ Gpc, $256^{3}$ resolution simulation. Underdense regions with $\delta\leq-0.1$ expand $4-5$\% faster than the mean at $z=0$, while overdense regions with $\delta\geq0.1$ expand $2-8$\% slower.

\subsubsection{Hubble parameter}
Figure~\ref{fig:dHvsdelta} shows the relation between the density, $\delta$, of a spherical domain and the corresponding deviation in the Hubble parameter $\delta H_\mathcal{D}/\bar{H}_\mathcal{D}=(H_\mathcal{D}-\bar{H}_\mathcal{D})/\bar{H}_\mathcal{D}$; the expansion rate of that sphere. We show the density and variation in the Hubble parameter for averaging radii $r_\mathcal{D}=20,40$, and 80 Mpc, (left to right) respectively. Points in each panel show individual measurements within 1000 randomly placed spheres of the same radius. The solid line of the same colour in each panel is the linear best-fit for the data points, with slope indicated in each panel. 

Linear perturbation theory predicts the relation between the average density, $\langle\delta\rangle$, of a spherical perturbation and the deviation from the Hubble flow of that spherical region, $\delta \mathcal{H_D}/\mathcal{H}_\mathrm{all}$, to be \citep{lahav1991}
\begin{equation} \label{eq:lindHvsdelta}
	\langle\delta\rangle = -3\,F\,\frac{\delta \mathcal{H_D}}{\mathcal{H}_\mathrm{all}},
\end{equation}
where $F=\Omega_m^{0.55}$ is the growth rate of matter \citep{linder2005}, which for our global average $\Omega_m\approx1$ is $F=1$. This in turn implies that the growth rate of structures in our simulations is larger than in the $\Lambda$CDM Universe where $\Omega_m\approx0.3$ \citep[e.g.][]{DESCollab2017a,bonvin2017,planck2016params,bennett2013}. The black dashed line in each panel of Figure~\ref{fig:dHvsdelta} is the relation \eqref{eq:lindHvsdelta}, a slope of -3. On 20 Mpc scales the line of best fit is 10\% larger than this prediction, on 40 Mpc scales it is 2.2\% larger, and on 80 Mpc scales is 0.9\% smaller.

\section{Discussion}\label{sec:discussion}
We have presented simulations of nonlinear structure formation with numerical relativity, beginning with initial conditions drawn from the CMB matter power spectrum. These simulations allow us to analyse the effects of large density contrasts on the surrounding spacetime, and consequently on cosmological parameters. We calculate the cosmological parameters $\Omega_m$, $\Omega_R$, $\Omega_Q,$ and $\Omega_L$, together describing the content of the Universe, for spherical subdomains embedded within a 256$^{3}$ resolution, $L=1$ Gpc simulation. We vary the averaging radius between $20\leq r_\mathcal{D}\leq250$ Mpc, representing scales both below and above the measured homogeneity scale of the Universe.

Our results were obtained using simulations sampling the matter power spectrum down to scales of two grid points. Quantifying the errors in such a calculation is difficult because structure formation occurs fastest on small scales, implying different physical structures at different resolutions. This is a known problem in cosmological simulations, not unique to General Relativistic cosmology \citep[see e.g.][]{schneider2016}. To correctly quantify such errors, we must maintain the same density gradients between several simulations at different computational resolution. This becomes difficult when the perturbations themselves are nonlinear. Even with identical initial conditions, we see a different distribution of structures at redshift $z=0$ when sampling nonlinear scales at different resolutions. To approximate the errors on our main results, we instead analyse a set of test simulations in which we simulate a fixed amount of large-scale structure (see Appendix~\ref{appx:convergence}). This allows for a reliable Richardson extrapolation of the solution to approximate the error in our main results at redshift $z=0$.

Regardless of this, the main result of this chapter is that we find $\Omega_m\approx1$ in all simulations we analyse here. Any unquantified errors are unlikely to significantly shift this result,  and all global effects of backreaction and curvature are likely to remain small with an improved sampling of small scales.

\subsection{Global averages}
We find global cosmological parameters consistent with a matter-dominated, flat, homogeneous, isotropic universe, and therefore no global backreaction. 
The evolution of the effective scale factor $a_\mathcal{D}$, evaluated over the whole domain, coincides with the corresponding FLRW model, as shown in Figure~\ref{fig:aDaFLRW}. The $<10^{-3}$ discrepancy between the two solutions does not correlate with the onset of nonlinear structure formation, indicating that this difference is most likely computational error. 

We find a globally flat geometry in our simulations with $\Omega_R \approx 10^{-8}$. This could be a result of our treatment of the matter as a fluid. We cannot create virialised objects and so any ``clusters'' will continue to collapse towards infinite density. In reality, a dark matter halo or galaxy cluster would form, be supported by velocity dispersion, and stop collapsing. The surrounding voids would continue to expand and potentially contribute to a globally negative curvature \citep[see e.g.][]{bolejko2017b,bolejko2018a}. Without a particle description for dark matter alongside numerical relativity we cannot properly capture this effect.

Any contribution from backreaction, $\mathcal{Q_D}$ or $\mathcal{L_D}$, is due to variance in the expansion rate and shear. The left panel of Figure~\ref{fig:GRstuff} shows the matter expansion rate $\theta$, where collapsing regions (yellow, orange, and red) balance the expanding regions (green) due to our treatment of matter. While we see spatial variance in $\theta$, there is no global contribution from backreaction under our assumptions. Therefore, in our chosen gauge and under the caveats described in Section~\ref{sec:caveat} below, backreaction from structure formation is unlikely to explain dark energy.

\subsection{Local averages}
We find strong positive curvature on scales below the homogeneity scale of the Universe. Variations in measured cosmological parameters are up to 31\% based purely on location in an inhomogeneous matter distribution. Our result is similar to that of \citet{bolejko2017b} on small scales, but with larger variance in $\Omega_R$ because of increased small-scale density fluctuations due to our fluid treatment of dark matter.

On the approximate homogeneity scale of the Universe we find mean cosmological parameters consistent with the corresponding FLRW model to $\sim1\%$. Aside from this, we find the parameters can deviate from these mean values by 4-9\% depending on physical location in the simulation domain. This implies that, although on average these coincide with a flat, homogeneous, isotropic Universe, an observers interpretation may differ by up to 9\% based purely on her position in space. 

As we approach larger averaging radii within a 1 Gpc$^3$ volume, we begin to move away from independent spheres, and each sphere begins to overlap with others; effectively sampling the same volume. Due to this, the confidence intervals contract, and eventually at $r_\mathcal{D}\approx400$ Mpc most spheres become indistinguishable from the mean. The beginning of this is evident in Figure~\ref{fig:omegas_larger} as we approach $r_\mathcal{D}=250$ Mpc. This transition appears to be due to overlapping spheres, although could in part be due to the statistical homogeneity of the matter distribution at these scales.

Local observations of type 1a supernovae generally probe scales of $75-450\,h^{-1}$Mpc \citep{wuhuterer2017}. Nearby objects are excluded from the data in an effort to minimise cosmic variance on the result \citep{riess2016,riess2018a,riess2018b}. In this work, we cannot meaningfully sample scales above 250 Mpc because our maximum domain size is only 1 Gpc$^3$. In order to sample all scales used in nearby SN1a surveys, we would need a domain size of $L\gtrsim 10\,h^{-1}$Gpc, with a resolution up to $1024^{3}$. Current computational constraints, and the overhead of numerical relativity, currently restrict us to domain sizes and resolutions used in this work. To address scales as similar as possible to those used in local surveys, we consider $75<r_\mathcal{D}<180\,h^{-1}$Mpc. On these scales we find $\Omega_m=1.002\pm0.06$, $\Omega_R=0.002\pm0.04$, and $\Omega_{Q}+\Omega_{L}=0.001\pm0.02$, where variances are the 68\% confidence intervals due to local inhomogeneity. This implies based on an observers physical location, measured deviations from homogeneity on these scales could be up to 6\%. We expect this variance to decrease when including the full range of observations; including radii up to $450\,h^{-1}$Mpc. We investigate this further, including extrapolation to larger scales, in our companion paper \citet{macpherson2018b}.

While the global effective scale factor demonstrates pure FLRW evolution, we find inhomogeneous expansion within spheres of 100 Mpc radius. Figure~\ref{fig:aDr100} shows the expansion rate differs by $2-8\%$ depending on the relative density of the region sampled. These differences agree with linear perturbation theory, to within 1\%, on $\gtrsim80$ Mpc scales, with smaller scales showing differences of up to 10\%. These differences are most likely due to the nonlinearity of the density field on these scales, although, in addition, could involve General-Relativistic corrections. To properly test this we would require an equivalent Newtonian cosmological simulation to compare this relation at nonlinear scales, which we leave to future work.

\subsection{Caveats}\label{sec:caveat}
\begin{enumerate}
	\item Our treatment of dark matter as a fluid is the main limitation of this work. Under this assumption, we are unable to form bound structures supported from collapse by velocity dispersions. In cosmological N-body simulations, particle methods are adopted so as to capture the formation of galaxy haloes, and local groups of galaxies as bound structures. Adopting a fully General-Relativistic framework in addition to particle methods would allow us to adopt a proper treatment of dark matter in parallel with inhomogeneous expansion. 
	\item We take averages over purely spatial volumes. In reality, an observer would measure her past light cone, and hence the evolving Universe. Our results can thus be considered an upper limit on the variance due to inhomogeneities, since any structures located in the past light cone will be more smoothed out.
	\item Our results are explicitly dependent on the chosen averaging hypersurface. The result of averaging across different hypersurfaces has been investigated \citep{adamek2019a,giblin2018}, and the results can show significant differences. It is clear the physical choice of hypersurface can be important for quantifying the backreaction effect.
	\item We assume $\Lambda=0$, and begin our simulations assuming a flat, matter dominated background cosmology with small perturbations. Throughout the evolution, on a global scale, we find the average $\Omega_m\approx1$; consistent with this model. It is extremely well constrained that our Universe is best described by a matter content $\Omega_m\approx0.3$ \citep[e.g.][]{DESCollab2017a,bonvin2017,planck2016params,bennett2013}. The growth rate of cosmological structures in our simulations will therefore be amplified relative to the $\Lambda$CDM Universe. 
	\item Given our limited spatial resolution, we underestimate the amount of structure compared to the real Universe. In addition, we resolve structures down to scales of two grid points, which means these structures may be under resolved.
\end{enumerate}

\section{Conclusions} \label{sec:conclude}

We summarise our findings as follows:
\begin{enumerate}
	\item We find no global backreaction under our assumptions. Over the entire simulation domain we have $\Omega_m \approx 1$, $\Omega_R \approx 10^{-8}$, and $\Omega_Q + \Omega_L \approx 10^{-9}$, in our chosen gauge; consistent with a matter-dominated, flat, homogeneous, isotropic universe.
	\item We find strong deviation from homogeneity and isotropy on small scales. Below the measured homogeneity scale of the Universe ($r_\mathcal{D}\lesssim 100\,h^{-1}$Mpc) we find deviations in cosmological parameters of $6-31\%$ based purely on an observers physical location.
	\item Above the homogeneity scale of the universe ($100<r_\mathcal{D}<180\,h^{-1}$Mpc) we find mean cosmological parameters coincide with the corresponding FLRW model, with potential $2-5\%$ deviations due to inhomogeneity.
	\item We find agreement with linear perturbation theory within 1\% on $\geq80$ Mpc scales for the relation between the density of a spherical region and its corresponding deviation from the Hubble flow. However, these few percent deviations on smaller scales may prove important in forthcoming cosmological surveys.
\end{enumerate}
While we find no global backreaction in our cosmological simulations, our numerical relativity calculations show significant contributions from curvature and other nonlinear effects on small scales.

\chapter[The Trouble with Hubble]{The Trouble with Hubble: Local versus Global Expansion Rates in Inhomogeneous Cosmological Simulations with Numerical Relativity} %

\label{Chapter5} %

\vspace{10mm}
Published in:\\
\citet{macpherson2018b}. The Astrophysical Journal Letters \textbf{856, L4}.

\section*{Abstract}
In a fully inhomogeneous, anisotropic cosmological simulation performed by solving Einstein's equations with numerical relativity, we find a local measurement of the effective Hubble parameter differs by less than 1\% compared to the global value. This variance is consistent with predictions from Newtonian gravity. We analyse the averaged local expansion rate on scales comparable to Type 1a supernova surveys, and find that local variance cannot resolve the tension between the \citet{riess2018b} and \citet{planck2018a} measurements.

\section*{A note on notation}
We have altered the notation throughout this chapter to be consistent with Chapters~\ref{Chapter1} and \ref{Chapter2}, unless explicitly stated otherwise. For these exceptions, we maintain the notation of the publication for consistency with figures in their published form. Aside from these changes, this chapter is consistent with the accepted version of \citet{macpherson2018b}.

\pagebreak

\section{Introduction} \label{sec:intro_chp5}
Recently, the tension in the locally measured value of the Hubble parameter, $H_0$ \citep{riess2011,riess2016} and that inferred from the cosmic microwave background \citep[CMB;][]{planck2018a} has reached $3.6\sigma$ \citep{riess2018a,riess2018b}. This tension has both motivated the search for extensions to the standard cosmological model, and for the improvement of our understanding of systematic uncertainties \citep[e.g.][]{efstathiou2014,addison2016,dhawan2018a}. The higher local expansion rate \citep{riess2018a,riess2018b} suggests we may live in a void \citep{cusin2017,sundell2015}, consistent with local $\sim20-40\%$ underdensites that have been found in the supernovae Type 1a (SN1a) data \citep{zehavi1998,jha2007,hoscheit2018}. 

In an attempt to address this tension, we perform cosmological simulations of nonlinear structure formation that solve Einstein's equations directly with numerical relativity. In this letter we quantify local fluctuations in the Hubble parameter based purely on physical location in an inhomogeneous, anisotropic universe. Further details of our simulations are given in \citet[][see Chapter~\ref{Chapter4}]{macpherson2019a}, including a quantification of backreaction of inhomogeneities on globally averaged quantities.

Local fluctuations in the expansion rate due to inhomogeneities have been analysed using Newtonian and post-Friedmannian N-body cosmological simulations \citep[e.g.][]{shi1998,wojtak2014,odderskov2014,odderskov2016,adamek2019a}, second-order perturbation theory \citep{ben-dayan2014}, and exact inhomogeneous models \citep[e.g.][]{marra2013}. These approaches predict local fluctuations in the Hubble parameter of up to a few percent. Inhomogeneities have also been proposed to have an effect on the globally measured expansion rate \citep[e.g.][]{buchert2015,roy2011}, with analytical approaches showing this can contribute to an accelerated expansion \citep[e.g.][]{rasanen2006b,rasanen2008,ostrowski2013}. Under the ``silent universe'' approximation, a globally, non-flat geometry has been shown to fully alleviate the Hubble tension \citep{bolejko2017b,bolejko2018b}. These works are important steps towards fully quantifying the effects of inhomogeneities on the Hubble expansion, although simplifying assumptions about the inhomogeneities themselves limit the ability to make a strong statement.

Considering a fully inhomogeneous, anisotropic matter distribution in General Relativity allows us to analyse the effects of inhomogeneities without simplifying the structure of the Universe. Simulations of large-scale structure formation with numerical relativity have been shown to be a viable way to study inhomogeneities \citep{giblin2016a,bentivegna2016a,macpherson2017a,giblin2017a,east2018}, although fluctuations in the Hubble parameter have not yet been considered. In this work we attempt to quantify the discrepancy between local and global expansion rates using cosmological simulations performed without approximating gravity or geometry.

We present our computational and analysis methods in Section~\ref{sec:method_chp5}, and outline our method for calculating the Hubble parameter in Section~\ref{sec:hubble_chp5}. We present results in Section~\ref{sec:results_chp5} and discuss them in Section~\ref{sec:discuss_chp5}.

Redshifts quoted throughout this chapter are based purely on the change in conformal time, and are stated as a guide to the reader, rather than corresponding to an observational measurement. We adopt geometric units with $G=c=1$, unless otherwise stated. Greek indices run from 0 to 3, and Latin indices run from 1 to 3, with repeated indices implying summation.

\section{Method} \label{sec:method_chp5}
We have simulated the growth of large-scale cosmological structures using numerical relativity. Our initial conditions were drawn from temperature fluctuations in the CMB radiation, using the Code for Anisotropies in the Microwave Background \citep[\textsc{CAMB};][]{lewis2002}. The initial density perturbation is a Gaussian random field drawn from the matter power spectrum of the CMB\footnote{To create a Gaussian random field following a particular power spectrum, we use the Python module \texttt{c2raytools}: https://github.com/hjens/c2raytools}, and the corresponding velocity and spacetime perturbations were found using linear perturbation theory. We use the free, open-source \textsc{Einstein Toolkit} along with our thorn \texttt{FLRWSolver} \citep{macpherson2017a} for defining initial perturbations. In a previous paper we benchmarked our computational setup for homogeneous and linearly perturbed cosmological solutions to Einstein's equations, achieving precision within $\sim10^{-6}$ \citep[see][]{macpherson2017a}. We refer the reader to Chapter~\ref{Chapter4} for full details of our computational methods (see also Section~\ref{sec:NR}), including generation of initial conditions and derivations of the appropriate equations (see also Section~\ref{subsec:FLRWSolver}), and details of gauge conditions (see also Section~\ref{subsec:our_code_setup}).

\begin{figure*}
	\includegraphics[width=\textwidth]{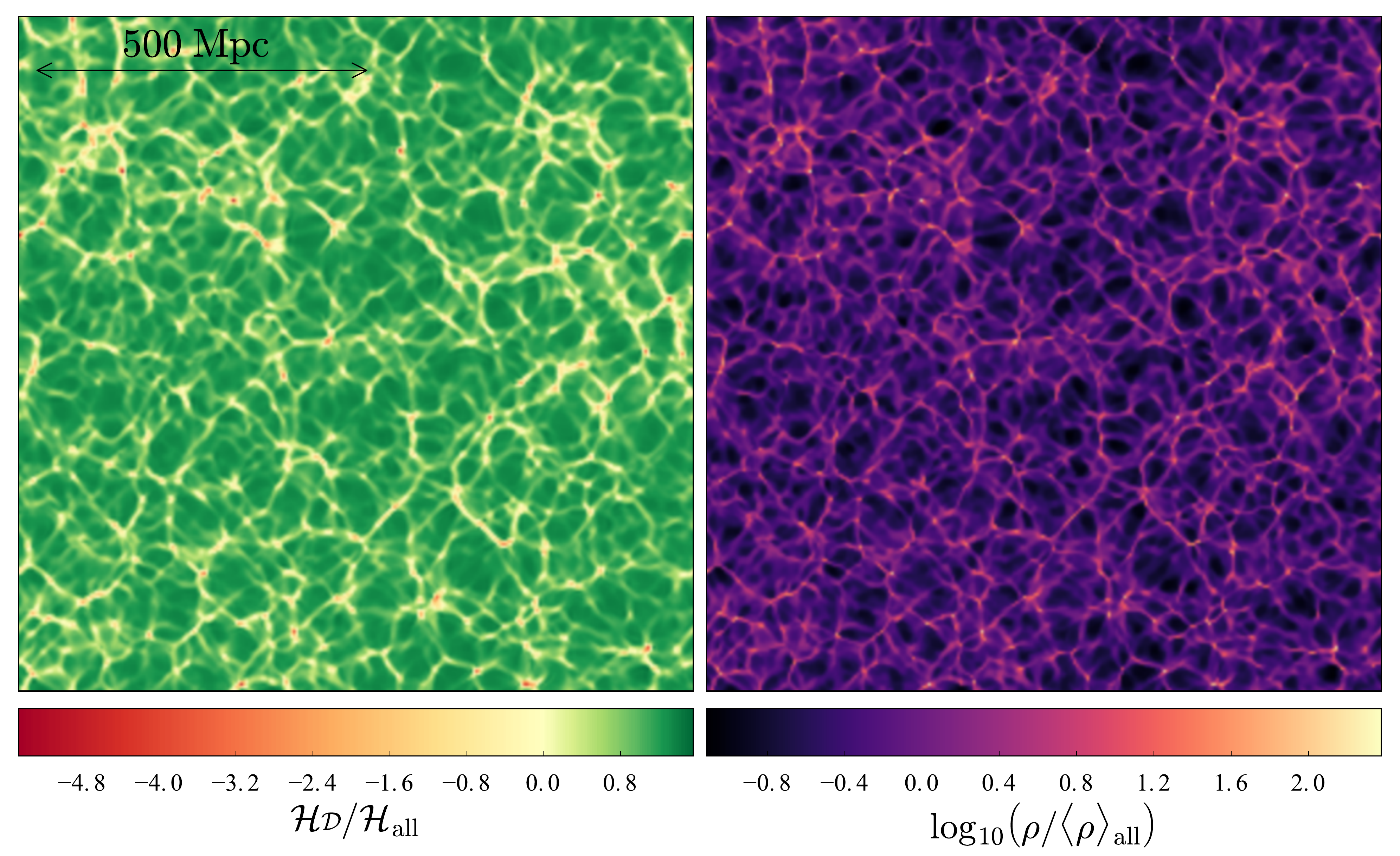}
    \caption{Expansion rate and density of an inhomogeneous, anisotropic universe. Left panel shows the deviation in the Hubble parameter relative to the global mean $\mathcal{H}_\mathrm{all}$. Right panel shows the density distribution relative to the global average, $\langle\rho\rangle_\mathrm{all}$. Both panels show a slice through the midplane of a $256^{3}$ resolution simulation with $L=1$ Gpc.}
    \label{fig:thetarho_chp5}
\end{figure*}

We evolve Einstein's equations in full, with no assumed background cosmology, beginning in the longitudinal gauge from $z=1100$, through to $z=0$. Since we have not yet implemented a cosmological constant in the \textsc{Einstein Toolkit}, we assume $\Lambda=0$, and a matter-dominated ($P\ll\rho$) universe. This implies the age of our model universe will differ from the Universe where $\Lambda\neq0$. We simulate a range of resolutions and domain sizes, detailed in \citet{macpherson2019a}. Here we analyse a $256^{3}$ resolution, $L=1$ Gpc simulation, where the total volume is $L^{3}$. 
Length scales are quoted under the assumption $h=0.704$ \citep[see][]{macpherson2019a}, and we use periodic boundary conditions in all simulations. The right panel of Figure~\ref{fig:thetarho_chp5} shows the density distribution at $z=0$, showing a two-dimensional slice through the midplane of the domain, normalised to the global average density, $\langle\rho\rangle_\mathrm{all}$\footnote{The energy-density $\rho$ used throughout this Chapter is the total \emph{rest-frame} energy-density, and is equivalent to $\rho_R$ used in Chapters~\ref{Chapter1} and \ref{Chapter2}.}. We evolve the matter distribution on a grid, treating dark matter as a fluid. This implies we cannot form virialised structures, and any dense regions will continue to collapse towards infinite density. This is a current limitation of any fully General-Relativistic cosmological simulation, since numerical relativity N-body codes for cosmology currently do not exist\footnote{Since the publication of this Letter, several codes for N-body numerical-relativity cosmology have been developed. See \citet{giblin2018,daverio2019,barrera-hinojosa2019}}.

\subsection{Averaging}
It is common to compare the evolution of global averages in an inhomogeneous, anisotropic universe \citep{buchertehlers1997,buchert2000} to the evolution of a homogeneous, isotropic universe. However, the correct choice of averaging time-slice remains ambiguous due to the presence of nonlinearities. We adopt the averaging scheme of \citet{buchert2000}, generalised to any hypersurface of averaging \citep{larena2009b,brown2009a,brown2009b,clarkson2009,gasperini2010,umeh2011}. The average of a scalar function $\psi$ over a domain $\mathcal{D}$, located within the chosen hypersurface, is
\begin{equation}
	\langle\psi\rangle = \frac{1}{V_{\mathcal{D}}} \int_{\mathcal{D}}\psi \sqrt{\gamma}\;d^{3}x,
\end{equation}
where $V_{\mathcal{D}} = \int_{\mathcal{D}} \sqrt{\gamma} \,d^{3}X$ is the volume of the domain, with $\gamma$ the determinant of the spatial metric $\gamma_{ij}$.  We define our averaging hypersurfaces by observers with four-velocity $n_{\mu}=(-\alpha,0,0,0)$, where $\alpha$ is the lapse function, and we set the shift vector $\beta^{i}=0$. The four-velocity of these observers differs from the four-velocity of the fluid $u^{\mu}\equiv dx^{\mu}/d\tau$, where $\tau$ is the proper time.

\subsection{Measuring the Hubble parameter} \label{sec:hubble_chp5}
The local expansion rate of the fluid projected onto our averaging hypersurface is
\begin{equation} \label{eq:theta_chp5}
	\theta\equiv \gam^{\mu\nu}\nabla_{\mu}u_{\nu},
\end{equation}
where $\gam_{\mu\nu}\equiv g_{\mu\nu} + n_{\mu}n_{\nu}$, and $\nabla_{\mu}$ is the covariant derivative associated with the metric tensor $g_{\mu\nu}$. 
We define the effective Hubble parameter in a domain $\mathcal{D}$ to be
\begin{equation} \label{eq:hubbledef_chp5}
	\mathcal{H_D} \equiv \frac{1}{3}\langle\theta\rangle.
\end{equation}
In a Friedmann-Lema\^{i}tre-Robertson-Walker spacetime, \eqref{eq:hubbledef_chp5} reduces to the usual conformal Hubble parameter $\mathcal{H}=a'/a$, where $'$ represents a derivative with respect to conformal time. 

The local expansion rate is not necessarily what the observer measures. Observations of SN1a \citep{riess2018a,riess2018b} measure the distance-redshift relation, and it is unclear how this relates to the local expansion rate. Recreating what an observer measures in an inhomogeneous Universe ultimately requires ray tracing \citep[see][]{giblin2016b,east2018}, which we leave to future work. 

\subsection{Averaging in subdomains}
In order to quantify $\mathcal{H_D}$ on different physical scales, we calculate averages over spherical subdomains placed randomly within the volume shown in Figure~\ref{fig:thetarho_chp5}. This allows us to analyse the effect of inhomogeneities independent of boundary effects. We calculate $\theta$ for each grid cell, and calculate $\mathcal{H_D}$ by averaging over subdomains of various radii $r_\mathcal{D}$.

Observations of SN1a in the local universe span a redshift range of $0.023\lesssim z\lesssim0.15$ \citep{riess2011,riess2016,riess2018a,riess2018b}, corresponding to distances of $75\lesssim r_\mathcal{D} \lesssim 450\,h^{-1}$ Mpc \citep{wuhuterer2017,odderskov2014}. Local SN1a with $z\lesssim0.023$ are excluded from the analysis in attempt to minimise cosmic variance; their inclusion results in a 3\% higher $H_0$, suggesting we are located in a void \citep{jha2007}. 

\begin{figure}
	\includegraphics[width=\columnwidth]{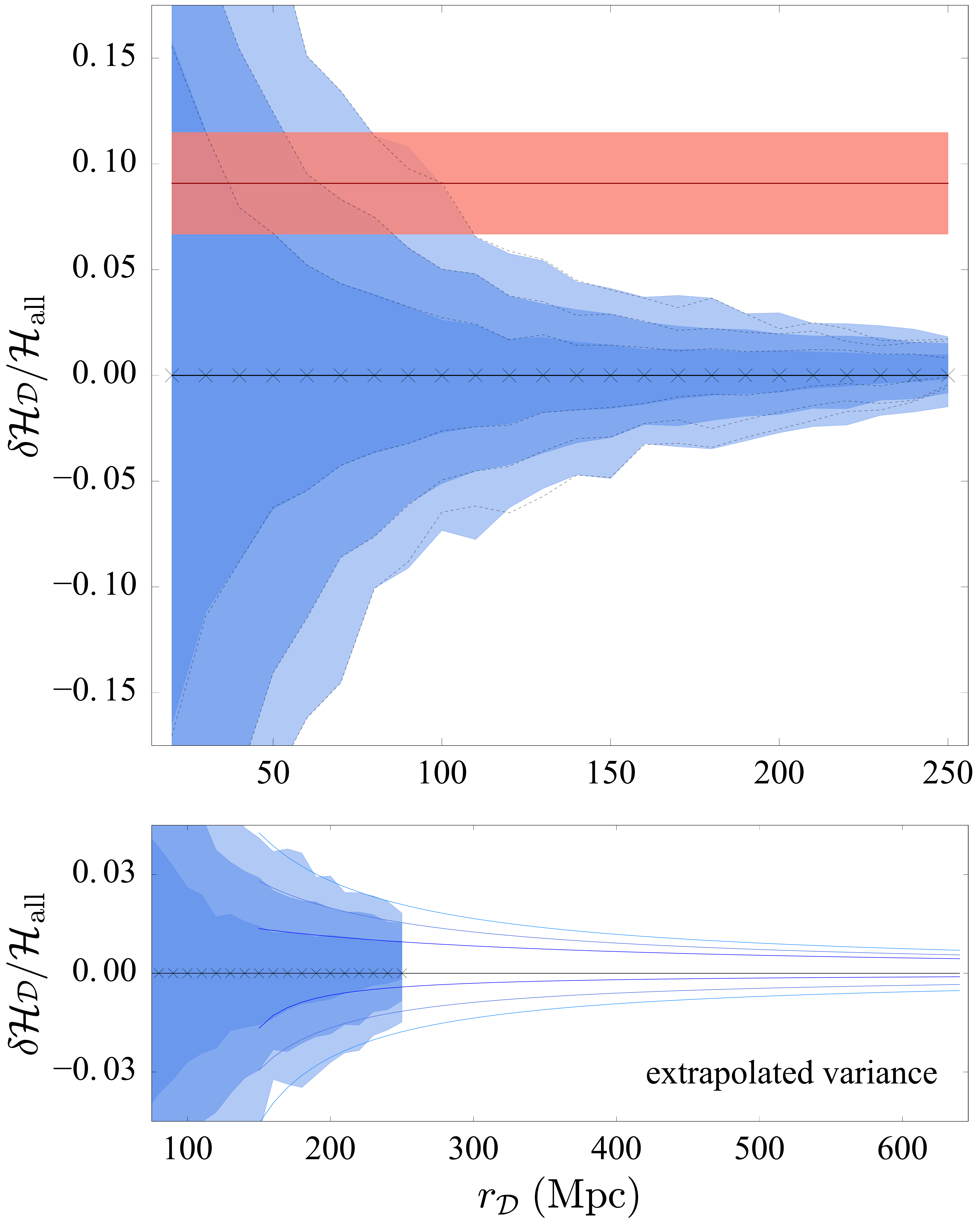}
    \caption{A General-Relativistic measurement of $\mathcal{H_D}$ at $z=0$. Top panel is the fractional deviation measured in any one sphere from the average over the whole domain, $\mathcal{H}_\mathrm{all}$, as a function of averaging radius $r_{\mathcal{D}}$. Progressively lighter blue shaded regions are the 68\%, 95\% and 99.7\% confidence intervals, respectively. The red line is the measurement from \citet{riess2018b}, and the shaded region represents the 1$\sigma$ uncertainty. Dashed curves represent 68\%, 95\%, and 99.7\% confidence intervals for the same sample of spheres weighted as a function of redshift in accordance with the SN1a sample used in \citet{riess2018a,riess2018b} \citep{wuhuterer2017,camarena2018}. Bottom panel shows the variance extrapolated to the full sample range \citep{riess2018a,riess2018b}. Progressively lighter blue curves are the extension of the 68\%, 95\%, and 99.7\% confidence intervals, respectively.}
    \label{fig:hubble_chp5}
\end{figure}

We approximate a measurement of the Hubble expansion using SN1a by calculating the average local expansion rate over a variety of scales. We sample spherical regions with radii up to $r_\mathcal{D}=250$ Mpc to ensure individual spheres are sufficiently independent within our $L=1$ Gpc domain. We therefore calculate $\mathcal{H_D}$ on scales $75<r_\mathcal{D}<180\,h^{-1}$Mpc, corresponding to an effective survey range of $0.023\lesssim z\lesssim0.06$. The reduced range is due to the computational overhead of numerical relativity currently limiting us to domain sizes and resolutions of this order. We extrapolate to $r_\mathcal{D}=450\,h^{-1}$ Mpc to estimate the variance over the full range adopted in \citet{riess2018a,riess2018b}. We perform this extrapolation by fitting a function of the form $\delta\mathcal{H_D}/\mathcal{H}_\mathrm{all}\propto 1/r_\mathcal{D}$ using our calculated variance at $r_\mathcal{D}\geq150$ Mpc, to minimise the effect of small-scale fluctuations (see lower panel of Figure~\ref{fig:hubble_chp5}). To properly test the full range of observations, a larger simulation volume and resolution would be required. 

\section{Results} \label{sec:results_chp5}
The left panel of Figure~\ref{fig:thetarho_chp5} shows deviations in the Hubble parameter, relative to the global mean $\mathcal{H}_\mathrm{all}$, at $z=0$. We show a two-dimensional slice through the midplane of the $L=1$ Gpc domain. Green regions are expanding ($\theta>0$), while yellow to red regions are collapsing ($\theta<0$). This expansion is strongly correlated with the density field shown in the right panel, which displays filaments, voids, knots, and clusters. Due to our fluid treatment of dark matter, collapsing regions will continue to do so towards infinite density, implying all regions in the left panel of Figure~\ref{fig:thetarho_chp5} will average to the corresponding homogeneous expansion.

The top panel of Figure~\ref{fig:hubble_chp5} shows the deviation in the Hubble parameter as a function of averaging radius $r_\mathcal{D}$. Crosses represent the radii at which our calculations were done, and progressively lighter blue shaded regions represent the 65\%, 98\%, and 99.7\% confidence intervals over 1000 randomly placed spheres with the corresponding radius $r_\mathcal{D}$. The red line and shaded region show the mean and $1\sigma$ deviation of the \citet{riess2018b} measurement from the \citet{planck2018a} measurement, respectively. The bottom panel of Figure~\ref{fig:hubble_chp5} shows the 68\%, 95\%,and 99.7\% confidence contours (dark to light blue curves, respectively) extrapolated to the full redshift range used in \citet{riess2018a,riess2018b}. 

Considering our averaging spheres as a survey volume including SN1a at redshifts $0.023\lesssim z\lesssim 0.06$, and assuming an isotropic distribution of objects across the sky with equal numbers of SN1a at all redshift, we estimate the expected variance in a local $H_0$ measurement due to inhomogeneities as the variance in $\mathcal{H_D}$. We calculate the $\pm1\sigma$ variance in a measurement as the $84^{\mathrm{th}}$ and $16^{\mathrm{th}}$ percentiles of the full distribution of spheres sampled over the effective survey range, and similarly for the $2-3\sigma$ variance. Sampling all scales in the top panel of Figure~\ref{fig:hubble_chp5}, including local SN1a with $z\lesssim0.023$, results in a $1\sigma$ variance of $\pm\,2.1\%$. Excluding these local SN1a the variance drops to (+1.2,-1.1)\%. 
We extrapolate to the full survey range $0.023\lesssim z\lesssim 0.15$ (bottom panel of Figure~\ref{fig:hubble_chp5}) by fitting a function $\delta\mathcal{H_D}/\mathcal{H}_\mathrm{all}\propto 1/r_\mathcal{D}$ to each confidence contour in Figure~\ref{fig:hubble_chp5}. While not intended to be a precise measure of the variance at large scales, we estimate a $1\sigma$ variance of (+0.8,-0.4)\%.

\begin{figure*}
	\includegraphics[width=\textwidth]{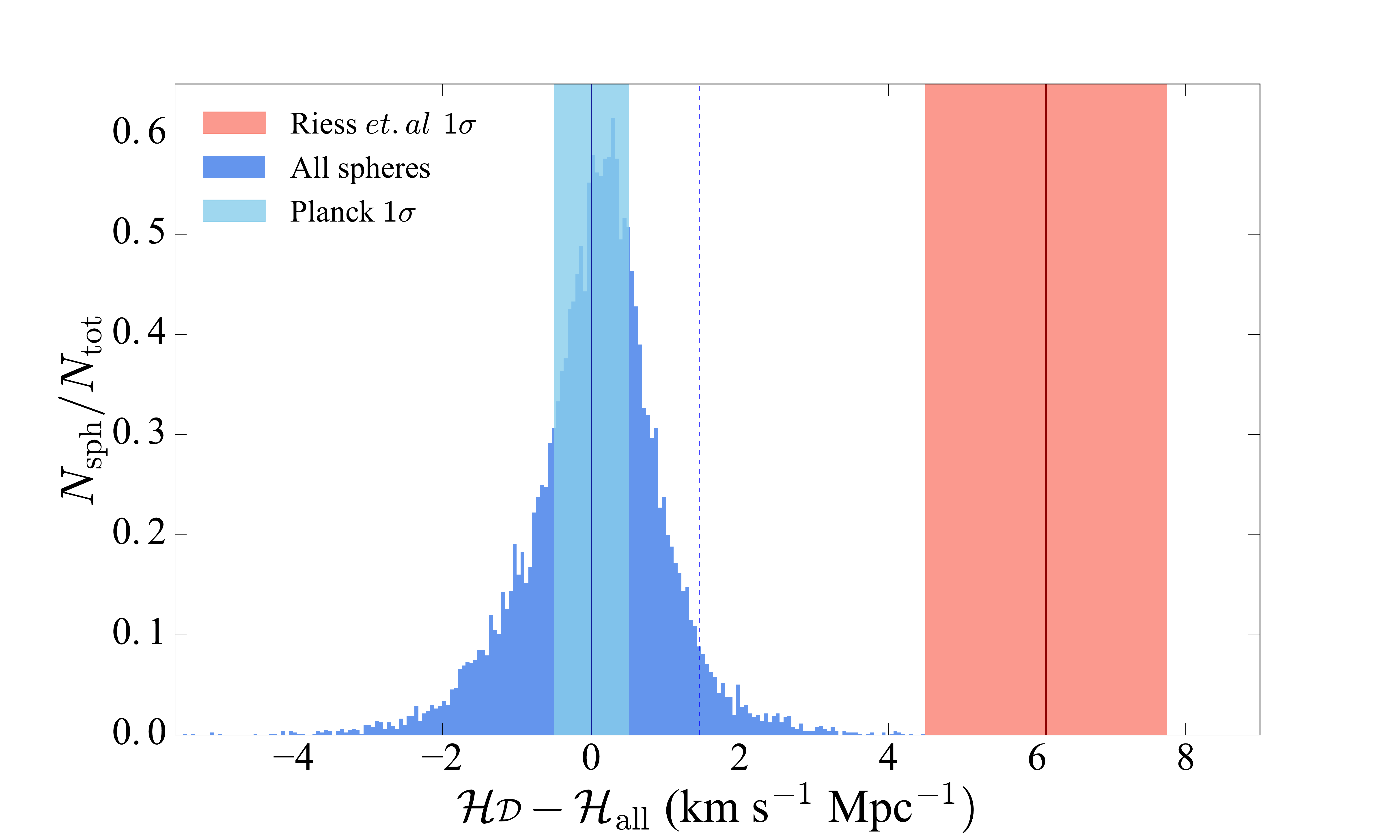}
    \caption{Local deviations in the Hubble parameter due to inhomogeneities. We show the full distribution of all spheres in the range $75< r_\mathcal{D} <180\,h^{-1}$ Mpc in blue. The dashed blue lines represent the $1\sigma$ deviation of the inhomogeneous distribution. The blue shaded region represents the $1\sigma$ uncertainties on the \citet{planck2016params} measurement, while the solid red line and shaded region represent the mean and $1\sigma$ deviation in the \citet{riess2018b} measurement, respectively.}
    \label{fig:hubble_dist_chp5}
\end{figure*}

The blue distribution in Figure~\ref{fig:hubble_dist_chp5} shows the local deviation in the Hubble parameter relative to the global mean, versus the fraction of total spheres with that deviation, $N_\mathrm{sph}/N_\mathrm{tot}$. We show the full sample of spheres in the range $0.023\lesssim z\lesssim0.06$, with the corresponding $1\sigma$ variations shown as dashed lines. The blue line and shaded region represent the \citet{planck2018a} measurement and $1\sigma$ uncertainties, respectively, while the red line and shaded region shows the \citet{riess2018b} measurement and the $1\sigma$ uncertainties, respectively. 

The Supercal SN1a compilation \citep{scolnic2015}, used by \citet{riess2016}, does not contain equal numbers of SN1a at all redshifts; a larger number of objects are sampled at low redshifts. Weighting our results in line with the redshift distribution of the sample \citep[as shown in][]{wuhuterer2017,camarena2018} we find the variance in the Hubble parameter increases to $(+1.5,-1.6)\%$ over our reduced redshift range. Dashed curves in the top panel of Figure~\ref{fig:hubble_chp5} show the variance as a function of averaging radius for the weighted sample. We proceed using the weighted sample for further analysis.

Extending to the $3\sigma$ variance over $0.023\lesssim z\lesssim 0.06$ we find a local Hubble constant can be up to 6.2\% larger than the mean. Taking the \citet{planck2018a} measurement of $67.4\pm 0.5$ km s$^{-1}$Mpc$^{-1}$ as the global mean expansion rate, this implies that if an observers position in the cosmic web is relatively underdense, she may measure a Hubble parameter up to $4.2$ km s$^{-1}$Mpc$^{-1}$ larger. Hence a local measurement using SN1a could reach $H_0=71.6 \pm 1.62$ km s$^{-1}$Mpc$^{-1}$, assuming the same statistical uncertainties as \citet{riess2018b}. This measurement would then be in $2.5\sigma$ tension with \citet{planck2018a}. 

In order to completely resolve the tension between a local measurement and the global value, we must restrict our sample range to $60<r_\mathcal{D}<180\,h^{-1}$ Mpc, or $0.02\lesssim z\lesssim0.06$. Over these scales, our $3\sigma$ variance in the Hubble parameter implies a local $H_0$ measurement could be up to 8.7\%, or $5.9$ km s$^{-1}$Mpc$^{-1}$, larger than the global expansion. Again taking the \citet{planck2018a} value as the global expansion, a local measurement could reach $H_0=73.3 \pm 1.62$ km s$^{-1}$Mpc$^{-1}$ purely based on the observers location in an inhomogeneous universe. This is consistent with the \citet{riess2018b} measurement within $1\sigma$.

\section{Discussion}\label{sec:discuss_chp5}

The variance in the effective Hubble parameter shown in Figure~\ref{fig:hubble_chp5} cannot resolve the tension between the \citet{planck2018a} and \citet{riess2018b} measurements. Excluding local SN1a with $z\lesssim0.023$ we find the variance in the Hubble parameter due to inhomogeneities is (+1.5,-1.6)\% over a reduced redshift range. We find an observer can only measure a local Hubble parameter up to 8.7\% higher than the global value when \textit{further} reducing the survey range to $0.02\lesssim z\lesssim0.06$. The restricted range required for such a measurement emphasises that it is unlikely to completely resolve the tension by local variance in expansion rate. Extrapolating our results to the full survey range results in an expected variance below percent-level, however, as the precision of cosmological surveys continues to improve, variations of this size can be significant.

In Chapter~\ref{Chapter4} and \citet{macpherson2019a}, we analysed the effects of inhomogeneities on globally averaged quantities. We found global averages coincide with the equivalent homogeneous, isotropic model, with negligible backreaction effects on the global expansion. These results are subject to several caveats, which we outline below.

In our simulations we treat dark matter as a fluid, implying we cannot form virialised structures. Any structures that should have formed dark matter haloes will continue to collapse to a single point, eventually growing towards infinite density. Ideally, a particle method would be used for simulating dark matter as dust. We cannot directly compare our simulations to Newtonian N-body simulations due to this difference, in addition to gauge differences, however we can check for consistency of results. On scales $r_\mathcal{D}=50, 75$ and $100\,h^{-1}$Mpc we find variations of $\pm\,4.3\%$, $\pm\,2.4\%,$ and $(+1.1,-0.6)\%$, respectively. These are consistent with Newtonian predictions, also sampling observers randomly located in space, from \citet{wojtak2014} and \citet{odderskov2016} to within $\lesssim1\%$. However, to address whether this difference is due to General-Relativistic effects or computational differences, we ultimately require a particle treatment of dark matter alongside numerical relativity.

Our results may be considered an upper limit for the variance in the Hubble parameter over the scales we sample for several reasons. We assume averages over a purely spatial volume, when in reality an observer would measure their past light cone. As we look back in time, structures are more smoothed out, which would reduce the overall variance. In addition, we evolve our simulations assuming $\Lambda=0$; a matter-dominated universe at the initial instance. We do not fix $\Omega_m=1$ over the course of the simulation, however, globally we find $\Omega_m=1$ to within computational error for all time \citep{macpherson2019a}. This implies the growth rate, $f$, of structures in our simulation will be larger than in $\Lambda$CDM, since $f=\Omega_m^{0.55}$ \citep{linder2005}, resulting in a larger density contrast in general. This will also increase our variance in the Hubble parameter relative to that measured in the Universe where $\Omega_m\approx0.3$ is well constrained \citep[e.g.][]{DESCollab2017a,bonvin2017,planck2018a,bennett2013}.

The effects of inhomogeneities can be dependent on the choice of observers. \citet{adamek2019a} used weak-field relativistic N-body simulations to study variance in the Hubble parameter in the comoving synchronous gauge and the Poisson gauge. In the comoving gauge the variance in the Hubble parameter reached 10\% at $z=0$, while the Poisson gauge remained below 0.01\%. A direct comparison to this work is not possible due to different definitions of the local expansion, however it outlines the importance of carefully choosing the averaging hypersurface. The comoving gauge is often used to represent observers on Earth, however this gauge breaks down at low redshifts due to shell crossings, and so it has been suggested the Poisson gauge --- similar to the gauge used here --- is better suited to study the effects of inhomogeneities in the nonlinear regime with simulations \citep{adamek2019a}.

\section{Conclusions}
We have investigated the effects of inhomogeneities on local measurements of the Hubble parameter. Using numerical relativity we have simulated the growth of density fluctuations drawn from the CMB through to $z=0$. We have calculated the expansion rate of dark matter within randomly placed spheres of various radii from a $256^{3}$ resolution simulation with domain size $L=1$ Gpc. Our conclusions are:
\begin{enumerate}
	\item We measure a (+1.5,-1.6)\% variance in the local expansion rate due to inhomogeneities over $0.023\lesssim z\lesssim0.06$ with a weighted sample of averaging spheres. 
	\item Estimating an extension to our results over $0.023\lesssim z\lesssim0.15$ reduces the variance to (+0.8,-0.4)\%. This is consistent with predictions from Newtonian N-body simulations. 
	\item Our $3\sigma$ variance in the Hubble parameter of 6.2\%, over $0.023\lesssim z\lesssim0.06$, could reduce the tension between a local and global measurement to $2.5\sigma$.
	\item When restricting the survey range to include more nearby SN1a, the tension is resolved. Over scales $0.02\lesssim z\lesssim0.06$, a local calculation of $\mathcal{H_D}$ can be up to 8.7\% larger than the global value. However, since the \citet{riess2018a,riess2018b} measurement considers a significantly wider survey range, we conclude that the tension cannot be explained by local inhomogeneities under our assumptions.
\end{enumerate}

\chapter{Conclusions} %

\label{Chapter6}

In this thesis, we have presented simulations of cosmological structure formation that solve Einstein's equations of GR directly. In Chapter~\ref{Chapter1} we outlined the current status of cosmological theory and observations, including the standard cosmological model; the \lcdm\, model, cosmological perturbation theory, and proposed extensions to \lcdm\, based on some current tensions with observational data. In Chapter~\ref{Chapter2} we derived the 3+1 foliation of Einstein's equations, specifically the BSSN formalism, for evolving arbitrary spacetimes numerically. We also discussed several common coordinate choices and an improvement to the BSSN formalism in terms of constraint violation management via the CCZ4 formalism. We then described the \textsc{Einstein Toolkit}; the computational framework used for the numerical-relativity simulations presented in this thesis, along with our initial-condition thorn \flrwsolver, and detailed \mesc; the post-processing analysis code used to extract our results from these simulations. 

In this final chapter we summarise the main findings from Chapters~\ref{Chapter3}, \ref{Chapter4}, and \ref{Chapter5}, and suggest directions for future work to further investigate and build on these results.

\section{Summary}

In Chapter~\ref{Chapter3} we presented two important code tests to ensure the validity of our computational setup. We initialised a homogeneous, isotropic FLRW metric with \flrwsolver, and evolved the flat, dust spacetime with numerical relativity using the \textsc{Einstein Toolkit}. We evolved over a change in the scale factor (and hence redshift) of $\Delta a \approx 100$, and matched the analytic solutions to the Friedmann equations for the scale factor \eqref{eq:Fried_dust_asoln} and density \eqref{eq:Fried_dust_rhosoln} to within $10^{-6}$ in a simulation with $80^3$ grid cells. We demonstrated the expected fourth-order convergence, for the time integrator, of the $L_1$ error in our solutions, in addition to the violation in the Hamiltonian constraint. Next, we initialised small perturbations to the background FLRW spacetime in the density, velocity, and metric. We chose a single-mode, sinusoidal form for the metric perturbation, and related this to the corresponding density and velocity perturbations using the solutions found in linear perturbation theory, \eqref{eq:growingmode_density} and \eqref{eq:growingmode_velocity}, respectively. We chose $\Phi_0 = 10^{-8} \ll 1$, with corresponding amplitudes of the density and velocity perturbations of $10^{-5}$ and $10^{-7}$, respectively, so that our assumption of linear perturbations was valid for the initial conditions. We evolved over a change in scale factor of $\Delta a \approx 500$, and found agreement with linear theory to within $10^{-3}$ for the growth of the density and velocity perturbations. We demonstrated the expected second-order convergence, for the spatial integrator, of the $L_1$ error of these solutions. Beginning with slightly larger perturbations, $\Phi_0 = 10^{-6}$, we simulated the growth of these perturbations into the nonlinear regime, and found the gravitational slip --- the difference between the temporal and spatial metric perturbations --- was nonzero, with an amplitude of $\sim 4\times 10^{-6}$ at $z\approx2$. We found the tensor perturbation, zero in linear theory, grew in the over-dense region of the simulation and was smoothed out in the under-dense region. This work was an important proof-of-concept test that cosmological simulations into the nonlinear regime of structure formation --- inaccessible with analytic methods --- is possible using numerical relativity, specifically with the open-source \textsc{Einstein Toolkit}. 

The main aim of this thesis was to complete the first steps towards a full investigation of GR effects in our own Universe. In Chapter~\ref{Chapter4} we therefore extended the work performed in Chapter~\ref{Chapter3} to a more realistic matter distribution. We drew a spectrum of Gaussian density perturbations from the anisotropies in the CMB using the matter power spectrum from \textsc{CAMB} \citep{seljak1996}, and found the corresponding metric and velocity perturbations using linear perturbation theory. We generated these initial conditions at a number of different resolutions and physical box sizes, with the main results presented for a 1 Gpc$^3$ domain with $256^3$ grid points. The simulations were each evolved over a total change in scale factor of $\Delta a \approx 1100$, i.e. from the CMB to $z\approx 0$. We used \citeauthor{buchert2000a} averaging for a general foliation \citep{larena2009} in this simulation to compute the global contributions to the total energy-density from matter, curvature, and backreaction, and compared with the equivalent FLRW model. We also calculated these contributions within sub-domains in the simulation to assess the affects from curvature and backreaction on small scales. Globally, we found the effect from curvature and backreaction to be negligible, with $\om{R}\approx10^{-8}$ and $|\om{Q}+\om{L}|\approx10^{-9}$, respectively, with the contribution from matter dominating; a match to the equivalent FLRW model. On small scales, we found the contribution from backreaction and curvature could be significant, anywhere between $\sim 1\%$ on $\geq 80$ Mpc scales, up to $\sim 30\%$ on $\lesssim 100$ Mpc scales. These results showed that, while backreaction from structures is unlikely to explain the accelerating expansion rate (under our assumptions), percent-level effects are possible, and could be relevant in upcoming precision cosmological surveys.

In Chapter~\ref{Chapter5} we used the simulations presented in Chapter~\ref{Chapter4} to investigate the effect of local inhomogeneities on a measurement of the Hubble parameter. Our aim was to address the recent tension between locally measured values of $H_0$ using SN1a \citep{riess2018b}, and that inferred from the CMB \citep{planck2018a}, which has been suggested to be caused by local inhomogeneities (see Section~\ref{subsec:curiosities}). We defined the effective Hubble parameter within a chosen domain to be the averaged expansion rate of the fluid within that domain, and calculated the variance as a function of domain size. From this we estimated the expected variance on a local measurement of $H_0$ using SN1a within a particular redshift interval, purely due to an observers physical location in an inhomogeneous universe. Due to computational limits on resolution, and therefore the physical box size of our simulations, we could not sample the full redshift range, $0.023\lesssim z \lesssim 0.15$, used for the local SN1a measurement in \citet{riess2018b}. We instead extrapolated our smaller-scale results out to the full range, and found the variance to be $(+0.8,-0.4)\%$; not sufficient to explain the current $3.7\sigma$ tension. It is therefore unlikely that the tension can be explained purely due to local inhomogeneities. This result is subject to several caveats --- and the importance of these remains to be investigated --- including periodic boundary conditions, the fluid treatment of dark matter, and the validity of using averaging to approximate a measurement of the Hubble parameter. 
Even small variances on our measurements could be important in upcoming precision survey data, and therefore investigation into the effect of these caveats is necessary.

\section{Future work}

Understanding the role of GR on our observations is imperative as we move into the era of precision cosmology. Upcoming surveys are expected to produce data at percent-level precision, and to ensure we correctly interpret these data we must first validate the accuracy of the underlying assumptions of our cosmological model. The work presented in this thesis was a step towards this goal, providing essential tests of the required computational framework and early quantifications of the effect of structure formation on the large-scale evolution of the Universe. This truly is only the beginning of this field, and many aspects of the work presented here can be improved on to solidify and extend our results.

\subsection{Improved general foliation averaging} \label{subsec:FW_improved_averaging}

Recently, \citet{buchert2018b} pointed out some potential issues in the averaging formalism of \citet{larena2009}, which we used to analyse our simulations in Chapters~\ref{Chapter4} and \ref{Chapter5}. Similar issues are also present in the averaging procedures of \citet{brown2009b,gasperini2010} \citep[see also][]{umeh2011}. 

The main issue with these formalisms is that the domain of averaging is non-conservative; the fluid is free to flow into and out of the domain of averaging because it is propagated along the normal to the hypersurfaces, rather than the fluid normal. These averaging schemes are therefore based on an ``extrinsic approach'' of studying the averaged fluid quantities as seen by observers located in the spatial hypersurfaces defined by the normal vector (which does not coincide with the fluid four velocity). Instead, \citet[][and in a forthcoming publication]{buchert2018b} derive a coordinate-independent averaging formalism for general foliations of spacetime to study average fluid quantities as seen by fluid observers; an ``intrinsic approach'', which we outlined in Section~\ref{subsubsec:improved_general_averaging}. The domains of averaging in this formalism are mass conserving, since they are propagated along the fluid four velocity vector. 

Including the lapse function in the definition of $\HD^h$, e.g. in \eqref{eq:HDh_def}, is technically arbitrary \citep[see][]{umeh2011}, however, including it ensures a covariant expression for the Hubble parameter. Since the effective fluid scale factor is subsequently defined from this, a transformation of time will therefore give a vastly different scale factor if $\alp$ is not included in the definition \citep[recently pointed out by][and in a forthcoming publication via private communication]{buchert2018b}. We include the lapse in the relation between the fluid and volume scale factors for the analysis presented in Chapter~\ref{Chapter4}. However, we do not include it when calculating the Hubble parameter or the cosmological parameters. We do not expect this to significantly affect the \emph{magnitude} of the cosmological parameters, although the inhomogeneous nature of the lapse could have an effect on the kinematic backreaction term itself. 

It is important to verify the effect of including the lapse in \emph{all averages} on the cosmological parameters, kinematic backreaction, and average curvature. We leave this investigation, along with a general improvement of our averaging scheme to the newly suggested intrinsic approach, to future work.

\subsection{Ray tracing} \label{subsec:FW_raytracing}

While the averaging procedures outlined in Section~\ref{subsec:backreaction} are useful for studying the large-scale evolution of our Universe compared to the homogeneous, isotropic equivalent, they are explicitly dependent on the chosen averaging domain. \citet{adamek2019a} showed backreaction can differ by 3-5 orders of magnitude depending on the spatial hypersurface chosen for averaging. However, whether the authors are actually measuring backreaction here, as opposed to \emph{cosmic variance}, is a point of contention \citep[see][]{buchert2018b}.

Connecting the results of averaging to our observations is not so clear, since for cosmology, the fact that our observations are made along our past light cone --- and \emph{not} on a purely spatial hypersurface --- can become important \citep[see e.g.][]{buchert2012}. While the global expansion rate will affect the redshift and distance of objects, we really must study the past light cone in an inhomogeneous Universe to determine the full, measurable effect. 

Light propagation in inhomogeneous cosmology is not a new field \citep[e.g.][]{zeldovich1964,tomita1998,rose2001,kostov2010,bolejko2011b,fleury2013}, however, application of this work to numerical simulations has only recently emerged. \citet{giblin2016b} studied the effects on the Hubble diagram in their inhomogeneous cosmological simulations \citep[see][for details]{mertens2016}, and found overall agreement with FLRW. The structures considered in \citet{giblin2016b} are of large wavelengths only (with small amplitudes), and therefore the lensing effects are well approximated by perturbation theory, with no significant deviations from FLRW. Extending this work to more realistic matter distributions, i.e. including small-scale structures that can have large amplitudes --- and therefore result in significant lensing --- is essential to determine the expected scatter on the Hubble diagram we measure. This scatter has been estimated in the context of N-body simulations adopting the weak-field approximation \citep{adamek2013,adamek2016a}, and was found to be small. In addition, a percent-level bias was found on the measured curvature parameter due to relativistic effects \citep{adamek2018b}. These simulations are however dependent on the weak-field approximation, and perhaps more importantly on the assumption of a background, flat FLRW cosmology, and so it is important to verify these results in the context of full GR for a complex matter distribution. 

The results presented in Chapter~\ref{Chapter5} approximated a local measurement of the Hubble parameter as the averaged expansion rate within the spatial volume encompassed by the SN1a measured. To validate the main findings of this chapter, we must re-perform this analysis by ray tracing. We can create synthetic observations of SN1a within the redshift range considered by \citet{riess2018a} to place a better constraint on the expected variance under a full treatment of GR.

\subsection{N-body simulations with numerical relativity}

Throughout this thesis we have adopted a fluid approximation for the matter content of the Universe, as in most numerical-relativity simulations. This approximation usually presents no issue in the context of most astrophysical phenomena in GR \citep[see, e.g.][]{font2008}, however, it can be problematic for cosmology. The reason for this is the common adoption of the \emph{comoving gauge} in theoretical relativistic cosmology, which is now used in numerical-relativity simulations \citep{giblin2016a,bentivegna2016a}. Simulations of nonlinear structure formation in this gauge will fail because shell-crossing singularities form as the structures collapse; since the coordinates coincide with the fluid flow lines (see Section~\ref{subsubsec:geodesic_slicing}). These singularities can be avoided with the addition of a small amount of pressure \citep[e.g.][]{bolejkolasky2008}, or by simply adopting a different gauge for cosmological simulations \citep{macpherson2017a,east2018}. However, the assumption of a continuous fluid itself also means we cannot properly capture the process of structure formation. As a galaxy or galaxy cluster collapses, it will reach a point at which its internal velocity dispersion prevents further collapse; and the structure becomes \emph{virialised}. This is an important aspect of structure formation in the Universe which cannot be captured naturally in the case of a continuous fluid. 

An \emph{effective} virialisation technique was implemented in \citet{bolejko2018a}, where regions were no longer evolved once they began to collapse, i.e. when they had $\Theta<0$. In this case, voids continue expanding and an overall, global negative curvature arises, which has been shown to explain the dimming of SN1a observations and the tension in the Hubble parameter \citep{bolejko2018b}. This suggests that virialisation is an important aspect of inhomogeneous cosmology \citep[see also][]{roukema2017}. 

Discretising the fluid with particles, as is done in Newtonian cosmological simulations (see Section~\ref{subsec:Nbody_sims}), allows virialised objects to form naturally. Early work adopting collisionless particles in numerical-relativity simulations focused on stellar collapse and black-hole or singularity formation \citep{shapiro1985,shapiro1986,shapiro1991,shibata1999}. More recent applications include black-hole formation from gravitational waves \citep{pretorius2018} and non-spherical gravitational collapse \citep{yoo2017}.

In a cosmological context, a particle description of matter has been adopted in the weak-field limit \citep{adamek2013,adamek2014b}, and also recently incorporated into numerical-relativity simulations \citep[][with each of these codes in the active development and testing phase]{giblin2018,daverio2019,barrera-hinojosa2019}. Ensuring we have a collection of independent codes with which to perform comparisons and validate results is essential for the advancement of this field. Extending already widely-used, open-source software, such as the \textsc{Einstein Toolkit}, for this purpose will allow for contribution to this field from the wider community. 

\subsection{Further code tests}

In this work we used the BSSN formalism of Einstein's equations for our cosmological simulations. As discussed in Section~\ref{subsec:CCZ4}, there are alternative formalisms that have been developed to damp the growth of constraint violation by evolving the constraint variables themselves. Any constraint violating modes can therefore be propagated off the grid. The conformal and covariant Z4 system \citep[CCZ4;][]{alic2012} is an example we discussed in Section~\ref{subsubsec:CCZ4}. This has been implemented in the numerical relativity formalism of \citet{giblin2018} used for cosmological simulations, and in the \textsc{GRChombo} numerical relativity package \citep{clough2015}; used for simulations of inflationary cosmology \citep[see, e.g.][]{clough2017b}. Comparing the BSSN formalism and the CCZ4 system for the simulations presented here is beyond the scope of this thesis. However, future work comparing the amplitude and evolution of constraint violations in the \textsc{Einstein Toolkit} for both of these formalism is important in choosing the best evolution system for cosmological simulations. 

The use of periodic boundary conditions is common in cosmological simulations in Newtonian gravity \citep[e.g.][]{boylan-Kolchin2009,genel2014,potter2017}, and in the case of numerical relativity \citep[e.g.][]{giblin2016a,bentivegna2016a,macpherson2017a,daverio2017,barrera-hinojosa2019}. Choosing boundary conditions inherently defines the topology of the domain, with periodic boundary conditions corresponding to a three-torus in Cartesian coordinates. In 2+1 dimensional GR the chosen topology sets the averaged curvature evolution, and in the case of periodic boundaries the averaged curvature will always tend towards zero; and hence towards a homogeneous, FLRW expansion. However, in 3+1 dimensional GR the connection between topology and curvature remains unclear \citep[see, e.g.][]{buchert2012}. The globally zero curvature and homogeneous expansion seen in Chapter~\ref{Chapter4} could be a consequence of our chosen topology. To properly test this, a numerical relativity code utilising a different coordinate system, e.g. spherical coordinates, must be used. %

\appendix %

\chapter{Newtonian Gauge} %

\label{appx:newt_gauge} %

\begin{figure*}[!ht]
	\includegraphics[width=\textwidth]{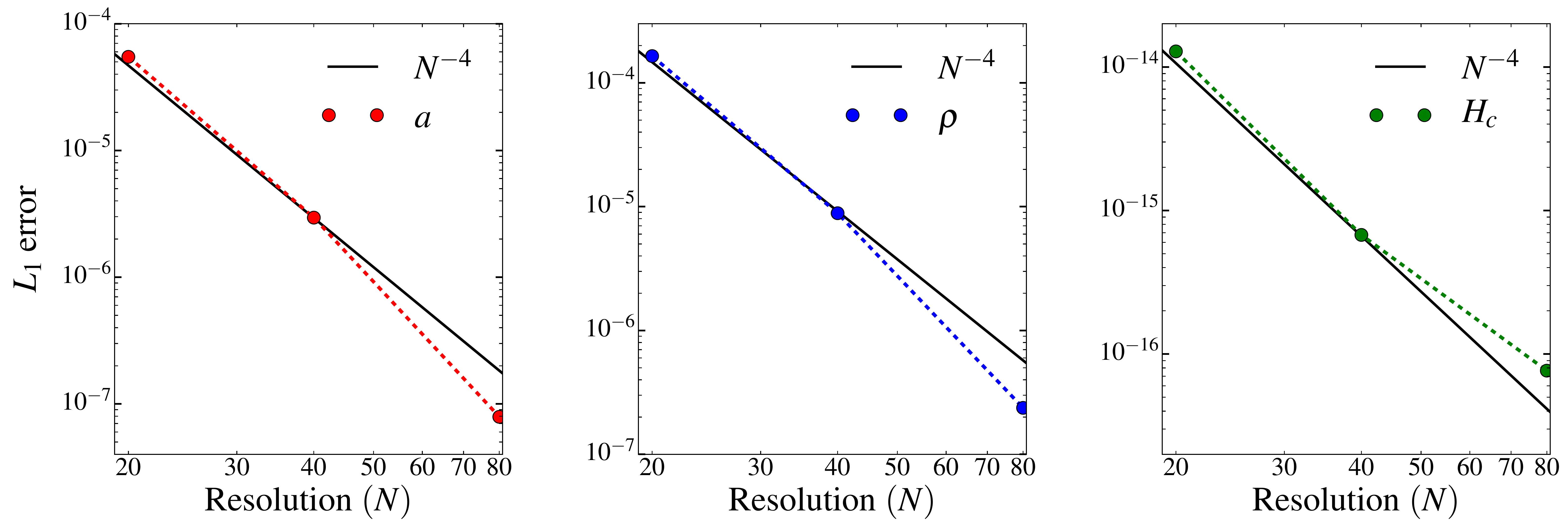}
	\caption{\label{fig:FLRW_RMS_newt}Fourth-order convergence of the FLRW solutions analysed in the Newtonian gauge. We show $L_{1}$ error as a function of resolution for the scale factor (left), density (middle), and Hamiltonian constraint (right). $N$ refers to the number of grid points along one spatial dimension. Filled circles indicate data points from the simulations, and black solid lines indicate the expected $N^{-4}$ convergence.}
\end{figure*}

Throughout Chapter~\ref{Chapter3} we work in the longitudinal gauge. For completeness, we show here the equivalent background and perturbation equations in the Newtonian gauge. In this Appendix we use geometric units with $G=c=1$. 
The flat FLRW metric is
\begin{equation}
	ds^{2} =  - d\hat{t}^{2} + a^{2}(\hat{t})dx^{i}dx^{j}\delta_{ij},
\end{equation}
which gives the Friedmann equations for a dust ($P\ll\rho$)\footnote{As in Chapter~\ref{Chapter3}, the density $\rho$ here is the total \emph{rest-frame} energy-density, and is equivalent to $\rho_R$ used in Chapters~\ref{Chapter1} and \ref{Chapter2}.} universe to be
\begin{subequations}\label{eq:Friedmann_newt}
	\begin{align}
		\left(\frac{\dot{a}}{a}\right)^{2} &= \frac{8\pi\rho}{3}, \label{eq:friedmann_1_newt}\\
		\dot{\rho} &= -3\frac{\dot{a}}{a}\rho, \label{eq:friedmann_2_newt}
	\end{align}
\end{subequations}
where a dot represents $d/d\hat{t}$. Solutions to these equations give the familiar time dependence of the scale factor,
\begin{equation}
		\frac{a}{a_{\mathrm{init}}} = s^{2/3},\quad
		\frac{\rho}{\rho_{\mathrm{init}}} = s^{-2},
\end{equation}
where 
\begin{equation}
	s\equiv1+\sqrt{6\pi\rho^{*}}\hat{t}.
\end{equation}
We match our numerical evolution to this alternative set of solutions by instead making the coordinate transform $t=t(\hat{t})$. With this we see the expected fourth-order convergence and maximum errors in the scale factor and density of $\sim10^{-7}$ for our highest resolution ($80^{3}$) simulation. Figure~\ref{fig:FLRW_RMS_newt} shows the convergence of the scale factor (left), density (middle) and Hamiltonian constraint (right) for analysis performed in this gauge.

The linearly perturbed FLRW metric in this gauge, including only scalar perturbations, is
\begin{equation}
	ds^{2} = -(1+2\psi_N)d\hat{t}^{2} + a^{2}(\hat{t})(1-2\phi_N)\delta_{ij}dx^{i}dx^{j}, \label{eq:perturbed_metric_newt}
\end{equation}
where $\psi_N, \phi_N$ are not the usual gauge-invariant Bardeen potentials (which are defined in the longitudinal gauge).
Solving the perturbed Einstein equations \eqref{eq:perturbed_einstein_chp3} in this gauge using the time-time, time-space, trace and trace free components gives
\begin{subequations} \label{eqs:perturbed_einstein_newt}
	\begin{align}
		\partial^2\phi_N - 3a\dot{a}\left(\dot{\phi}_N + \frac{\dot{a}}{a} \psi_N\right) &= 4\pi  \bar{\rho}\,\delta a^{2}, \label{eq:einstein_1_newt} \\ 
		\frac{\dot{a}}{a} \partial_{i}\psi_N + \partial_{i}\dot{\phi}_N &= -4\pi \bar{\rho} \,a^{2} \delta_{ij}\delta \hat{v}^{j}, \label{eq:einstein_2_newt} \\ 
		\ddot{\phi}_N + \frac{\dot{a}}{a}\left(\dot{\psi}_N + 3\dot{\phi}_N \right) &= \frac{1}{3 a^{2}}\partial^2(\phi_N - \psi_N), \label{eq:einstein_3_newt} \\ 
		\partial_{\langle i}\partial_{j\rangle} \left(\phi_N - \psi_N\right) &= 0, \label{eq:einstein_4_newt}
	\end{align}
\end{subequations}
in the linear regime. Solving these equations we find the form of the potential $\phi_N$ to be
\begin{equation}
	\phi_N = f(x^{i}) - \frac{3}{5}s^{-5/3}\,g(x^{i}),
\end{equation}
where $f, g$ are functions of the spatial coordinates. From this we find the density and velocity perturbations\footnote{The velocity here is $\hat{v}^i\equiv dx^i/d\hat{t}$, which differs from the velocity used in Chapter~\ref{Chapter3}.} to be, respectively,
\begin{subequations}
	\begin{align}
		\delta &= C_{1}\, s^{2/3}\,\partial^2f(x^{i}) - 2\,f(x^{i}) \\
		\phantom{\delta}&\phantom{=C_{1}\,} + 3\,C_{2}\, s^{-1}\,\partial^2g(x^{i}) - \frac{9\,a_{\mathrm{init}}^{3}}{5}s^{-5/3}\,g(x^{i}), \nonumber \\
		\delta \hat{v}^{i} &= C_{3}\,s^{-1/3}\,\partial^{i}f(x^{i}) + 3\,C_{4} \,s^{-2}\,\partial^{i}g(x^{i}), 
	\end{align}
\end{subequations}
where the $C_i$ were defined in \eqref{eq:c1c2} and \eqref{eq:c3c4}. 
We set $g(x^{i})=0$ to extract only the growing mode of the density perturbation, giving exact solutions to be
\begin{subequations}
	\begin{align}
		\phi_N &= f(x^{i}),\\
		\delta &= C_{1}\, s^{2/3}\,\partial^2f(x^{i}) - 2\,f(x^{i}), \label{eq:delta_ex_newt}\\  
		\delta \hat{v}^{i} &= C_{3}\,s^{-1/3}\,\partial^{i}f(x^{i}).\label{eq:deltav_ex_newt}
	\end{align}
\end{subequations}
We note that these solutions give equivalent initial conditions to those found in Section \ref{sec:linear_setup} since, initially, $s=\xi=1$. 

\begin{figure*}[!ht]
	\begin{centering}
	\includegraphics[width=\textwidth]{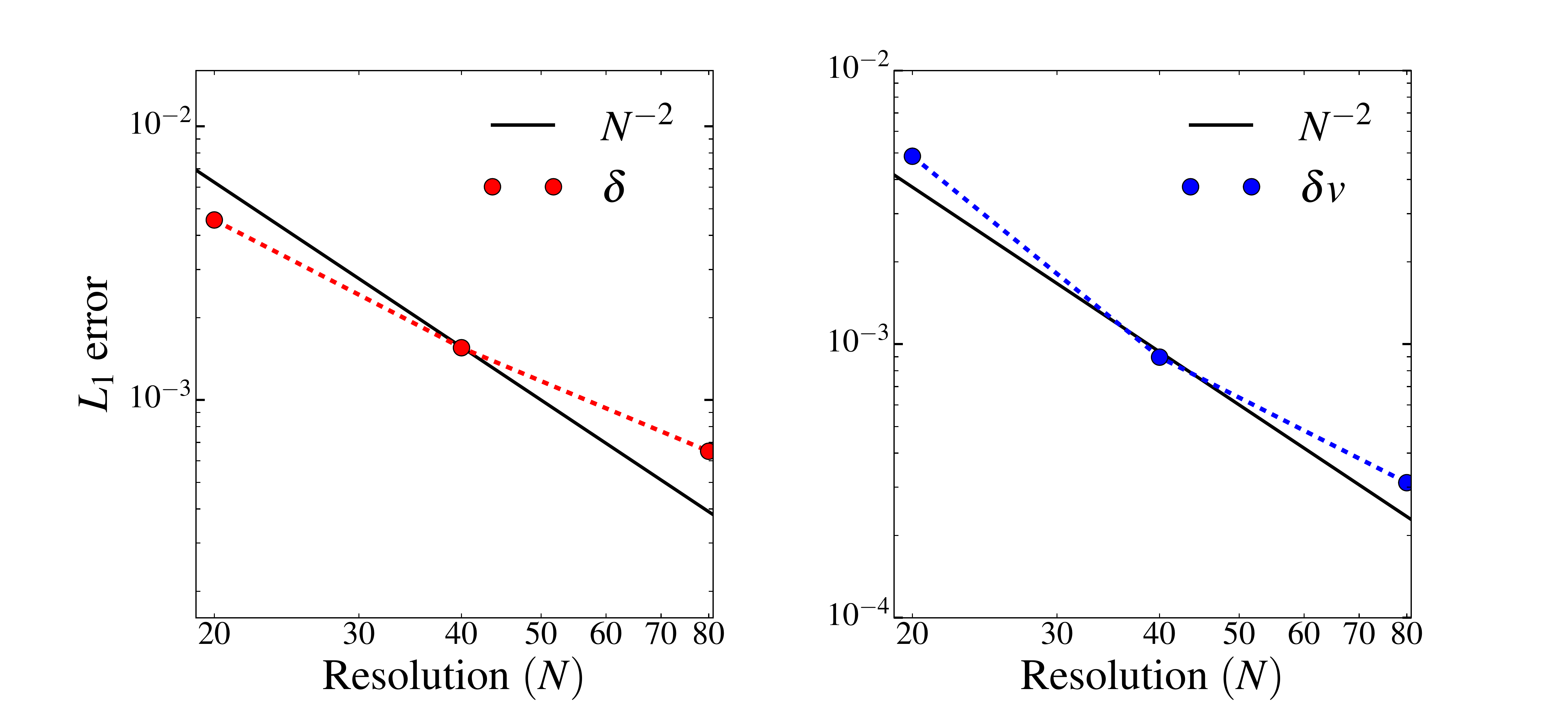}
	\caption{\label{fig:perturb_converge_newt}Second-order convergence of the numerical solutions for a linearly perturbed FLRW spacetime, analysed in the Newtonian gauge. We show $L_{1}$ errors in the density (left) and velocity perturbations (right). $N$ refers to the number of grid points along one spatial dimension. Filled circles indicate data points from our simulations, and black solid lines indicate the expected $N^{-2}$ convergence.}
	\end{centering}
\end{figure*}

We compare our numerical relativity solutions to the exact solutions for linear perturbations in this gauge using the coordinate transform $t=t(\hat{t})$. We find the expected second-order convergence with maximum errors in the density and velocity perturbations of $\sim10^{-3}$ for our highest resolution ($80^{3}$) simulation. Figure~\ref{fig:perturb_converge_newt} shows the convergence of the density (left) and velocity (right) perturbations when analysed in the Newtonian gauge.

\chapter{Averaging in the non-comoving gauge} %

\label{sec:appx_avg} %

Averaging Einstein's equations in a non-comoving gauge results in the averaged Hamiltonian constraint
\begin{equation}
	6\mathcal{H_D}^{2} = 16\pi\langle W^{2}\rho\rangle - \mathcal{R_D} - Q_\mathcal{D} + \mathcal{L_D},
\end{equation}
where we define
\begin{align}
	\mathcal{R_D} &\equiv \langle W^{2}\mR\rangle, \label{eq:RD}\\
	Q_\mathcal{D} &\equiv \frac{2}{3} \left( \langle\theta^{2}\rangle - \langle\theta\rangle^{2}\right) - 2\langle\sigma^{2}\rangle, \label{eq:QD}\\
	\mathcal{L_D} &\equiv 2\langle\sigma_{B}^{2}\rangle - \frac{2}{3}\langle\theta_{B}^{2}\rangle - \frac{4}{3}\langle\theta\theta_B\rangle. \label{eq:LD}
\end{align}
Here, $W=1/\sqrt{1-v^{i}v_{i}}$ is the Lorentz factor, $\mR\equiv \gamma^{ij}\mR_{ij}$ is the three-dimensional Ricci curvature of the averaging hypersurfaces, with $\mR_{ij}$ the spatial Ricci tensor. In this Appendix we work in geometric units with $G=c=1$. Here 
\begin{equation} \label{eq:sigma2}
	\sigma^{2}=\frac{1}{2}\sigma^{i}_{\phantom{i}j}\sigma^{j}_{\phantom{j}i},
\end{equation}
where $\sigma_{ij}$ is the shear tensor, defined as
\begin{equation} \label{eq:sigma_def}
	\sigma_{\mu\nu}\equiv h^{\alpha}_{\phantom{\alpha}\mu}h^{\beta}_{\phantom{\beta}\nu} \nabla_{(\alpha} u_{\beta)} - \frac{1}{3}\theta h_{\mu\nu}.
\end{equation}
As in \citep{umeh2011}, we introduce for simplification
\begin{subequations} \label{eqs:tensor_defs}
	\begin{align}
		\sigma_B^{2} &= \frac{1}{2}\sigma^{i}_{\phantom{i}Bj}\sigma^{j}_{\phantom{j}Bi} + \sigma_{ij}\sigma_{B}^{ij} \\
		\sigma_{Bij} &\equiv -W \beta_{ij} - W^{3}\left(B_{(ij)} - \frac{1}{3}Bh_{ij}\right) \\
		\theta_B &\equiv - W \kappa - W^3 B \\
		\beta_{\mu\nu} &\equiv h^{\alpha}_{\phantom{\alpha}\mu}h^{\beta}_{\phantom{\beta}\nu}\nabla_{(\alpha}v_{\beta)} - \frac{1}{3}\kappa h_{\mu\nu} \\
		B_{\mu\nu} &\equiv \frac{1}{3}\kappa(v_\mu n_\nu + v_\mu v_\nu) + \beta_{\alpha\mu}v^{\alpha}n_\nu + \beta_{\alpha\mu}v^{\alpha}v_\nu \\
				&+ M_{\alpha\mu}v^{\alpha}n_\nu + M_{\alpha\mu}v^{\alpha}v_\nu,		
	\end{align}
\end{subequations}
where we also define
\begin{subequations}\label{eqs:kapWB_defs}
    \begin{align} 
    	\kappa &\equiv h^{\alpha\beta}\nabla_{\alpha}v_{\beta}, \quad M_{\mu\nu} \equiv h^{\alpha}_{\phantom{\alpha}\mu}h^{\beta}_{\phantom{\beta}\nu}\nabla_{[\alpha}v_{\beta]}, \\
    	B &= \frac{1}{3}\kappa v^{\alpha}v_{\alpha} + \beta_{\mu\nu}v^{\mu}v^{\nu}.
    \end{align}
\end{subequations}
For a given tensor $A_{\mu\nu}$ we adopt the notation $A_{(\mu\nu)} = \frac{1}{2}(A_{\mu\nu} + A_{\nu\mu})$ and $A_{[\mu\nu]} = \frac{1}{2}(A_{\mu\nu} - A_{\nu\mu})$.

\chapter{Effective scale factors} %

\label{sec:appx_expn} %

The effective expansion of an inhomogeneous domain can be defined by
\begin{align}
	\frac{\partial_\eta \,a_\mathcal{D}^{V}}{a_\mathcal{D}^{V}} &\equiv \frac{1}{3}\frac{\partial_\eta V_\mathcal{D}}{V_\mathcal{D}}, \\
	\Rightarrow a_\mathcal{D}^{V} &= \left( \frac{V_\mathcal{D}(\eta)}{V_\mathcal{D}(\eta_\mathrm{init})} \right)^{1/3},
\end{align}
where $V_\mathcal{D}(\eta)$ is the volume of the domain $\mathcal{D}$ at a given conformal time. The physical interpretation of this scale factor depends on the chosen hypersurface of averaging. If we choose the averaging surface to be comoving with the fluid; a surface with normal $u^{\mu}$, then the scale factor $a_\mathcal{D}^{V}$ describes the effective expansion of the fluid averaged over the domain. We define the averaging surface to be comoving with a set of observers with normal $n^{\mu}$; \textit{not} coinciding with $u^{\mu}$. In this case, $a_\mathcal{D}^{V}$ describes the expansion of the volume element, not of the fluid itself.

We define the Hubble parameter as the expansion of the fluid projected into the gravitational rest frame; the frame of observers with normal $n^{\mu}$. From this we define the effective scale factor of the fluid, $a_\mathcal{D}$\footnote{The fluid scale factor $\aD$ here is equivalent to $\aDh$ used in Chapter~\ref{Chapter1} and Section~\ref{sec:mescaline}.} in \eqref{eq:aDdef}. We can relate the two scale factors by first considering the rate of change of the volume (with $\beta^{i}=0$) in the (3+1) formalism \citep{larena2009b}\footnote{After the publication of this paper, we noticed an error in \eqref{eq:appxC_detaVD}, \eqref{eq:aDVlog}, and \eqref{eq:aDfromaDV} from \citet{larena2009b}. In Appendix~\ref{sec:appx_aD_typo} we show this error has negligible effect on our results.},
\begin{equation} \label{eq:appxC_detaVD}
	\frac{\partial_\eta V_{\mathcal{D}}}{V_{\mathcal{D}}} = \langle \frac{\alp}{W}(\theta - \kappa)\,\rangle.
\end{equation}
Now, with $\partial_\eta a_\mathcal{D}/a_\mathcal{D} = \partial_\eta \mathrm{ln}(a_\mathcal{D})$, we can write
\begin{align}
	\partial_\eta \mathrm{ln}(a_{\mathcal{D}}) &= \frac{1}{3}\langle\alpha\theta\rangle, \label{eq:aDlog} \\
	\partial_\eta \mathrm{ln}(a_{\mathcal{D}}^{V}) &= \frac{1}{3}\langle \frac{\alp}{W}(\theta - \kappa)\,\rangle, \label{eq:aDVlog}
\end{align}
subtracting \eqref{eq:aDVlog} from \eqref{eq:aDlog} we arrive at the relation
\begin{equation} \label{eq:aDfromaDV}
	a_{\mathcal{D}} = a_{\mathcal{D}}^{V} \;\mathrm{exp}\left(-\frac{1}{3}\int \langle \frac{\alp}{W} (\theta - \kappa) - \alpha\theta\,\rangle \;d\eta\right).
\end{equation}
Here, $a_{\mathcal{D}}^{V}$ is found by calculating the volume of the domain relative to the initial volume. Figure~\ref{fig:aDaFLRW} shows the evolution of \eqref{eq:aDfromaDV} (blue solid curve) as a function of redshift for a $256^{3}$ simulation over an $L=1$ Gpc domain, relative to the equivalent FLRW solution (purple dashed curve).

\chapter{Constraint violation} %

\label{appx:constraints} %

\begin{centering}
\begin{figure*}
\begin{subfigure}{0.5\textwidth}
  \includegraphics[width=\textwidth]{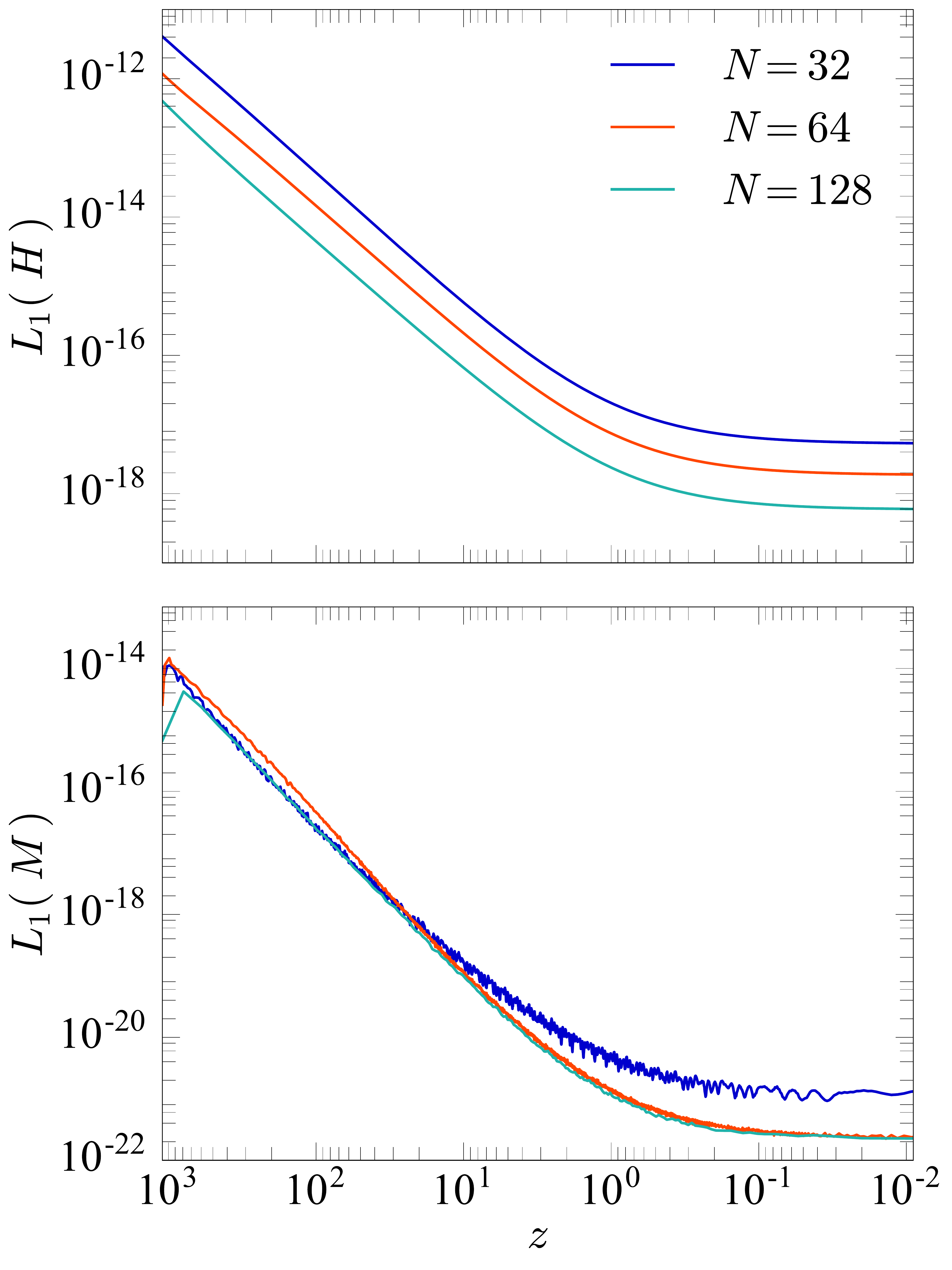}
  \label{fig:constraints_controlRaw}
\end{subfigure}%
\begin{subfigure}{0.5\textwidth}
  \includegraphics[width=\textwidth]{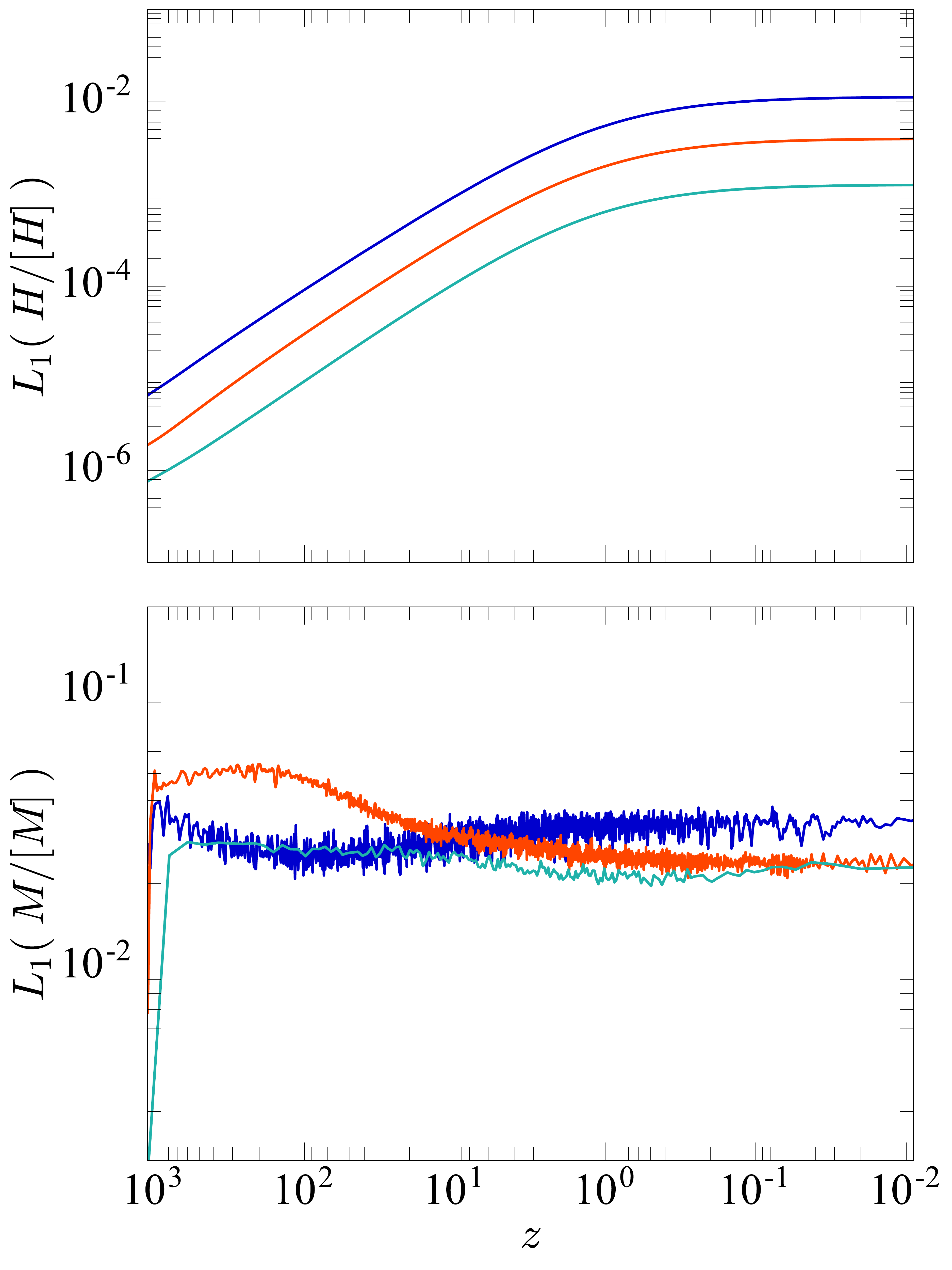}
   \label{fig:constraints_controlRel}
\end{subfigure}
\caption{Relative (right panels) and raw (left panels) constraint violation as a function of effective redshift calculated using \eqref{eq:l1Herror_rel} and \eqref{eq:l1Herror_raw}, respectively. We show violations for the simulations with a controlled number of physical modes. Top panels show the Hamiltonian constraint violation, and bottom panels show the momentum constraint violation. Colours show different resolutions as indicated by the legend.}
 \label{fig:constraints_control}
\end{figure*}
\end{centering}

\begin{figure*}
	\begin{centering}
	\begin{subfigure}{\textwidth}
     		\includegraphics[width=\textwidth]{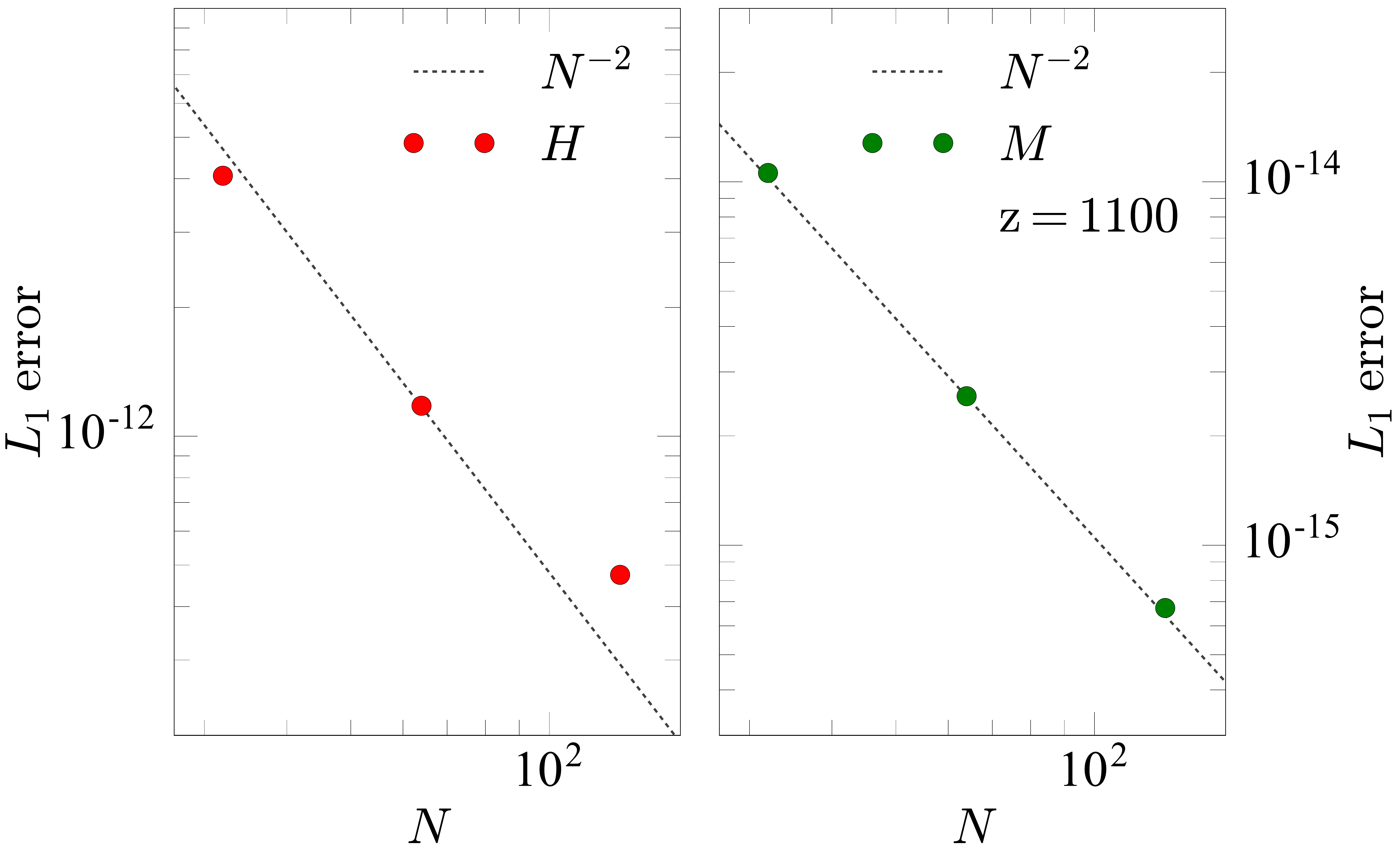}
	\end{subfigure}

	\begin{subfigure}{\textwidth}
	     \includegraphics[width=\textwidth]{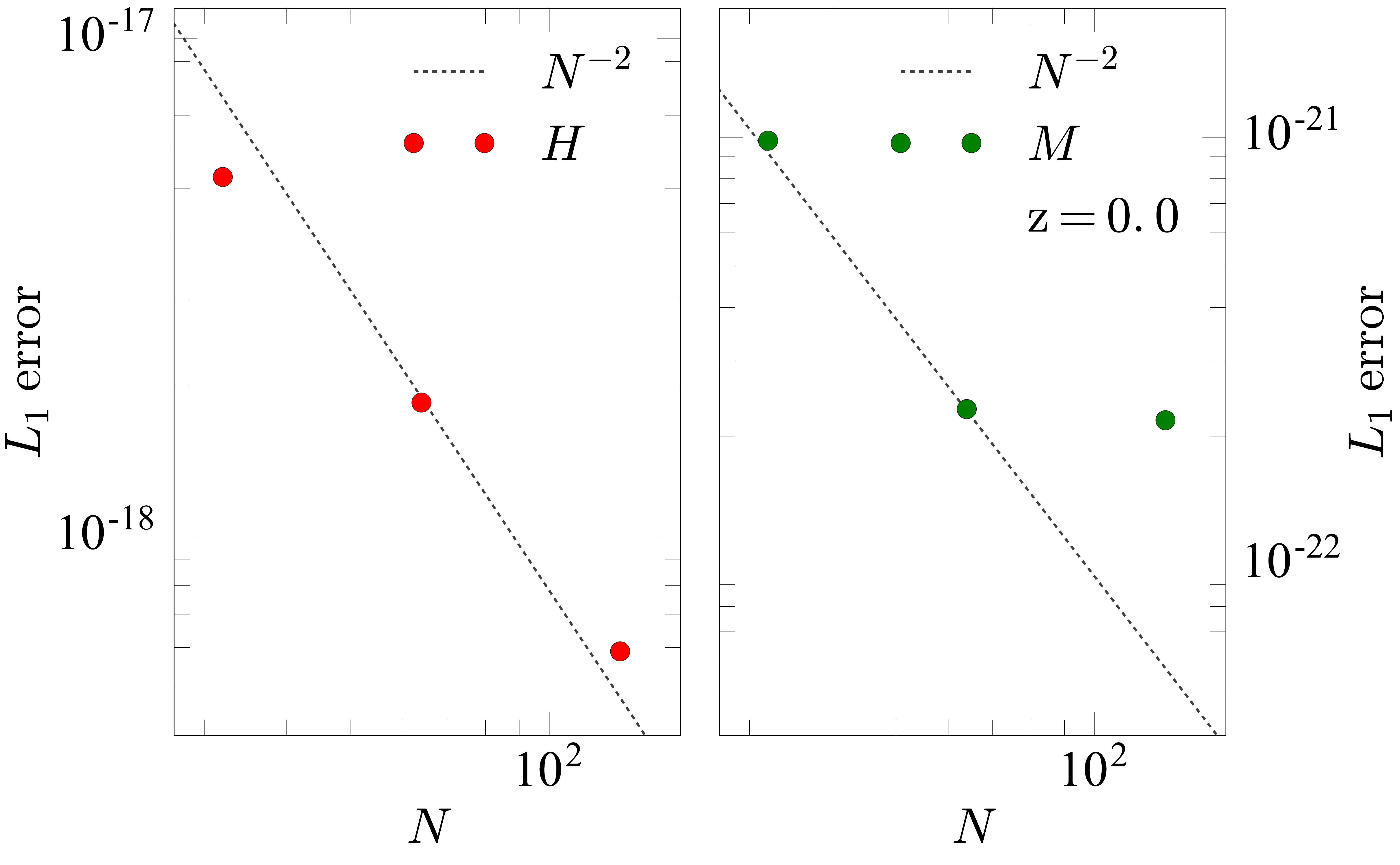}
	\end{subfigure}
    	\caption{Second order convergence of the Hamiltonian and momentum constraints for the set of simulations with a fixed number of physical modes. We show the $L_1$ error calculated using \eqref{eq:l1Herror_raw} for both constraints. The top two panels show the $L_1$ error for the initial conditions, at $z=1100$, and the bottom two panels show the $L_1$ error for $z=0$.}
   	 \label{fig:10dx_HMconverge_z0_z1100}
	\end{centering}
\end{figure*}

In numerical relativity, the error can be quantified by analysing the violation in the Hamiltonian and momentum constraint equations, defined by
\begin{subequations} \label{eqs:HamMom}
	\begin{align}
		H &\equiv \mR + K^{2} - K_{ij}K^{ij} - 16\pi\rho = 0, \label{eq:ham} \\
		M_{i} &\equiv D_{j} K^{j}_{\phantom{j}i} - D_{i}K - 8\pi S_{i} = 0, \label{eq:mom}
	\end{align}
\end{subequations}
respectively, where $S_{i}\equiv-\gam_{i\alpha}n_{\beta}T^{\alpha\beta}$ and $D_{i}$ represents the covariant derivative associated with the three-metric $\gamma_{ij}$. We define the magnitude of the momentum constraint to be $M=\sqrt{M_i M^i}$. In this Appendix we adopt geometric units with $G=c=1$. An exact solution to Einstein's equations will identically satisfy \eqref{eqs:HamMom}. Since we are solving Einstein's equations numerically, we expect some non-zero violation in the constraints. We use the \textsc{mescaline} code, described in Section~\ref{sec:mescaline} and \ref{subsec:mescaline_chp4}, to calculate the constraint violation as a function of time.

For the simulations we present in this work, we do not expect (in general) to see convergence of the constraint violation. At each different resolution we are sampling a different section of the power spectrum, and hence a different physical problem. In order to see convergence of the constraints at the correct order, we must analyse a controlled case in which the gradients are kept constant between resolutions. We perform three simulations at resolutions $32^3,64^3$, and $128^3$ inside an $L=1$ Gpc domain. We generate the initial conditions for the $32^3$ simulation using CAMB; restricting the minimum sampling wavelength to be $\lambda_{\rm min}=10\Delta x_{32} = 312.5$ Mpc. We use linear interpolation to generate the same initial conditions at $64^3$ and $128^3$. 

Figure~\ref{fig:constraints_control} shows the violation in the Hamiltonian (top panels) and momentum (bottom panels) constraints, for the set of simulations with a controlled number of physical modes, as a function of effective redshift. Left panels show the raw $L_1$ error for the violation, which for the Hamiltonian constraint we define as
\begin{equation}\label{eq:l1Herror_raw}
	L_1(H) = \frac{1}{N}\sum_{a=1}^{N} |H_a|,
\end{equation}
where $N$ is the total number of grid cells, and $H_a$ is the Hamiltonian constraint violation at grid cell $a$, and similarly for the momentum constraint. To quantify the "smallness" of this violation, we normalise the constraint violations to their relative "energy scales". Similar to \citet{mertens2016,giblin2017a}, we define
\begin{subequations} \label{eqs:energyscale}
	\begin{align}
		[H] &\equiv \sqrt{\mR^{2} + (K^{2})^{2} + (K_{ij}K^{ij})^{2} + (16\pi\rho)^{2}}, \label{eq:enham} \\
		[M] &\equiv \sqrt{(D_{j} K^{j}_{\phantom{j}i})(D_{k} K^{ki}) + (D_{i}K)(D^{i}K) + (8\pi)^2 S_i S^i}.\label{eq:enmom}
	\end{align}
\end{subequations}
Right panels in Figure~\ref{fig:constraints_control} show the relative $L_1$ error for each constraint violation, which we define as
\begin{equation}\label{eq:l1Herror_rel}
	L_1(H/[H]) = \frac{ \frac{1}{N}\sum_{a=1}^{N} |H_a| }{ \frac{1}{N} \sum_{a=1}^{N} [H]_a},
\end{equation}
where $[H]_a$ is the energy scale calculated at grid cell $a$, and similarly for the momentum constraint. We take the positive root of both $[H]_a$ and $[M]_a$.

Figure~\ref{fig:10dx_HMconverge_z0_z1100} shows the raw $L_1$ error for the same set of simulations, for the initial data ($z=1100$; top two panels) and for the data at $z=0$ (bottom two panels). The left panels show the $L_1$ error for the Hamiltonian constraint, and the right panels show the $L_1$ error for the momentum constraint. We see the expected second order convergence for the \textsc{Einstein Toolkit}, with the exception of the $N=128$ simulation's violation in the momentum constraint. Our initial speculation was that this was roundoff error, given the smallness of the quantities involved. The top panels of Figure~\ref{fig:10dx_HMconverge_z0_z1100} show that this issue is not due to the non-convergence of our initial data. Whether or not this is roundoff error remains unclear, and cannot be clarified without re-performing our simulations in quad precision.

For the simulations with a controlled number of modes, at $z=0$ for resolution $N=128$ the relative Hamiltonian constraint violation is $L_1(H/[H]) = 1.3\times10^{-3}$, the momentum constraint is $L_1(M/[M]) = 2.3\times10^{-2}$. For the simulations with full power spectrum sampling, at $z=0$ and resolution $N=256$ we find $L_1(H/[H]) = 4.4\times10^{-1}$, and $L_1(M/[M]) = 5.4\times10^{-2}$.

\chapter{Convergence and errors} %

\label{appx:convergence} %

\begin{figure*}
	\begin{centering}
     \includegraphics[width=\textwidth]{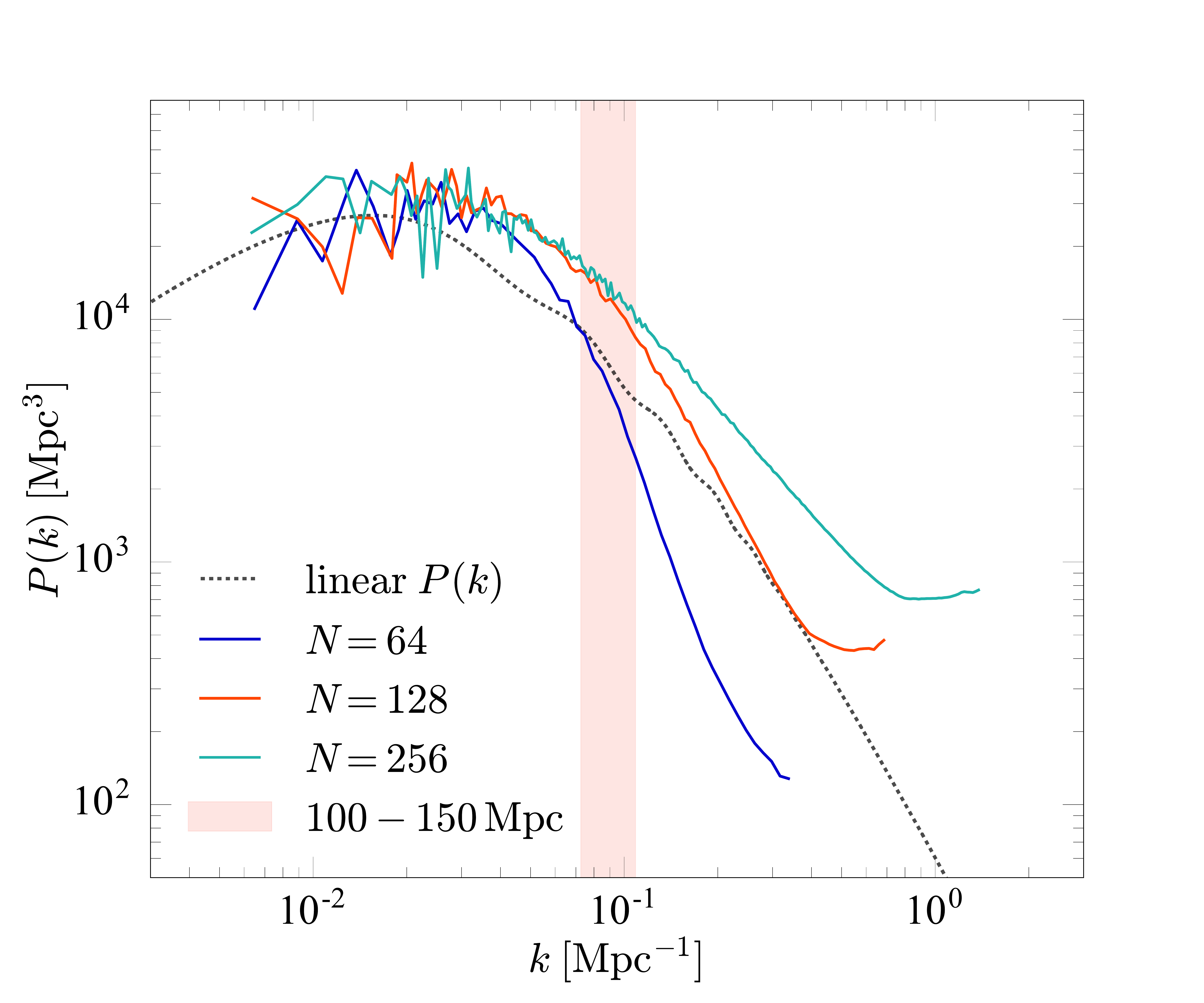}
    \caption{Power spectrum of fractional density fluctuations $\delta$ at $z=0$. Solid coloured curves show $P(k)$ for three simulations with $64^3,128^3,256^3$, each for an $L=1$ Gpc domain, and the dashed curve shows the linear power spectrum at $z=0$. The pink shaded region represents one-dimensional scales of $80-100$ Mpc. }
    \label{fig:pkz0}
    \end{centering}
\end{figure*}

\begin{figure*}
	\begin{centering}
	\begin{subfigure}{0.5\textwidth}
	 	 \includegraphics[width=\textwidth]{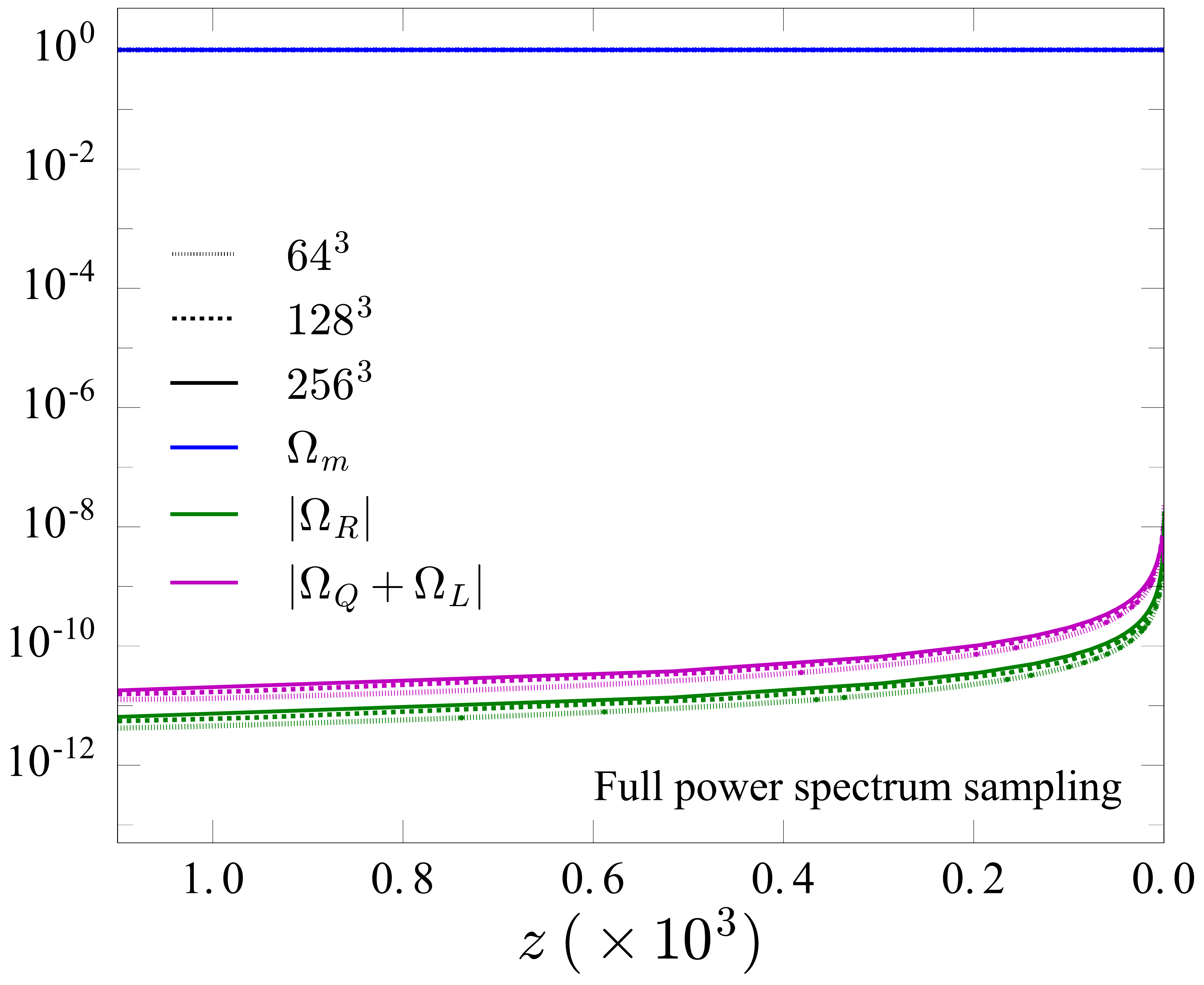}
  		\label{fig:omegas_z_noncontrol}
	\end{subfigure}%
	\begin{subfigure}{0.5\textwidth}
	  	\includegraphics[width=\textwidth]{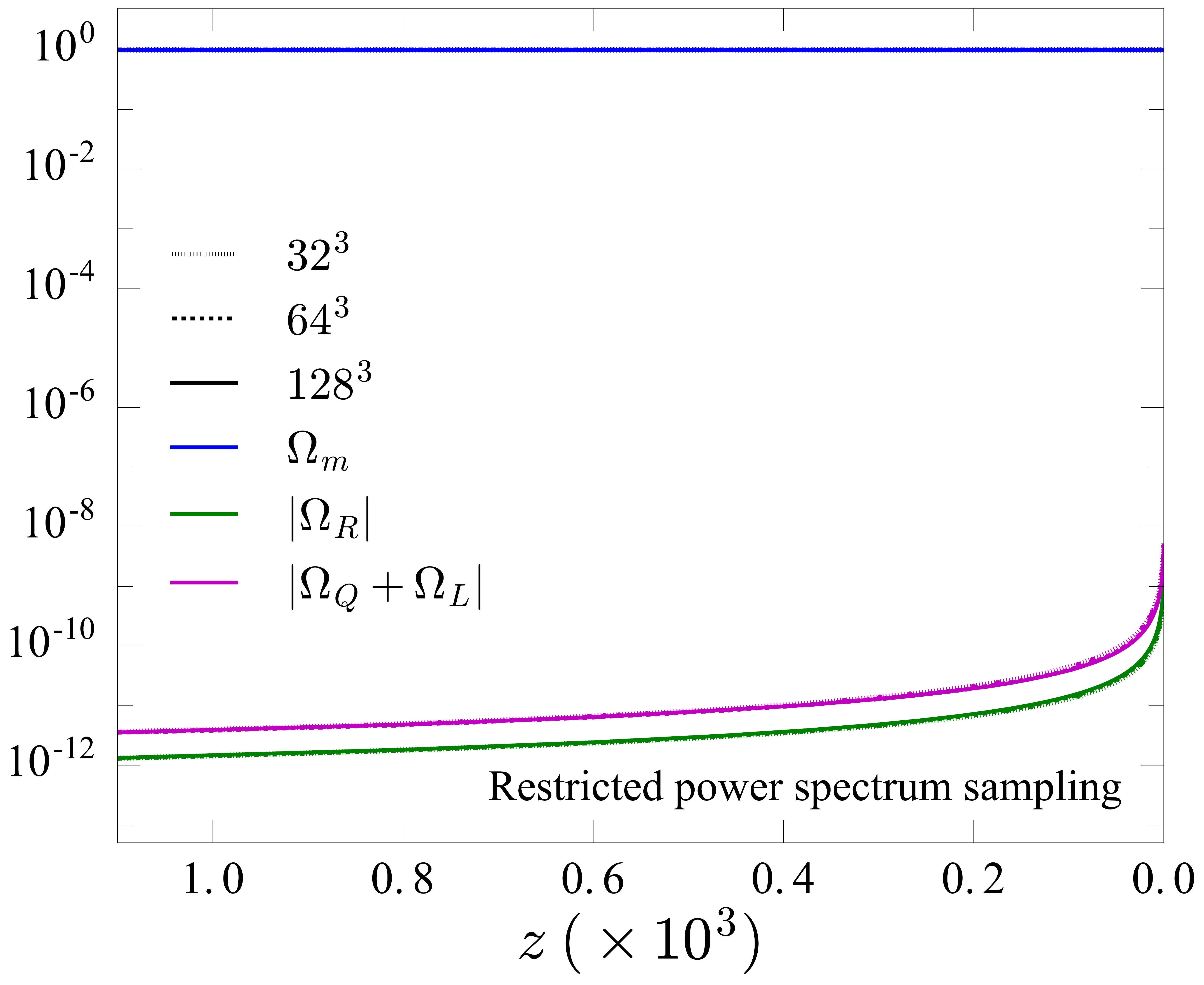}
   		\label{fig:omegas_z_control}
	\end{subfigure}
	\caption{Global cosmological parameters as a function of effective redshift, for simulations with full power spectrum sampling (left panel) and a controlled number of physical modes (right panel). Dotted, dashed, and solid curves show resolutions as indicated in each seperate legend. Blue curves show $\Omega_m$, green curves show $|\Omega_R|$, and purple curves show $|\Omega_Q + \Omega_L|$.}
 \label{fig:omegas_z}
\end{centering}
\end{figure*}

\begin{figure*}
     \includegraphics[width=\textwidth]{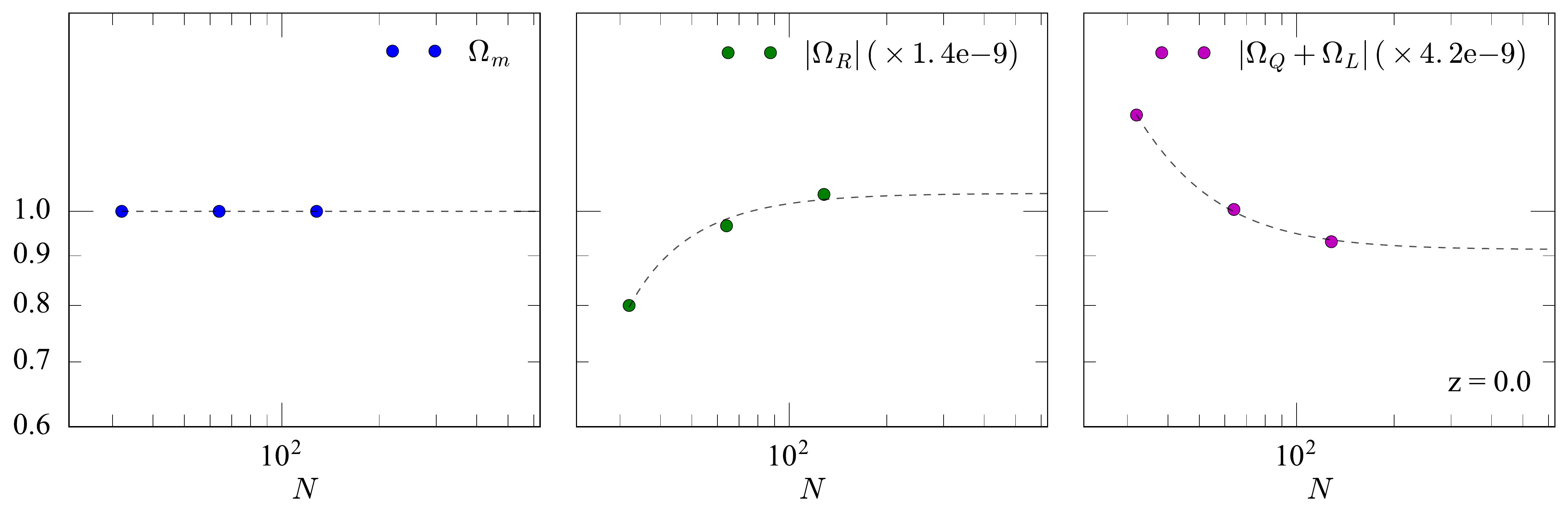}
    \caption{Cosmological parameters measured at $z=0$ for simulations with a controlled number of physical modes. Coloured points show $\Omega_m$ (left), $|\Omega_R|$ (middle), and $|\Omega_Q+\Omega_L|$ (right) for resolutions $N=32, 64,$ and $128$. Dashed curves are the convergence fit for each parameter, detailed in the text. We use these curves for a Richardson extrapolation to calculate the true value of the parameters and hence the errors on our measurements. }
    \label{fig:omegas_curvefit}
\end{figure*}

\begin{figure*}
	\begin{centering}
     \includegraphics[width=\textwidth]{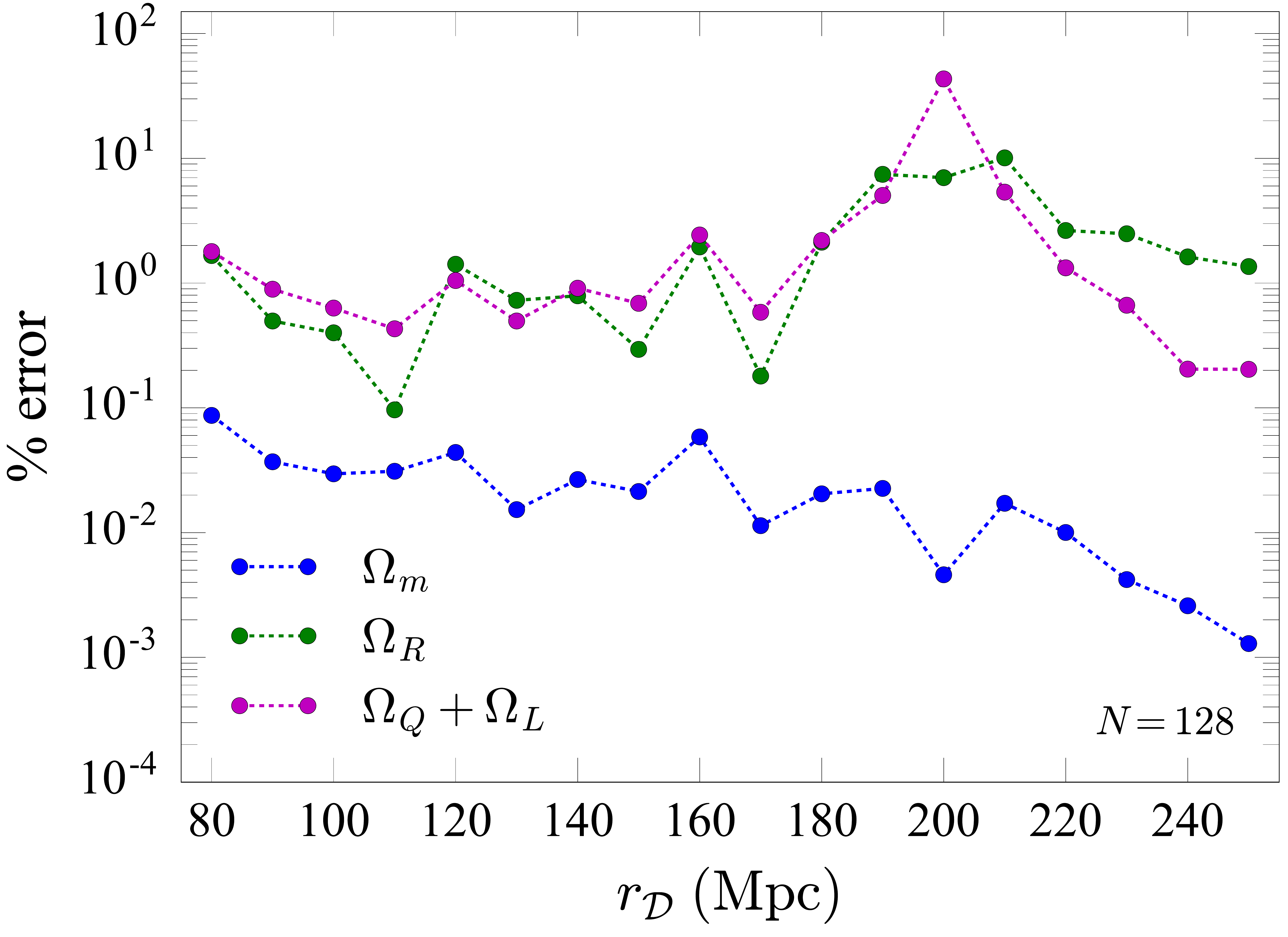}
    \caption{Richardson extrapolated errors for cosmological parameters calculated within subdomains, for the $128^3$ resolution controlled simulation, as a function of averaging radius $r_\mathcal{D}$. Blue points show the percentage error for $\Omega_m$, green points for $\Omega_R$, and purple points for $\Omega_Q+\Omega_L$.}
    \label{fig:omegas_rD_errors}
	\end{centering}
\end{figure*}

In Appendix~\ref{appx:constraints} we discussed the convergence of the constraint violation, which for the main simulations presented in Chapter~\ref{Chapter4} we do not expect to reduce with resolution, since the physical problem is changing. Regardless of this, we expect the power spectrum of fractional density fluctuations, $\delta$, to be converged in these simulations. Coloured curves in Figure~\ref{fig:pkz0} show the $z=0$ power spectrum of the fractional density fluctuations for resolutions $64^3,128^3$, and $256^3$, each within an $L=1$ Gpc domain. The dashed curve shows the linear power spectrum at $z=0$ calculated with CAMB. The pink shaded region shows one-dimensional scales of $100-150$ Mpc, roughly corresponding to scales which are converged. Below these scales we therefore underestimate the growth of structures, and hence underestimate the contribution from curvature and backreaction in our calculations. 

Figure~\ref{fig:omegas_z} shows the global cosmological parameters as a function of redshift for simulations sampling the full power spectrum (left panel), and for those with a restricted sampling of the power spectrum; our controlled case discussed in the previous section (right panel). Dotted, dashed, and solid curves show different resolutions as indicated in each seperate legend. Blue curves show the density parameter $\Omega_m$, green curves show the curvature parameter $|\Omega_R|$, and purple curves show the backreaction parameters $|\Omega_Q+\Omega_L|$. As resolution increases in the left panel --- as we add more small-scale structure --- the contributions from curvature and backreaction increase, however still remain negligible relative to the matter content, $\Omega_m$, and are unlikely to grow large enough to be significant when reaching a realistic resolution. In the right panel, we have kept the physical problem constant and varied only the computational resolution, and so we see all global parameters converged towards a single value. Comparing the left panel to the right panel, the value of the curvature and backreaction parameters differ by almost an order of magnitude. This is due to the restricted power spectrum sampling for the simulations in the right panel, in which we only sample structures down to $\lambda_{\rm min}=10\Delta x_{32} = 312.5$ Mpc. These simulations should therefore not be considered the most realistic representation of our Universe. From this comparison we see that adding more structure results (in general) in a larger contribution from curvature and backreaction. 

We calculate the errors in the cosmological parameters using a Richardson extrapolation, which requires the gradients between resolutions to remain the same. This is not the case for the simulations with full power spectrum sampling, however it is the case for the controlled simulations with a restricted mode sampling. We therefore use the controlled simulations to approximate the errors for our main calculations. Figure~\ref{fig:omegas_curvefit} shows the values of the globally averaged cosmological parameters at $z=0$ for the controlled simulations. Coloured points show $\Omega_m$ (left panel), $|\Omega_R|$ (middle panel), and $|\Omega_Q+\Omega_L|$ (right panel) at resolutions $N=32,64,$ and $128$. We use the function \texttt{curve\_fit} as a part of the SciPy\footnote{https://scipy.org} Python package to fit each set of points with a curve of the form $\Omega_i(N) = \Omega_{\rm inf} + E\times N^{-2}$, where $\Omega_{\rm inf}$ is the value of the relevant cosmological parameter at $N\rightarrow\infty$, and $E$ is a constant. Black dashed curves in each panel of Figure~\ref{fig:omegas_curvefit} show the best-fit curves.  

The best-fit value for $\Omega_{\rm inf}$ provides an approximation of the correct value of each cosmological parameter for this set of test simulations. The residual between our calculations and $\Omega_{\rm inf}$ gives the error in our calculations. For the controlled simulation with $128^3$ resolution, the errors in the global cosmological parameters are $10^{-8}$, $4\times10^{-12}$, and $7\times10^{-11}$ for $\Omega_m$, $\Omega_R$, and $\Omega_Q+\Omega_L$, respectively. Expressed as a percentage error, these are  $10^{-6}\%$, 0.27\%, and 1.9\%.

We follow the same procedure to estimate the errors on the cosmological parameters calculated within subdomains. Figure~\ref{fig:omegas_rD_errors} shows the percentage error in each parameter as a function of averaging radius of the subdomain, $r_\mathcal{D}$, for the controlled simulation with $128^3$ resolution. Blue points show the error for $\Omega_m$, green points show $\Omega_R$, and purple points show $\Omega_Q+\Omega_L$. The jump in errors evident at $\sim 200$ Mpc in $\Omega_R$ and $\Omega_Q+\Omega_L$ is due to a change in sign of the curvature and backreaction parameters.

\chapter[Effective scale factors part 2]{Effective scale factors part 2: investigation into an error} %

\label{sec:appx_aD_typo} %

In Section~\ref{subsubsec:general_foliation} we introduced the averaging formalism of \citet{larena2009b} for general foliations of spacetime. Here, we detail an error in  \citeauthor{larena2009b}'s equation (30) for the rate of change of $\sqrt{h}$, subsequently carried through into equation (31) for the rate of change of the volume, and equation (34) relating the fluid and volume scale factors --- used for our analysis in Chapter~\ref{Chapter4}. Here we re-analyse the simulations presented in Chapter~\ref{Chapter4} and find the error makes a difference of $\approx 10^{-8}$ globally, and $\approx 10^{-10}$ on 100 Mpc scales, both at $z\approx0$. In this Appendix, we work in units with $c=1$.

Expressing the evolution equation for the spatial metric \eqref{eq:gamij_evolution} in terms of the fluid variables defined in Section~\ref{subsubsec:general_foliation}, we arrive at equation (24) in \citet{larena2009b},
\begin{equation}\label{eq:larena_eq24_appx}
	\frac{W}{\alp} \pdt h_{ij} = \frac{2}{3} \left(\theta + \theta_B \right) h_{ij} + 2\left(\sigma_{ij} + \sigma_{Bij} \right) + \frac{2W}{\alp} D_{(i} \beta_{j)},
\end{equation}
and taking the trace of this results in an evolution equation for $\sqrt{h}$
\begin{equation}\label{eq:larena_typo_fixed_appx}
	\frac{1}{\sqrt{h}} \pdt \sqrt{h} = \frac{\alp}{W} \left( \theta + \theta_B \right) + D_i \beta^i,
\end{equation}
where we have used \eqref{eq:dt_lndetgam} and the fact that both $\sigma_{ij}$ and $\sigma_{Bij}$ are traceless. The equivalent to equation \eqref{eq:larena_typo_fixed_appx} in \citeauthor{larena2009b}'s paper, equation (30), reads
\begin{equation}\label{eq:larena_typo_appx}
	\frac{1}{\sqrt{h}} \pdt \sqrt{h} = \frac{\alp}{W} \left( \theta - \kappa \right) + D_i \beta^i.
\end{equation}
Comparing \eqref{eq:larena_typo_fixed_appx} and \eqref{eq:larena_typo_appx} we can see there is a difference of $ \theta_B \rightarrow -\kappa$. In \citet{larena2009b}, this error is propagated into equation (31) for the evolution of the volume, and subsequently into equation (34) for the relation between the effective volume and fluid scale factors, which reads
\begin{equation} \label{eq:larena_typo_aDh_aDV}
	\aD^V = \aDh \, {\rm exp} \left( \int_{t_{\rm init}}^t \avgh{ \frac{\alp}{W} \left(\theta - \kappa \right) - \alp\theta + D_i\beta^i} dt \right).
\end{equation}
We find this relation, instead using the evolution of the volume derived from \eqref{eq:larena_typo_fixed_appx}, to be
\begin{equation} \label{eq:larena_typo_fixed_aDh_aDV}
	\aD^V = \aDh \, {\rm exp}\left( \frac{1}{3} \int_{t_{\rm init}}^t \avgh{ \frac{\alp}{W} \left(\theta + \theta_B \right) - \alp\theta + D_i \beta^i } dt \right),
\end{equation}
see Section~\ref{subsubsec:general_foliation} for the derivation. 

We used \eqref{eq:larena_typo_aDh_aDV} for our analysis in Chapter~\ref{Chapter4}, however, \emph{only} the calculation of the fluid scale factor is affected, and all other analysis remains valid. We also note the missing factor of $1/3$ in the exponential in \eqref{eq:larena_typo_aDh_aDV} that \emph{was} corrected in our analysis in Chapter~\ref{Chapter4}. 

Starting from the definition of $\theta_B$ in \eqref{eq:thetaB}, and using the trace of $B_{\mu\nu}$ \eqref{eq:Bmunu_larena} as $B=\frac{1}{3} \kappa v^i v_i + \beta_{ij} v^i v^j$, we find
\begin{align}
	\theta_B &= -W \kappa - \frac{W^3}{3}\kappa v_i v^i - W^3 \beta_{ij} v^i v^j, \\
		&= -W\kappa - W^3 v^i v^j \nabla_i v_j, \label{eq:thetaB_vs_kappa_line2}
\end{align}
where we have substituted $\beta_{ij}$ from \eqref{eq:betaij_larena}. Even for small velocities, with $W\approx1$, it is not clear that $\theta_B\approx-\kappa$ since the \emph{gradient} of the velocity may not be small, and so the second term in \eqref{eq:thetaB_vs_kappa_line2} is not obviously negligible. It is therefore important to verify whether this makes a difference to the calculation of the fluid scale factor.

We perform the same analysis outlined in Chapter~\ref{Chapter4}, on the same simulations, but instead using \eqref{eq:larena_typo_fixed_aDh_aDV} to calculate the effective fluid scale factor, $\aDh$\footnote{We note that in Chapter~\ref{Chapter4} we denote the fluid scale factor by $\aD$, equivalent to $\aDh$ in this Appendix, and in Chapter~\ref{Chapter1}.}, from the volume scale factor, $\aD^V$. We average over the whole domain as well as within subdomains, i.e. we reproduce the calculations in Figures~\ref{fig:aDaFLRW} and \ref{fig:aDr100}, respectively, and compare the evolution with and without the error. 

\begin{figure*}[!ht]
	\begin{centering}
	\includegraphics[width=\textwidth]{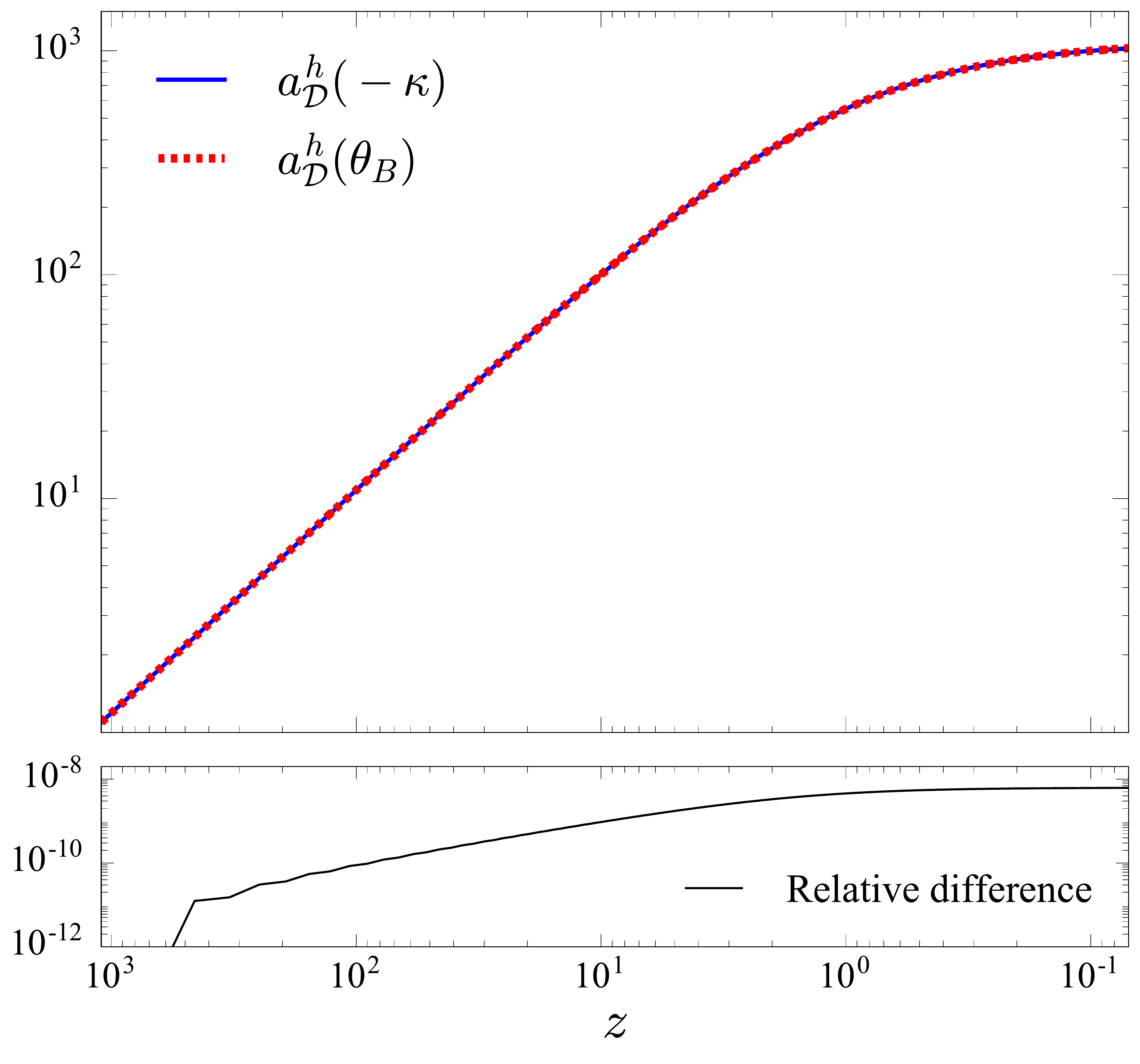} %
	\caption{\label{fig:aDh_typo_vs_notypo} Globally-averaged, effective fluid scale factors with the error described in this section present, $\aDh(-\kappa)$; black solid curve, and with the error corrected, $\aDh(\theta_B)$; red dashed curve. Top panel shows the evolution as a function of effective redshift, and the bottom panel shows the relative difference \eqref{eq:appxF_rel_diff}, which remains below $10^{-8}$ for the entire simulation.}
	\end{centering}
\end{figure*}

The top panel of Figure~\ref{fig:aDh_typo_vs_notypo} shows $\aDh$ calculated over the entire domain as a function of effective redshift, for $\aDh (-\kappa)$ (error present; solid black curve), and $\aDh (\theta_B)$ (error corrected; dashed red curve). The bottom panel shows the relative difference, i.e. 
\begin{equation}\label{eq:appxF_rel_diff}
	\aDh (-\kappa) / \aDh (\theta_B) - 1,
\end{equation}
which remains below $10^{-8}$, even at $z\approx0$.

\begin{figure*}[!ht]
	\begin{centering}
	\includegraphics[width=\textwidth]{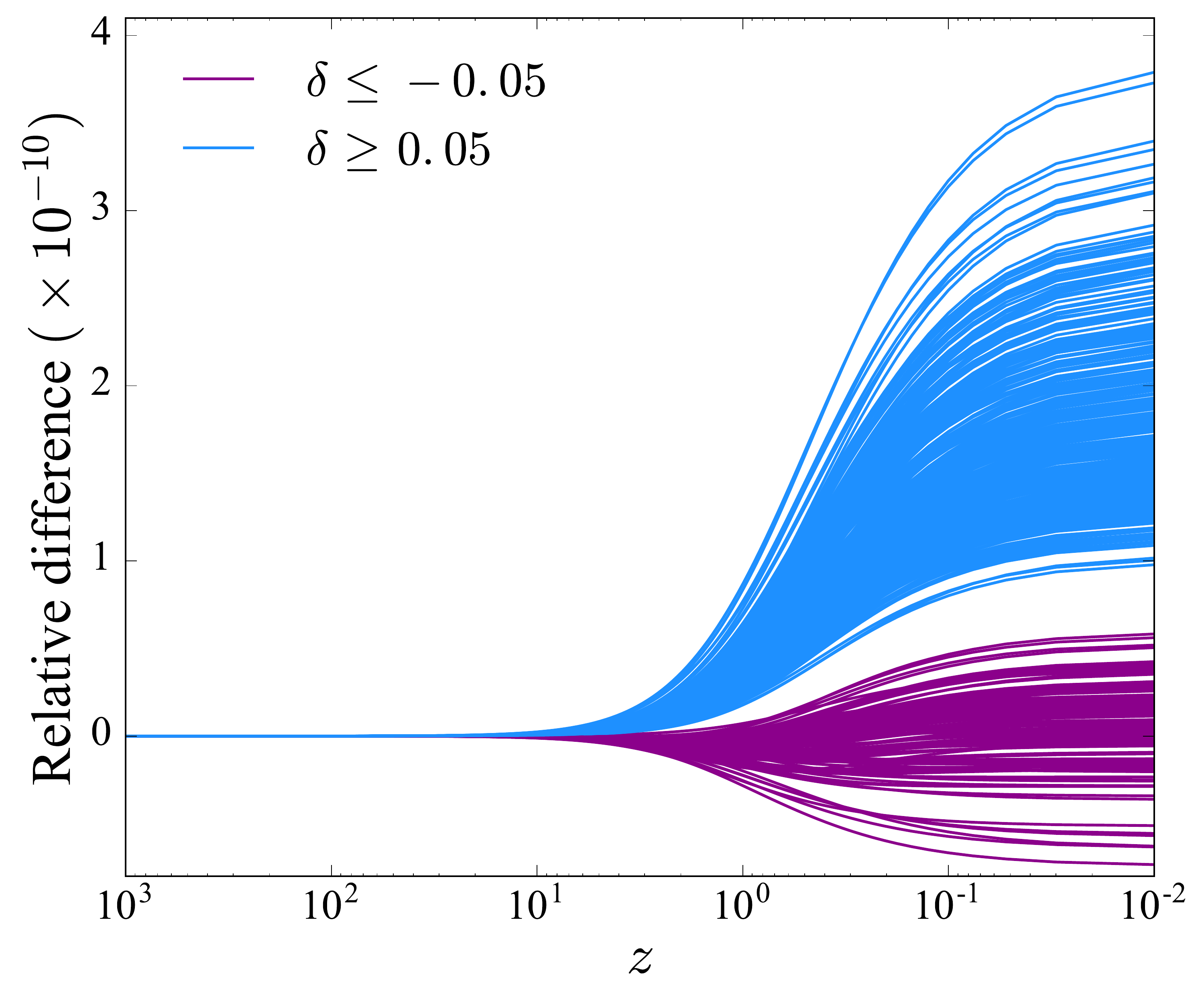}
	\caption{\label{fig:aDh_spheres_z_typo_vs_notypo} Relative difference \eqref{eq:appxF_rel_diff} between effective fluid scale factors (with and without the error corrected) calculated in spheres of radius $r_\mD=100$ Mpc. Each curve shows the difference for an individual sphere, either over- or under-dense as indicated by the legend, as a function of effective redshift. The difference remains of order $\approx 10^{-10}$ throughout the simulation.}
	\end{centering}
\end{figure*}

\begin{figure*}[!ht]
	\begin{centering}
	\includegraphics[width=\textwidth]{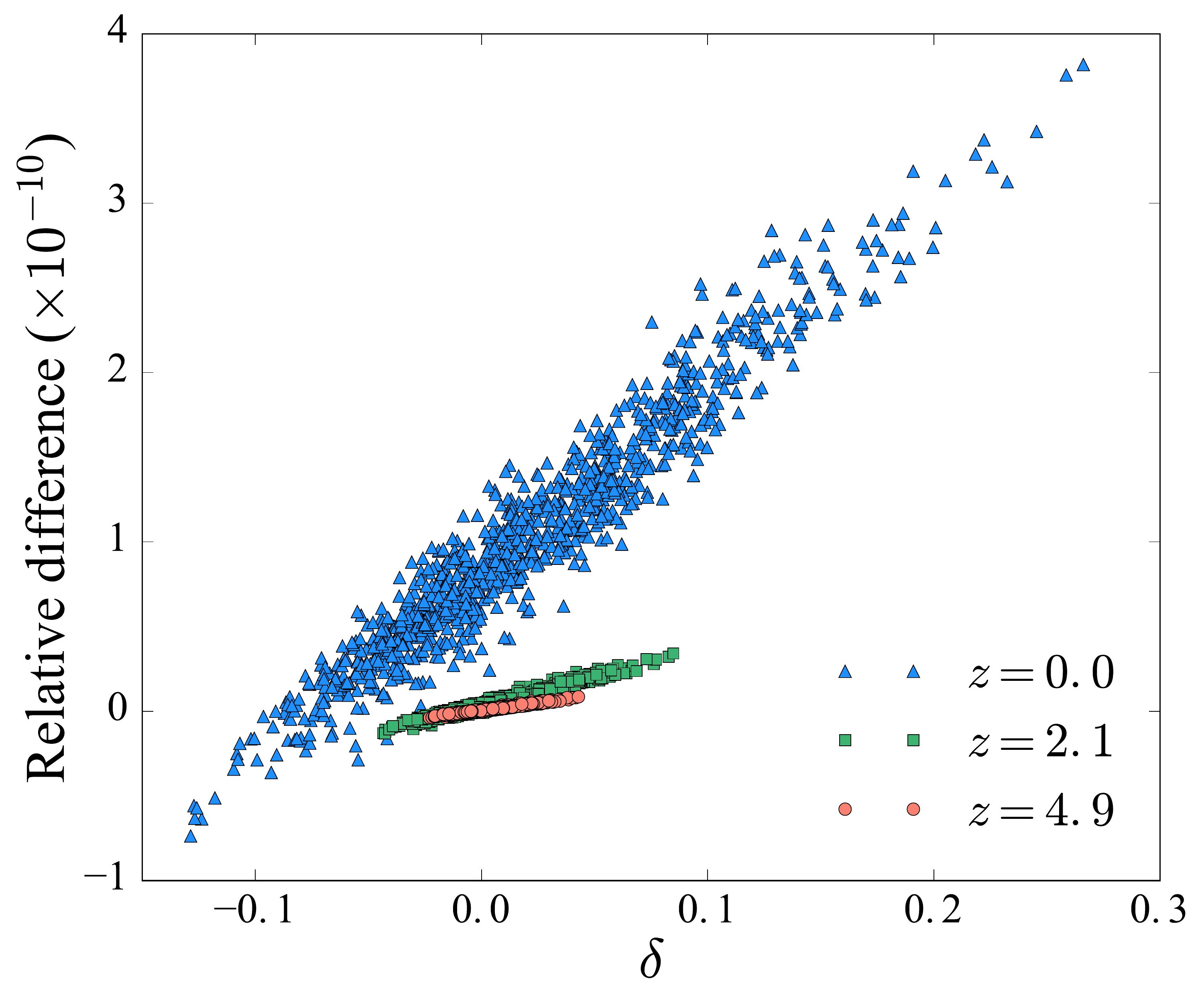}
	\caption{\label{fig:aDh_spheres_delta_typo_vs_notypo} Relative difference \eqref{eq:appxF_rel_diff} between effective fluid scale factors (with and without the error corrected) calculated in 1000 spheres of radius $r_\mD=100$ Mpc. Each point shows an individual sphere measured at redshifts $z=0.0, 2.1, 4.9$ as indicated by the legend. As density contrasts grow, so too does the difference between the effective scale factors. The difference remains of order $\approx 10^{-10}$ throughout the simulation.}
	\end{centering}
\end{figure*}

Each curve in Figure~\ref{fig:aDh_spheres_z_typo_vs_notypo} shows the relative difference \eqref{eq:appxF_rel_diff} for an individual sphere of radius $r_\mD=100$ Mpc as a function of effective redshift. We have coloured the curves depending on their averaged density contrast at $z\approx0$, with over-dense regions in blue and under-dense regions in purple. The curves differ from one another because each sphere contains different structures, and therefore will contain different velocity gradients. Figure~\ref{fig:aDh_spheres_delta_typo_vs_notypo} shows the relative difference \eqref{eq:appxF_rel_diff} for all 1000 spheres sampled with $r_\mD=100$ Mpc at three different redshifts $z=0.0,2.1$, and $4.9$, as a function of the averaged density contrast. At high redshift, we see the spread in $\delta$ is less, implying smoother structures and therefore smaller velocity gradients. At $z\approx0$ the maximum difference still remains below $4\times 10^{-10}$. 

We have detailed an error we found in the derivation of \citeauthor{larena2009b}'s generalised averaging formalism, which was used in our calculation of the fluid scale factor. We found maximum differences between the globally-averaged and subdomain-averaged fluid scale factors, with and without the error fixed, of $10^{-8}$ and $4\times10^{-10}$, respectively. These differences are negligible compared with our estimated errors (see Appendix~\ref{appx:convergence}).

\printbibliography[heading=bibintoc]

\end{document}